\documentclass[preprint2,dvipsnames,usenames,longtable]{aastex63p}

\usepackage{amsmath, bm}

\usepackage[utf8]{inputenc}
\usepackage{natbib}
\usepackage{textcomp} %JC

\submitjournal{Astrophysical Journal Supplements}
\received{February 20, 2020}
\revised{April 27, 2020}
\accepted{April 28, 2020}

\newcommand{\xte}{{\it RXTE}}
\newcommand{\sax}{{\it BeppoSAX}}
\newcommand{\igr}{{\it INTEGRAL}}
\newcommand{\xspec}{{\sc XSpec}}
\newcommand{\eps}{{\rm erg\, s^{-1}}}
\newcommand{\epcs}{{\rm erg\, cm^{-2}\, s^{-1}}}
\newcommand{\epc}{{\rm erg\, cm^{-2}}}

\newcommand{\cspcu}{\mathrm{count\,s^{-1}\,PCU^{-1}}}
\newcommand{\fper}{F_\mathrm{per}}

\newcounter{magicrownumbers}
\newcommand\rnum{\stepcounter{magicrownumbers}\arabic{magicrownumbers}}

\newcommand{\sources}{115}  % minbar_sources.fits v2.6, now (2020 Apr) including Swift J1858.6-0814 and IGR J17591-2342
\newcommand{\minbarsources}{85} % 11.9.17
\newcommand{\pubdatacutoffdate}{2012 May 3}
\newcommand{\pubdatacutoffmjd}{56050}
\newcommand{\pubdatacutoffrev}{1166}
\newcommand{\observations}{118848} % following corrections, 2020 Aug
\newcommand{\bursts}{7111}  % updated 17.1.20
\newcommand{\ubursts}{7083}  % updated 17.1.20
\newcommand{\obstotxte}{17901}
\newcommand{\exptotxte}{46.08} % Ms  
\newcommand{\obstotsax}{14545}
\newcommand{\exptotsax}{224.1} % Ms  
\newcommand{\obstotigr}{245340}
\newcommand{\exptotigr}{605.7} % Ms  
\newcommand{\burstssax}{2203}   % checked 17.1.20
\newcommand{\sourcessax}{54}
\newcommand{\burstsxte}{2288}    % checked 17.1.20
\newcommand{\sourcesxte}{60}
\newcommand{\burstsigr}{2620}   % updated 17.1.20
\newcommand{\sourcesigr}{63}

\newcommand{\burstdupes}{28}    % 18.11.2019

\newcommand{\expsax}{133.6~Ms}
\newcommand{\expigr}{268.7~Ms}
\newcommand{\expxte}{42.71~Ms}

\newcommand{\rexpthresh}{1.629}

\newcommand{\alloscbursts}{1042}    %{1036} updated by laura 19.11.2018
\newcommand{\oscburstsexcluded}{91} %{78} updated by laura 19.11.2018
\newcommand{\oscbursts}{950}        %{938} updated by laura 19.11.2018

\shorttitle{The Multi-INstrument Burst ARchive}%: DR 1} 
\shortauthors{Galloway et al.}

\begin{document}

\title{The Multi-INstrument Burst ARchive (MINBAR)}

\correspondingauthor{D. K. Galloway}
\email{duncan.galloway@monash.edu}

\author[0000-0002-6558-5121]{Duncan K. Galloway}
\affiliation{School of Physics \& Astronomy,  Monash University,  
  Clayton VIC 3800,  Australia}

\author[0000-0002-4363-1756]{Jean in 't Zand}
\affiliation{SRON Netherlands Institute for Space Research, Sorbonnelaan 2, 3584 CA Utrecht, The Netherlands}

\author[0000-0002-4397-8370]{J\'er\^ome Chenevez}
\affiliation{DTU Space,
Technical University of Denmark, Elektrovej 327-328, DK-2800 Lyngby, Denmark}

\author[0000-0002-5042-9070]{Hauke W\"orpel}
\affiliation{Leibniz-Institut f\"ur Astrophysik, Potsdam, An der Sternwarte 16, 14482 Potsdam, Germany}

\author{Laurens Keek}
\affiliation{Department of Astronomy, University of Maryland, College Park, MD 20742, USA}

\author[0000-0002-2800-8309]{Laura Ootes}
\affiliation{Anton Pannekoek Institute for Astronomy, University of Amsterdam, Science Park 904, 1090GE Amsterdam, the Netherlands}

\author[0000-0002-1009-2354]{Anna L.~Watts}
\affiliation{Anton Pannekoek Institute for Astronomy, University of Amsterdam, Science Park 904, 1090GE Amsterdam, the Netherlands}

\author{Luis Gisler}
\affiliation{School of Physics \& Astronomy,  Monash University,  
  Clayton VIC 3800,  Australia}

\author[0000-0002-0778-6048]{Celia Sanchez-Fernandez }
\affiliation{European Space Agency, ESAC, 28691 Villanueva de la Cañada, Madrid, Spain}

\author[0000-0002-5790-7290]{Erik Kuulkers}
\affiliation{European Space Agency, ESTEC, 2201 AZ Noordwijk, The Netherlands}

\begin{abstract}
We present the largest sample of type-I (thermonuclear) X-ray bursts yet assembled, 
comprising \ubursts\ bursts from \minbarsources\ bursting sources. 
The sample is drawn from observations with 
Xenon-filled proportional counters on the long-duration satellites {\it RXTE}, %the {\it Rossi X-ray Timing Explorer},%jc190917
{\it BeppoSAX}  and {\it INTEGRAL}, between 1996 February 8, and 2012 May 3.
\replaced{We assembled a catalog of \sources\ burst sources,
from which the sample burst sources are drawn.}{The burst sources were drawn from a comprehensive catalog of \sources\ burst sources, assembled from 
earlier catalogs and the literature.}
We carried out a consistent analysis for each  burst lightcurve (normalised to the relative instrumental effective area),
and provide measurements of rise time, peak intensity, 
\replaced{exponential decay timescale, burst $\tau$-value}{burst timescale}, 
and fluence.
For bursts observed with the \xte/PCA and \sax/WFC we also provide time-resolved spectroscopy, including estimates of bolometric peak flux and fluence, and spectral parameters at the peak of the burst.
For \oscbursts\ bursts observed with the PCA from sources with previously detected burst oscillations, we include an analysis of the high-time resolution data, providing information on the detectability and amplitude of the oscillations, as well as where in the burst they are found.
We also present analysis of  \observations\ observations of the burst sources within the sample timeframe.
We extracted 3--25~keV X-ray spectra from most observations, and 
(for observations meeting our signal-to-noise criterion), we provide measurements of the flux, spectral colours, and for selected sources, the position on the colour-colour diagram, for the best-fit spectral model.
We present a description of the sample, a summary of the science investigations completed to date, and suggestions for further studies.
\end{abstract}


\tableofcontents

\newpage

\section{Introduction}

\noindent
Type-I (thermonuclear) X-ray bursts 
are flashes in the few-keV X-ray sky that typically %jc190917
last for about one minute, and rival the brightest 
cosmic objects
in intensity. They were discovered in the mid 1970s \citep{grindlay76,bce76}, although already observed in 1969 \citep{bce72,kuul09c}. 
Thanks to earlier theoretical work  
\citep{hvh75,woosley1976},
it was soon realised
that these events arise
from
unstable ignition of accreted hydrogen and/or helium on neutron stars. %jc190917
X-ray bursts are thus
the neutron-star equivalent of classical novae, thermonuclear shell flashes that occur instead on white dwarfs \citep{mc77,joss77,lamblamb78}.
Here
we provide a brief overview of the knowledge about type-I X-ray bursts; for more detail, we refer to 
comprehensive reviews by \cite{lew93}, \cite{sb03} and \cite{gal17b}.

The fuel for thermonuclear bursts is provided from a %jc1900917low-mass 
companion star via Roche-lobe overflow in a low-mass X-ray binary (LMXB).
The bursts occur when the hot, dense matter at the base of the accumulated layer 
ignites unstably. Thermonuclear burning then proceeds to engulf the entire neutron-star surface in less than $10$~s, converting most of the accreted hydrogen and helium to heavy-element ashes. At the peak of the burst, the luminosity can reach the Eddington limit of  $\approx3\times10^{38}\ \eps$ \cite[for a $1.4\ M_\odot$ neutron star; e.g. ][]{lew93}. Subsequent accretion builds a new fuel layer, which is then ignited, and the process repeats every few hours or longer, mainly depending on the mass accretion rate. %jc190917
The basic physics of this process has been understood for many years, %\cite[e.g.][]{fhm81}, 
although there are several observational aspects that have not yet been satisfactorily explained.

\begin{figure}
	\includegraphics[width=\columnwidth]{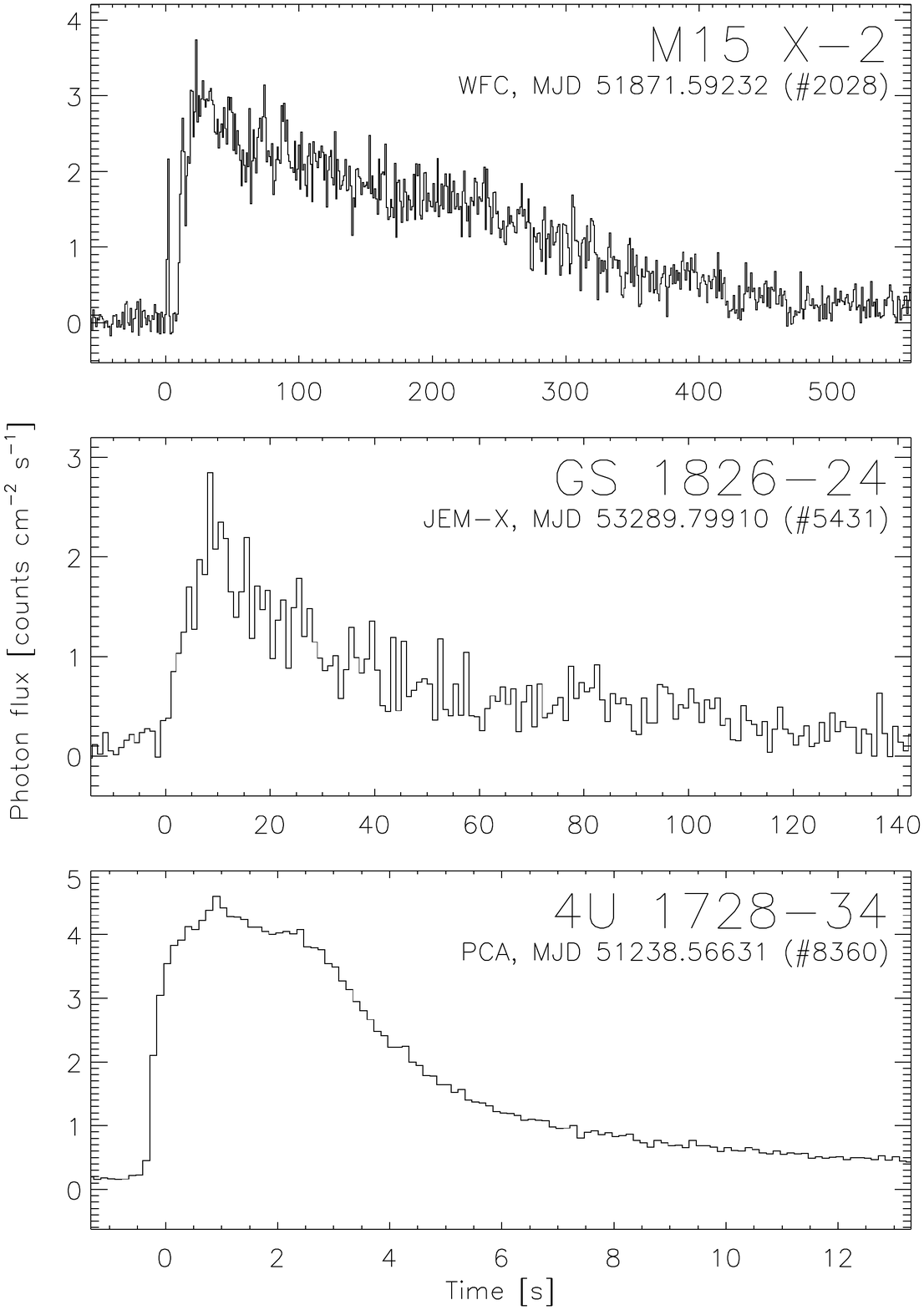}
	\caption{Example bursts from the MINBAR sample, demonstrating the range of durations and intensities. From top to bottom, we show an intermediate duration burst observed with \sax/WFC
	% added in response to the referee's comments
	% \added{(in the energy range 2--30~keV)} 
	(in the energy range 2--30~keV)
	from 
	% 2S~0918$-$614; 
	M15~X-2; a mixed H/He burst observed with \igr/JEM-X 
	% \added{(3--25~keV)} from GS~1826$-$24; and a H-deficient burst observed with \xte/PCA \added{(2--60~keV)}
	(3--25~keV) from GS~1826$-$24; and a H-deficient burst observed with \xte/PCA (2--60~keV)
	from 4U~1728$-$34.
	% \added{The $y$-axis is common to all three panels, and represents the flux normalised according to the relative effective areas of the three instruments determined in \S\ref{sec:area}.}
	The $y$-axis is common to all three panels, and represents the flux normalised according to the relative effective areas of the three instruments determined in \S\ref{sec:area}.
	\label{fig:examples}}
\end{figure}

X-ray bursts are commonly classified according to their duration. The
``classical'' bursts, discovered in the 1970s \citep{grindlay76} and early
1980s 
\citep[e.g.,][]{lew93}
have a duration of order 1~min (Fig. \ref{fig:examples}). They are
 frequent, with wait times of order 1~hr. With the advent of wide field
X-ray imaging through the {\it BeppoSAX} Wide Field Cameras (WFCs) in the 1990s, as much as half the LMXB population could be covered in a single observation and rare kinds of X-ray bursts were picked up, such as 
``intermediate-duration'' bursts, %  was also identified, 
lasting $\sim0.5$~hr and with strong radiation pressure effects \citep[e.g.,][]{zand11a}. These events are thought to result from ignition of a pure helium layer accreted at low rates, that has one to two orders of magnitude more mass than for classical X-ray bursts \citep{zand05a,cumming06}. % They are usually seen in ultracompact X-ray binaries (UCXBs) with low mass accretion rates 
A third class of 
``superbursts'' was also identified \citep{corn00},
lasting 
$\sim10$~hr instead of 1~min. 
These events are thought to result from thermonuclear ignition not of helium and hydrogen, but of carbon at column depths 10$^3$ times larger \citep{cumming01,stroh02}.

The nuclear burning of hydrogen and helium on the neutron star surface proceeds via four main channels (e.g. \citealt{gal17b}; see also  \citealt{meisel18b}). 
Prior to ignition, 
hydrogen burns primarily through the CNO cycle,
at a rate that depends
on the abundance of CNO nuclei. When the temperature is above $7\times10^7$~K, this process becomes  stable (the ``hot CNO'' cycle). Once a burst is triggered, 
helium burns primarily through the $3\alpha$ process, independent of the 
fuel composition (but not the temperature). 
Additional burning channels include the 
$\alpha$,p process, arising from captures of He-nuclei onto light elements, and the ``$rp$ process'', that involves rapid proton captures followed by beta decay of heavy nuclei that are produced during all nuclear burning. The 
$rp$ process is 
particularly complex, and can produce hundreds of unstable isotopes with a wide range of decay times, up to many seconds \citep[e.g.][]{schatz01,fis08}.

The accretion rate largely sets the temperature of the layer prior to ignition, due to 
heating processes arising from pycnonuclear reactions and electron captures in the neutron star crust 
\cite[]{brown00,hz08}.
At sufficiently high temperatures,
the helium fuel also burns stably prior to ignition, because the $T$-dependence of the nuclear power weakens and becomes similar to 
that
of the cooling, 
and no runaway will occur \cite[e.g.][]{bil98a}. 

At the lowest accretion rates, any accreted hydrogen burns stably 
via hot CNO burning
at a constant rate per gram (for a certain column depth), 
and the time to ignite the burst may be long enough that all the hydrogen is exhausted at the base. In that case, 
a pure helium layer grows and subsequently ignites. At higher accretion rates, bursts occur so frequently that hydrogen does not have the time to burn completely and a mixed hydrogen/helium burst may occur. 
These two are the most common ignition regimes of a growing number
\citep{fhm81,keek16}, which contribute to the diversity in the observed bursts.

Bursting LMXBs can be subdivided into two groups: those with orbital periods shorter or longer than 80~min \citep{rjw82,nrj86}. The former class is referred to as ``ultracompact'' X-ray binaries (UCXBs). The orbits are too small to fit the hydrogen envelope of the companion star and what remains is the extinguished core in the form of a white dwarf. The
implication for the X-ray bursts on the 
neutron star in such systems is that they occur in a hydrogen-poor environment.
The lack of hydrogen 
strongly influences the appearance of the bursts;
the faster triple-$\alpha$ burning leads to shorter burst rise times, 
and the absence of stable hydrogen burning
delays ignition,
yielding larger accumulated fuel layers. As a result, the total nuclear energy is larger \cite[despite the yield per gram being smaller; e.g.][]{goodwin19a}, and such bursts are thus more likely to reach the Eddington limit, 
preferentially (but not exclusively)
producing so-called ``photospheric radius-expansion'' (PRE). 

Burst sources can be further discriminated on the basis of the typical range of accretion rates. 
UCXBs typically exhibit lower accretion rates ($\lesssim0.01\ \dot{M}_{\rm Edd}$,
where $\dot{M}_{\rm Edd}$ is the accretion rate corresponding to the Eddington luminosity limit, or roughly 
$2\times10^{-8}\ M_\odot\,{\rm yr}^{-1}$), 
resulting in long wait times to energetic bursts. The highest accretion rates are found in the so-called Z sources \cite[so-named for the shape of their X-ray ``colour-colour'' diagrams;][]{hvdk89}, including Cyg~X-2 and GX~17+2. 
The most prolific burst sources exhibit  wait times of a few hours, resulting from intermediate accretion rates (i.e., a few percent of $\dot{M}_{\rm Edd}$). Exceptionally short wait times (of order minutes) are seen 
in one unusual source at high accretion rates (IGR~J17480$-$2446; \citealt{linares12a}), or in systems accreting H-rich fuel after incomplete burning of the available fuel buffer \citep{keek10,keek17a}.

About 20\% of burst sources exhibit ``burst oscillations'' intermittently during some bursts \cite[e.g.][]{watts12a}. These oscillations are detected at a few percent fractional amplitude, at frequencies that are characteristic for each source, corresponding to the neutron star spin  \cite[]{chak03a}. Oscillations typically exhibit a slight (few Hz) drift to higher frequencies while they are present, and may occur during the burst rise, peak, or even into the burst decay, and in some cases all three. Burst oscillations are not found in every burst of sources that exhibit them, but tend to occur in bursts at high accretion rates \cite[e.g.][]{muno01,ootes17a}.
The details of the mechanism that give rise to the oscillations remains unknown.

\paragraph{Outstanding questions}

Although X-ray bursts are fairly well understood, important science questions remain, concerning the details of the ignition conditions, thermonuclear burning and interaction with the environment. 
For many burst sources, 
at higher mass accretion rates bursts  become {\it less} frequent, contrary to the predictions of numerical models 
\cite[e.g.][]{corn03a,bcatalog}. 
All exceptions have slow ($<400$~Hz) neutron-star spin frequencies, where known; 
notably, one of these (IGR~17480$-$2446) %jc190917 \citealt{gal18a,Linares12a}
has a spin rate that is at least 20 times slower than any other bursting neutron star with a measured spin rate \cite[e.g.][]{linares12a}. 
The decreasing burst rates for rapidly spinning sources at high accretion rates may be explained by a burst regime where stable helium burning coexists with unstable, and further influenced by 
systematic drifts of the ignition location to higher latitudes (\citealt{cavecchi17,gal18a}; see also \citealt{zand03b,keek14b}).

The burning occurs through a complex nuclear chain involving hundreds of isotopes and thousands of reactions that are intimately dependent on each other and often difficult to study in the laboratory. These reactions have a noticeable effect on the light curve of the X-ray burst \citep{woos04,cyburt16}. 
Detailed measurements of bursts may thus be used to constrain the rates of individual nuclear reactions \cite[e.g.][]{meisel19}.

X-ray bursts are the brightest phenomena that we can observe from the surfaces of neutron stars,
and thus offer
a unique probe of quantum chromodynamics under dense and cool circumstances. Accurately constraining the average density of neutron stars (ergo, measuring their mass and radius) is a prime goal of studying these objects. % modified in response to referee's comment
\added{Observations of} X-ray bursts 
and burst oscillations
are considered a promising 
\replaced{opportunity for that}{approach to achieve such constraints} (e.g. \citealt{vp79, damen90, ozel06, wbs06, lp07, ozel16b, watts16, nattila17a}). % jc190917: has added Weinberg, Bildsten & Schatz 2006 
\added{However, such constraints rely critically on assumptions regarding the dynamics during PRE bursts \cite[e.g.][]{slb10}, the detailed shape of the X-ray spectrum \cite[e.g.][]{suleimanov17}, and the ability to separate the burst emission from other components including reflection \cite[e.g.][]{ball04,keek14a}.}

The persistent emission, arising from accretion, is usually much fainter than the emission during the bursts, but is clearly not completely independent of that phenomenon.  A fraction of the burst photons may be reprocessed by the accretion flow and scattered in or out of the line of sight \citep{vp86,zhangma11,chen12,zand11a}.
Alternatively, photons and matter ejected by the burst may disturb the accretion flow and temporarily change its spectrum \citep{worpel13a} or geometry \citep{zand11a}. For a recent review, see 
% fixed duplicate in proofs, Q11
% \cite{degenaar18}. 
\cite{deg18a}. 

\paragraph{Motivation for a new burst sample}

In order to make progress in answering the  science questions posed above and elsewhere, and to stimulate further work in the wider community, we 
assembled the Multi-INstrument Burst ARchive (MINBAR) from data acquired by three instruments (\sax/WFC, \xte/PCA and \igr/JEM-X).
These three instruments have accumulated the largest sets of observations of burst sources, amongst the 
$\approx20$ satellite-based instruments that have contributed to 
many thousands of events observed in total since their discovery  (see 
\S\ref{sec:sources}).
Conveniently, these three instruments all comprise Xenon-filled proportional counter detectors, with similar 
spectral response curves
which makes their data readily comparable. 
Additionally, the instruments offer complementary properties; the high effective area (and hence sensitivity) of the PCA is offset by the lack of imaging and the relatively narrow field of view, while WFC and JEM-X offer moderate sensitivity imaging observations across wide fields of view, ideally suited to collecting large burst samples including rare types of bursts.

This paper describes the assembly and content of the MINBAR sample. 
This work is an extension of previous studies of large databases, such as that based on 
\xte/PCA observations through 2008 \citep[][hereafter G08]{bcatalog} or on all observations with the \sax/WFCs \citep{corn03a}. The extension results in a more than doubling of the sample size, the provision of an online facility to query the database, and the inclusion of additional observational features such as  persistent spectral analyses and burst oscillations. 

This paper is organized as follows. In \S\ref{sec:sources} we describe the selection of the burst sources for which we select observations.
In \S\ref{sec:source-data} we describe the selection criteria to identify the data from each of the instruments.
\S\ref{sec:analysis} provides comprehensive details of the analysis procedures, both for the bursts and the persistent spectra, as well as the searches for burst oscillations
and the instrumental cross-calibration. % from 2019 Dec
In \S\ref{sec:assembly} we describe the 
analysis steps undertaken to combine the data from the different instruments, including 
establishment of a uniform luminosity scale and determination of burst timescales and energetics, bolometric corrections and spectral colours.
\S\ref{minbar} describes the burst sample itself, including all analysis parameters for each event detected within the observation sample.
In \S\ref{osc_table} we describe the results of the burst oscillation search, covering selected sources with bursts observed by \xte/PCA.
\S\ref{minbar-obs} describes the observation table, which includes the analysis results for each observation used as a source for the sample. In this and the following two sections, we also present a broad overview of the data comprising each table.
Finally, in \S\ref{sec:discussion} we present a summary of the results  already arising from the sample, and provide some suggestions for future extensions of this work.

\section{A catalog of thermonuclear burst sources}
\label{sec:sources}

We assembled a 
complete list of thermonuclear burst sources
by first cross-matching the \igr\/ source catalog \cite[]{bird10}\footnote{\url{http://www.isdc.unige.ch/integral/science/catalogue}} with the burst sources (source type code ``B') in the catalog of LMXBs of \cite{lmxb07}. % resulting in an initial table of 102 sources (18 without a listing in the earlier sample). 
To ensure completeness, we cross-matched our original list with a separate catalog,
assembled following
a systematic search through the literature and incorporating new discoveries since the mid 1990s\footnote{\url{http://www.sron.nl/~jeanz/bursterlist.html}}.

The resulting sample includes \sources\ known LMXBs which have exhibited
type-I (thermonuclear) bursts\footnote{See also 
\url{http://burst.sci.monash.edu/sources}} (Table \ref{tab:bursters}). 
The corresponding columns in the FITS file which we also provide as part of the MINBAR sample (see \S\ref{sec:assembly}) are described in Table \ref{tab:fitscolumns}.

The instrument and year in which the burst behaviour was first detected is summarized in column 2 ({\tt disc} in the FITS file). %
The three-letter acronym gives the spacecraft and instrument name, as described in Table \ref{tab:instruments}; the 
year in which the first burst from the source was detected
is given as the two-digit number following. 
The source type \cite[column 3, adapted from][ {\tt type} in the FITS table]{lmxb07}, gives the basic properties of each object.
The source type in the MINBAR table includes additional flags ``C'', ``I'', ``O'', and ``S'' compared to the previous authors (see below for an explanation); we omit the ``U'' (ultra-soft X-ray spectrum) flag.
We reviewed the literature to obtain the most precise known position  for each source, as summarised in columns 4--6 (columns {\tt RA\_OBJ}, {\tt DEC\_OBJ}, {\tt ERR\_RAD} and {\tt ERR\_CONF} in the FITS table). 
The binary orbital period, where known, is given in column 7 ({\tt Porb} in the FITS table), and the measured (or estimated) line-of-sight hydrogen column density is given in column 8 ({\tt NH}). This latter value is adopted for all spectral fits for both persistent and time-resolved burst spectra (see \S\ref{trsanl} and \S\ref{obsanl}).
The estimated total exposure for the three instruments comprising the MINBAR sample is given in column 9 ({\tt exp} in the FITS table). 

The number of {\it unique} bursts detected from each source through \pubdatacutoffdate\ (MJD~\pubdatacutoffmjd) is given in
column 10 ({\tt nburst}). Because we list analysis results independently for each detection by each instrument, the burst table (see \S\ref{minbar}) includes duplicate events observed by multiple instruments (see also \S\ref{sec:crosscal}).
\minbarsources\ of the listed sources have one or more bursts observed with 
the PCA, WFC or JEM-X
during the sample interval, which includes
the entire mission durations for \sax\/ and \xte.
At the time of writing \igr\/ is continuing to observe; the data analysed here is through revolution \pubdatacutoffrev.
\added{We note that because of the lack of imaging capability of PCA/\xte, there is some uncertainty about the origin for bursts observed in fields with more than one active burst source within the $1^\circ$~field of view (FOV; see also \S\ref{xteburstid}).}

For those sources with no bursts in MINBAR, we adopt the convention of listing a 0 in Table \ref{tab:bursters} for bursters known at the cutoff date, and an ellipsis for sources for which the burst activity was discovered after that date.
These \minbarsources\ sources form the sample we adopt for this paper; we explicitly exclude from our analysis sources where the first discovery of bursts was after the cutoff date, even if that discovery was made with \igr/JEM-X. There have been 
15 % as at 2019 Nov
new burster discoveries since. 

The corresponding mean burst rate (or limit) averaged over all the observations is listed in column 11 ({\tt rate}). For sources with bursts in MINBAR, the burst rate is calculated from the exposure while the source is active, as described in \S\ref{sec:burstrates}.
For sources with no bursts in MINBAR, we calculate an estimated 95\% upper limit
assuming Poisson statistics
\cite[$\approx3$ bursts over the total observation period, neglecting any corrections for source activity;][]{gehrels86}.
Finally, in column 12 we list the references from which the 
first detection of bursts, 
the position, orbital period, and $N_H$ values are drawn. 
These references are taken from the {\tt NH\_bibcode}, {\tt Porb\_bibcode}, {\tt pos\_bibcode} and {\tt disc\_bibcode} columns of the FITS table.

Our objective in assembling this list is a complete sample, but there remains some uncertainty primarily due to long intervals between burst activity, and uncertainty of localisation by some instruments.
There is evidence for a few additional burst sources that may exist,
for example the burst event detected by {\it MAXI}, localised to a relatively large region including the known burst source, RX~J1718.4$-$4029 \cite[]{iwakiri18}. Although the known burst source (with two events in the MINBAR sample) is the most likely origin for the event, it is also possible that a previously unknown source is the origin.
Conversely, some distinct entries in the list of bursters may actually be the same object.
A single (unusual) burst was observed in 1995 from a poorly-localised source in the globular cluster M28, designated 
AX~J1824.5$-$2541 \cite{gk97}. It seems possible that the burst origin may be the same object as IGR~J18245$-$2452, just $75\farcs9$ away, with a much improved localisation thanks to the identification of an optical counterpart \cite[]{pallanca13}.
There remains the possibility that other clusters host multiple burst sources that have not been positionally separated during past activity intervals due to limited instrumental spatial resolution.

% Created using srctable.pro, Thu Jun 18 13:14:05 2020
%
%%%%%%%%%%%%%%%%%%%%%%%%%%%%%%%%%%%%%%%%%%%%%%%%%%%%%%%%%%%%%%%%%%%%%%%%%%%%%%
%
%            THIS FILE IS AUTOMATICALLY GENERATED - DO NOT EDIT
%
%%%%%%%%%%%%%%%%%%%%%%%%%%%%%%%%%%%%%%%%%%%%%%%%%%%%%%%%%%%%%%%%%%%%%%%%%%%%%%
%
\begin{longrotatetable} % AASTeX 6.3 environment
\begin{deluxetable}{lccccccccccl}
\tabletypesize{\scriptsize}
\tablecaption{Known LMXBs exhibiting type-I X-ray bursts and their representation in the MINBAR sample
  \label{tab:bursters}
}
\tablewidth{0pt}
\tablehead{
  \colhead{}
 & 
 & 
 & 
 & 
 & \colhead{Error}
 & \colhead{$P_{\rm orb}$}
 & \colhead{$N_{\rm H}$}
 & \colhead{Time}
 & 
 & \colhead{Mean burst}
 & 
\\
  \colhead{Source}
 & \colhead{Disc.}
 & \colhead{Type\tablenotemark{a}}
 & \colhead{RA}
 & \colhead{Dec}
 & \colhead{(conf.)}
 & \colhead{(hr)}
 & \colhead{($10^{22}\ {\rm cm^2}$)}
 & \colhead{(Ms)\tablenotemark{b}}
 & \colhead{$n_{\rm burst}$}
 & \colhead{rate (hr$^{-1}$)\tablenotemark{c}}
 & \colhead{Ref.}
}
\colnumbers
\startdata
\object{IGR J00291+5934} & XRT'15 & PT & $00^\mathrm{h}29^\mathrm{m}03\fs050$ & $+59^\circ34\arcmin18\farcs91$ & $0\farcs04$ & 2.46 & \nodata & 11.9 & \nodata & $<0.00091$ & [1,2] \\
4U 0513$-$40 & UHU'72 & CG & $05^\mathrm{h}14^\mathrm{m}06\fs48$ & $-40^\circ02\arcmin38\farcs8$ & $0\farcs6$(90\%) & 0.283 & 0.0300 & 8.73 & 35 & 0.043 & [3,4,5] \\
\object{4U 0614+09} & OS8'75 & ACRS & $06^\mathrm{h}17^\mathrm{m}07\fs35$ & $+09^\circ08\arcmin13\farcs4$ & $0\farcs1$ & 0.855? & 0.380 & 8.46 & 2 & 0.0018 & [6,7,8,9] \\
EXO 0748$-$676 & EXO'85 & DEOT & $07^\mathrm{h}48^\mathrm{m}33\fs70$ & $-67^\circ45\arcmin07\farcs9$ & $0\farcs6$(90\%) & 3.82 & 0.800 & 22.9 & 357 & 0.19 & [10,11,12,13] \\
4U 0836$-$429 & GIN'90 & T & $08^\mathrm{h}37^\mathrm{m}23\fs6$ & $-42^\circ54\arcmin02\arcsec$ & 10$''$(90\%) & \nodata & 2.20 & 12.9 & 82 & 0.16 & [14,15] \\
2S 0918$-$549 & XTE'00 & C & $09^\mathrm{h}20^\mathrm{m}26\fs473$ & $-55^\circ12\arcmin24\farcs47$ & $0\farcs06$ & 0.290 & 0.350 & 11.2 & 7 & 0.0065 & [16,17,18,9] \\
4U 1246$-$588 & WFC'97 & C & $12^\mathrm{h}49^\mathrm{m}39\fs364$ & $-59^\circ05\arcmin14\farcs68$ & $0\farcs05$(90\%) & \nodata & 0.500 & 12.5 & 4 & 0.0037 & [19,20,21] \\
4U 1254$-$69 & OPT'79 & DS & $12^\mathrm{h}57^\mathrm{m}37\fs15$ & $-69^\circ17\arcmin21\farcs0$ & $0\farcs6$(90\%) & 3.93 & 0.320 & 19.4 & 34 & 0.019 & [22,23,24,25] \\
SAX J1324.5$-$6313 & WFC'97 & \nodata & $13^\mathrm{h}24^\mathrm{m}30\fs30$ & $-63^\circ13\arcmin50\farcs0$ & $0\farcs7$ & \nodata & 1.50 & 14.4 & 1 & 0.0036 & [26,27] \\
4U 1323$-$62 & EXO'84 & D & $13^\mathrm{h}26^\mathrm{m}36\fs3$ & $-62^\circ08\arcmin10\arcsec$ & $1''$ & 2.93 & 2.42 & 14.2 & 99 & 0.065 & [28,29,30,31] \\
MAXI J1421$-$613 & JEM'14 & T & $14^\mathrm{h}21^\mathrm{m}38\fs0$ & $-61^\circ36\arcmin25\arcsec$ & $2''$(90\%) & \nodata & \nodata & 10.8 & \nodata & $<0.0010$ & [32,33] \\
\object{Cen X-4} & VEL'69 & RT & $14^\mathrm{h}58^\mathrm{m}21\fs92$ & $-31^\circ40\arcmin07\farcs4$ & $0\farcs5$ & 15.1 & \nodata & 3.30 & 0 & $<0.0033$ & [34,35,36] \\
\object{Cir X-1} & EXO'84 & ADMRT & $15^\mathrm{h}20^\mathrm{m}40\fs87$ & $-57^\circ10\arcmin00\farcs3$ & $0\farcs6$ & 398 & 0.660 & 10.9 & 14 & 0.0071 & [37,38,39,40] \\
4U 1543$-$624 & MAX'18 & C & $15^\mathrm{h}47^\mathrm{m}54\fs69$ & $-62^\circ34\arcmin05\farcs4$ & $0\farcs6$ & 0.300 & \nodata & 10.9 & \nodata & $<0.0010$ & [41,42] \\
\object{UW CrB} & ASC'97 & DE & $16^\mathrm{h}05^\mathrm{m}45\fs872$ & $+25^\circ51\arcmin45\farcs20$ & $0\farcs06$(68\%) & 1.85 & \nodata & 4.06 & 0 & $<0.0027$ & [43,44,45] \\
4U 1608$-$522 & VEL'69 & AOST & $16^\mathrm{h}12^\mathrm{m}43\fs0$ & $-52^\circ25\arcmin23\arcsec$ & $1''$ & 12.9? & 0.891 & 11.7 & 145 & 0.087 & [46,47,48] \\
MAXI J1621$-$501 & NUS'17 & T & $16^\mathrm{h}20^\mathrm{m}22\fs0$ & $-50^\circ01\arcmin12\arcsec$ & $3''$ & \nodata & \nodata & 11.7 & \nodata & $<0.00092$ & [49] \\
4U 1636$-$536 & OS8'76 & AOS & $16^\mathrm{h}40^\mathrm{m}55\fs57$ & $-53^\circ45\arcmin05\farcs2$ & $0\farcs3$(90\%) & 3.80 & 0.250 & 11.4 & 664 & 0.26 & [50,51,52,53] \\
MAXI J1647$-$227 & XRT'12 & T & $16^\mathrm{h}48^\mathrm{m}12\fs32$ & $-23^\circ00\arcmin53\farcs6$ & $0\farcs2$(68\%) & \nodata & \nodata & 11.4 & \nodata & $<0.00094$ & [54,55] \\
XTE J1701$-$462 & BAT'08 & R?TZ & $17^\mathrm{h}00^\mathrm{m}58\fs46$ & $-46^\circ11\arcmin08\farcs6$ & $0\farcs6$(90\%) & \nodata & 2.00 & 14.8 & 6 & 0.0053 & [56,57,58] \\
XTE J1701$-$407 & XTE'07 & T & $17^\mathrm{h}01^\mathrm{m}44\fs33$ & $-40^\circ51\arcmin30\farcs1$ & $0\farcs6$(90\%) & \nodata & 3.10 & 12.5 & 1 & 0.0018 & [59,60,61] \\
MXB 1658$-$298 & SAS'76 & DEOT & $17^\mathrm{h}02^\mathrm{m}06\fs53$ & $-29^\circ56\arcmin44\farcs3$ & $0\farcs1$ & 7.11 & 0.200 & 8.80 & 27 & 0.031 & [62,63,64,65] \\
4U 1702$-$429 & SAS'77 & AO & $17^\mathrm{h}06^\mathrm{m}15\fs31$ & $-43^\circ02\arcmin08\farcs7$ & $0\farcs6$ & \nodata & 1.87 & 13.3 & 284 & 0.13 & [66,67,68] \\
IGR J17062$-$6143 & BAT'12 & CPT & $17^\mathrm{h}06^\mathrm{m}16\fs3$ & $-61^\circ42\arcmin41\arcsec$ & $4''$ & 0.633 & \nodata & 7.66 & \nodata & $<0.0014$ & [69,70] \\
4U 1708$-$23 & SAS'76 & \nodata & $17^\mathrm{h}08^\mathrm{m}23\fs0$ & $-22^\circ48\arcmin12\arcsec$ & 40$''$ & \nodata & \nodata & 1.12 & 0 & $<0.010$ & [71] \\
4U 1705$-$32 & WFC'00 & C & $17^\mathrm{h}08^\mathrm{m}54\fs27$ & $-32^\circ19\arcmin57\farcs1$ & $0\farcs6$(90\%) & \nodata & 0.400 & 11.0 & 1 & 0.0019 & [72] \\
4U 1705$-$44 & EXO'85 & AR & $17^\mathrm{h}08^\mathrm{m}54\fs47$ & $-44^\circ06\arcmin07\farcs4$ & $0\farcs5$(68\%) & \nodata & 1.90 & 13.1 & 267 & 0.12 & [73,74,75] \\
XTE J1709$-$267 & WFC'97 & CT & $17^\mathrm{h}09^\mathrm{m}30\fs40$ & $-26^\circ39\arcmin19\farcs9$ & $0\farcs6$ & \nodata & 0.440 & 9.31 & 11 & 0.027 & [76,77,78] \\
XTE J1710$-$281 & XTE'01 & DET & $17^\mathrm{h}10^\mathrm{m}12\fs53$ & $-28^\circ07\arcmin51\farcs0$ & $0\farcs1$ & 3.28 & 0.400 & 10.4 & 47 & 0.072 & [79,80,81] \\
4U 1708$-$40 & NFI'99 & \nodata & $17^\mathrm{h}12^\mathrm{m}23\fs83$ & $-40^\circ50\arcmin34\farcs0$ & $0\farcs6$ & \nodata & \nodata & 12.2 & 0 & $<0.00088$ & [82] \\
SAX J1712.6$-$3739 & WFC'99 & CT & $17^\mathrm{h}12^\mathrm{m}37\fs1$ & $-37^\circ38\arcmin40\arcsec$ & $5''$(90\%) & \nodata & 1.34 & 12.5 & 2 & 0.0015 & [83,84,85] \\
2S 1711$-$339 & WFC'98 & T & $17^\mathrm{h}14^\mathrm{m}19\fs78$ & $-34^\circ02\arcmin47\farcs3$ & $0\farcs6$(90\%) & \nodata & 1.50 & 12.4 & 21 & 0.036 & [26,86] \\
RX J1718.4$-$4029 & WFC'96 & C & $17^\mathrm{h}18^\mathrm{m}24\fs1$ & $-40^\circ29\arcmin30\arcsec$ & 20$''$ & \nodata & 1.32 & 11.8 & 2 & 0.0032 & [87,72] \\
1H 1715$-$321 & SAS'76 & T & $17^\mathrm{h}18^\mathrm{m}47\fs02$ & $-32^\circ10\arcmin13\farcs5$ & $0\farcs4$(68\%) & \nodata & \nodata & 14.1 & 0 & $<0.00077$ & [71,88] \\
IGR J17191$-$2821 & XRT'07 & OT & $17^\mathrm{h}19^\mathrm{m}15\fs1$ & $-28^\circ17\arcmin57\arcsec$ & $4''$ & \nodata & 0.300 & 14.0 & 5 & 0.019 & [89,90] \\
XTE J1723$-$376 & XTE'99 & T & $17^\mathrm{h}23^\mathrm{m}38\fs7$ & $-37^\circ39\arcmin42\arcsec$ & 30$''$ & \nodata & 7.94 & 12.4 & 12 & 0.023 & [91,92] \\
IGR J17254$-$3257 & JEM'06 & CT & $17^\mathrm{h}25^\mathrm{m}25\fs5$ & $-32^\circ57\arcmin17\arcsec$ & $2''$(68\%) & \nodata & 1.79 & 15.7 & 11 & 0.017 & [93,94] \\
4U 1722$-$30 & OS8'75 & ACG & $17^\mathrm{h}27^\mathrm{m}32\fs9$ & $-30^\circ48\arcmin08\arcsec$ & $2''$ & \nodata & 0.780 & 18.9 & 97 & 0.028 & [95,5,96] \\
4U 1728$-$34 & SAS'76 & ACOR & $17^\mathrm{h}31^\mathrm{m}57\fs6782$ & $-33^\circ50\arcmin01\farcs547$ & $0\farcs007$(68\%) & 0.179? & 2.60 & 18.9 & 1173 & 0.27 & [97,98,99,100] \\
MXB 1730$-$335 & SAS'77 & DGRT & $17^\mathrm{h}33^\mathrm{m}24\fs61$ & $-33^\circ23\arcmin19\farcs8$ & $0\farcs1$ & \nodata & 1.66 & 19.9 & 126 & 0.054 & [101,102,103] \\
KS 1731$-$260 & TTM'89 & OST & $17^\mathrm{h}34^\mathrm{m}13\fs46$ & $-26^\circ05\arcmin18\farcs6$ & $0\farcs2$(99\%) & \nodata & 1.30 & 18.8 & 366 & 0.20 & [104,105,106] \\
Swift J1734.5$-$3027 & BAT'13 & \nodata & $17^\mathrm{h}34^\mathrm{m}24\fs2$ & $-30^\circ23\arcmin53\arcsec$ & $1''$(90\%) & \nodata & \nodata & 18.8 & \nodata & $<0.00057$ & [107,108] \\
1RXH J173523.7$-$354013 & BAT'08 & \nodata & $17^\mathrm{h}35^\mathrm{m}23\fs0$ & $-35^\circ40\arcmin13\arcsec$ & $4''$ & \nodata & \nodata & 10.3 & 0 & $<0.0011$ & [109] \\
SLX 1732$-$304 & HAK'80 & GRT & $17^\mathrm{h}35^\mathrm{m}47\fs26$ & $-30^\circ28\arcmin55\farcs3$ & $0\farcs6$(90\%) & \nodata & 1.63 & 21.5 & 1 & 0.00073 & [110,5,111] \\
IGR J17380$-$3749 & IBI'04 & T & $17^\mathrm{h}37^\mathrm{m}58\fs8$ & $-37^\circ46\arcmin20\arcsec$ & $1''$(90\%) & \nodata & \nodata & 7.26 & 0 & $<0.0015$ & [112,113] \\
SLX 1735$-$269 & WFC'97 & CS & $17^\mathrm{h}38^\mathrm{m}17\fs12$ & $-26^\circ59\arcmin38\farcs6$ & $0\farcs6$(90\%) & \nodata & 1.50 & 21.0 & 23 & 0.0073 & [114,115,86] \\
4U 1735$-$444 & SAS'77 & AR?S & $17^\mathrm{h}38^\mathrm{m}58\fs3$ & $-44^\circ27\arcmin00\arcsec$ & $1''$ & 4.65 & 0.140 & 9.6 & 71 & 0.036 & [116,51,117] \\
XTE J1739$-$285 & JEM'05 & T & $17^\mathrm{h}39^\mathrm{m}53\fs95$ & $-28^\circ29\arcmin46\farcs8$ & $0\farcs6$(90\%) & \nodata & 2.01 & 22.5 & 43 & 0.021 & [118,119,56] \\
SLX 1737$-$282 & WFC'00 & C & $17^\mathrm{h}40^\mathrm{m}42\fs83$ & $-28^\circ18\arcmin08\farcs4$ & $0\farcs6$(90\%) & \nodata & 1.90 & 23.7 & 3 & 0.0011 & [120,121] \\
IGR J17445$-$2747 & JEM'17 & T & $17^\mathrm{h}44^\mathrm{m}30\fs4$ & $-27^\circ46\arcmin00\arcsec$ & $1''$(68\%) & \nodata & \nodata & 21.3 & \nodata & $<0.00051$ & [122,123] \\
KS 1741$-$293 & TTM'89 & T & $17^\mathrm{h}44^\mathrm{m}51\fs1$ & $-29^\circ21\arcmin17\arcsec$ & $1''$(90\%) & \nodata & 33.0 & 25.5 & 29 & 0.0095 & [124,125,126] \\
XMM J174457$-$2850.3 & BAT'12 & T & $17^\mathrm{h}44^\mathrm{m}57\fs3$ & $-28^\circ50\arcmin20\arcsec$ & $4''$ & \nodata & \nodata & 9.6 & \nodata & $<0.0011$ & [127,128] \\
GRS 1741.9$-$2853 & WFC'96 & OT & $17^\mathrm{h}45^\mathrm{m}02\fs32$ & $-28^\circ54\arcmin49\farcs6$ & $0\farcs2$ & \nodata & 11.3 & 25.4 & 27 & 0.0090 & [129,130,131] \\
AX J1745.6$-$2901\tablenotemark{e} & ASC'94 & ET & $17^\mathrm{h}45^\mathrm{m}35\fs4$ & $-29^\circ01\arcmin34\arcsec$ & $3''$ & 8.36 & \nodata & 25.5 & \nodata & \nodata & [132,133] \\
1A 1742$-$289 & SAS'76 & RT & $17^\mathrm{h}45^\mathrm{m}37\fs19$ & $-29^\circ01\arcmin04\farcs7$ & $0\farcs4$(90\%) & \nodata & 10.0 & 25.5 & 3 & 0.0010 & [134,135,136] \\
1A 1742$-$294 & SAS'76 & \nodata & $17^\mathrm{h}46^\mathrm{m}05\fs2$ & $-29^\circ30\arcmin53\arcsec$ & $1''$ & \nodata & 1.16 & 25.7 & 794 & 0.15 & [137,135,138] \\
SAX J1747.0$-$2853 & WFC'98 & ST & $17^\mathrm{h}47^\mathrm{m}02\fs60$ & $-28^\circ52\arcmin58\farcs9$ & $0\farcs7$ & \nodata & 8.80 & 25.7 & 113 & 0.033 & [139,140,141] \\
IGR J17464$-$2811 & JEM'05 & CT? & $17^\mathrm{h}47^\mathrm{m}16\fs16$ & $-28^\circ10\arcmin48\farcs0$ & $0\farcs5$(90\%) & \nodata & 8.90 & 25.2 & 2 & 0.00079 & [142,143,144] \\
IGR J17473$-$2721 & AGI'08 & T & $17^\mathrm{h}47^\mathrm{m}18\fs08$ & $-27^\circ20\arcmin38\farcs7$ & $0\farcs5$ & \nodata & 3.80 & 22.9 & 61 & 0.027 & [145,146,147] \\
SLX 1744$-$299\tablenotemark{e} & GRA'99 & CT? & $17^\mathrm{h}47^\mathrm{m}25\fs89$ & $-30^\circ00\arcmin01\farcs6$ & $0\farcs4$(90\%) & \nodata & \nodata & 24.4\tablenotemark{d} & \nodata & \nodata & [148,149] \\
SLX 1744$-$300 & SLX'85 & T? & $17^\mathrm{h}47^\mathrm{m}26\fs01$ & $-30^\circ02\arcmin41\farcs8$ & $0\farcs7$(90\%) & \nodata & 4.50 & 24.4 & 304 & 0.068 & [150,151,149] \\
\object{GX 3+1} & HAK'80 & AS & $17^\mathrm{h}47^\mathrm{m}56\fs096$ & $-26^\circ33\arcmin49\farcs35$ & $0\farcs09$(68\%) & \nodata & 1.59 & 21.7 & 204 & 0.038 & [152,153,154] \\
IGR J17480$-$2446 & JEM'10 & GOPT & $17^\mathrm{h}48^\mathrm{m}04\fs819$ & $-24^\circ46\arcmin48\farcs90$ & $0\farcs06$ & \nodata & 0.500 & 18.3\tablenotemark{d} & 303 & 1.9 & [155,156,157] \\
EXO 1745$-$248 & HAK'80 & DGST & $17^\mathrm{h}48^\mathrm{m}05\fs23$ & $-24^\circ46\arcmin47\farcs7$ & $0\farcs2$(68\%) & \nodata & 3.80 & 18.3 & 25 & 0.018 & [5,111,158] \\
Swift J174805.3$-$244637 & XRT'12 & GT & $17^\mathrm{h}48^\mathrm{m}05\fs41$ & $-24^\circ46\arcmin38\farcs0$ & $0\farcs2$ & \nodata & \nodata & 18.3\tablenotemark{d} & \nodata & $<0.00059$ & [159,160] \\
1A 1744$-$361 & TTM'89 & A?DRT & $17^\mathrm{h}48^\mathrm{m}13\fs15$ & $-36^\circ07\arcmin57\farcs0$ & $0\farcs3$(68\%) & 1.62? & 0.410 & 14.0 & 4 & 0.012 & [161,162,163,164] \\
SAX J1748.9$-$2021 & WFC'98 & AGIT & $17^\mathrm{h}48^\mathrm{m}52\fs16$ & $-20^\circ21\arcmin32\farcs4$ & $0\farcs2$ & \nodata & 0.470 & 10.6 & 46 & 0.073 & [165,5,166] \\
Swift J1749.4$-$2807 & BAT'06 & PT & $17^\mathrm{h}49^\mathrm{m}31\fs73$ & $-28^\circ08\arcmin05\farcs1$ & $0\farcs6$ & \nodata & 3.00 & 10.5 & 1 & 0.0020 & [167,168] \\
IGR J17498$-$2921 & JEM'11 & OPT & $17^\mathrm{h}49^\mathrm{m}55\fs34$ & $-29^\circ19\arcmin19\farcs7$ & $0\farcs1$ & \nodata & 1.28 & 8.61 & 7 & 0.036 & [169,170,171] \\
4U 1746$-$37 & SAS'77 & ADG & $17^\mathrm{h}50^\mathrm{m}12\fs73$ & $-37^\circ03\arcmin06\farcs5$ & $0\farcs4$ & 5.16 & 0.260 & 12.1 & 37 & 0.019 & [172,173,5,174] \\
SAX J1750.8$-$2900 & WFC'97 & A?OT & $17^\mathrm{h}50^\mathrm{m}24\fs42$ & $-29^\circ02\arcmin15\farcs4$ & $0\farcs6$(90\%) & \nodata & 0.900 & 23.4 & 24 & 0.0092 & [137,175,176] \\
EXO 1747$-$214 & EXO'85 & T & $17^\mathrm{h}50^\mathrm{m}24\fs52$ & $-21^\circ25\arcmin19\farcs9$ & $0\farcs6$ & \nodata & 0.190 & 12.0 & 1 & 0.0021 & [177,178] \\
GRS 1747$-$312 & XTE'01 & DEGT & $17^\mathrm{h}50^\mathrm{m}46\fs86$ & $-31^\circ16\arcmin28\farcs9$ & $0\farcs4$(95\%) & 12.4 & 1.39 & 22.7 & 21 & 0.0089 & [5,179,180] \\
IGR J17511$-$3057 & XRT'09 & OPT & $17^\mathrm{h}51^\mathrm{m}08\fs66$ & $-30^\circ57\arcmin41\farcs0$ & $0\farcs6$ & 3.47 & 0.600 & 11.9 & 16 & 0.030 & [181,182,183,184] \\
SAX J1752.3$-$3138 & WFC'99 & T & $17^\mathrm{h}52^\mathrm{m}24\fs0$ & $-31^\circ37\arcmin42\arcsec$ & $2\farcm9$(99\%) & \nodata & 0.490 & 21.8 & 2 & 0.0012 & [137,185,186] \\
SAX J1753.5$-$2349 & WFC'96 & T & $17^\mathrm{h}53^\mathrm{m}31\fs90$ & $-23^\circ49\arcmin14\farcs9$ & $0\farcs6$(90\%) & \nodata & 0.880 & 16.4 & 2 & 0.0015 & [137,187,188] \\
AX J1754.2$-$2754 & JEM'07 & \nodata & $17^\mathrm{h}54^\mathrm{m}14\fs50$ & $-27^\circ54\arcmin35\farcs6$ & $0\farcs5$(90\%) & \nodata & 2.70 & 21.4 & 2 & 0.0013 & [189,190,191] \\
IGR J17591$-$2342 & JEM'19 & PRT & $17^\mathrm{h}59^\mathrm{m}02\fs856$ & $-23^\circ43\arcmin08\farcs19$ & $0\farcs03$ & 8.80 & \nodata & 21.4 & \nodata & $<0.00051$ & [192,193,194] \\
IGR J17597$-$2201 & XTE'03 & D & $17^\mathrm{h}59^\mathrm{m}45\fs53$ & $-22^\circ01\arcmin39\farcs2$ & $0\farcs1$ & \nodata & 2.84 & 12.4 & 16 & 0.021 & [195,196] \\
1RXS J180408.9$-$342058 & JEM'12 & T & $18^\mathrm{h}04^\mathrm{m}08\fs37$ & $-34^\circ20\arcmin51\farcs4$ & $0\farcs5$ & \nodata & 0.480 & 7.34 & 1 & 0.072 & [197,198,199] \\
SAX J1806.5$-$2215 & WFC'96 & T & $18^\mathrm{h}06^\mathrm{m}32\fs168$ & $-22^\circ14\arcmin17\farcs32$ & $0\farcs03$(68\%) & \nodata & 0.97 & 11.8 & 9 & 0.012 & [137,200,188] \\
2S 1803$-$245 & WFC'98 & ART & $18^\mathrm{h}06^\mathrm{m}50\fs72$ & $-24^\circ35\arcmin28\farcs6$ & $0\farcs8$(68\%) & \nodata & 0.630 & 14.3 & 3 & 0.0031 & [137,201,202] \\
\object{MAXI J1807+132} & NIC'19 & T & $18^\mathrm{h}08^\mathrm{m}07\fs54$ & $+13^\circ15\arcmin05\farcs4$ & $0\farcs2$ & \nodata & \nodata & 14.3 & \nodata & $<0.00076$ & [203,204] \\
SAX J1808.4$-$3658 & WFC'96 & OPRT & $18^\mathrm{h}08^\mathrm{m}27\fs60$ & $-36^\circ58\arcmin43\farcs9$ & $0\farcs5$ & 2.01 & 0.120 & 10.9 & 12 & 0.018 & [205,206,207,208] \\
XTE J1810$-$189 & XTE'08 & T & $18^\mathrm{h}10^\mathrm{m}20\fs86$ & $-19^\circ04\arcmin11\farcs2$ & $0\farcs6$ & \nodata & 4.20 & 4.34 & 19 & 0.036 & [176,209,210] \\
SAX J1810.8$-$2609 & WFC'98 & OT & $18^\mathrm{h}10^\mathrm{m}44\fs47$ & $-26^\circ09\arcmin01\farcs2$ & $0\farcs6$ & \nodata & 0.350 & 14.4 & 16 & 0.015 & [211,212,213] \\
XMMU J181227.8$-$181234 & XTE'08 & CT & $18^\mathrm{h}12^\mathrm{m}27\fs8$ & $-18^\circ12\arcmin34\arcsec$ & $2''$(68\%) & \nodata & 12.8 & 9.18 & 7 & 0.057 & [214,215] \\
XTE J1814$-$338 & XTE'03 & OPT & $18^\mathrm{h}13^\mathrm{m}39\fs04$ & $-33^\circ46\arcmin22\farcs3$ & $0\farcs2$(90\%) & 4.27 & 0.160 & 11.5 & 28 & 0.14 & [216,217,218,219] \\
\object{GX 13+1} & GIN'89 & ADR & $18^\mathrm{h}14^\mathrm{m}31\fs08$ & $-17^\circ09\arcmin26\farcs1$ & $0\farcs6$ & 578 & 3.40 & 10.8 & 1 & 0.00041 & [220,221,222,223] \\
4U 1812$-$12 & HAK'82 & AC & $18^\mathrm{h}15^\mathrm{m}06\fs15$ & $-12^\circ05\arcmin46\farcs7$ & $0\farcs3$(68\%) & \nodata & 1.55 & 9.7 & 25 & 0.018 & [224,19,225] \\
\object{GX 17+2} & EIN'80 & RSZ & $18^\mathrm{h}16^\mathrm{m}01\fs39$ & $-14^\circ02\arcmin10\farcs6$ & $0\farcs1$(90\%) & \nodata & 1.90 & 10.3 & 43 & 0.019 & [226,227,228] \\
Swift J181723.1$-$164300 & BAT'17 & T & $18^\mathrm{h}17^\mathrm{m}23\fs2$ & $-16^\circ43\arcmin00\arcsec$ & $4''$ & \nodata & \nodata & 9.9 & \nodata & $<0.0011$ & [229,230] \\
\object{SAX J1818.7+1424} & WFC'97 & T & $18^\mathrm{h}18^\mathrm{m}44\fs0$ & $+14^\circ24\arcmin12\arcsec$ & $2\farcm9$(99\%) & \nodata & 0.100 & 3.69 & 2 & 0.034 & [26,27] \\
4U 1820$-$303 & ANS'75 & ACGRS & $18^\mathrm{h}23^\mathrm{m}40\fs5029$ & $-30^\circ21\arcmin40\farcs088$ & $0\farcs007$(68\%) & 0.190 & 0.160 & 11.0 & 67 & 0.029 & [231,98,232,5] \\
AX J1824.5$-$2451 & ASC'95 & G & $18^\mathrm{h}24^\mathrm{m}30\fs0$ & $-24^\circ51\arcmin00\arcsec$ & 40$''$(95\%) & \nodata & 1.50 & 9.30 & 1 & 0.0044 & [233] \\
IGR J18245$-$2452 & XRT'13 & GPRT & $18^\mathrm{h}24^\mathrm{m}32\fs50$ & $-24^\circ52\arcmin07\farcs8$ & $0\farcs2$(90\%) & 11.0 & \nodata & 9.30 & \nodata & $<0.0012$ & [234,235,236] \\
4U 1822$-$000 & MAX'16 & \nodata & $18^\mathrm{h}25^\mathrm{m}22\fs02$ & $-00^\circ00\arcmin43\farcs0$ & $0\farcs6$ & 3.18 & \nodata & 2.64 & \nodata & $<0.0041$ & [237,238,239] \\
SAX J1828.5$-$1037 & WFC'01 & S & $18^\mathrm{h}28^\mathrm{m}34\fs0$ & $-10^\circ36\arcmin59\arcsec$ & $4''$(90\%) & \nodata & 1.90 & 10.0 & 1 & 0.0066 & [26,240] \\
GS 1826$-$24 & WFC'97 & T & $18^\mathrm{h}29^\mathrm{m}28\fs2$ & $-23^\circ47\arcmin49\arcsec$ & $2''$ & 2.09 & 0.400 & 8.91 & 455 & 0.28 & [241,242,243,244] \\
XB 1832$-$330 & WFC'96 & CG & $18^\mathrm{h}35^\mathrm{m}43\fs65$ & $-32^\circ59\arcmin26\farcs8$ & $0\farcs6$(68\%) & 0.727 & 0.0500 & 7.89 & 19 & 0.021 & [245,246,5,247] \\
\object{Ser X-1} & OS8'75 & ARS & $18^\mathrm{h}39^\mathrm{m}57\fs55$ & $+05^\circ02\arcmin09\farcs5$ & $0\farcs1$ & \nodata & 0.380 & 4.18 & 55 & 0.063 & [50,248,249] \\
Swift J185003.2$-$005627 & BAT'11 & T & $18^\mathrm{h}50^\mathrm{m}03\fs3$ & $-00^\circ56\arcmin23\arcsec$ & $2''$(90\%) & \nodata & \nodata & 3.81 & 0 & $<0.0028$ & [250] \\
4U 1850$-$086 & SAS'78 & ACGR? & $18^\mathrm{h}53^\mathrm{m}04\fs88$ & $-08^\circ42\arcmin20\farcs0$ & $0\farcs4$ & 0.343 & 0.390 & 5.00 & 4 & 0.010 & [251,252,5,253] \\
Swift J1858.6$-$0814 & NIC'20 & DT & $18^\mathrm{h}58^\mathrm{m}34\fs92$ & $-08^\circ14\arcmin16\farcs0$ & $0\farcs7$ & 21.8? & \nodata & \nodata & \nodata & \nodata & [254,255] \\
HETE J1900.1$-$2455 & HET'05 & IOT & $19^\mathrm{h}00^\mathrm{m}09\fs77$ & $-24^\circ54\arcmin04\farcs3$ & $0\farcs1$ & 1.39 & 0.160 & 6.88 & 10 & 0.027 & [256,257,258,259] \\
\object{XB 1905+000} & SAS'76 & CT & $19^\mathrm{h}08^\mathrm{m}27\fs0$ & $+00^\circ10\arcmin08\arcsec$ & $5''$ & \nodata & \nodata & 7.54 & 0 & $<0.0014$ & [260,261] \\
\object{Aql X-1} & SAS'76 & ADIORT & $19^\mathrm{h}11^\mathrm{m}16\fs047$ & $+00^\circ35\arcmin05\farcs85$ & $0\farcs08$ & 18.9 & 0.400 & 7.49 & 96 & 0.10 & [262,17,261,263] \\
XB 1916$-$053 & OS8'76 & ACD & $19^\mathrm{h}18^\mathrm{m}47\fs87$ & $-05^\circ14\arcmin17\farcs1$ & $0\farcs6$(90\%) & 0.834 & 0.320 & 3.52 & 36 & 0.079 & [264,265,266,267] \\
Swift J1922.7$-$1716 & BAT'11 & T & $19^\mathrm{h}22^\mathrm{m}36\fs99$ & $-17^\circ17\arcmin01\farcs1$ & $0\farcs6$(90\%) & \nodata & \nodata & 1.78 & 0 & $<0.0061$ & [268,269] \\
XB 1940$-$04 & HAK'81 & \nodata & $19^\mathrm{h}42^\mathrm{m}37\fs9$ & $-03^\circ52\arcmin51\arcsec$ & $1.^\circ$ & \nodata & \nodata & 2.66 & 0 & $<0.0041$ & [225] \\
XTE J2123$-$058 & XTE'98 & AET & $21^\mathrm{h}23^\mathrm{m}14\fs54$ & $-05^\circ47\arcmin53\farcs2$ & $0\farcs6$ & 5.96 & 0.0700 & 1.52 & 6 & 0.13 & [270,271,272,273] \\
\object[4U 2129+12]{M15 X-2} & GIN'88 & CGR? & $21^\mathrm{h}29^\mathrm{m}58\fs13$ & $+12^\circ10\arcmin02\farcs6$ & $0\farcs5$ & 0.376 & 0.0300 & 2.11 & 8 & 0.028 & [274,5,275,276] \\
\object{XB 2129+47} & EIN'78 & E & $21^\mathrm{h}31^\mathrm{m}26\fs19$ & $+47^\circ17\arcmin24\farcs7$ & $0\farcs1$(68\%) & 5.24 & \nodata & 10.8 & 0 & $<0.0010$ & [277,278,279] \\
\object{Cyg X-2} & EIN'80 & RZ & $21^\mathrm{h}44^\mathrm{m}41\fs15$ & $+38^\circ19\arcmin17\farcs1$ & $0\farcs2$ & 236 & 0.0500 & 8.66 & 70 & 0.050 & [50,17,280,281] \\
\object{SAX J2224.9+5421} & WFC'99 & \nodata & $22^\mathrm{h}24^\mathrm{m}49\fs7$ & $+54^\circ23\arcmin10\arcsec$ & $2''$(90\%) & \nodata & 0.500 & 11.3 & 1 & 0.017 & [26,282] \\
\tableline
 Total (115 sources) & & & & & & & & 7083 \\
 \enddata
%
% All codes present over all sources: PTGCARSDEOMZ?I
\tablenotetext{a}{Source type, adapted from \cite{lmxb07}; 
  A = atoll source, 
  C = ultracompact X-ray binary (including candidates), 
  D = ``dipper'', 
  E = eclipsing, 
  G = globular cluster association, 
  I = intermittent pulsar, 
  M = microquasar,  
  O = burst oscillation, 
  P = pulsar,
  R = radio-loud X-ray binary, 
  S = superburst, 
  T = transient, 
  Z = Z-source. 
We omit the ``B''designation indicating a burst source.}
\tablenotetext{b}{For sources with a neighbor within $1\arcdeg$, 
we combine all \xte\/ observations which include this source within the
field of view (possibly including observations of the neighbor).}
\tablenotetext{c}{For systems with no bursts detected in the
MINBAR sample, we calculate the 95\% upper limit on the
average burst rate, assuming Poisson-distributed number of
bursts.}
\tablenotetext{d}{The observation table entries for these systems
are not complete, and may be attributed to their nearby neighbor.
We thus adopt the maximum exposure for any of the nearby sources
as the common value for the group}
\tablenotetext{e}{These sources are indistinguishable from the next nearest
source, and so we cannot separate the bursts; we attribute all the observed
events to the neighbor (SLX~1744$-$300 in the case of SLX~1744$-$200, and 
1A~1742$-$289 in the case of AX~J1745.6$-$2901.}
\tablerefs{
  1. \cite{gal05a}; 
  2. \cite{kuin15}; 
  3. \cite{fiocchi11}; 
  4. \cite{fj76}; 
  5. \cite{kuul03a}; 
  6. \cite{Shahbaz08}; 
  7. \cite{migliari10}; 
  8. \cite{swank78}; 
  9. \cite{ucb01}; 
 10. \cite{hwvdb03}; 
 11. \cite{parmar85c}; 
 12. \cite{parmar86}; 
 13. \cite{torres08b}; 
 14. \cite{belloni93}; 
 15. \cite{makino90}; 
 16. \cite{Zhong11}; 
 17. \cite{cutri03}; 
 18. \cite{jon01}; 
 19. \cite{bassa06}; 
 20. \cite{piro97}; 
 21. \cite{zand08a}; 
 22. \cite{bp03}; 
 23. \cite{c3p86}; 
 24. \cite{iaria07}; 
 25. \cite{mason80}; 
 26. \cite{corn02b}; 
 27. \cite{corn02c}; 
 28. \cite{church05}; 
 29. \cite{parmar89}; 
 30. \cite{smale95}; 
 31. \cite{vdk84}; 
 32. \cite{Bozzo14}; 
 33. \cite{kennea14}; 
 34. \cite{bce72}; 
 35. \cite{canizares80}; 
 36. \cite{chevalier89}; 
 37. \cite{iaria05}; 
 38. \cite{iaria08}; 
 39. \cite{khbs76}; 
 40. \cite{tfs86a}; 
 41. \cite{Serino18}; 
 42. \cite{Wang04}; 
 43. \cite{morris90}; 
 44. \cite{mukai01}; 
 45. \cite{sdss7}; 
 46. \cite{bce76}; 
 47. \cite{keek08a}; 
 48. \cite{wachter02}; 
 49. \cite{bult17}; 
 50. \cite{asai00}; 
 51. \cite{casares06}; 
 52. \cite{russell12}; 
 53. \cite{swank76c}; 
 54. \cite{garnavich12}; 
 55. \cite{kennea12}; 
 56. \cite{krauss06}; 
 57. \cite{lin09}; 
 58. \cite{mark08b}; 
 59. \cite{falanga09}; 
 60. \cite{homan07}; 
 61. \cite{kaplan08}; 
 62. \cite{cw89}; 
 63. \cite{lhd76}; 
 64. \cite{oo01b}; 
 65. \cite{wachter98}; 
 66. BeppoSAX standard result on 1999 observation; 
 67. \cite{mr77}; 
 68. \cite{wachter05}; 
 69. \cite{degenaar12c}; 
 70. \cite{stroh18}; 
 71. \cite{hoff78b}; 
 72. \cite{zand05b}; 
 73. \cite{disalvo05}; 
 74. \cite{piraino07}; 
 75. \cite{szt85b}; 
 76. \cite{cocchi98}; 
 77. \cite{jonk04}; 
 78. \cite{jonker03}; 
 79. \cite{jain11}; 
 80. \cite{ratti10}; 
 81. \cite{ybs09}; 
 82. \cite{mig03}; 
 83. \cite{cocchi99b}; 
 84. \cite{cummings14}; 
 85. \cite{fiocchi08}; 
 86. \cite{wilson03}; 
 87. \cite{kaptein00}; 
 88. \cite{jonker07}; 
 89. \cite{klein-wolt07}; 
 90. \cite{kw07}; 
 91. in 't Zand fit of PCA spectrum (wa comptt, chi=4.772); 
 92. \cite{marshall99b}; 
 93. \cite{brandt06a}; 
 94. \cite{chenevez07}; 
 95. \cite{grindlay80}; 
 96. \cite{swank77}; 
 97. \cite{dai06}; 
 98. \cite{diaz17}; 
 99. \cite{gal10b}; 
100. \cite{lcd76}; 
101. \cite{frogel95}; 
102. \cite{hoff78c}; 
103. \cite{moore00}; 
104. \cite{cackett06}; 
105. \cite{sun89}; 
106. \cite{zurita10}; 
107. \cite{Kennea13}; 
108. \cite{bozzo15b}; 
109. \cite{degenaar10a}; 
110. \cite{cackett06c}; 
111. \cite{mak81}; 
112. \cite{chelov10a}; 
113. \cite{krimm08b}; 
114. \cite{bazz97}; 
115. \cite{david97}; 
116. \cite{august98}; 
117. \cite{lewin77}; 
118. \cite{brandt05}; 
119. in 't Zand fit to XRT spectrum (wa po, chi=1.056); 
120. \cite{tomsick07}; 
121. \cite{zand02}; 
122. \cite{chakrabarty17}; 
123. \cite{mereminsky17}; 
124. \cite{decesare07}; 
125. \cite{marti07}; 
126. \cite{zand91}; 
127. \cite{degenaar12b}; 
128. \cite{sakano05}; 
129. \cite{cocchi99a}; 
130. \cite{lin12}; 
131. \cite{sakano02}; 
132. \cite{degenaar09}; 
133. \cite{maeda96}; 
134. \cite{brand76}; 
135. \cite{lhd76b}; 
136. \cite{muno09}; 
137. FTOOLS; 
138. \cite{wij06b}; 
139. \cite{werner04}; 
140. \cite{wij02a}; 
141. \cite{zand98b}; 
142. \cite{brandt06b}; 
143. \cite{degenaar07}; 
144. \cite{sidoli04}; 
145. \cite{altamirano08a}; 
146. \cite{delmonte08}; 
147. \cite{juett05}; 
148. \cite{pgs94}; 
149. \cite{zolotukhin11}; 
150. \cite{mori05}; 
151. \cite{skin90}; 
152. \cite{mak83}; 
153. \cite{oo01}; 
154. \cite{vandenberg14}; 
155. \cite{bozzo10}; 
156. \cite{chenevez10a}; 
157. \cite{riggio12}; 
158. \cite{tremou15}; 
159. \cite{altamirano12b}; 
160. \cite{bahramian14}; 
161. \cite{bhatt06b}; 
162. \cite{emelyanov01}; 
163. \cite{gsb12}; 
164. \cite{rupen03}; 
165. \cite{cadelano17}; 
166. \cite{zand98e}; 
167. \cite{jonker13}; 
168. \cite{wij09a}; 
169. \cite{falanga12}; 
170. \cite{ferrigno11}; 
171. \cite{vandenberg11}; 
172. \cite{balucinska04}; 
173. \cite{homer02}; 
174. \cite{lc77}; 
175. \cite{bazz97b}; 
176. \cite{chakrabarty08}; 
177. \cite{parmar85b}; 
178. \cite{tomsick05}; 
179. \cite{zand00a}; 
180. \cite{zand03}; 
181. \cite{bozzo09}; 
182. \cite{paizis12}; 
183. \cite{papitto10}; 
184. \cite{riggio11}; 
185. \cite{cocchi01b}; 
186. \cite{cocchi99c}; 
187. \cite{chakrabarty10}; 
188. \cite{zand99c}; 
189. \cite{armas13}; 
190. \cite{bassa08}; 
191. \cite{chelov07b}; 
192. \cite{ferrigno18}; 
193. \cite{kuiper20}; 
194. \cite{shaw18}; 
195. \cite{chaty08}; 
196. \cite{markwardt03c}; 
197. \cite{baglio15}; 
198. \cite{chenevez12a}; 
199. \cite{degenaar16a}; 
200. \cite{kaur17}; 
201. \cite{hjellming98}; 
202. \cite{muller98b}; 
203. \cite{Arzou19}; 
204. \cite{Kennea17}; 
205. \cite{chak98d}; 
206. \cite{gaensler99}; 
207. \cite{wang01}; 
208. \cite{zand98c}; 
209. \cite{krimm08a}; 
210. \cite{mss08a}; 
211. \cite{jonker04}; 
212. \cite{nat00b}; 
213. \cite{ubert98}; 
214. \cite{cackett06b}; 
215. \cite{zand17}; 
216. FTOOLS (H map); 
217. \cite{krauss05}; 
218. \cite{papitto07}; 
219. \cite{stroh03a}; 
220. \cite{corbet03a}; 
221. \cite{dai14}; 
222. \cite{matsuba95}; 
223. \cite{sidoli02}; 
224. \cite{barret03}; 
225. \cite{murakami83}; 
226. \cite{ftf07}; 
227. \cite{lmxb07}; 
228. \cite{oda81}; 
229. \cite{barthelmy17}; 
230. \cite{parikh17b}; 
231. \cite{anderson97}; 
232. \cite{grindlay75}; 
233. \cite{gk97}; 
234. \cite{pallanca13}; 
235. \cite{papitto13a}; 
236. \cite{papitto13b}; 
237. \cite{asai16}; 
238. \cite{juett05b}; 
239. \cite{shahbaz07}; 
240. \cite{hands04}; 
241. \cite{barret95}; 
242. \cite{homer98}; 
243. \cite{ubert97}; 
244. \cite{zand99}; 
245. \cite{dma00}; 
246. \cite{heg01}; 
247. \cite{zand98}; 
248. \cite{migliari04}; 
249. \cite{swank76b}; 
250. \cite{beardmore11}; 
251. \cite{hoff80}; 
252. \cite{homer96}; 
253. \cite{lehto90}; 
254. \cite{Kennea18}; 
255. \cite{buisson20b}; 
256. \cite{campana05}; 
257. \cite{kaaret05b}; 
258. \cite{rupen05}; 
259. \cite{vand05a}; 
260. \cite{cic85}; 
261. \cite{lhd76c}; 
262. \cite{campana03b}; 
263. \cite{wry00}; 
264. \cite{becker77}; 
265. \cite{chou00}; 
266. \cite{church98}; 
267. \cite{iaria06b}; 
268. \cite{barthelmy11}; 
269. \cite{deg11a}; 
270. \cite{hynes01}; 
271. \cite{tomsick04}; 
272. \cite{tomsick99}; 
273. \cite{ts98}; 
274. \cite{dieball05}; 
275. \cite{vpd90}; 
276. \cite{wa01}; 
277. \cite{garcia87}; 
278. \cite{mcclintock81}; 
279. \cite{nowak02}; 
280. \cite{kg84}; 
281. \cite{premach16}; 
282. \cite{degenaar14b}.} 
\end{deluxetable}
\end{longrotatetable} % AASTeX 6.3 environment
 % label tab:bursters

\begin{deluxetable*}{ccccp{10cm}}
\tabletypesize{\scriptsize}
\tablecaption{MINBAR sources FITS table selected columns, formats and description
  \label{tab:fitscolumns}}
\tablehead{
\colhead{Column} & \colhead{Label}& \colhead{Format} & \colhead{Units} &
\colhead{Description}  }
\startdata
1  & {\tt NAME } &  A22  & & Burster name   \\
6  & {\tt Type } &  A7  &  & Source type, as for Table \ref{tab:bursters}  \\
8  & {\tt RA\_OBJ } &  E  &  deg & Source right ascension \\
9  & {\tt DEC\_OBJ } &  E  &  deg & Source declination \\
10  & {\tt ERR\_RAD } &  E  &  deg & Positional error \\
16  & {\tt NH } &  E  &  $10^{22}\ {\rm cm^{-2}}$ & Adopted neutral column density $N_H$ along the line of sight \\
17  & {\tt NH\_err } &  D  &  $10^{22}\ {\rm cm^{-2}}$ & $1\sigma$ uncertainty on $N_H$ \\
18  & {\tt NH\_bibcode } &  A21  & & Reference bibliographic code for $N_H$ value, providing a link to the publication on NASA's ADS\tablenotemark{a}   \\
20  & {\tt COMMENTS } &  A132  &  & Aliases, host clusters, and other information  \\
23  & {\tt Porb } &  D  &  h & Orbital period, where known \\
24  & {\tt Porb\_flag } &  A1  & & A ``?'' indicating cases for which the orbital period is a candidate only   \\
25  & {\tt Porb\_bibcode } &  A21  & & Bibliographic code for orbital period\tablenotemark{a} \\
57  & {\tt Vmag } &  E  &  mag & $V$-magnitude, where measured (from \citealt{lmxb07}) \\
60  & {\tt nburst } &  I  &  & Number of bursts in MINBAR  \\
61  & {\tt nobs } &  I  &  & Number of observations in MINBAR  \\
62  & {\tt exp } &  E  &  ks & Total exposure \\
63  & {\tt ERR\_CONF } &  E  & & Confidence level for positional error   \\
64  & {\tt pos\_method } &  A20  & & Method by which source position determined   \\
65  & {\tt pos\_bibcode } &  A21  &  & Reference for source position\tablenotemark{a}  \\
67  & {\tt disc } &  A6  &  & Instrument and year of discovery of burst activity \\
68  & {\tt disc\_bibcode } &  A19  & & Reference bibliographic code for burst discovery\tablenotemark{a} \\
70 & {\tt rate} & E & hr$^{-1}$ & Burst rate (or limit) measured in the MINBAR sample
\enddata
\tablenotetext{a}{The URL for the publication can be accessed via \url{https://ui.adsabs.harvard.edu/abs/[bibcode]} where 
[bibcode] is one of the entries in columns {\tt NH\_bibcode}, {\tt Porb\_bibcode}, {\tt pos\_bibcode} or {\tt disc\_bibcode}.}
\tablecomments{The FITS table also includes several columns copied from the \igr\/ source table, which for the sake of brevity we omit here.}
\end{deluxetable*}

The first bursts observed from these sources were discovered with instruments on 19 platforms, %(see column 2),
spanning almost fifty years.
The earliest detection was with Vela 5B in 1969; 
\sax\/  identified 25 sources in the late 1990s; 
\xte\/ (10 sources), % as at 2019 Dec 18
{\it Swift} (15 sources to date) % what date?
and 
\igr\/ 
(11 sources so far) % 1 IBI & 10 JEM, as at 2019 Dec 18
have each made notable contributions. % as of what date
Most of the recent discoveries are transient LMXBs
that have not previously been detected,
although the MAXI instrument \cite[onboard the {\it International Space Station};][]{maxi11} first detected bursts from 4U~1822$-$000 \citep{asai16} and 4U~1543$-$624 
\citep{Serino18},
both persistently accreting sources known since discovery in the early 1970s \cite[]{2ucat}. The rate of discovery has dwindled since 2014 to approximately one a year, which suggests that the knowledge of the bursting nature of all presently known LMXBs is nearing completion. 

\begin{deluxetable*}{llc}
\tablecaption{Summary of burster discoveries by instrument
  \label{tab:instruments}
}
\tablewidth{0pt}
\tablehead{
  \colhead{Acronym}
 & \colhead{Spacecraft/instrument}
 & \colhead{No. bursters}
}
\startdata
WFC & \sax\/ Wide Field Camera (WFC) & 24 \\
NFI & \sax\/ Narrow Field Instruments (NFI) & 1 \\
XTE & \xte\/ Proportional Counter Array (PCA) & 10 \\
JEM & \igr\/ Joint European X-Ray Monitor (JEM-X) & 10 \\
IBI & \igr\/ Imager on Board the \igr\/ Satellite (IBIS) & 1 \\
SAS & Small Astronomy Satellite 3 (SAS-3) & 13 \\
BAT & {\it Swift}\/ Burst Alert Telescope (BAT) & 9 \\
XRT & {\it Swift}\/ X-Ray Telescope (XRT) & 6 \\
EXO & {\it European X-Ray Observatory Satellite} ({\it EXOSAT}) & 5 \\
OS8 & {\it Orbiting Solar Observatory 8} ({\it OSO-8}) & 5 \\
HAK & {\it Hakucho} & 5 \\
TTM & {\it Mir-Kvant} Coded Mask Imaging Spectrometer (COMIS) & 3 \\
GIN & {\it Ginga} & 3 \\
EIN & {\it Einstein X-ray Observatory} & 3 \\
ASC & {\it Advanced Satellite for Cosmology and Astrophysics} ({\it ASCA}) & 3 \\
VEL & {\it Vela} & 2 \\
MAX & {\it ISS} Monitor of All-Sky X-ray Image (MAXI) & 2 \\
NIC & {\it ISS} Neutron star Interior Composition Explorer  (NICER) & 2 \\ % bumped to 2 with discovery of Swift J1858.6-0814
NUS & {\it Nuclear Spectroscopic Telescope Array} ({\it NuSTAR})  & 1 \\
SLX & Spacelab-2 & 1 \\
AGI & {\it Astro-rivelatore Gamma a Immagini Leggero} ({\it AGILE}) & 1 \\
ANS & {\it Astronomical Netherlands Satellite} ({\it ANS}) & 1 \\
GRA & {\it Granat}\/ Astrophysical Roentgen Telescope (ART-P) & 1 \\
HET & {\it High-Energy Transient Explorer 2} ({\it HETE-2}) & 1 \\
UHU & {\it Uhuru} & 1 \\
\enddata
\tablecomments{Bursts from just one source, 4U~1254$-$69, were discovered in high-time resolution optical photometry \cite[]{mason80}; this source is labeled OPT}
\end{deluxetable*}

The source demographics include several significant sub-groups.
Sixteen burst sources reside in thirteen globular clusters (labelled ``G''), with --- notably ---
three in a single cluster (Terzan 5). 
One third of the burst sources have been persistently accreting for more than 10 years, while the remainder are flagged as transient (label ``T'').
Eleven
burst sources are confirmed ultracompact X-ray binaries \cite[i.e., with measured orbital periods below 80~min; e.g. ][]{rjw82}, and a further 
fourteen
are candidate ultracompact X-ray binaries \cite[based on the ratio of optical to X-ray luminosity, or on the persistent nature combined with the low mass accretion
% fixed duplicate in proofs, Q11
% rate;][]{zjm07}.
 rate;][]{zand07}.
Both cases are labelled ``C''.
Three sources (HETE~J1900.1$-$2455 SAX~J1748.9$-$2021 and Aql X-1) are flagged ``I'' indicating intermittent pulsations (as distinct from the six sources flagged ``P'' which show pulsations consistently when they are in outburst). 
Eighteen sources are flagged ``O'' as having burst oscillations; see \S\ref{osc_summary}.
Fourteen sources are flagged ``S'' as having exhibited at least one candidate superburst 
\cite[e.g.][]{zand17c}. 
There are 
eight % from check of table numbers, 2020 Apr
eclipsing sources (flagged ``E'') which, therefore, are viewed edge on (with inclinations greater than about 80~degrees).
Seventeen % including Swift J1858.6-0814, added 2020 Apr, & checked numbers
sources are ``dippers'' (flagged ``D''), exhibiting incomplete and/or irregular reductions in X-ray flux at particular phases in the binary orbit, suggestive of somewhat lower inclinations than the eclipsing sources 
($i\gtrsim 70^\circ$; \citealt{white95}, although see \citealt{gal16a}).

The most precise positions for burst sources come not from X-ray observations, but from observations of a radio counterpart, for 13 sources. Long-running observational target-of-opportunity campaigns have resulted in precise ($\approx1''$) X-ray positions with {\it Chandra}\/ for 44 sources, which (provided the extinction along the line of sight is not too great) have also allowed identification of the optical counterpart.
A further 15 sources have positions for the X-ray sources known to a few arcseconds thanks to {\it Swift}/XRT or {\it XMM-Newton}\/ observations.
For the remainder of the sources which do not have known optical counterparts, the X-ray positions may be known only to tens of arcseconds, or even as poorly as within a degree \cite[in the case of XB~1940$-$04;][]{murakami83}. For these systems, in the absence of any new transient activity, any cross-identification with optical or X-ray catalogs must be viewed with extreme caution.

The total exposure is calculated for most sources in the sample
simply as the sum of exposures of all the observations of that source. However, the $1\deg$ field-of-view (FOV) of \xte/PCA, combined with the lack of imaging capability and the high source density in certain sky regions (particularly the Galactic centre) makes the contribution of those observations more complex (see also \S\ref{xteobs}). 
For PCA observations covering multiple sources, we added the exposure to the total for each source within the FOV,
since only one entry per pointing is present in the observation table (see \S\ref{minbar-obs}).
For the WFC and JEM-X, we instead list each source detectable in each pointing in the observation table. %jc190917 removed () 

The variation in the range of accretion rates for different sources have the result that the
average
burst rates of the  sample span a wide range.
Several
sources (including 4U 0614+09, 2S 0918$-$549, 1A 1246$-$588, 
4U~1705$-$32, RX~J1718.4$-$4029 % added
SLX 1737$-$282, 4U 1850$-$086, and M15 X-2) 
have less than 10 bursts detected in MINBAR, despite having been persistently accreting for at least 10 years.
This paucity is in remarkable contrast to 
prolific sources like 4U~1636$-$536 and 4U~1728$-$34.
On the other extreme of the accretion rate range are the so-called ``Z'' sources (GX 17+2, Cyg X-2 and Cir X-1) which are thought to accrete near the Eddington limit. The burst behaviour of these sources is difficult to reconcile with burst theory, 
particularly for GX~17+2 which shows a mix of long- and short-duration bursts at high accretion rates \cite[e.g.][]{kuul02a}.
We note that the average burst rates may have significant systematic errors, for sources with only one or a few bursts (e.g. 1RXS~J180408.9$-$342058), or for those ``burst-only'' sources where the persistent emission is so weak that it is typically not detectable by \sax/WFC \cite[e.g. SAX~JJ1818.7+1424, SAX~J2224.9+5421;][]{corn02b}. For these (and similar cases) we would expect the burst rate determined from the MINBAR sample to substantially overestimate the typical rate.

\section{Data selection and reduction}
\label{sec:source-data}

Here we describe the characteristics and treatment of the data for constructing the burst and observation catalogs, for each instrument (as summarised in Table \ref{tab:properties}).

The selection criteria for each instrument were adopted to achieve a sample that was 
as complete as possible over the interval for the study, covering the \xte\/ launch through to \igr\/ revolution \pubdatacutoffrev\ (MJD~\pubdatacutoffmjd; \pubdatacutoffdate).
For \xte\/ and \sax, completeness is in principle achievable, because 
the cutoff date is beyond the end of the mission.
For \igr, observations are 
continuing, but we defer their analysis to future data releases.

We calculated the exposure for each observation based on the ``good-time'' intervals adopting standard screening criteria. 

\begin{deluxetable*}{lcccccccc}
\tabletypesize{\scriptsize}
\tablecaption{Summary of the properties of instruments contributing to MINBAR
  \label{tab:properties}
}
\tablewidth{0pt}
\tablehead{
 & & \colhead{Mission}
 & \colhead{Effective}
 & \colhead{FoV}
 & \colhead{FWHM\tablenotemark{b}}
 & \colhead{$\Delta E/E$}
 & \colhead{Total}
 & \colhead{No.}
 \\
 \colhead{Spacecraft/instrument}
 & \colhead{Launched}
 & \colhead{duration (yr)}
 & \colhead{area\tablenotemark{a} (cm$^2$)}
 & \colhead{(deg)}
 & \colhead{(arcmin)}
 & \colhead{@ 6~keV}
 & \colhead{exp. (Ms)}
  & \colhead{bursts}
}
\startdata
\xte/PCA & 1995-12-30 & 16.0 & 1400\tablenotemark{c} & 1\tablenotemark{d} & \nodata & 17\% & \exptotxte & \burstsxte \\
\sax/WFC & 1996-04-30 & 6.0 & 140 & $40\times40$\tablenotemark{e} & 5 & 20\% & \exptotsax & \burstssax  \\
\igr/JEM-X & 2002-10-17 & ongoing & 64\tablenotemark{f} & 6.6\tablenotemark{d} & 3 & 17\% & \exptotigr & \burstsigr\tablenotemark{g} \\
\enddata
\tablenotetext{a}{For the PCA and JEM-X, these values are determined empirically as described in \S\ref{sec:area}, and also include corrections for the different energy bands of the instruments}
\tablenotetext{b}{Spatial resolution, full width at half maximum (FWHM)}
\tablenotetext{c}{For the PCA, the quoted effective area is per PCU; with all five operational, the total area is $\approx7000\ {\rm cm^2}$}
\tablenotetext{d}{Radius to zero response}
\tablenotetext{e}{Full width to zero response}
\tablenotetext{f}{Effective area adopted for the persistent emission. For the  spectrum typical of bursts, a larger relative effective area of $\approx100$~cm$^2$ is consistent with the comparison of bursts observed with both PCA and JEM-X; see \S\ref{ss:jemxpca_obs} and \S\ref{ss:jemxpca_burst}.}
\tablenotetext{g}{Through revolution \pubdatacutoffrev, \pubdatacutoffdate (MJD \pubdatacutoffmjd)}
\end{deluxetable*}

\subsection{Rossi X-ray Timing Explorer Proportional Counter Array} %jc190917: RXTE-PCA
\label{obs-xte}

The {\it Rossi X-ray Timing Explorer}\/ (\xte) was launched into an approximately $90$-min low-Earth orbit
on 1995 December 30 and operated until the end-of-mission on  2012 January 3.
The spacecraft featured three science instruments: the All-Sky Monitor \cite[ASM;][]{asm96}, the High-Energy X-ray Timing Experiment \cite[HEXTE;][]{hexte98}, and the Proportional Counter Array \cite[PCA;][]{xte96}. We used PCA data for the principal analysis for this paper; the key properties of this instrument are summarised in Table \ref{tab:properties}.

The PCA is comprised of five identical proportional counter units (PCUs), sensitive to X-ray photons in the energy range 2--60~keV and with total 
geometric collecting area of about $8000\ {\rm cm}^2$ \cite[]{xtecal06}.
Each PCA is fitted with a passive collimator admitting photons within a $1^\circ$ radius of the pointing direction, with an approximately linear response as a function of off-axis angle.
The spectral resolution is approximately 17\% at 6~keV, improving to 8\% at 22~keV, as measured from ground calibration sources.
Gradual degradation of the PCUs over the mission lifetime led to a mission-wide average number of active PCUs of roughly three. For most observations PCU\#2 was active, with the other units rotated in and out of service to maintain operation for as long as possible. 

Photons can be time-tagged to a precision of $\sim1\ \mu$s, and are collected in a variety of data modes adopted for each of five event analyzers. The principal modes used for the MINBAR analysis are the ``Standard-1'' modes, with time resolution of 0.125~s but no spectral resolution; ``Standard-2'' mode, with 16~s time resolution and 129 spectral channels; and ``Event'' modes, typically with time resolution of $125\ \mu$s and 64 spectral channels.

The PCA sensitivity is primarily a function of the number of PCUs on, and the instrumental background rate, which varies over the orbit. For a source observed on-axis with all 5 PCUs, the $3\sigma$ sensitivity over a 1-s time bin is roughly
0.01~count~s$^{-1}$~cm$^{-2}$, or $5\times10^{-11}\ \epcs$ (3--25~keV).

The {\it RXTE}\/ PCUs are subject to a short ($\approx 10\ \mu$s) interval of inactivity following the detection of each X-ray photon. This ``deadtime'' reduces the detected count rate below what is incident on the detector (by approximately 3\% for an incident rate of 
400~count~s$^{-1}$~PCU$^{-1}$).

In mid-2000 PCU \#0 developed a leak in the propane veto layer, used to exclude charged particles, with the pressure dropping to zero within a day. The PCU remained operational, although with a higher background rate and different gain. PCU \#1 experienced a similar issue in late 2006, dropping to a similar level of performance to PCU \#0. 

We also used HEXTE spectra, covering 16--250~keV, to measure the persistent spectrum beyond the PCA range. HEXTE consists of two independent clusters each with four NaI(Tl)/CsI(Na) phoswich scintillation counters, covering a circular field of view of $1^\circ$ full-width at half maximum (FWHM). %jc190917
The photon collecting area is approximately 1600~cm$^2$, and the energy resolution is $\approx15$\% at 60~keV. Each cluster can ``rock'' on and off-source to provide background measurements, with one cluster designed to cover the target at any given time.

Beginning in 2006, cluster A experienced intermittent failures of the rocking mechanism, and late in that year was set to stare permanently at the source, to avoid being stuck instead in the off-source position. Modulation of cluster B failed some years later, in early 2010, leaving it stuck in the off-source position. For the last years of the mission the source spectrum could be measured with cluster A, and background estimated from cluster B.

We used public ASM data, consisting of 90~s dwells covering the entire X-ray sky a few times per day, to assess the activity of sources that fell within a single PCA field, as described below.
The three ASM cameras each have a position-sensitive proportional counter offering an effective area of at most $\approx30~{\rm cm^2}$ at 5~keV, and covering the 0.5--12~keV energy range.
We used lightcurves of daily averages provided by the MIT ASM team\footnote{\url{http://xte.mit.edu}}.

The analysis approach for the \xte\/ observations and bursts was based on that adopted for 
G08, with a few exceptions (as described in \S\ref{xtetrse} and \S\ref{trsanl}).

\subsubsection{Observations}
\label{xteobs}

Pointed PCA observations were made as a mix of scheduled, target-of-opportunity (TOO) and monitoring observations as part of the guest observer (GO) program over the mission lifetime. The shortest observations have typical durations of $\approx2$~ks, corresponding to one orbit, but observations can last up to 3~days, for a maximum exposure of $\approx150$~ks. The total exposure time for each source depends only weakly on the sky position, but is boosted for sources around the Galactic centre, where a single pointing can span multiple sources. The total exposure for most sources was less than 1~Ms, but up to 3.8~Ms for the best-studied example (4U~1636$-$536). 

We  selected all observations including burst sources within 
the full field of view, 
and the resulting sample totals \exptotxte~Ms (from \obstotxte\ individual observations). 

The PCA instrument collects photons from any source within the FOV, so that persistent spectra may include contributions from more than one source. %jc190917: rephrased
We attempted to flag spectra so affected by testing for ASM detections of each source in such fields, close to the time of the PCA observation. Where this information is available, we flagged those observations in which the count rate and persistent spectrum 
is
contaminated by sources other than the target (see \S\ref{minbar-obs}).
Where contemporaneous ASM dwells were not available, we also indicated this via the observation flag (see \S\ref{obsanl}).

\subsubsection{Source lightcurves and burst searches}
\label{xteburstid}

We extracted lightcurves from the PCA data, 
covering the full energy range 2--60~keV,
using Standard-1 mode data
(0.125~s resolution, no energy resolution, PCUs resolved). For
a few
cases, these data were absent and we instead employed event-mode data (available with a range of time resolutions typically $\ll1$~s).
We normalized the light curves to the number of
active PCUs and the collimator response, and then to a photon collecting area
of 1400~cm$^2$ 
as determined from the cross-calibration described in \S\ref{sec:crosscal}.
No deadtime
correction was applied to the lightcurves. 
The collimator was modeled with a simple triangular function peaking at the optical axis, and decreasing to zero at an off-axis angle of $1.0\deg$. The response was
calculated as 
$1-\theta$, where $\theta$ is the off-axis angle of the source in degrees. This simple model introduces a systematic error of 5--10\% in normalized count rates for off-axis angles up to 0\fdg5.

We 
searched for bursts by selecting excess measurements within the lightcurve. 
This search was confounded by ``breakdown'' events in individual PCUs, which manifest as a short-lived burst of X-rays, similar in some cases to the profile of a thermonuclear burst. We identified such events by reviewing the lightcurves calculated from individual PCUs around the time of each candidate event.
A second source of confusing events arises from gamma-ray bursts, which may be observed even from outside the FOV of the instrument. These events may be identified by an extremely hard X-ray spectrum, rising towards the upper energy limit for the PCA. 
Both types of events are relatively easy to identify from the PCA data, and we do not expect any remaining examples are present in our sample.

\added{Where multiple bursting sources were active in the FOV during an observation, we assigned the burst origin following the procedure of G08. We measured intensity variations between PCUs (where available), which arise in part from slight differences in the pointing direction of each unit; and also matched the burst properties to each source. It remains possible that some of the PCA events are attributed incorrectly, which will also introduce systematic errors into the burst rates (see \S\ref{sec:burstrates}) and possibly also the measured Eddington flux (\S\ref{fluxEdd}).}

A total of \burstsxte\ type-I X-ray bursts from \sourcesxte\ sources were found.

\subsubsection{Time-resolved spectral extraction}
\label{xtetrse}

Where available, we utilised Event mode data to extract time-resolved spectra in the range 2--60~keV covering the burst.
For a small number of bursts the Event mode data was unavailable, and we instead used ``Binned'' mode data to extract the spectra.
We set the interval for spectral extraction initially at 0.25~s during the burst rise and peak. For fainter bursts, we began with 0.5~s bins or as long as 1~s. The %bin jc190917
size of the time intervals was gradually increased into the burst tail to maintain roughly the same signal-to-noise level \added{($\gtrsim50$)}. 
A spectrum taken from a 16-s interval prior to the burst was adopted as the background. 

We estimated the deadtime correction using the Standard-1 mode data\footnote{following the recipe at \url{http://heasarc.gsfc.nasa.gov/docs/xte/recipes/pca\_deadtime.html}} and applied the correction by 
calculating an effective exposure, depending upon the measured count rate, which takes into account the deadtime fraction. The largest deadtime fraction we found in our analysis is 23\%, for the brightest bursts from SAX J1808.4-3658.

We generated a response matrix specifically for each burst, but incorporating the contribution from each active PCU, otherwise as for the persistent spectra (see \S\ref{xtepersspec}). 
The spectral fitting approach for the time-resolved spectra is described in \S\ref{trsanl}.

\subsubsection{Persistent spectra}
\label{xtepersspec}

We extracted observation-averaged PCA spectra separately from each PCU from Standard-2 mode data, binned every 16~s. In contrast to the treatment for the WFC and JEM-X, we excluded an interval beginning 32~s before and ending 256~s after each burst, to avoid contamination from the burst emission. % formally not enough (think of long bursts GX 17+2) but lets leave it -- jz
We estimated the instrumental background for each PCU over each interval in which it was active using the all-mission background model file appropriate for ``bright'' sources ($\gtrsim40$~counts~s$^{-1}$~PCU$^{-1}$).
We calculated instrumental responses appropriate for the epoch of each observation using
the revised PCA response matrices,  v11.7\footnote{see 
% replaced in proof, Q2
% \url{http://www.universe.nasa.gov/xrays/programs/rxte/pca/doc/rmf/pcarmf-11.7}
\url{https://heasarc.gsfc.nasa.gov/docs/xte/pca/doc/rmf/pcarmf-11.7} }.
We estimated the effects of deadtime as for the time-resolved burst spectra. The correction factor for the persistent spectra was typically 1.02, or 1.13 for the highest-intensity spectrum.
The analysis of the observation-averaged spectra is described in \S\ref{obsanl}.

\subsubsection{Selection of data for burst oscillation search}
\label{xteoscillationbursts}

We provide burst oscillation properties for 
16
sources for which the detection of burst oscillations is considered to be robust \cite[labeled as ``O'' in Table \ref{tab:bursters}; see the discussion in][]{watts12a,Bilous19pub}, and for which sufficiently high-quality \xte/PCA data is available.
There are currently 
19  % added SAX J1810.8-2609, Bilous et al. 2018
known burst oscillation sources\footnote{see \url{https://staff.fnwi.uva.nl/a.l.watts/bosc/bosc.html} for an up to date list};
the three that are omitted from this analysis are IGR~J17480$-$2446 \citep{cavecchi11}, which rotates too slowly, see discussion below; 
IGR~J17498$-$2921 \citep{linares2011b,chakraborty12}, for which one of the two bursts observed with \xte\/ was eliminated because it was flagged with label h (see 
% modified ref to preserve table ordering, as per proofs
% Table \ref{pca-code-table}), %jc190918
\S\ref{subsec:minbar})
and the other one because it did not pass the burst count limit; and
IGR~J18245$-$2452 \citep{patruno2013,papitto13b}, which was not observed with \xte.

We first selected all the bursts observed by \xte/PCA from the candidate sources. The resulting sample includes \alloscbursts\ candidate events; see \S\ref{osc_table} for the description of the table parameters.

From this sample, we discarded some bursts, using the following criteria:

\begin{itemize}
\item{We eliminated bursts that are marked with flags e, f, g, h (see 
% modified ref to preserve table ordering, as per proofs
% Table \ref{burst-analysis-flag-table}). 
\S\ref{subsec:minbar}).
This excludes very faint bursts, bursts where there are problems with the background subtraction, bursts that were only partly observed, and bursts that were not covered by the high time resolution PCA data modes 
(see \S\ref{minbar} for more details).}
\item{We set a minimum background-subtracted burst count of 5000~counts within the first 16 seconds of the burst, to ensure that each burst can be divided into at least one full time bin (see \S\ref{burstosc}).}
\item{We excluded bursts with data gaps lasting for $\gtrsim 1$ second, to avoid eliminating one or more full time bins (as defined in \S\ref{burstosc}) from our analysis. 
In such events
there is a significant chance that the time bin with the strongest signal 
will be excluded from the analysis, which would affect the outcomes.}
\item{We excluded bursts that are not fully observed by \xte.
Some of these bursts have the flag g, but for some others 
without this flag 
the PCA data does not include (part of) the last phase before the start of the burst, or the burst decay.
Since we determine the background count rate based on the 17~s before the start of the burst, or up to 16~s after, we also eliminated bursts that were not fully observed in these windows.}
\end{itemize}
These criteria exclude \oscburstsexcluded\ events; one more burst was excluded because it lacked the high-time resolution data necessary for the search.
The resulting sample includes \oscbursts\ bursts.

\subsection{BeppoSAX Wide Field Camera} %jc190917: BeppoSAX-WFC
\label{obs-sax}

The \sax\/ broad-band X-ray observatory \citep{boella97a} was launched on April 30th, 1996. It became operational two months later and remained active until May 1st, 2002. \sax\/ comprised four sets of instruments,
including a pair of identical Wide Field Cameras  \cite[WFCs;][]{jager97,zand04b},  operating on the principle of coded aperture imaging 
\cite[]{dicke68}.
The two cameras pointed in opposite directions with fields of view of $40\times40$~square~degrees,  encompassing 4\% of the sky each. The imaging was provided by the combination of a coded mask and a position-sensitive large-area proportional counter,  enabling an on-axis angular resolution of 5~arcmin (FWHM). The net photon-collecting detector area of the data is highest on-axis at 140~cm$^2$ and drops linearly to zero at the edge of the field of view. The spectral resolution is 20\% FWHM at 6~keV, in a 2--30 keV bandpass. 
The WFC is the primary instrument adopted for MINBAR, with properties summarised in Table \ref{tab:properties}.

The WFC sensitivity is a strong function of the source position within the field of view, and the total flux from all sources contained within the FWHM of the field of view  from the source position. For the on-axis position,  the 3$\sigma$ sensitivity %at how many sigmas? -jc
on a time scale of 1~s is at best about 0.4~count~s$^{-1}$cm$^{-2}$ or 4$\times10^{-9}$~erg~s$^{-1}$~cm$^{-2}$ (3--25 keV), and at worst about 4 times higher.

\subsubsection{Observations}
\label{saxobs}

The WFC observations lasted between 10$^3$~s and 9~d, typically about 1~d. We searched all observations with a fixed pointing for X-ray bursts. The total net exposure time over the whole \sax\/  mission is a strong function of sky position, and  
for most sources is in the range 3--5~Ms.

The WFC angular resolution is generally sufficient to separate all close pairs of burst sources, except for SLX~1744$-$299 and SLX~1744$-$300 (separated by only $2\farcm4$),
and AX~1745.6$-$2901 and 1A~1742$-$289 ($5\farcm7$). 
For observations covering these pairs of sources, we report the observation and burst parameters as coming from the latter of the pair 
for convenience. % dkg (2019 Nov)
There are other cases of close pairs of bursters, but those concern transient sources whose active periods were always disjoint so that bursts could be attributed to the active source of the pair. There may be an incidental burst from a persistent globular cluster source that could be from an undocumented burster in the same cluster. This possibility applies to JEM-X and PCA as well. 

We analysed a total of \obstotsax\ \sax/WFC observations, totalling 
\exptotsax~Ms. 
We performed blind searches for bursts (see \S \ref{saxburstid}) and, therefore, included searches for sources that were discovered to be burst sources after the mission.

\subsubsection{Source lightcurves and burst searches}
\label{saxburstid}

We generated 2--30~keV light curves from the WFC data by fitting the point-spread function (PSF),
with the source position
and PSF shape fixed, and only the intensity left free. The
source position was determined 
as follows: 
\begin{enumerate}
\item we identified the burst
time and duration in a light curve of the complete
detector;
\item we reconstructed an image for this time frame;
\item we identified the
burst source in this image;
\item we generated an ``imaged'' light curve
for this source using the initial position;
\item we identified the optimum time frame within this light
curve to achieve the best
signal-to-noise ratio;
\item we calculated a new image for the newly-identified
time frame;
\item we determined the most accurate source position from
this image. 
\end{enumerate}
Note that these light curves are subtracted for diffuse and
particle-induced background, but not for the source's persistent emission.
The flux was normalized to the photon collecting area.

We searched each light curve for X-ray bursts in two ways: first, 
with a burst-search algorithm 
\cite[e.g.][]{bagnoli15}
applied to the 1-s light curve for each
active burst source in each observation,
with confirmation by visual inspection.

Second, we generated light curves 
of all photons over the whole detector,  as well as the four quarters of the detector,  at various time resolutions between 1~s and 500~s. These light curves were searched for bursts with the same algorithm as for the ``imaged'' light curves, as well as by eye. This second search finds X-ray bursts from sources that are unknown as low-mass X-ray binaries,  sometimes even from previously unknown sources \cite[e.g.,][]{corn02c}. There is some confusion with gamma-ray bursts,  but most of those can be distinguished by atypical light curve shapes (lacking the fast rise and exponential-like decay) and spectra (being much harder than 2--3~keV black bodies).

The ``Rapid Burster'' (MXB~1730$-$335) is unique as it shows both type-I and type-II bursts during active periods \cite[e.g.][]{bagnoli15}. The latter events are thought to arise from quasi-regular ``bursts'' of accretion onto the neutron star, and the energy generation processes are distinct from the type-I events. In moderate signal-to-noise data it is difficult to separate the type-II events from the (less frequent) type-I events, and therefore we excluded all the burst events detected by the WFC from this source.

We identified a total of \burstssax\ type-I X-ray bursts from \sourcessax\ sources observed with the WFC. 

\subsubsection{Time-resolved spectral extraction}
\label{saxtrse}

The time-resolved spectral extraction for the bursts observed with \sax/WFC first involved defining the limits of each time interval. 
We fitted the full-bandpass time profiles with an exponential function to determine the start time
and the exponential e-folding decay time over the complete bandpass, 
among other parameters
(for details see \S\ref{sec:burst-lightcurves}). We set the time intervals (beginning with the burst start time) by the requirement that the significance of the burst signal in that time interval
be $\geq10$, significance being defined as the total flux divided by its 1-$\sigma$ uncertainty.  Experience showed that this criterion reveals spectra that have sufficient quality to allow a meaningful fit with a blackbody model. 

We extracted spectra for each burst time interval % \jz{defined how? -- dkg / sorry I don't understand; the derivation of the time interval has been explained 2 sentences up? -- jz} 
by extracting images for each channel in 
the required
time interval and fitting a  model point-spread function (PSF) appropriate for the channel energy and field-of-view position
to the source image. This series of flux measurements constitutes the spectrum. The WFC spectral response is a strong function of field-of-view position and mission time, and was calculated for every spectrum separately. 

We also corrected each WFC spectrum (both burst and persistent emission) for instrumental deadtime,
calculated from rate meters that count the events triggering the front-end electronics before and after anti-coincidence criteria are applied.
The accuracy of the deadtime measurements was  about 1\%; the resulting correction factors 
were at most about 35\%. 
The spectrum of the non-burst emission during the burst was estimated by taking the  observation-averaged persistent spectrum (see \S\ref{saxpersspec}) and normalizing it to match the background flux determined from the time profile fit of the burst. 

Analysis of the resulting spectra is described in \S\ref{trsanl}.

\subsubsection{Persistent spectra}
\label{saxpersspec}

The WFC 2--30 keV bandpass was read out in 32 channels. We generated spectra 
through the same procedure as for the burst spectrum explained above, including a correction for instrumental deadtime and a separate response matrix for each time-interval and source.
We extracted a spectrum covering each observation of each burster,
including any bursts that 
occurred\footnote{Any burst contribution would constitute less than 1\% of the fluence of the whole observation, which is negligible considering the WFC's sensitivity}. For about half of all observations the source was not detected, up to our detection significance threshold (based on the count rate) of 3.

The analysis of the observation-averaged spectra is described in \S\ref{obsanl}.

\subsection{INTEGRAL JEM-X} %jc190917: "INTErnational Gamma RAy Laboratory"?
\label{obs-jemx}

The hard X-ray and $\gamma$-ray observatory \igr\/ 
\citep{integral03} % \citep{w03} 
has been
orbiting the Earth about every three days since launch on 2002 October 17. %\jc{INTEGRAL's orbit was actually changed in 2015. Now it makes 3 revolutions in 8 days, so I would maybe write that INTEGRAL orbit is about 3 days. - cs}
The satellite carries, besides an optical monitor camera, three % high energy 
coded-mask instruments operating simultaneously 
and covering different energy bands from 3~keV up to 10~MeV. 

In the present work we use data from the X-ray monitor JEM-X \citep{lund03}, with properties summarised in Table \ref{tab:properties}. 
The twin X-ray cameras JEM-X~1 and JEM-X~2 contain each a micro-strip xenon gas chamber 
located at a distance of 3.4~m from the
coded mask 
to observe the same $\simeq6.6^\circ$-radius (to zero response) FOV % (FWZR)
and provide good imaging capabilities 
at about 
3~arcmin (FWHM)
angular resolution. 
Like {\it BeppoSAX}/WFC,  the sensitivity of JEM-X  strongly depends on the source angle in the FOV 
with an on-axis effective photon-collecting detector area of $64~{\rm cm}^2$ %\jc{do we know this any more precisely? cf. with the value of 64 cm$^2$ derived for the cross-calibration -- dkg
per instrument at 10~keV, 
dropping by a factor of two at
$3^\circ$ off-axis. 
The spectral resolution as a function of energy  is roughly $0.4\left(1/E({\rm keV})+1/60\right)^{1/2}$, corresponding to 17\% at 6~keV.

With no 
confounding sources producing a background %\jc{change OK? --- dkg} OK - jc190917
stronger than 0.1~Crab\footnote{Note that 1 Crab unit, which is the flux of the Crab nebula plus pulsar, is 
equivalent to roughly
$3\times10^{-8}\ \epcs$ (3--25~keV)} % 3.02 to be more precise, according to dkg's model
in the FOV, the 3-$\sigma$ on-axis sensitivity for each JEM-X unit is about 0.3~Crab %\jc{removed ``/s'' here, OK? --- dkg} OK - jc190917
or 0.5~count~s$^{-1}$cm$^{-2}$ equivalent to 9$\times10^{-9}$~erg~s$^{-1}$~cm$^{-2}$ (3--25 keV) for a timescale of 1 second. These numbers must be multiplied by a factor $\approx2$ if the total background is about 1~Crab. %\jc{should this be mCrab? -- dkg. No -- jc190917}
Bursts observed simultaneously with other instruments have shown that the burst detection below 1~Crab drops for an off-axis position $\geqslant 3.5^\circ$, but burst peaks above 2~Crab can be detected up to $4.5^\circ$ off-axis.

As for every instrument aboard \igr, JEM-X data are reduced with 
the standard Off-line Science Analysis 
\citep[OSA;][]{isdc03} % \citep[OSA;][]{c03} %distributed by the INTEGRAL Science Data Centre
software version 10.1.  
JEM-X data are thus corrected for vignetting effects of the collimator,  
dead-time effects of the detector, as well as calibration effects 
due to short and long term variations of the detector gain and sensitivity.

\subsubsection{Observations}
\label{jemxobs}

For the first eight years of the \igr\/ mission, the two JEM-X units were 
operated independently, with only one unit switched on at any given time.  JEM-X 2 operated alone from launch through to satellite revolution 170 (2004 March 5), when it was switched off and JEM-X 1 was switched on. The instruments were swapped back at the end of revolution 861 (end of October 2009). Since revolution 976 (2010  October 10),  both 
JEM-X units have been operating simultaneously (apart from short periods due to technical reasons).

 Thanks to its elliptical orbit with an apogee at $\simeq150,000$~km, \igr\/ can perform nearly uninterrupted observations that commonly last from several hours to days. 
During a typical observation the satellite slews around a predefined target or sky area following a given pattern, consisting of a number of stable pointings separated by $\simeq2^\circ$ slews lasting two minutes (e.g. \citealt{jensen03}). %\jc{Reference? - jz}.
Each pointing is referred to as a science window (ScW), with a typical duration of $0.5$~hr up to $1$~hr %\jc{should this be a range, or really discrete values? -- dkg},  
depending on the actual observation. %\jc{what determines this? -- dkg}. \cs{unless specifically requested by the observer the pointing duration is determined by the ISOC software, to provide a complete number of patterns for the requested exposure. Pointing durations can have any value between 1800 and 3600 sec. This is to reply your question, but actually I do not think all this detail is necessary.}
\igr\/ datasets are thus identified by their ScW number in each of the satellite revolutions.

We selected every %publicly available %jc190917: all data involved here have indeed being public for years now.
observation of burst sources through \igr\ revolution \pubdatacutoffrev\ (\pubdatacutoffdate; MJD~\pubdatacutoffmjd), totalling \exptotigr~Ms (\obstotigr\ observations). Because we only searched for bursts through revolution \pubdatacutoffrev, we exclude from the observation table those sources with bursts first detected after this date (see \S\ref{sec:sources}). 

We  analysed all ScWs containing any of the 
target sources 
inside the zero-response FOV.
Our selection includes bursts observed at angles $>5^\circ$, 
but these data must be viewed with some caution, as 
the sensitivity of the JEM-X detectors gets so low 
that only very strong sources (and therefore the brightest bursts) can be detected \citep[see][]{brandt03}.

As for the WFC (see \S\ref{saxobs}), the spatial resolution of JEM-X is sufficient to separate almost all bursters, %jc190917
apart from those in globular clusters and the close pairs of sources SLX~1744$-$299/300 as well as AX~1745.6$-$2901 and 1A~1742$-$289. For those two pairs, we report observations as coming only from the latter sources, respectively. %SLX~1744$-$300.

During the first two years following the launch of \igr, the JEM-X instruments could adopt an alternative ``restricted imaging'' data-tacking mode, which was automatically activated to reduce the telemetry in case of increased count rates. This restricted mode, with only eight energy channels, was abandoned in 2004 and has not been supported by the OSA software since 2006. Therefore, any observations or bursts that occurred in this restricted mode 
could not be analysed for MINBAR.
We identified 114 science windows (through revolution 163) that were taken in ``restricted imaging'' mode, of which 99 included at least one burster. 

We estimated the total exposure lost to the restricted imaging mode data by  
counting all the burst sources within $5^\circ$ of the aimpoint of each affected science window, and multiplying by the median length of science windows in the MINBAR observation catalog of 2~ks, 
to give 4.4~Ms (about 
0.7\% of the total \exptotigr~Ms 
of JEM-X observations that were analysed).
For individual sources, the fraction of observations in this mode may have been as high as a few percent, but does not factor in the transient source activity, so may not have meant any significant loss of bursts. We further explore the effect of this data taking mode on the completeness of the burst sample in \S\ref{completeness}.

\subsubsection{Source lightcurves and burst searches}
\label{jemxburstid}

The source light curve extraction in JEM-X is based on an algorithm where each detected photon 
in the energy range 3--25~keV
is back-projected through the mask so its contribution to the source signal is computed using the expected pixel illumination fraction (PIF) of each detector pixel from a given sky position. 
The energy-dependent PIF-map on the detector, obtained from knowledge of the source position relative to the instrument mask, collimator and detector geometry, as well as the physics of photon interaction, depends strongly on the off-axis angle of the source direction. This vignetting effect is therefore corrected so the source light curve is obtained as if the source were observed on-axis, 
although the uncertainties will typically be higher with increasing off-axis angle. 
Other sources in the FOV are expected to yield a poor contribution to the source PIF and are considered as background, which is subtracted bin by bin during the light curve extraction. Since this procedure is basically a matter of counting and scaling events, statistical uncertainties in derived count rates are estimated 
assuming Poisson
statistics for the counts. 
We normalized the light curves to a photon collecting area
of 64~cm$^2$,
as determined from the cross-calibration
appropriate for the persistent emission,
as described in \S\ref{sec:crosscal}.

We thus generated  3--25 keV source light curves at 1~s resolution 
for each of the known X-ray bursters 
included in our source catalog,
for every ScW where the source position intercepted 
the JEM-X FOV. 

Due to incomplete coding of the field of view by JEM-X, 
the deconvolution of the coded detector data to  the decoded sky image 
can be affected by some crosstalk between sources inside the same field of view (for a review of coded aperture imaging, see e.g. \citealt{caroli87}). 
This effect may result in some cases in a bright burst from one given source 
showing up in more than one source light curve. 
In order to alleviate this degeneracy we have systematically produced sky images 
of the FOV inside a radius of $5^\circ$ during (typically) a 30~s time interval around 
every burst detected in the source light curves. 
These burst images are then automatically screened so as to detect the most significant 
group of pixels and identify it with the bursting source.
A double check with the corresponding source light curve is eventually performed 
to confirm the source origin of the burst.
Solar flares and particle events may also affect the instruments and produce 
excesses in the light curves that may be interpreted as X-ray bursts. 
Also in such cases the imaging verification does make it possible to rule out a real 
X-ray burst if the whole sky image actually has a low S/N.

As for \sax, we excluded all the burst events from the ``Rapid Burster'', MXB~1730$-$335, as it was not possible to distinguish the type-I events from the (much more frequent) type-II (see also \S\ref{saxburstid}).

A total of \burstsigr\ type-I X-ray bursts from \sourcesigr\ sources were found up to \igr\/  revolution 
\pubdatacutoffrev\ (\pubdatacutoffdate, or MJD \pubdatacutoffmjd)

\subsubsection{Time-resolved spectral extraction}
\label{igrtrse}

The  version of the OSA adopted for our analysis (see \S\ref{obs-jemx}) cannot extract source spectra on intervals shorter than $10$~s. As this limit is  longer than the typical timescale for spectral variation in bursts, it has not been possible to perform time-resolved spectroscopy of bursts detected with JEM-X. 
Thus, we defer spectroscopy of JEM-X bursts to future MINBAR releases. 

\subsubsection{Persistent spectra}
\label{jemxpersspec}

We extracted average persistent spectra, with 16 energy bins covering the range 3--25~keV, %\jc{the standard binning is 16 energy bins in the 3--35~keV energy range, did you use non-starndard binning? -- CS}
for each burster less than $5^\circ$ off-axis in each ScW. 
As for \sax/WFC, the spectral response is strongly dependent on mission time and source position within the FOV, so it is calculated together with each source spectrum, which is also corrected for dead time based on the infalling count rate on the whole detector.

The analysis of the observation-averaged spectra is described in \S\ref{obsanl}.

\section{Data analysis and calibration}
\label{sec:analysis}

Here we describe the analysis procedures that were used to derive the  properties for each burst and observation, from each instrument.

\subsection{Light curve analysis}
\label{sec:burst-lightcurves}
\label{sec:lcs:modeling} % now that we merged these two sections

We determined
the basic  parameters describing each burst, including the start time, observed peak (photon) flux, total photon fluence and duration, from the 
instrumental light curves, normalised using empirical coefficients taking into account the differences in effective area and energy band. 
We chose these ``instrumental'' lightcurves
rather than the bolometric flux measurements (as adopted by G08), because the latter were not available for faint bursts observed by WFC, or the bursts observed by JEM-X (see \S\ref{igrtrse}). Furthermore, the burst lightcurves
do not depend on spectral models
or detailed assumptions about the detector response, and are not subject to systematic
errors arising from spectral analysis,
in contrast to 
the estimated peak bolometric flux or fluence. 

Prior to performing the analysis, we corrected the lightcurves for instrument-to-instrument differences in energy range and effective area, as described in \S\ref{sec:area}.
The energy range over which the lightcurves were extracted for the three instruments was not consistent.
Specifically, the full range of \xte/PCA, of 2--60~keV, is substantially wider than the bandpass used for \sax/WFC (2--30~keV), or \igr/JEM-X (3--25~keV).
Additionally, the effective area of the three instruments are substantially different, with the \xte/PCA almost two orders of magnitude higher than \sax/WFC and \igr/JEM-X (e.g. Table \ref{tab:properties}).

Since all three devices are proportional counters based on the same photon detection
principle --- photo-ionization of Xenon atoms in a gas chamber and signal amplification in a strong electrical field ($E\sim1$ kV~cm$^{-2}$) --- with similar quantum efficiencies, we expect that the relative normalisation between the instruments would be only weakly dependent upon the source spectrum.
We thus took the approach of 
comparing light curves between devices and adopting a linear cross-correlation relation (accounting also for the differences in photon-collecting area) for each pair of instruments WFC-PCA and PCA-JEM-X, as described in \S\ref{sec:area}. The coefficients for the linear relations are given in the {\tt count} row of Table \ref{tab:crosscal}.
We then determined the burst parameters from the lightcurves rescaled with those relations.
As we describe in \S\ref{crosscal:bursts}, the corrections are in fact inconsistent between the typical burst and persistent spectrum, notably for JEM-X, so an additional correction is applied to give the final values quoted in the burst table (see \S\ref{minbar}).

For each instrument, we extracted light curves 
for the full instrumental bandpasses with a resolution
of 1~s, as described in \S\ref{xteburstid}, \ref{saxburstid} and \ref{jemxburstid}. This resolution samples the light curves sufficiently for
accurate determination of peak flux and other parameters. The PCA data allow
in principle for much higher resolution while preserving statistically meaningful
data points, but we require the same resolution for all
three devices in order to make comparisons between instruments as fair
as possible and are, thus, limited by the capabilities of the instruments with smaller detector areas, 
WFC and JEM-X. The basis for the extraction was a list of burst onset
times resulting from the burst searches explained in \S\ref{sec:source-data}.
Typically, a time window beginning 50~s before the burst start, and extending to 300~s after the onset
time was extracted. For  longer duration bursts (particularly from GX
17+2), these time frames were enlarged on a burst-by-burst case.

In order to derive the duration and photon fluence of a burst in an instrument-independent way, it is appropriate to adopt a model for the decay phase of the burst, when the flux drowns in the noise sooner for WFC and JEM-X than for PCA data. 
Times were
redefined to be measured with respect to the burst onset time.

The following steps were followed to model the light curves and
determine the basic burst parameters:

\begin{enumerate}

\item Determine flux 
and time of the peak in the light curve.

\item Subtract the background level, calculated from the measurements between -50 and -15 s from the onset time.  If there are no data in that time frame, no background
   is subtracted and the burst is flagged.

\item Re-calculate the burst onset time, in two stages. The first stage 
searches backward in time from the burst peak, for the first bin 
that drops below a certain threshold value\footnote{For PCA we adopt a threshold of 5\% of the peak flux above the pre-burst level; for the other two instruments, we choose the pre-burst level itself as the threshold. The difference in treatment is to accommodate the difference in sensitivity between the instruments.}. 
The 
search
is continued 
for another 15~s earlier to test for
data points $\geq3.5\sigma$ above the background level, and at least 10\%
of the peak flux;
such measurements may indicate a
superexpansion burst (see
\S\ref{sec:superexpansion}). If found, the time of the earliest excess
is taken as the burst start time. 

\item The
second step involves a visual verification of the start time in the light curve of the bursts. In a few ($\approx1$\%) cases, the automatically determined start time was off by up to 10 s because the persistent flux of the source varied on similar time scales as the burst.
\item Estimate the burst rise time, as the number of 1-s bins since the first bin that is above the 
   threshold of 5\% of the peak flux.

\item Identify the first data point for the model fit of the decay phase, as the first data point following the peak where the flux drops below 75\% of the peak flux.
 %  \jc{How is "the end of the burst" determined? I suggest rewriting this part so it reads chronologically from the burst peak with time increasing: the first data-point where the (bkg-subtracted) cflux is lower than 75\% of the peak... - done - jz}. 
   The 75\% threshold was determined by trial and error.

\item Fit each of three model functions to the decay phase beyond the start
   point defined in the latter step:

\begin{itemize}

   \item an exponential function, with normalization, 
      $e$-folding decay time and background level as free parameters;

   \item a power-law decay, with normalization and
      power law decay index as free parameters;

   \item a power-law decay plus a one-sided Gaussian function,
      introducing two additional free
      parameters: the Gaussian width (standard deviation) and normalization. 
      %\ek{For an alternative way, see Kuuttila et al. 2017, A\&A 604, A77. jz: I think this 
      %would be a confusing statement here. I believe Kuutila et al. do not employ a model for 
      %function for the decay. Note that Kuutila is cite in \S \ref{sec:cooling}}

\end{itemize}

\end{enumerate}

\begin{figure}
	\begin{center}
	\includegraphics[width=\columnwidth]{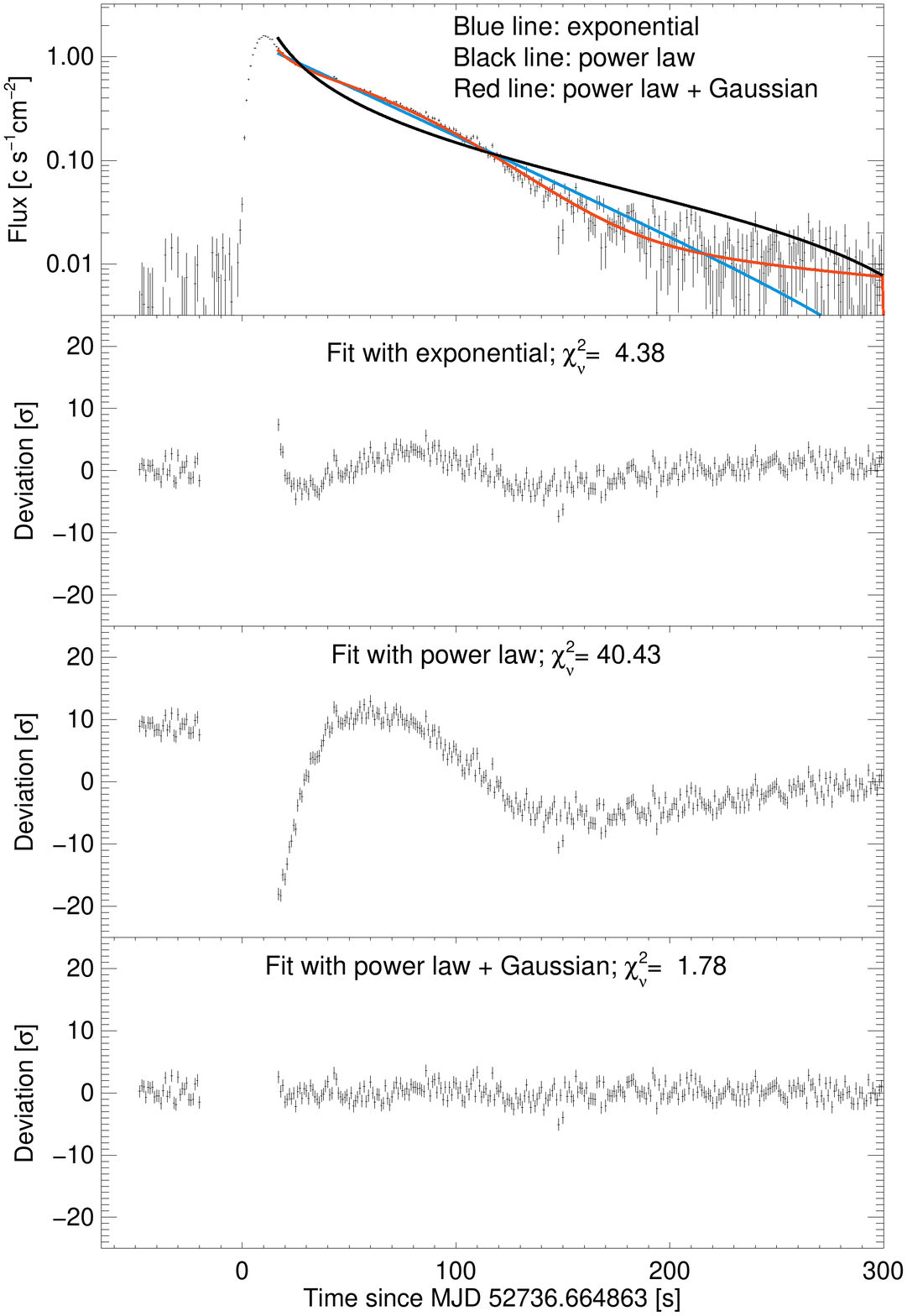}
	\end{center}
	\caption{Example of model fits to light curves,
	% . This is 
	for a burst observed on MJD~52736 with \xte/PCA from GS~1826$-$24 (burst id \#3076). Top panel: light curve data with 3 model fits.
	Second panel: deviations
	of data with respect to exponential model. Third panel: deviations with respect to power-law model.
	Fourth panel: deviations with respect to power law plus Gaussian model. \label{gs1826lcfits}}
\end{figure}

The light curve models and burst onset times were visually inspected for all bursts.
We estimate the accuracy of the start times to be at best 1~s for bursts where the ratio
of the peak flux to its error was higher than about 50 (this involves mostly PCA-detected
bursts) and at worst 5-10 s for the least significant bursts.

We find 
   that all WFC and JEM-X data can satisfactorily be modeled
   with the exponential decay or power law, while the PCA data are generally
   better modeled with either the pure power law or the power law plus Gaussian.
   In other words, only the better-quality data of the PCA show the ``hump'',
   likely reflecting the contributions of $rp$-burning \cite[see][]{zand17b}.
   This contribution is modeled by a
      one-sided Gauss function with the centroid fixed to the burst
      onset.
% added in proof, Q1
An example is shown in Fig. \ref{gs1826lcfits}.

   % (see also in 't Zand et al., 2017, A\&A 606, A130).

The following basic burst parameters for the catalog are determined from these 
light curves and
model fits; we also list the burst table parameters, as described further in \S\ref{minbar}:

\begin{enumerate}

   \item Burst start time ({\tt time} in the MINBAR table), given in MJD

   \item Peak photon flux and uncertainty ({\tt pflux, pfluxe}), measured on 1-s time scale, in count~s$^{-1}$~cm$^{-2}$.

   \item Burst duration and uncertainty ({\tt dur, dure}), in seconds. We define duration
      as the time interval when the flux is above a threshold 
      % flux which is defined as 
      value of 
      5\% of the peak flux. The interval end time
      is determined from the model fit and the start time as that of the first
      data point surpassing the threshold. The 5\% threshold is chosen as a
      compromise between accuracy (lower values are more uncertain since one
      reaches the noise level in WFC and JEM-X data) and best
      representative of the burst duration.

   \item Photon fluence and uncertainty ({\tt fluen, fluene}), in count~cm$^{-2}$. This quantity is the sum of
      the integral under the best fit model curve (until infinity) and 
      % the sum of 
      all earlier
      data points 
      % before that until the 
      beginning with the burst onset time. 
      %\ek{I may be missing something, but why include data points before the burst onset time? %jz: we don't. note tht it says untill the burst onset time}

   \item Exponential decay time and uncertainty ({\tt edt, edte}), in seconds. Although the exponential
      fit is bad for many bursts, particularly those detected with the PCA, this
      number provides easy comparison with the literature. The 1$\sigma$ error
      is multiplied with $\sqrt{\chi^2_\nu}$ when $\chi^2_\nu>1$ to account for the lack of fit.

\end{enumerate}

We cross-calibrate the measurements between the three instruments in \S\ref{crosscal:bursts}. 

\subsection{Burst time-resolved spectroscopy} 
\label{trsanl}

We describe in sections \S\ref{xtetrse} and \S\ref{saxtrse} how the \xte/PCA and \sax/WFC data were reduced to produce time-resolved spectra covering each burst.
We fit each burst spectrum  with 
an absorbed blackbody model
(model {\tt wabs*bbodyrad} in \xspec) in the range 3--25~keV for PCA and 2--30 keV for WFC,  after 
subtracting off the pre-burst (persistent) spectrum as background. Note that this is the ``traditional'' approach, as distinct from the 
variable persistent flux
method introduced by \cite{worpel13a}, 
which accounts for variations (typically an increase) in the contribution of the persistent emission during the burst.
The hydrogen column density $N_{\rm H}$
was fixed to values adopted for each source (Table \ref{tab:bursters}). The model fits yield 
black
body temperature $kT$ and emission area $K_{\rm bb}$ 
(the projected area of a sphere at 10~kpc) as a function of time.

For \xte, although we largely replicate the analysis of G08, there are a number of important differences with the earlier analysis.
We included the recommended systematic error of 0.5\% \citep{shaposhnikov12}, and adopted Churazov weighting \citep{churazov96} to resolve an issue with low-count rate bins that arose with the adoption of \xspec\ version 12 over version 11.

Indeed, %jc190918
many of the spectra from the sample of G08 had a minimum number of counts per bin $<10$ 
and for a significant sub-sample,  one or more of the bins within the energy range of interest had zero counts. Typically these spectra were in the tail of the bursts,  when the burst flux had dropped to low levels. These zero-count rate bins 
had no effect on 
the original analysis,  since \xspec\ version 11 and earlier arbitrarily attribute a statistical error (the {\tt STAT\_ERR} column adopted by {\sc XSpec}) 
of 1 to bins with zero counts. However,  version 12 only substitutes the 1-count minimum uncertainty when the combined variance of the data and background spectra is 0.0 (C. Gordon,  pers. comm). The effect of these low-count rate bins,  when re-fitting with \xspec\ version 12,  was that around 10\% of the spectra exhibited much higher $\chi^2$ values than for the original fits,  and typically had blackbody temperatures much lower than the previously fit values.

We found that Churazov weighting \cite[]{churazov96} for fits of simulated data performed the best in terms of agreement between the input and fitted spectral parameters. Furthermore, this weighting provided parameter uncertainties which encompassed the true (input) value of each spectral parameter in a fraction of spectra corresponding as closely as possible to the confidence level used (i.e. for 1$\sigma$,  68\%).

For each time interval we 
estimated the bolometric flux $F_i$ based on the best-fit spectral parameters $(T_{{\rm bb},i},K_{{\rm bb},i})$ at each timestep $i$, after G08:
\begin{eqnarray}
  F_i & = & \sigma T_{{\rm bb},i}^4 \left( \frac{R}{d} \right)_i^2 \nonumber \\
      & = & 1.076\times10^{-11}\ 
          \left(\frac{kT_{{\rm bb},i}}{1\ {\rm keV}}\right)^4 K_{{\rm bb},i}\ 
\epcs\\
\label{flux}
\end{eqnarray}
The bolometric burst fluence $E_b$ 
was calculated by integrating all flux measurements up to the last 
flux measurement,
supplemented with an estimate of the
fluence beyond that point as extrapolated by an exponential fit to
the light curve of the complete bandpass. Note that this approach is slightly different to that used for the photon fluence, as described in \S\ref{sec:lcs:modeling}, which adopts instead the best-fit model curve.

For very faint bursts, only one flux measurement could be made during the burst; these bursts are flagged to indicate the limited information available (see \S\ref{minbar}).
Even when multiple flux measurements were possible, the fluence measurements sometimes exhibited 
large uncertainties. These low precision values could arise because the time-resolved flux measurements were also low precision, or because the extrapolation of the decay beyond the burst tail was uncertain. 
The latter issue was particularly noticeable for bursts from Cyg~X-2, where for several events observed with \xte/PCA the steady flux level after the burst appeared to be significantly higher than before, making it difficult to distinguish from the burst emission. 

\subsection{Burst oscillation search}
\label{burstosc}

Burst oscillations are high frequency ($\sim$~kHz) X-ray timing phenomena, and to date have mostly been studied using the high time resolution data modes of \xte/PCA.  
Our analysis follows the procedure outlined in \citet{ootes17a}, which we summarize below.  For a more complete discussion of the methodology used, readers are referred to that paper.

We analyse each burst of the oscillation sources (for burst selection, see \S\ref{xteoscillationbursts}) individually to determine whether an oscillation can be detected. We searched for signals in the first 16~s of the burst, 
with a frequency within 5~Hz of the known oscillation frequency ($\nu_o\pm 5$ Hz) to account for any frequency drift. Although in most cases the frequency drift is only 1--3~Hz \citep{muno02b}, larger drifts have been reported in some sources \citep{wij01}. 
In case of a detection (see below), we compute the fractional root mean square (rms) amplitude of the signal.
For those bursts in which we do not detect an oscillation signal that passes the detection criterion, we compute an upper limit on the rms amplitude.

We carried out the search on each burst as follows. 
First we compute a burst start time for timing purposes ($t_0$)\footnote{The burst start time for timing purposes may be slightly different from that identified from the light curve fits; see \S\ref{sec:burst-lightcurves}. For 94\% of the bursts searched for oscillations, the calculated $t_0$ is within 1.5 seconds of the burst start time as defined in \S\ref{sec:lcs:modeling}.} and the background count rate. We estimate the background count rate using the count rate in the range 20--5~s prior to the MINBAR start time.
Note that we have chosen somewhat different pre-burst time intervals in the various analyses in this study. We anticipate that this has a negligible effect on results.

$t_0$ for the timing analysis is then defined as the time where the count rate exceeds for the first time 1.5 times the estimated background count rate.  This ensures that all the burst start times are defined by the same criterion.  The 
background count rate ($C_\mathrm{B}$) is then defined as the average count rate in the range 17 to 1 seconds before $t_0$. 
A time buffer of 1-s is maintained between the burst start time and the interval over which the background is calculated, to ensure that the background is not overestimated in bursts with a slow rise.

The first 16 seconds of the burst, measured from $t_0$ onward,
are then divided into non-overlapping time bins with 5000 counts each,
to ensure that
the error bars on each measurement are similar \citep{watts05}\footnote{Note that in a previous analysis by \citet{muno04a} equal {\it duration} time bins were used, so that error bars later in the burst were larger as the burst intensity decreased.}.   The number of time bins in our analysis thus depends on the strength of the burst and the underlying persistent intensity. We use non-overlapping time bins to ensure that each time bin is independent,
to simplify 
computing the number of trials to obtain a signal (see below).

We define 10 frequency bins (with frequencies in the range $\nu_\mathrm{o}\pm5$ Hz), to obtain a frequency resolution of 1~Hz. We thus create for each burst a two-dimensional grid of time-frequency bins in which we attempt to detect oscillation signals \citep[see Figure 2 of][for a visualisation of the grid]{ootes17a}.

For each time bin we compute the signal power for each of the 10 trial frequencies. We obtain the measured power for a signal with trial frequency $\nu$ by calculating the $Z^2$ statistic \citep[see][]{buccheri83, sm99}, defined as:

\begin{equation}
Z^2_n= \frac{2}{N_\gamma}\sum_{k=1}^n\left[\left(\sum_{j=1}^{N_\gamma} \cos{k\nu t_j}\right)^2 
+\left(\sum_{j=1}^{N_\gamma} \sin{k\nu t_j}\right)^2\right]
\label{zsquare}
\end{equation}
where $Z^2$ is the measured power of the signal, $n$ is the number of harmonics, $N_\gamma$ is the number of counts in the time bin, and $t_j$ the arrival time of the j-th count relative to some reference time. We only 
consider
the first harmonic of each signal, so $n=1$. By definition of the time bins $N_\gamma=5000$. 
Using this statistic, we obtain a power spectrum for each time bin in which the power of the oscillation signals is plotted as function of the 10 trial frequencies.  We do not attempt to compensate for any drifts in frequency that might occur during a given time bin.

For each individual time bin of a burst, we select the frequency bin with the largest measured power and determine whether or not the signal is considered a detection.  We assume a Poisson noise process, for which powers in the absence of a signal are distributed as $\chi^2$ with two degrees of freedom.  This assumption is reasonable at high frequencies, but not at low frequencies where the red noise contribution due to the burst light curve envelope becomes significant (see also the more extensive analysis\footnote{The analysis by \citet{Bilous19pub} also uses the MINBAR burst sample, but the analysis of burst oscillation amplitudes takes into account factors such as lightcurve shape and deadtime, which are not considered in \citet{ootes17a}.} in \citealt{Bilous19pub}). 
For this reason, we exclude sources from our analysis with signals below 50 Hz (at present only one source, see above).  Based on the assumption for noise distribution, we can then determine the chance that any measured power is produced by noise alone. We  then set a threshold for the measured power above which we define a signal to be significant. We set the detection criterion such that the chance that a signal was produced by noise is less than 1\% when taking into account the number of trials for each burst $(N)$. The number of trials is defined as the total number of time-frequency bins in which one looks for a signal; where $N=N_\mathrm{t}\times N_\nu$ with $N_\mathrm{t}$ the number of time bins and $N_\nu$ the number of frequency bins. 

The probability $P_{\rm noise}$ that a measured signal with noise chance $\delta$ was produced by noise for $N$ trials is given by:
\begin{equation}
P_{\rm noise}=N\delta(1-\delta)^{N-1}
\label{prob}
\end{equation}
We define three criteria \citep[similar to those used in][]{muno04a} 
to identify
a significant detection. 
We choose these criteria to ensure that, on average
each detection in a single burst 
has a 1\% chance of being a false detection. 
The specific criteria are:
\begin{enumerate}
\item The chance that a measured power $Z_m$ was produced by noise is less than $7\times10^{-5}$ in a single trial ($\delta\leq 7\times 10^{-5}$), assuming that a burst will on average consist of 16 individual time bins, such that $N=16\times10$. This corresponds for 1\% chance overall to a measured power criterion $Z_m^2\geq 19.4$.  

\item{A signal occurring in the first second of a burst has a single trial chance probability $\delta\leq10^{-3}$. This probability results in a measured power limit $Z_m^2\geq13.8$. At the burst onset, the difference in brightness between burning and non-burning material is largest, and therefore oscillation signals would be expected to be largest in the burst rise (first second). }

\item{A signal distributed over two adjacent time-frequency bins has a combined single trial noise chance probability $\delta_1 \times \delta_2\leq1.3\times10^{-6}$. We 
test for this possibility 
using the fact that this probability is similar to that associated with a measured power limit of the averaged signal in these two adjacent bins of $\bar{Z}_m^2\geq13.8$.
}
\end{enumerate}

Our motivation for the final criterion is as follows.
There is a significant chance that a signal does not peak exactly in one time-frequency bin, but is spread over multiple bins instead. Therefore, we select from each time interval the frequency bin with the largest measured power and compute the noise chance of the signal that is spread over the selected time-frequency bin and one of three directly adjacent time-frequency bins: the same time bin and one of two the adjacent frequency bins, or the same frequency bin and the next time bin.   The chance that both bins consist of noise alone is given by the product of the noise chance probabilities of the two individual bins ($P_{{\rm noise},1,2}=P_{\rm noise,1}(N_1,\delta_1)*P_{\rm noise,2}(N_2,\delta_2)$). To meet the detection criterion of the burst, the single trial probabilities of the two bins ($\delta_1$ and $\delta_2$) must satisfy the equation for $P_{\rm noise,1,2}$. Using an approximation for $P_{\rm noise,1,2}$ given by 
\begin{equation}
P_{\rm noise,1,2}\approx 3N_\mathrm{t}^2N_\nu\delta_1\delta_2\label{doubleprob}
\end{equation}
(taking into account that $N_2$ is reduced due to the fact that the second bin has to be selected from one of the three bins surrounding the first bin) yields the solution $\delta_1\delta_2=1.3\times 10^{-6}$ that adjacent bins must satisfy to meet the threshold burst probability $P_{\rm noise,1,2}=10^{-2}$.
We note that if one considers each of the three detection criteria as individual trials, the noise probability would increase to a 3\% chance that a detected oscillation is actually a false detection.

\begin{figure}
	\begin{center}
	\includegraphics[width=\columnwidth]{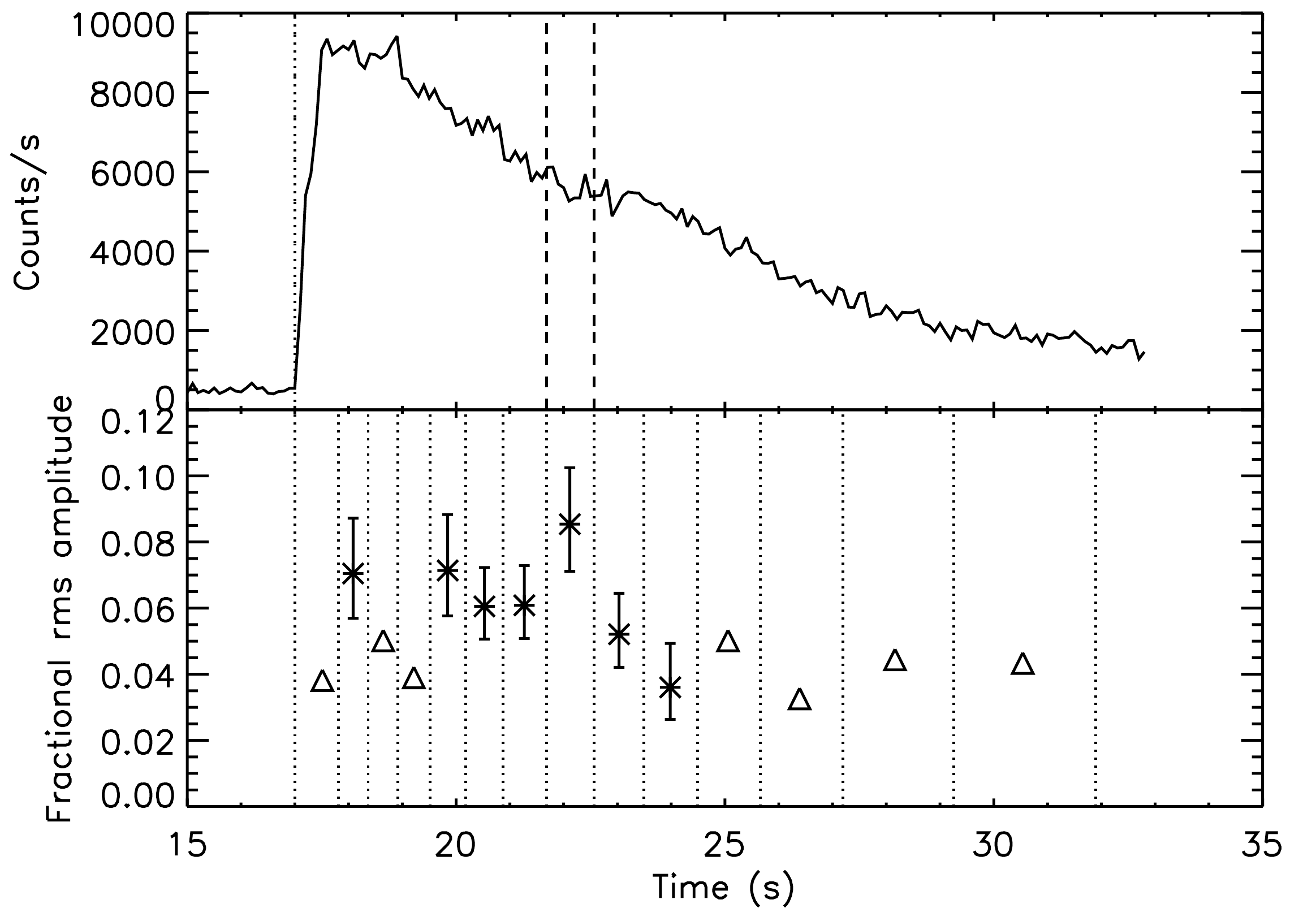}
	\end{center}
	\caption{Result of the analysis of a burst from 4U 1728-34 with observation ID 95337-01-02-00, and $t_0=$MJD~55474.1755 (\#3966). The upper panel shows the burst lightcurve, and the lower panel shows the limits of the time bins (dotted lines) and in each time bin the computed amplitude (asterisks with vertical error bars) or amplitude upper limit (triangles) in the case of a non-significant signal. In the upper panel the dotted line indicates the burst start time ($t_0$) and the dashed lines represent the time bin in which the oscillation signal with the largest signal power $Z_s^2$ was found. Figure from \citet{ootes17a}}\label{141}
\end{figure}

There are five possibilities to pass the detection criterion: one from the first criterion, one from the second and three from the third. For each time bin we select from the five options the signal with the largest (averaged) measured power that passed the detection criterion. The measured power consists of two components, the signal power and the noise power. The signal power is derived using the probability distribution $p_n$ of measured signals $Z_m$ for given signal power $Z_s$:

\begin{multline}
p_n(Z_m | Z_s)=\frac{1}{2}\exp{\left[-\frac{(Z_m+Z_s)}{2}\right]}\times\\ \left(\frac{Z_m}{Z_s}\right)^{(n-1)/2}
I_{n-1}\left(\sqrt{Z_mZ_s}\right)
\end{multline}
where $n$ is the number of harmonics (we always use $n=1$), and $I$ is a 
% modified in response to Anna's feedback on the proof
% first kind 
modified Bessel function
of the first kind. The computational procedure provides a signal power and 1$\sigma$ errors.  From the signal power of this oscillation, we compute the fractional rms amplitude of the oscillation ($A_\mathrm{rms}$) as follows:

\begin{equation}
A_\mathrm{rms}=\sqrt{\frac{Z_s^2}{N_\gamma}}\left(\frac{N_\gamma}{N_\gamma - B}\right)
\label{amp}
\end{equation}

The term in brackets in Equation \ref{amp} is the factor that corrects for the background emission, where $N_\gamma$ is the number of counts, and $B$ is the estimated number of background counts in the investigated time bin ($N_\gamma=5000$ and $B=C_B \Delta t$, with $\Delta t$ the time width of the bin(s) over which the signal is considered). We calculate the 1$\sigma$ error on the amplitude using linear error propagation of the independent parameters, for which the standard deviations of $N_\gamma$ and $B$ are calculated as the square root of the considered parameter.

If none of the detection criteria are passed, a $3\sigma$ upper limit on the oscillation amplitude is calculated.  
From the oscillation signals detected in a burst, we select the amplitude  of signal with the largest signal power to compare with the results from other bursts (see Figure \ref{141}). We thus select one specific time-frequency bin for each individual burst. If no oscillation signals are found throughout an entire burst, we select the upper limit found for the signal with the largest non-significant signal power.

If we detect burst oscillations during the burst, we also determine the burst phase in which the signal was found. First we determine the maximum intensity of the burst and define the boundary limit as 90\% of the peak luminosity, similar to 
G08. 
The peak of the burst is defined as the phase that exceeds this boundary limit. The time from the start of the burst until the start of the peak is defined as the rising phase, and similarly the time span from the end of the peak up to 16 seconds after the burst start time is defined as the burst tail. If the time bin of strongest oscillation signal falls on both sides of one of the boundaries, we select as burst phase of the signal the one in which the largest fraction of the time bin falls.

The results of the burst oscillation search are provided as a table with format described in \S\ref{osc_table}.

\subsection{Persistent spectral fitting} 
\label{obsanl}

We measured the persistent source flux $F_p$ in the energy range 3--25~keV for each 
observation in which a source was significantly detected
and for which a spectrum was available.
We set the detection criteria as when the source count rate averaged over the observation was greater than or equal to three times the uncertainty (roughly equivalent to a $\geq3\sigma$ detection).
Spectra were not available for every observation, and we indicate this with the {\tt flag} column in the catalog table (\S\ref{minbar-obs}).

We estimated the flux from the observation-averaged spectra generated as described for each instrument in \S\ref{xtepersspec} , \S\ref{saxpersspec} and \S\ref{jemxpersspec}.
We fit the net spectra using \xspec\ version 12 \cite[]{xspec12} over the range 
3--25 keV for the PCA and JEM-X, and 2--30 keV for the WFCs % set in obsflux.tcl
with an empirical model,
including the effects of neutral absorption by material (with solar abundances assumed) along the line-of-sight. 

The field of view of the \xte/PCA can cover multiple active sources, and (in contrast to the WFC or JEM-X) the lack of imaging makes it impossible to separate their fluxes. As described in \S\ref{xteobs}, for such observations we used the intensities of each source in the FOV as measured independently with the \xte/ASM (where available), to assess the level of source activity. We identified three possible cases for such observations, which are labelled with {\tt flag} values as 
% modified ref to preserve table ordering, as per proofs
% specified in Table \ref{analysis-flag-table}: 
described in \S\ref{osc_table_format}: 
``a'' for those observations where only the named source ({\tt name} attribute in the table)
% see \S\ref{minbar-obs}) 
is active; and ``b'' where more than one source was determined to be active, so that the measured spectrum contains contributions from multiple objects. A third category, labelled ``c'' indicates those observations where no ASM data covering the PCA measurement was available, and so no independent information about the source activity could be determined.
We thus excluded observations flagged ``b'' or ``c'' from our estimates of the accretion rate (as a fraction of the Eddington value) $\gamma$, as described in \S\ref{bolcor}.

Since the instruments for our source data offer fairly poor low-energy sensitivity, it is difficult to constrain the neutral absorption column density $N_{\rm H}$.
Thus, 
we fixed $N_{\rm H}$ to the value listed in Table \ref{tab:bursters} for each source in all the fits, for consistency. These values were taken
from the 
literature, and were determined
from high-sensitivity observations with other instruments encompassing sub-$2$~keV photon
energies.

For each observation, we calculated a preliminary fit with the simplest model, an absorbed power-law. 
For observations in which the detection significance was $<8$, we fixed the power law spectral index 
$\Gamma$ at 2, and otherwise left it free to vary.
If the reduced-$\chi^2$ ($\chi^2_\nu$) was in excess of 2
we added components successively until a good fit was achieved. For the majority of observations with the WFC (85.3\%) and JEM-X (97.0\%), the preliminary (power-law) model offered a sufficiently good fit. The remaining observations for those instruments were fit either with a {\tt compTT} model, or a blackbody and powerlaw, respectively. The {\tt compTT} model describes Comptonization in a homogeneous environment \cite[]{tit94}.

For \xte, the higher signal-to-noise meant that for many observations these simple phenomenological continuum %jc1901918
models were inadequate. 
Additionally, for many of the \xte\/ spectra,  residuals were present around 6.4~keV. 
Such residuals are common features of persistent spectra from LMXBs \cite[e.g.][]{cackett13}, and are usually interpreted as fluorescent Fe K$\alpha$ emission arising from the source or nearby environment, such as from reflection off the accretion disk
\cite[e.g.][]{fr10}
or the diffuse emission from the Galactic ridge  \cite[e.g.][]{val98}.
Where these residuals resulted in a reduced $\chi^2$ above the threshold, we added a Gaussian component to improve the fit. 
We allowed the line energy to range between 6.39--7.1~keV, and the line width up to 1.5~keV.
We describe the statistics of the various models over the sample in \S\ref{minbar-obs}.

We then integrated the model over the energy range 3--25~keV
to estimate the persistent flux for each observation. 
We note that the flux measurements were only weakly sensitive to the particular choice of models, although the model produced significant offsets in the spectral colours (see \S\ref{colours}).

We estimated the uncertainty on the flux using the {\tt error} flag of the {\tt flux} command in {\sc XSpec}. This routine draws values from the estimated probability density functions of the model parameters, and calculates the corresponding confidence range on the distribution of fluxes. Where the flux measurement was different from zero at less than the $3\sigma$ level, we re-fit the spectrum with the power-law model, fixing the spectral index $\Gamma$ at 2. 
See 
% modified ref to preserve table ordering; this one missed in the proofs
% Table \ref{tab:spec_summ} in 
\S\ref{minbar-obs} for the numbers of spectra affected.
In the cases where the resulting flux was still not significantly different from zero, we report the $3\sigma$ upper limit. % (see \S\ref{minbar-obs}).

The spectral fitting procedure necessitated some variations on the approach depending upon the specific instrument used.
For \xte, most observations were performed with more than one PCU functional, so (following G08), we analysed each PCU separately and then averaged the spectral parameters and flux. 
This approach simplifies the generation of response matrices, which otherwise would need to be summed from the matrices for the individual PCUs, with weights corresponding to the relative exposure times (which were not always identical). Later in the mission, the degradation of certain PCUs means that for best results we need to omit spectral fits from those PCUs. 
For PCU\#1,  we found that the background estimation was a poor match to the on-source spectrum, particularly from 2010 October 
through to the mission end. 
For some observations with PCU\#1 active in this time interval, we neglected the spectra from PCU\#1 in the fits. 

The WFCs were both operated essentially without changing the instrument settings. The gain changed gradually over the mission life time, and that was accounted for by calculating the response custom-wise for each observation.

As described in \S\ref{jemxobs}, since revolution 976 (October 10, 2010) both JEM-X cameras were on during all observations.
For those observations we carried out spectral fits independently, but then averaged the fluxes and spectral parameters between the two fits.

\subsection{Spectral colours}
\label{colours}

Here we describe the approach used to determine the spectral colours and hence the position on the colour-colour diagram, $S_\text{z}$.
Spectral colours are normally calculated as the ratio of count rates in different energy bands (e.g. G08), and so will vary between different instruments even for the same spectrum. As we wish to combine information from multiple instruments, we choose to calculate the colours instead as the ratios of fluxes within different bands, based on the fitted persistent spectral model.

We chose four energy bands to calculate ``soft'' and ``hard'' spectral colors, following G08: 
2.2--3.6~keV, 3.6--5.0~keV, 5.0--8.6~keV and 8.6--18.6~keV. %
These bands span the common 3--25~keV energy range of the spectral responses of the three instruments, although the JEM-X response is negligible below 3~keV.
We then calculated a soft and hard colour as the ratio of integrated fluxes (based on the spectral model) in each pair of low and high fluxes. 

For systems in which sufficient observations have been accumulated with a significant variation in the spectral colours, the colour measurements typically define an ``atoll'' or Z-shaped track in colour-colour space \cite[]{hvdk89}, with correlated behaviour reflected in the X-ray variability. Measuring the position $S_\text{z}$ 
along this track has been suggested
\cite[e.g.][]{vdk90,hertz92,kuul94}
as an alternative way to constrain the accretion rate, independently of the persistent flux
\cite[an equivalent quantity, $s_a$, has been defined for ``atoll'' sources; ][]{vs00}.

For the MINBAR sample of observations we take the same approach, also following G08; but our definition of the colours as based on the integrated model flux introduces biases between observations fitted with different models. We describe in \S\ref{sec:colour} the approach taken to correct these biases.

\subsection{Instrumental area correction}
\label{sec:area}

A critical requirement for combining  data gathered by multiple instruments is to quantify any variation in the absolute calibrations. Typically, the absolute calibration of most X-ray missions is guaranteed at only the level of a few tens of percent 
\cite[]{tsujimoto11}. %\jz{please check the year here -- you had 2015}
Making the situation worse is the fact that calibration sources are scarce, and even sources which have been used for many decades as ``standards'' like the Crab, have also been shown to vary in intensity by almost 10\%
\cite[]{crabvar11}.

The missions from which the MINBAR sample is assembled were active over periods which coincide, offering the ideal opportunity for verifying the consistency of the measurements between different instruments.
The entire \sax\ mission occurred during the longer active period for \xte\ 
(Table \ref{tab:properties}); and \igr\ and  \xte\ were both active between the launch of \igr\ 
and the end of mission for \xte.

We determined the relative effective area of \xte/PCA and \igr/JEM-X to \sax/WFC based on comparing the mean count rates from pairs of overlapping observations.
We assessed the relative effective area of the three instruments in pairs, as there was no common interval during which \sax\/ and \igr\/ were operational.

We first searched for overlapping observations for each pair of instruments, and set a minimum overlap fraction of $f=0.1$, where $f=\Delta t_\mathrm{overlap}/(t_\mathrm{max}-t_\mathrm{min})$, $\Delta t_\mathrm{overlap}$ is the time in common to both instruments, and $t_\mathrm{min}$, $t_\mathrm{max}$ represents the maximal extent of the two observations.

We adopt a linear relation, so that the mean count rates for contemporaneous measurements by both instruments are related by 
\begin{equation}
c_i = z_{{\rm off,c,}i} + k_{{\rm c,}i} \times c_\mathrm{PCA} 
\label{eq:sax-xte-count}
\end{equation}
where $c_\mathrm{i}$ is the mean count rate observed with the 
alternative instrument (WFC or JEM-X) and $c_\mathrm{PCA}$ for the PCA, in units of $\cspcu$.
We used the Bayesian method of \cite{kelly07} to derive the linear correlation coefficients; this method takes into account the errors in both $x$ and $y$ (in this case, the count rates or fluxes measured by each instrument), as well as any possible correlation of the errors.
We identified outliers (which we attribute to flux variability over timescales shorter than the observation duration) deviating by more than $3\sigma$ confidence from the trend, and excluded these from a subsequent fit. 

We list the results for each pair of instruments in Table \ref{tab:crosscal}, and describe the results in detail below.

\begin{deluxetable*}{lccccc}
\tablewidth{0pt}  % set to "natural" width
\tablecaption{Cross-calibration results for parameters derived from 
coincident bursts and observations
  \label{tab:crosscal} \label{tab:crosscal-bursts}}
\tablehead{
 &  & \multicolumn{2}{c}{WFC--PCA} & \multicolumn{2}{c}{JEM-X--PCA} \\
\colhead{Parameter} & \colhead{Units} & $k$ & $z_{\rm off}$\tablenotemark{a} & $k$ & $z_{\rm off}$\tablenotemark{a} }
\startdata
{\tt count}\tablenotemark{b} & various & 
$(7.14\pm0.10)\times10^{-4}$ & $<0.022$ & $(4.58\pm0.03)\times10^{-2}$ & $<0.67$ \\
\hline
\multicolumn{6}{c}{{\tt minbar}} \\
{\tt pflux} & count~cm$^{-2}$~s$^{-1}$ & 
$1.00\pm0.12$ & $<1.15$ & $1.6\pm0.2$\tablenotemark{c} & $<1.21$ \\
{\tt fluen} & count~cm$^{-2}$ & 
$0.82\pm0.09$ & $<9.3$ & $0.76\pm0.17$ & $<19$ \\
{\tt bpflux} & erg~cm$^{-2}$~s$^{-1}$ & 
$0.97\pm0.18$ & $<30$ \\
{\tt bfluen} & erg~cm$^{-2}$ & 
$0.69\pm0.12$ & $<0.25$ \\
\hline
\multicolumn{6}{c}{{\tt minbar-obs}} \\
{\tt flux} & $10^{-9}\ \epcs$ & $1.058\pm0.020$ & $<0.18$ & $0.866\pm0.010$ & $0.08\pm0.05$
 \\
{\tt sc} & & $0.91\pm0.06$ & $0.04\pm0.03$ & $0.18\pm0.10$ & $0.44\pm0.06$ \\
{\tt hc} & & $1.29\pm0.09$ & $-0.11\pm0.05$ & $0.72\pm0.07$ & $0.22\pm0.04$ \\
\enddata
\tablenotetext{a}{Note that the upper limits  are at 
3$\sigma$ % as of 2020 Feb
confidence.}
\tablenotetext{b}{We adopted the inverse of the correlation coefficient for the count rate averaged over each coincident observation, as the relative effective area for the \xte/PCA and \igr/JEM-X}
\tablenotetext{c}{Note that the JEM-X values for the MINBAR table attributes {\tt pflux}, {\tt pfluxe} (see \S\ref{minbar}) were subsequently rescaled by the inverse of the coefficient $k$ to bring them in line with those measured with the PCA}
\end{deluxetable*}

\begin{figure}[ht]
\includegraphics[width=\columnwidth]{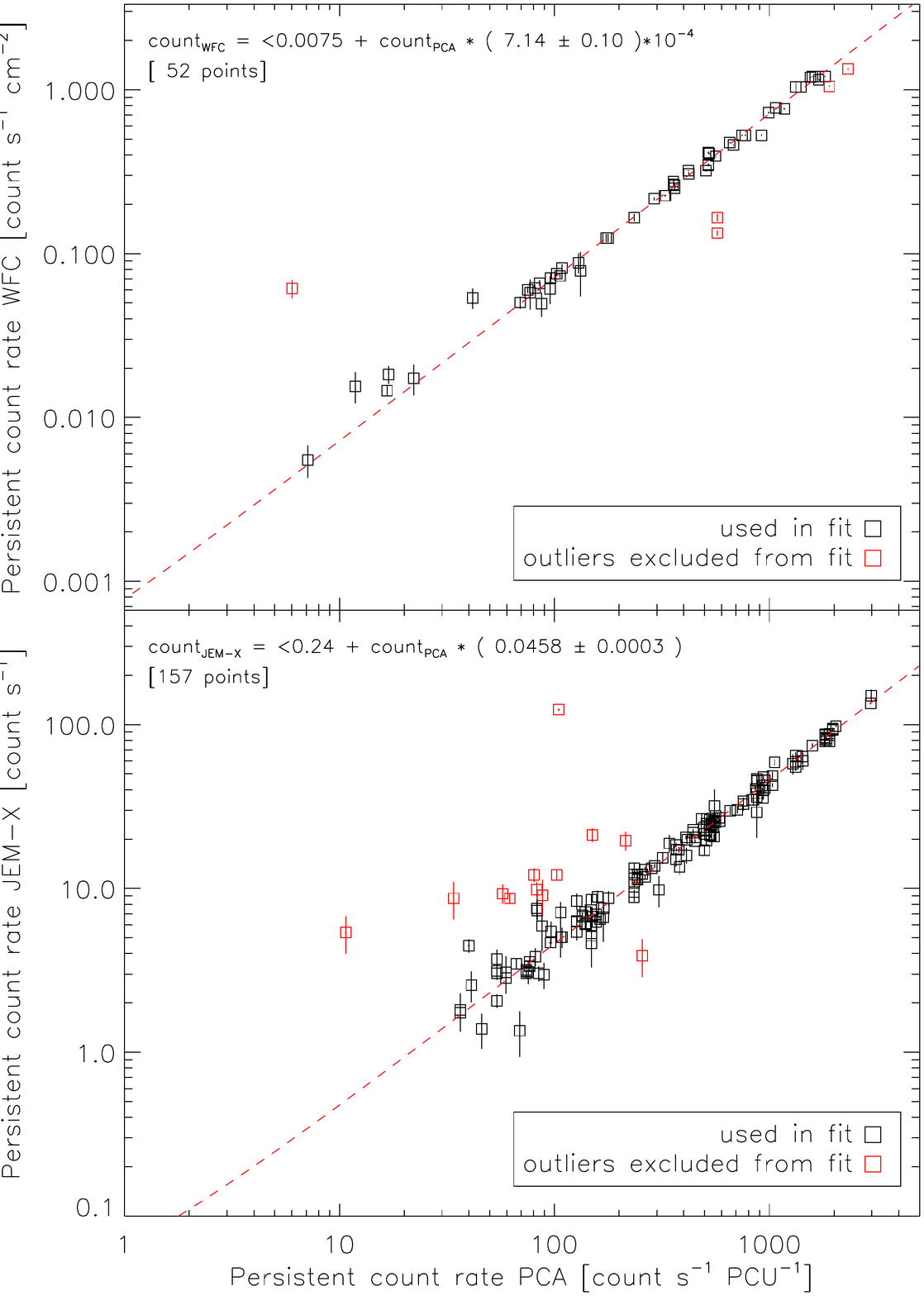}
\caption{
Cross-calibration for the mean (uncorrected) count rates for overlapping observations with the three instruments comprising the MINBAR
observation table.
The top panel shows the comparison between \sax/WFC and \xte/PCA, while the bottom panel shows \igr/JEM-X against \xte/PCA.
The points which are excluded from the fit are marked ({\it red symbols}); the line of best-fit is overplotted ({\it dashed red line}).
Note that the WFC measurements are already normalised to the area of the instrument, while the PCA and JEM-X are not; these fits are used to establish the effective area of the instruments, as described in \S\ref{sec:area}.
\label{fig:sax-xte-count} \label{fig:igr-xte-count}}
\end{figure}

\subsubsection{WFC-PCA}

The relative count rates between overlapping  WFC and PCA observations are shown in Figure \ref{fig:sax-xte-count}, top panel. 
We find an offset 
consistent with zero, and an (inverse) proportionality factor 
$1/k_{\rm c,WFC} = 1400\pm20$ (Table \ref{tab:crosscal}).
This value is roughly consistent with the approximate geometric area per PCU \cite[]{xtecal06}, as one would expect.
Moderate deviations from the measured geometric area of the detector might be attributed to the combined effects of the different effective area curves, convolved with the typical persistent spectral shape.
For the subsequent analysis, we adopt the value of $A_{\rm eff,PCU}=1400~{\rm cm}^2$ as the effective area of each PCU %
to convert $\mathrm{count\,s^{-1}\,PCU^{-1}}$ to $\mathrm{count\,cm^{-2}\,s^{-1}}$ for the observations and burst lightcurve analyses (see \S\ref{sec:burst-lightcurves}).

\subsubsection{JEM-X-PCA}
\label{ss:jemxpca_obs}

As for the \sax/WFC, we estimated the relative effective area of the \igr/JEM-X relative to the \xte/PCA by fitting a linear relation between the count rates measured for overlapping observations:
\begin{equation}
c_\mathrm{JEM-X} = z_{\rm off,c,JEM-X} + k_{\rm c,JEM-X} \times c_\mathrm{PCA} 
\label{eq:igr-xte-count}
\end{equation}
The cross-calibration for the mean count rate between overlapping  \igr/JEM-X and \xte/PCA observations is shown in Figure \ref{fig:igr-xte-count}, lower panel. The inverse coefficient for the count rate is 
$k_{\rm c,JEM-X}=21.84\pm0.15$.
When combined with the corresponding value for the WFC--PCA cross calibration, the measurements correspond to an approximate photon-collecting detector area of 64~cm$^2$, roughly consistent with the expected value for each of the JEM-X cameras (see also \citealt{brandt03}). 
As for the WFC--PCA comparison, we do not expect that this value would correspond exactly to the effective area, given the combined effects of different effective areas at different energies, convolved with the typical persistent spectra.

There is significantly more scatter in the measurements compared to the plot for the WFC. One contribution to this scatter is the short duration of the JEM-X pointings; each science window is just 30-60~min, compared to much longer WFC observations of typically 1~d (see \S\ref{saxobs}).
We note that this scatter is asymmetric, with the JEM-X outliers typically measuring a systematically higher count rate than would be expected given the average trend. This asymmetry is not obviously the result of large off-axis angles, since the distribution of angles is similar for both the outliers and the observations included in the fit.
We adopt the value of $A_{\rm eff,JEM-X}=64~{\rm cm}^2$ %jc190918
to convert JEM-X $\mathrm{count\,s^{-1}}$ to $\mathrm{count\,s^{-1}\,cm^{-2}}$ for the observations and the burst lightcurve analysis (see \S\ref{sec:burst-lightcurves}).

\subsection{Burst cross-calibration}
\label{crosscal:bursts}

As a result of the overlapping observations between the three instruments described in the previous section,
\burstdupes\ bursts were observed simultaneously by more than one instrument. These events are flagged in the table with the attribute {\tt mult=2} (see \S\ref{minbar}); the burst times and ID numbers are listed in Table \ref{tab:multbursts}.
Here we compare the analysis results for each pair of events, with the objective of determining any systematic bias for measurements from each instrument.

\begin{deluxetable*}{lcccc}
\tablewidth{0pt}  % set to "natural" width
\tablecaption{MINBAR bursts observed by multiple instruments
  \label{tab:multbursts}}
\tablehead{  & \colhead{Time} & \colhead{\sax/WFC} & \colhead{\xte/PCA} & \colhead{\igr/JEM-X} \\
\colhead{Source} & \colhead{(MJD)} & \colhead{\#ID} & \colhead{\#ID} & \colhead{\#ID}}
\startdata
 1A 1742$-$294           & 50527.71336 & 873 & 2267 & \nodata\\
 SLX 1744$-$300       &    50532.72233 & 1470 & 2270 & \nodata\\
 4U 1608$-$522        &    50899.58702 & 1600 & 2380 & \nodata\\
 KS 1731$-$260        &    51088.31732 & 586 & 2430 & \nodata\\
 KS 1731$-$260        &    51092.14551 & 589 & 2433 & \nodata\\
 4U 1705$-$44         &    51223.53390 & 1538 & 2487 & \nodata\\
 4U 1702$-$429        &    51230.99203 & 1404 & 2489 & \nodata\\
 4U 1702$-$429        &    51231.20562 & 1405 & 2490 & \nodata\\         
 4U 1636$-$536        &    51765.37284 & 1756 & 2659  & \nodata\\
 GS 1826$-$24         &    51813.66629 & 1248 & 2680  & \nodata\\
 GS 1826$-$24         &    51814.00691 & 1249 & 2679  & \nodata\\
 KS 1731$-$260        &    51816.26129 & 830 & 2683  & \nodata\\
 EXO 1745$-$248       &    51819.63818 & 2200 & 2689 & \nodata\\
 SAX J1747.0$-$2853   &    52179.46856 & 2140 & 2778 & \nodata\\
 SAX J1747.0$-$2853   &    52181.97975 & 2141 & 2807 & \nodata\\
 4U 1728$-$34         &    52896.15827 & \nodata & 8075 & 4461 \\     
 4U 1323$-$62         &    53370.46377 & \nodata & 3246 & 5467 \\
 4U 1636$-$536          &  53611.83666 & \nodata & 3315  & 5757 \\
 4U 1728$-$34           &  53986.15180 & \nodata & 3412 & 6168 \\
 4U 1728$-$34           &  54004.54993 & \nodata & 3430 & 6221 \\
 GS 1826$-$24           &  54167.28453 & \nodata & 3480 & 6302 \\
 IGR J17473$-$2721      &  54564.56700 & \nodata & 3684 & 6747 \\
 4U 1705$-$44           &  55074.12414 & \nodata & 3850 & 7545\\
 IGR J17511$-$3057      &  55091.27326 & \nodata & 3860 & 7653\\
 IGR J17511$-$3057      &  55091.61717 & \nodata & 3861 & 7656 \\
 SLX 1744$-$300         &  55792.52507 & \nodata & 8240 & 8506\\
 1A 1742$-$294          &  55792.58406 & \nodata & 8241 & 8507\\
 1A 1742$-$294          &  55793.44454 & \nodata & 8243 & 8512\\
\enddata
\end{deluxetable*}

We first compare the parameters derived from the 
burst lightcurves, as described in \S\ref{sec:lcs:modeling}. 
We identified 15 events observed by both \sax/WFC and \xte/PCA (Table \ref{tab:multbursts}), from 10 sources. Two events were observed by both instruments for 4U~1702$-$429, GS~1826$-$24, and SAX~J1747.0$-$2853, while three events from KS~1731$-$26 were observed by both missions.
We found 13 bursts detected by both \xte/PCA and \igr/JEM-X, from 9 sources. Two bursts each were observed for 1A~1742$-$294 and IGR~J17511$-$3057, with three bursts observed from 4U~1728$-$34 with both instruments.

\begin{figure}[ht]
\includegraphics[width=\columnwidth]{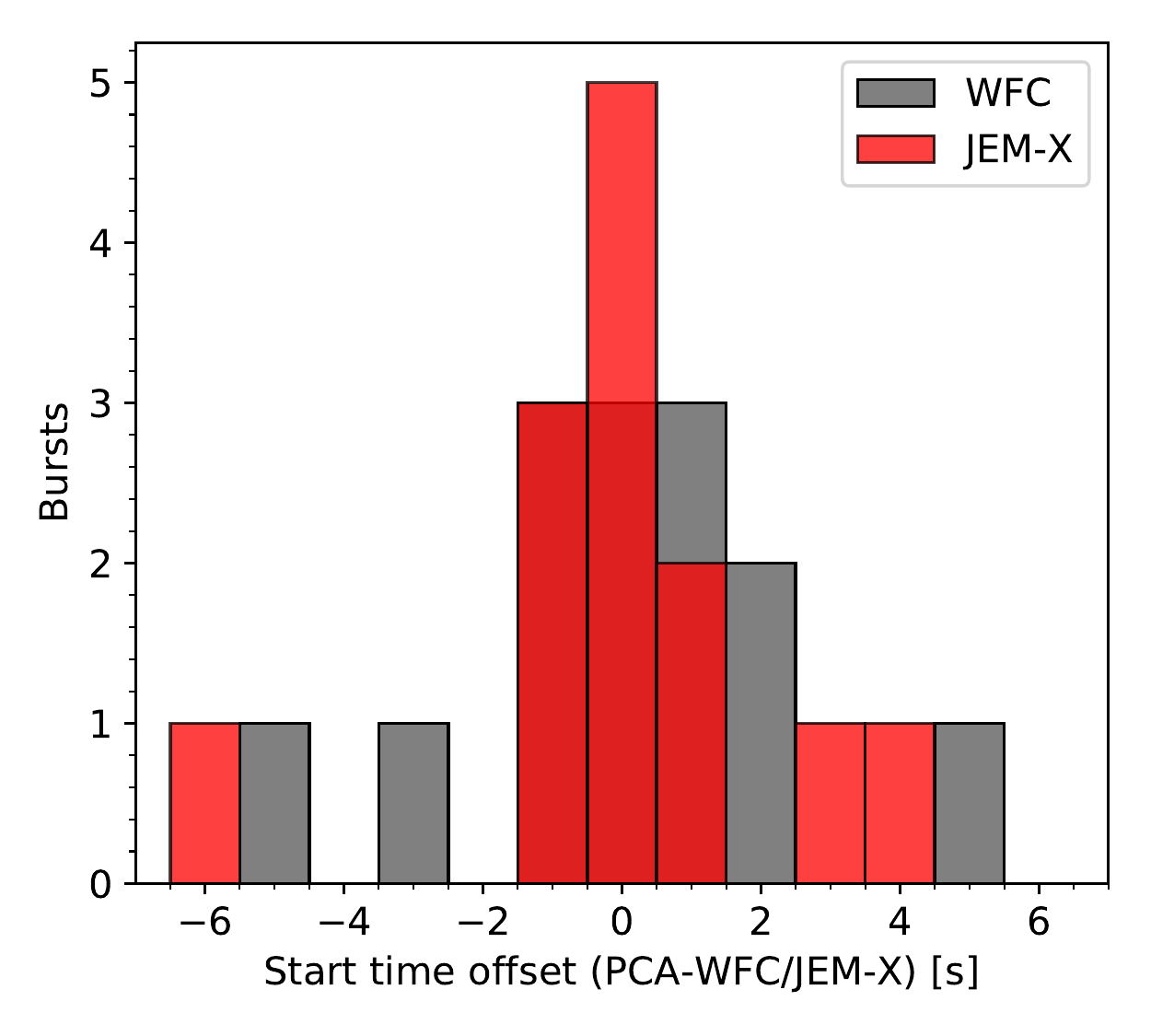}
\caption{Difference in the derived burst start time  measured for events observed with more than one instrument.
\label{fig:time_offset}}
\end{figure}

The discrepancy between the burst start times was typically limited to a few seconds between each pair of events (Fig. \ref{fig:time_offset}). For the combined set of observations the average offset was just 0.04~s, and the absolute time offset was 
$<1.47$~s ($<5.9$~s) at 68\% (95\%) confidence.

\subsubsection{WFC-PCA}%{\xte-\sax} 
\label{burstpcountxcal}

\begin{figure}[ht]
\includegraphics[width=\columnwidth]{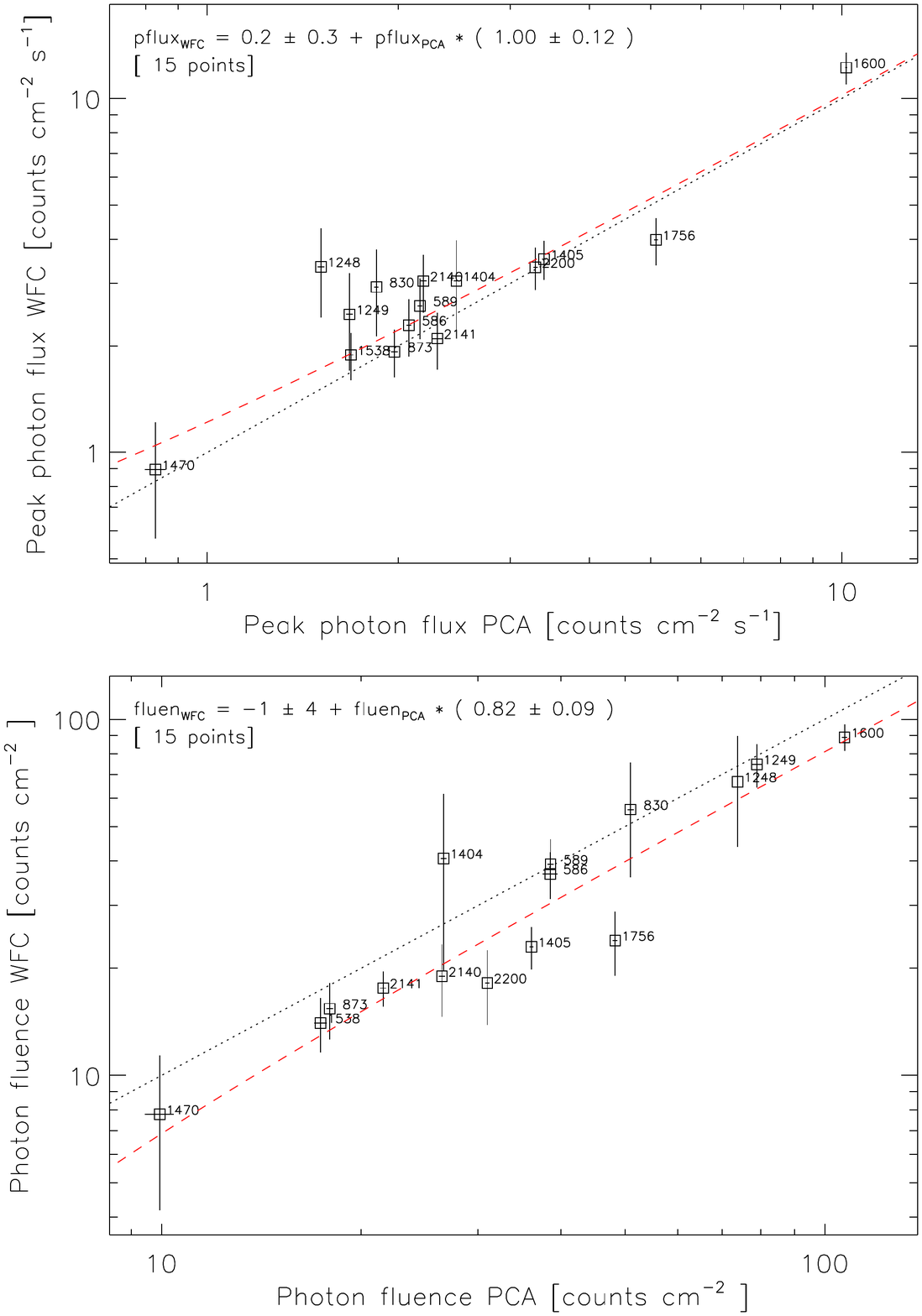}
\caption{Cross-calibration for the 
peak photon flux ({\it top panel}) and fluence ({\it bottom panel}) for bursts observed with both \sax/WFC and \xte/PCA in the MINBAR burst sample. The bursts are labelled with the WFC burst number. The dotted line shows a 1:1 correspondence; the dashed line shows the best fit linear correlation. The best fit parameters are listed in Table \ref{tab:crosscal-bursts}.
\label{fig:sax-xte-pflux} \label{fig:sax-xte-fluen}}
\end{figure}

We first compared the peak photon flux for each set of burst pairs, as shown in Fig. \ref{fig:sax-xte-pflux}, top panel.
Measurements by the two instruments generally correspond well, over a range of more than an order of magnitude. 
We calculated a linear correlation between the two sets of measurements, as for the observation-averaged counts in \S\ref{sec:area}.
We obtained the best-fit parameters as listed in Table \ref{tab:crosscal-bursts}.
The best-fit linear correlation indicates no significant bias between the two sets of measurements. % \notitie{useful to put a measure of the RMS difference here? -- dkg}

We next compared the photon fluences, calculated by integrating over the rescaled lightcurves, as described in \S\ref{sec:lcs:modeling}. While the two quantities exhibited reasonable correspondence, the fluence measured by \sax/WFC underestimated that from \xte/PCA by almost 20\% on average (Fig. \ref{fig:sax-xte-fluen}, botttom panel; Table \ref{tab:crosscal-bursts}).
We attribute this offset to the effect of the lower sensitivity of the WFC, which means that the burst emission cannot be measured as far into the burst decay.

\begin{figure}[ht]
\includegraphics[width=\columnwidth]{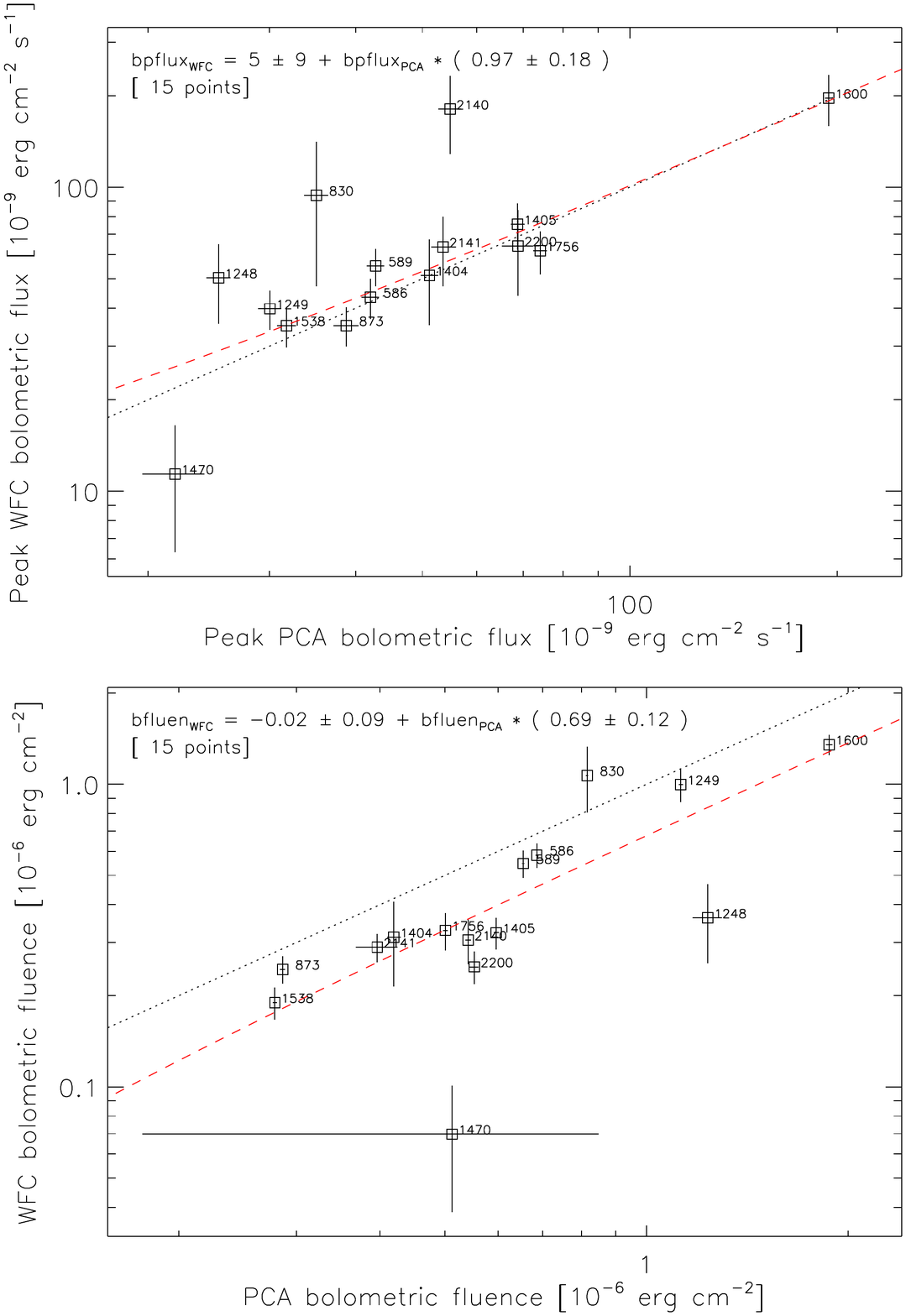}
\caption{Cross-calibration for the peak bolometric flux ({\it top panel}) and the bolometric fluence ({\it bottom panel}) for bursts observed with both \sax/WFC and \xte/PCA in the MINBAR burst sample. 
Other details as for Fig. \ref{fig:sax-xte-pflux}.
\label{fig:sax-xte-bpflux} \label{fig:sax-xte-bfluen}}
\end{figure}

We carried out time-resolved spectroscopy for bursts from \sax/WFC and \xte/PCA, as described in \S\ref{saxtrse} and \S\ref{xtetrse}. Consequently, we also compared the corresponding peak bolometric flux and fluence. As with the peak photon fluxes, we found a relatively good correspondence between the peak bolometric fluxes, although with more scatter about the line of best fit (Fig. \ref{fig:sax-xte-bpflux}, top panel).
The correlation coefficients are listed in Table \ref{tab:crosscal-bursts}.
We adopted the slope of the line of best fit between the peak bolometric fluxes measured by the two instruments, as a correction factor (equivalent to only a few \% difference) to apply to the peak fluxes of bursts detected by WFC, when calculating the mean peak flux of the PRE bursts (see \S\ref{fluxEdd}). 
We also compared the estimated bolometric fluence measured by \sax/WFC, which underestimates that measured by \xte/PCA on average by 30\%, and substantially more for weak bursts
(Fig. \ref{fig:sax-xte-bfluen}, bottom panel; Table \ref{tab:crosscal-bursts}). %jc190918

\subsubsection{JEM-X-PCA}%{\xte-\igr} 
\label{ss:jemxpca_burst}

\begin{figure}[ht]
\includegraphics[width=\columnwidth]{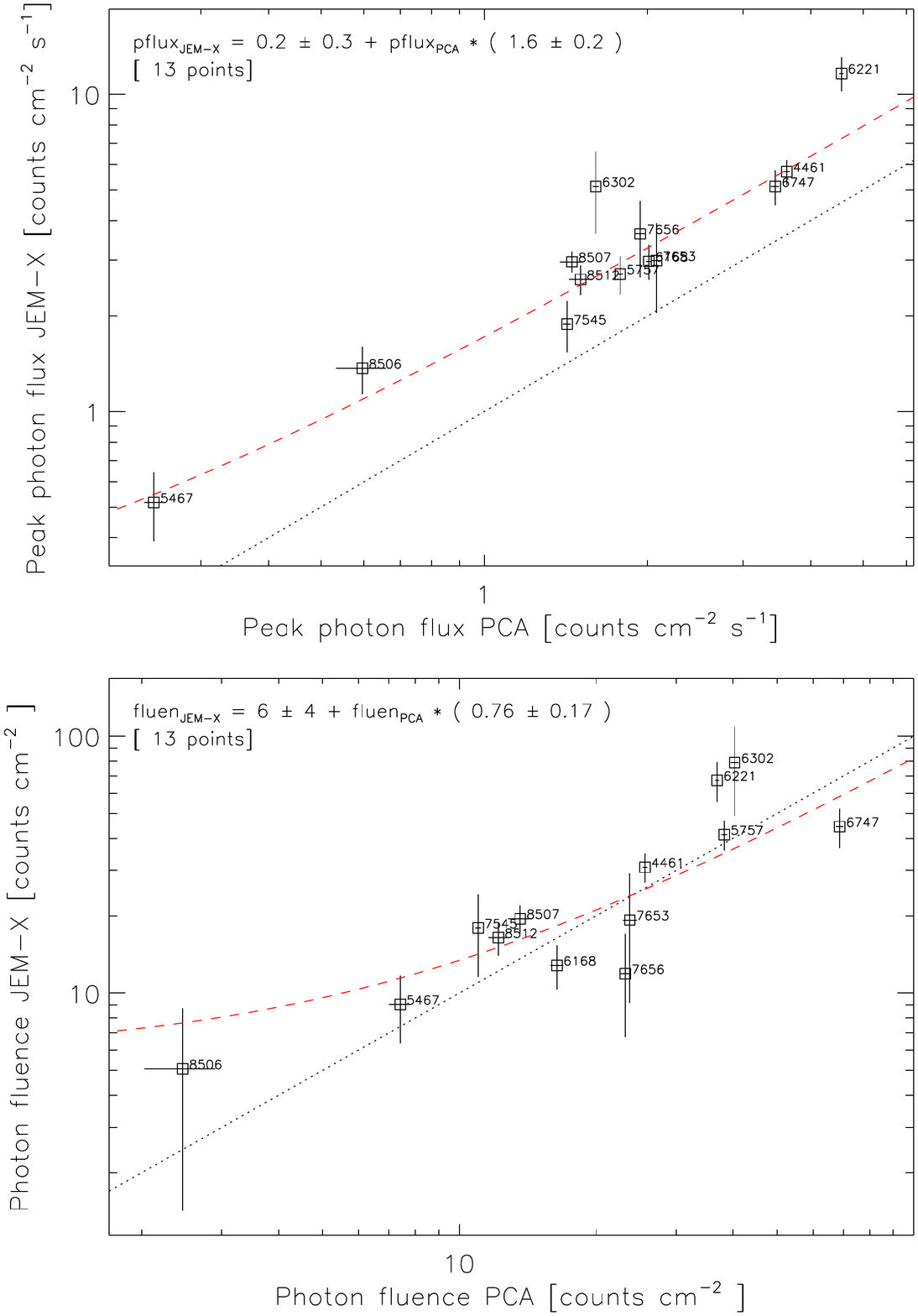}
\caption{Cross-calibration for the peak photon flux ({\it top panel}) and photon fluence ({\it bottom panel}) for bursts observed with both \igr/JEM-X and \xte/PCA in the MINBAR burst sample. 
Note the marked offset between the two sets of peak flux measurements; to ensure consistency between 
instruments, we reduced the JEM-X measurements for this parameter by a factor of 1.6 in the final MINBAR table.  
Other details as for Fig. \ref{fig:sax-xte-pflux}.
\label{fig:igr-xte-pflux} \label{fig:igr-xte-fluen}}
\end{figure}

In the absence of time-resolved spectroscopy for the bursts detected with JEM-X, we compared the measured peak photon flux and fluence between the two sets of measurements 
in the 3--25 keV energy band. %jc190918
The peak photon flux measured with \igr/JEM-X was consistently 56\% larger than the value measured by \xte/PCA, although with moderate scatter about the line of best fit (Fig. \ref{fig:igr-xte-pflux}, top panel). 

\begin{figure}[ht]
\includegraphics[width=\columnwidth]{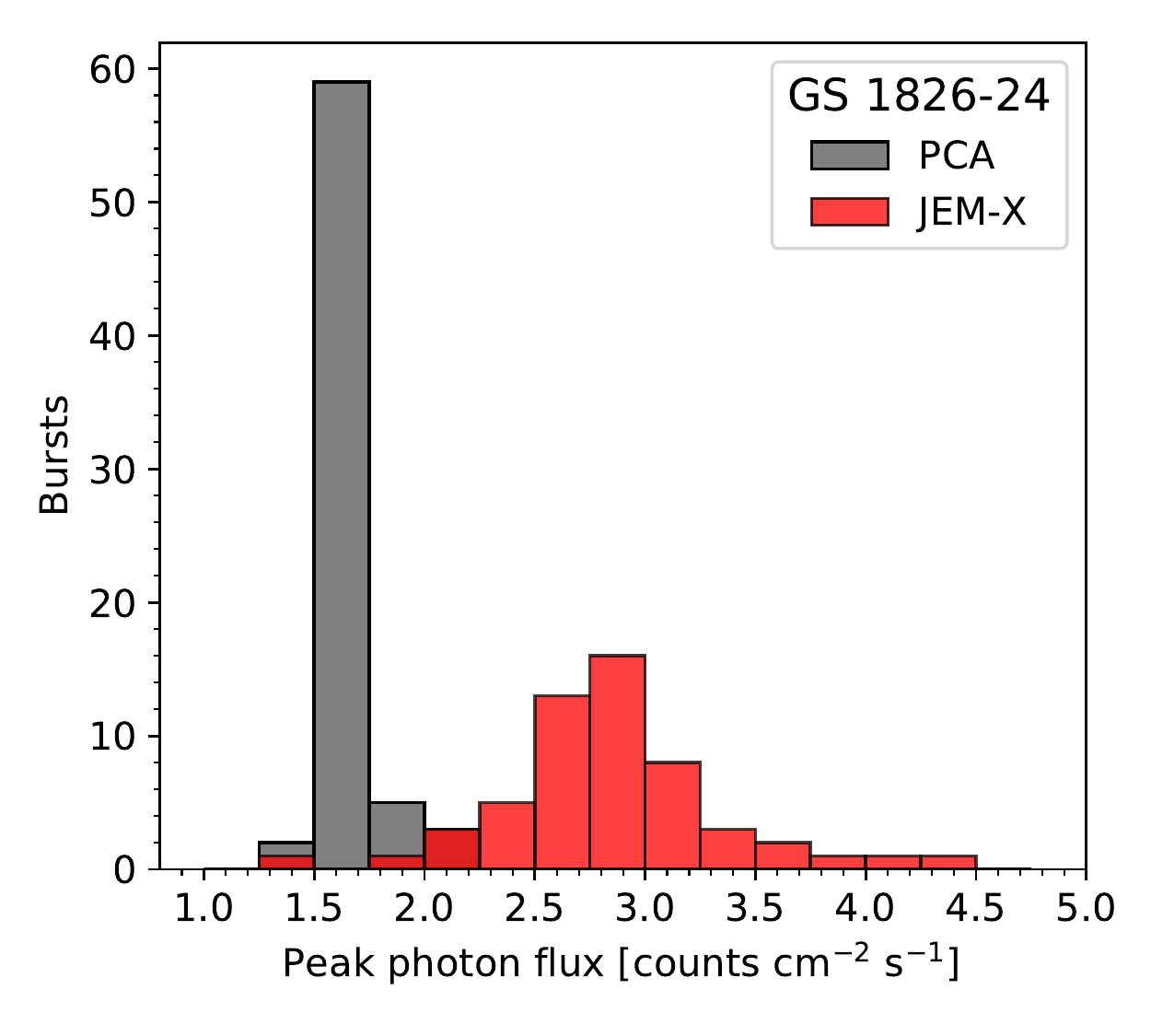}
\caption{Comparison of the distributions of raw (uncorrected) peak photon flux for bursts from GS~1826$-$24, observed with both \igr/JEM-X and \xte/PCA in the MINBAR burst sample. The JEM-X bursts plotted are those observed $<4^\circ$ from the camera aimpoint, to include only the most precise measurements (cf. with Fig. \ref{fig:JMX_offangle}). Note the marked offset in the two distributions, illustrating the bias in JEM-X towards higher peak fluxes. 
\label{fig:1826-igr-xte-pflux}}
\end{figure}

This offset in the peak fluxes is also apparent in distribution of peak fluxes from sources with highly consistent bursts, for example GS~1826$-$24 (Fig. \ref{fig:1826-igr-xte-pflux}). Similar results are found for other bursters with bursts reaching roughly consistent peak fluxes.

We attribute this discrepancy to the variation between the spectral response of the two instruments, that are more acute for the burst spectra than for the persistent spectra. 
We compared the predicted count rates for both JEM-X and PCA for a 2.5~keV blackbody and
for a power law with spectral index $\Gamma=2$, both affected by neutral absorption with $N_H = 10^{22}\ {\rm cm}^{-2}$. The ratio of PCA to JEM-X count
rates was 1.53 times larger for the blackbody than for the power
law. 
This value is roughly consistent with the discrepancy identified from the burst cross-correlation. We note that for
PCA/WFC, the corresponding factor is only 1.12, suggesting that JEM-X has a
qualitatively different effective area curve compared to PCA or WFC.

To correct this discrepancy for the JEM-X values, we reduced the burst peak count rates 
by a factor of 1.6 as determined from the cross correlation illustrated in
Fig. \ref{fig:igr-xte-pflux}. Strictly speaking, we should also correct for the WFC count rates for the same reason, but since the discrepancy is only a few percent, we neglect any correction for those values.

\begin{figure}
	\begin{center}
	\includegraphics[width=\columnwidth]{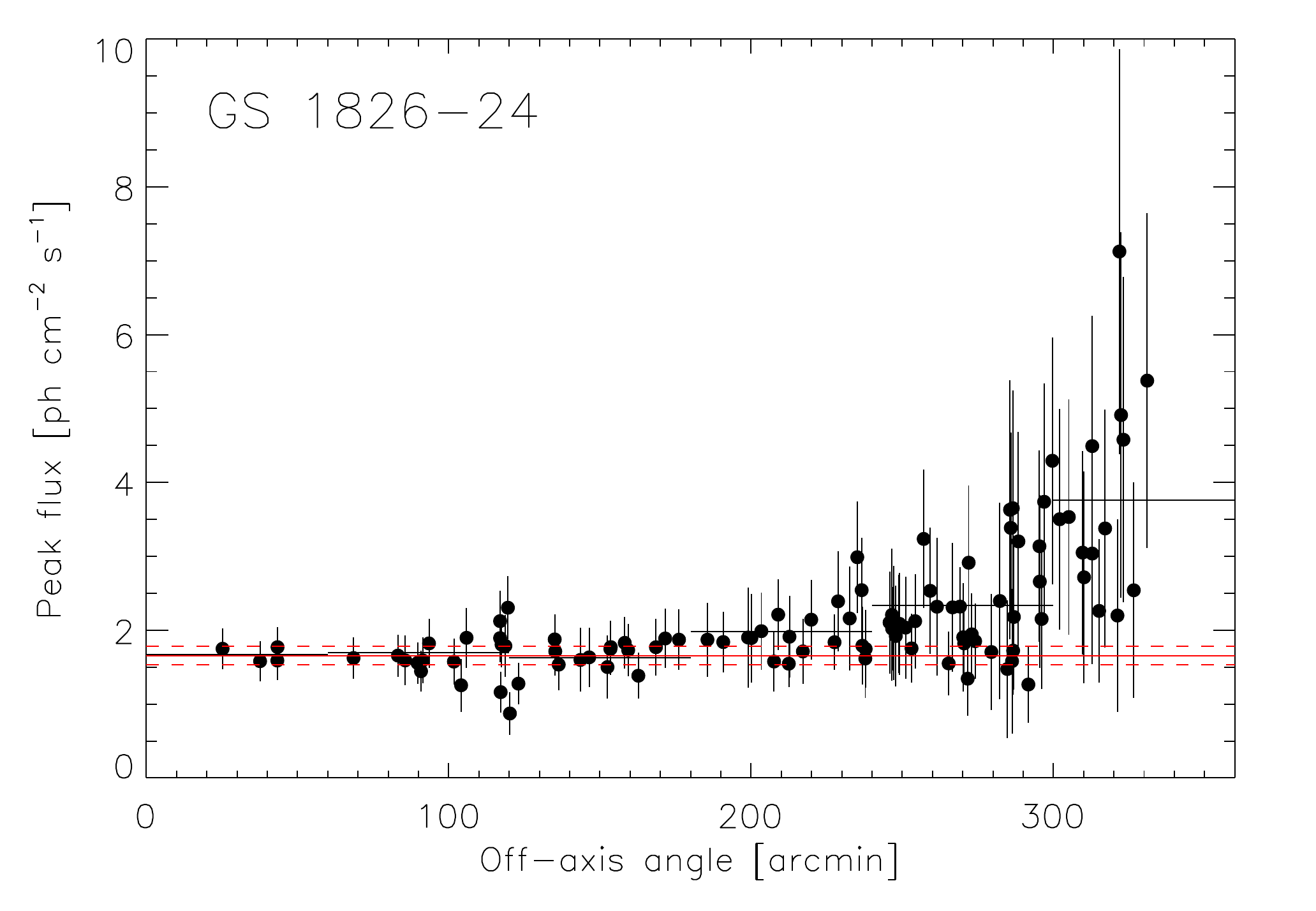}
	\end{center}
	\caption{Corrected burst peak fluxes as function of JEM-X off-axis angle for 
	% 559 bursts from 4U~1728-34. \dg{need to re-create this figure in IDL with proper axis labels etc.}
	107 bursts from GS~1826$-$24. The mean value calculated within each degree of off-axis angle is overplotted ({\it black horizontal lines}). For comparison, the mean peak flux and standard deviation for bursts observed with \xte/PCA is overplotted ({\it red lines}). Note the substantial bias on the mean values for JEM-X at high off-axis angles.
	\label{fig:JMX_offangle}}
\end{figure}

One likely contributing factor to the scatter between the cross-calibration for peak flux is the angle between the aimpoint and the source position in each observation. 
Since the JEM-X sensitivity drops with the source off-axis angle (see \ref{obs-jemx}), the measured fluxes are affected by increasing uncertainties with increasing angles and there is a positive bias for stronger fluxes at larger off-axis angle. 
As an example, Figure \ref{fig:JMX_offangle} shows corrected peak fluxes for all bursts detected from
GS~1826$-$24 with JEM-X, plotted as function of off-axis angle. Over the period that this source was covered by MINBAR, it produced highly consistent bursts with a narrow range of peak fluxes. The JEM-X data clearly show increased variation towards larger off-axis angles, as well as a marked bias to the mean values calculated within the angle bins.
Therefore, absolute flux values of bursts seen in JEM-X at off-axis angles larger than $4^\circ$ must be considered with caution, while the time and origin of the bursts are still valuable information.

The measured photon fluence between JEM-X and the PCA exhibits the opposite bias to the peak flux, with the value estimated from JEM-X underestimating, on average, the value measured with the PCA (Fig. \ref{fig:igr-xte-fluen}, bottom panel). This underestimate is despite the exaggerated count rates around the peak as measured with JEM-X. The opposite discrepancy in this case is likely due to similar effects as noted for the fluences measured with the WFC, which are more than sufficient to offset the excess count rate values.
The best-fit correlation parameters are listed in Table \ref{tab:crosscal-bursts}.

\subsection{Observation cross-calibration}
\label{sec:crosscal}

Following the same approach as for the bursts (in \S\ref{crosscal:bursts}), and using the overlapping observations identified in \S\ref{sec:area}, we cross-correlated the 
observation-averaged 3--25~keV flux over the pairs of observations, and the spectral colours for each pair of instruments. 
For the flux and the spectral colours, an additional potential source of variation is the choice of spectral model. 
We list the results for each pair of instruments in Table \ref{tab:crosscal}, and describe the results in detail below.

\subsubsection{WFC-PCA}

\begin{figure}[ht]
\includegraphics[width=\columnwidth]{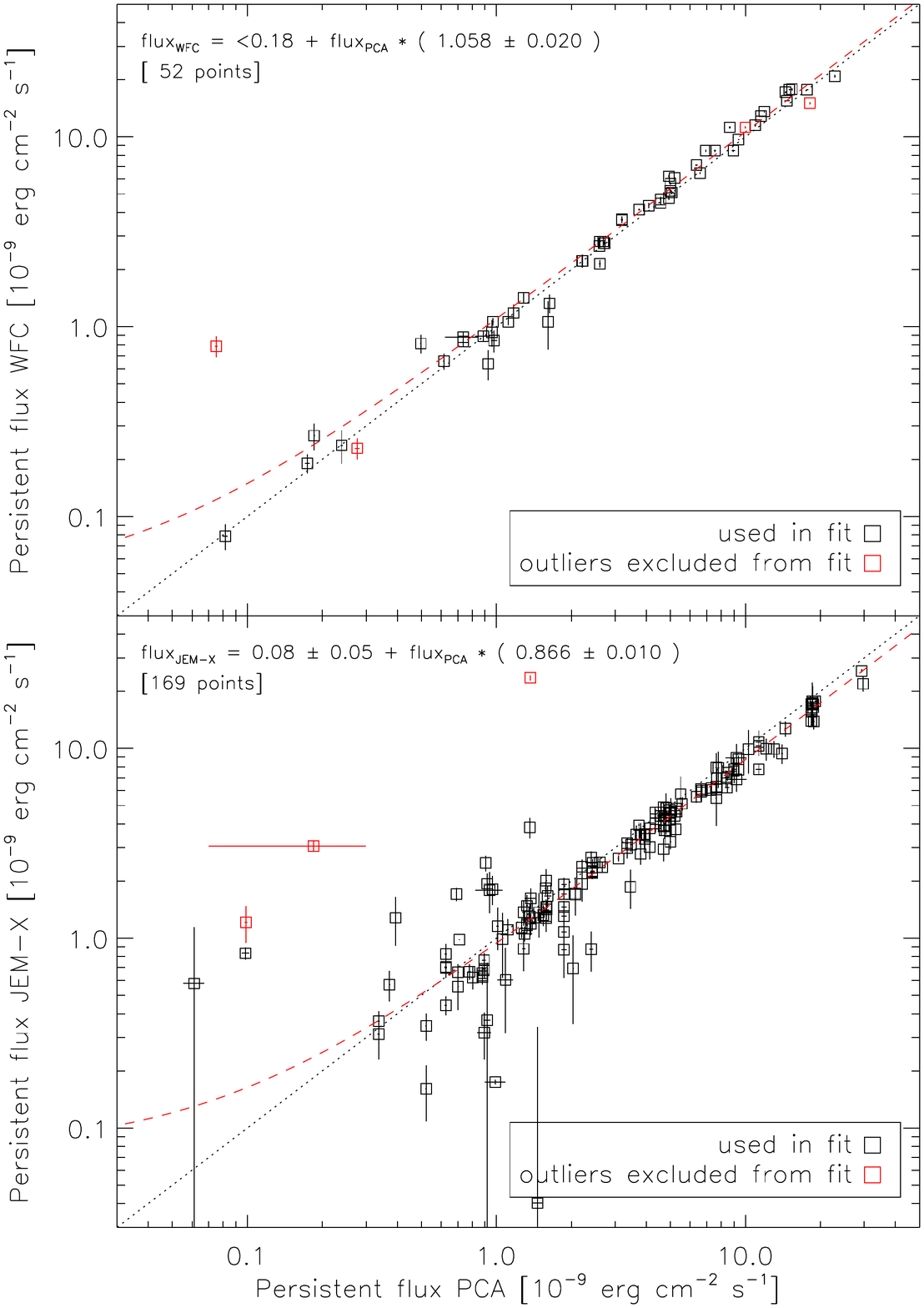}
\caption{Cross-calibration for the 3--25~keV flux for overlapping observations with instruments contributing to the MINBAR observation table. The top panel shows the comparison of measurements between \xte/PCA and \sax/WFC; the bottom panel shows PCA against \igr/JEM-X. The dotted black line in each panel shows a 1:1 correspondence; other elements as for Fig. \ref{fig:sax-xte-count}
\label{fig:sax-xte-flux} \label{fig:igr-xte-flux}}
\end{figure}

The best linear fit to the contemporaneous flux measurements with both the PCA and WFC gives
\begin{equation}
    F_\mathrm{WFC} = (0.04 \pm 0.14) + F_\mathrm{PCA} \times ( 1.058\pm0.020 )
\label{eqn:pca_wfc_crosscal}
\end{equation}
and is plotted in Figure \ref{fig:sax-xte-flux}, top panel. In general the results are highly consistent; the best-fit gradient deviates from 1 at  the 6\% level, and at weak significance, just over $3\sigma$.
This agreement is satisfactory given the typical estimated absolute precision for X-ray instruments, of up to a few tens of per cent.
We adopt the cross-calibration factor above as a correction for calculating the estimated $\gamma$-factor (proportional to the accretion rate) as described in \S\ref{bolcor}.

\subsubsection{JEM-X-PCA}
As for the WFC, we compare the fluxes measured by \igr/JEM-X and \xte/PCA in Figure \ref{fig:igr-xte-flux}, bottom panel. We find a best-fit cross-calibration factor of 
$0.866\pm0.010$,
with the JEM-X persistent flux values systematically below those measured by \xte/PCA by approximately 12\%. Again, this discrepancy appears reasonable given the typical absolute calibration uncertainties for X-ray instruments.
We adopt the cross-calibration factor above as a correction for calculating the estimated $\gamma$-factor (proportional to the accretion rate) as described in \S\ref{bolcor}.

\section{Catalog assembly}
\label{sec:assembly}

In the following sections we describe the tasks undertaken to assemble the available data from the primary sources into a uniform format for integration into the final tables. 
In \S\ref{sec:prebursts} we describe the classification and analysis of the radius-expansion bursts in the sample, including those events for which time-resolved spectroscopy was not available.
In \S\ref{fluxEdd} we use the bursts identified as PRE with high confidence, to measure the Eddington flux for each source and hence provide a uniform scale for the luminosity (and hence accretion rate) across all burst sources.
In \S\ref{sec:trec} we describe the determination of the burst recurrence times and the derived quantities $\tau$ and $\alpha$.
In \S\ref{bolcor} we describe the approach taken to correct the persistent flux, measured in the (typically) 3--25~keV energy range, to estimate the bolometric fluxes and hence the accretion rate.
In \S\ref{sec:colour} we describe the approach adopted to calculate hard and soft spectral colours for each observation.

\subsection{Radius-expansion bursts}
\label{sec:prebursts}

We took one of two approaches to identify photospheric radius-expansion (PRE) bursts in the sample, and hence measure the Eddington flux for each source. 
For the bursts observed with \xte/PCA, we adopted the same criteria as G08, specifically: the presence of a local maximum in the blackbody normalisation, coincident with a local minimum in the blackbody temperature. This classification could only be performed where there was time-resolved spectroscopy covering the burst peak, which was not the case for all bursts (see \S\ref{minbar}). Where these features are present, we classify the burst as radius expansion ({\tt rexp=2}); in cases where the radius maximum is only weakly significant ($<3\sigma$) compared to the subsequent values, we classify it as marginal ({\tt rexp=3}).

For \sax/WFC, we identified a small number of bright bursts as PRE based on the time-resolved spectroscopy, adopting similar criteria. If, in the early phases of a burst, the blackbody normalization was seen to increase in tandem with a decrease in temperature, followed by a reverse trend for both, the burst was qualified as PRE. % \jz{need more details here? -- dkg} I explained this now, is this what you meant? --jz
For weaker WFC bursts, and those detected by \igr/JEM-X, in the absence of time-resolved spectroscopy, we could not determine 
whether PRE was present.

We augmented the PRE flags for the bursts detected with WFC and JEM-X using a classification from a machine-learning algorithm trained on a subset of bursts with known (or assumed) PRE status. 
Using the Random Forest Classifier from the Python {\tt scikit-learn} library \cite[]{scikit-learn}, an estimate for the likelihood of PRE was established. The classifier was trained on a data set which was compiled using bursts matching the following criteria. 
First, sources that have exhibited predominately radius-expansion bursts or predominately non radius-expansion bursts, according to the \xte/PCA analysis, had the PRE flag values of all the bursts set to Y or N, respectively.
Second, if bursts were detected on multiple instruments the \xte/PCA analysis determined the PRE flag. %\luis{need to add training set information}
The training set so constructed consists of 779 bursts, and the composition is summarised in Table \ref{tab:training}.

\begin{deluxetable}{lcccc}
\tablecaption{Composition of the training set for PRE classification
  \label{tab:training}
}
\tablewidth{0pt}
\tablehead{
  \colhead{}
 & \multicolumn{2}{c}{WFC}
 & \multicolumn{2}{c}{JEM-X}
\\
  \colhead{Source}
 & \colhead{non-PRE}
 & \colhead{PRE}
 & \colhead{non-PRE}
 & \colhead{PRE}
}
\startdata
4U 0513$-$40           &   0 &   2 &   0 &   0 \\
EXO 0748$-$676         & 175 &   0 &  22 &   0 \\
4U 0836$-$429          &   0 &   0 &  61 &   0 \\
2S 0918$-$549          &   0 &   3 &   0 &   0 \\
4U 1246$-$588          &   0 &   4 &   0 &   0 \\
4U 1323$-$62           &   0 &   0 &   1 &   0 \\
4U 1608$-$522          &   0 &   4 &   0 &   0 \\
4U 1636$-$536          &   0 &   1 &   1 &   0 \\
4U 1702$-$429          &   2 &   0 &   0 &   0 \\
4U 1705$-$44           &   1 &   0 &   1 &   0 \\
RX J1718.4$-$4029      &   0 &   1 &   0 &   0 \\
4U 1722$-$30           &   0 &  24 &   0 &   0 \\
4U 1728$-$34           &   0 &   0 &   2 &   1 \\
KS 1731$-$260          &   2 &   1 &   0 &   0 \\
SLX 1737$-$282         &   0 &   1 &   0 &   0 \\
1A 1742$-$294          &   0 &   1 &   0 &   0 \\
SAX J1747.0$-$2853     &   0 &   2 &   0 &   0 \\
IGR J17473$-$2721      &   0 &   0 &   1 &   0 \\
SLX 1744$-$300         &   1 &   0 &   0 &   0 \\
EXO 1745$-$248         &   0 &   1 &   0 &   0 \\
IGR J17511$-$3057      &   0 &   0 &   2 &   0 \\
SAX J1808.4$-$3658     &   0 &   3 &   0 &   0 \\
4U 1812$-$12           &   0 &  18 &   0 &   7 \\
4U 1820$-$303          &   0 &  49 &   0 &   2 \\
GS 1826$-$24           & 272 &   0 & 107 &   0 \\
HETE J1900.1$-$2455    &   0 &   0 &   0 &   2 \\
M15 X-2                &   0 &   1 &   0 &   0 \\
\enddata
\end{deluxetable}

The Random Forest classifier operates by generating multiple ``decision trees'' to divide the sample based on the attributes. Each individual tree operates on a randomly-chosen subset of the data, and the final classification probability for each instance is based on the fraction of trees which classify in each category.
We chose 128 trees for the PRE classifier, and iterated using random seed values of 1--50. 
We verified the classifier by testing on each of 10 subsets consisting of 10\% of the training set, with the remaining instances used for training. This ``stratified 10 fold cross validation'' 
yielded an average accuracy of $0.98\pm0.01$ for the training set. 

\begin{figure}[ht]
	\includegraphics[width=\columnwidth]{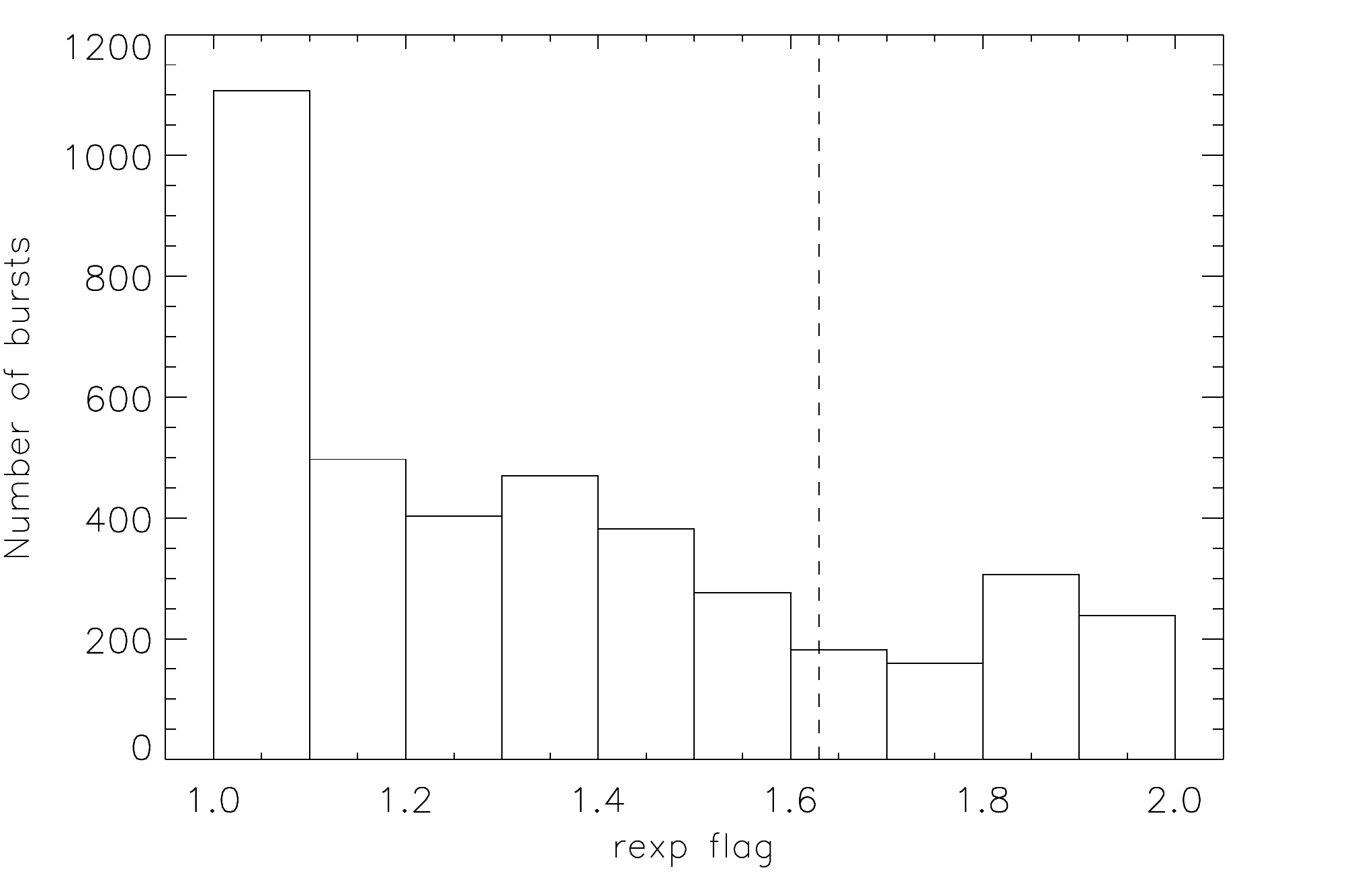}
	\caption{Distribution of {\tt rexp} flag values for those bursts for which the time-resolved spectroscopy was unavailable or ambiguous, such that the presence of PRE was determined via the classification algorithm.
    The vertical dashed line indicates the suggested threshold of {\tt rexp}=\rexpthresh\/ such that the fraction of PRE bursts in this sample matches that measured by \xte/PCA alone.
    % , 26.5\%.
	\label{fig:rexp} }
\end{figure}

We set the PRE flag (column {\tt rexp}; see \S\ref{minbar}) with the results from the classification as follows. For the PCA and WFC bursts (and the training set sample) we set non-PRE bursts as {\tt rexp=1}, and PRE bursts with {\tt rexp=2}. (A small number of bursts observed by \xte\/ are flagged with {\tt rexp=3}, indicating ``marginal'' radius expansion, following G08). 
We then set the flag for the remaining 4020 WFC and JEM-X bursts (which we refer to as the ``classification sample'') as $1+p_i$, where $p_i$ is the probability of radius-expansion according to the classification algorithm. 
The distribution of the {\tt rexp} values is shown in Fig. \ref{fig:rexp}.

With these values, one can set the desired confidence level for non-PRE or PRE bursts, or simply choose those events with integer values for the highest-confidence samples.
A suitable threshold value may be that which gives the same fraction of PRE bursts in the classification sample as for a set with confirmed PRE status, as observed by \xte/PCA. In this sample we find 
20.0\% % 2019 Nov
bursts with PRE (including marginal cases). 
The corresponding threshold required to have the same fraction of PRE bursts in the classification sample is \rexpthresh. 

\subsection{Eddington flux}
\label{fluxEdd}

For each Eddington-limited burst (that is, those with radius-expansion flag in the range $\rexpthresh\leq$~{\tt rexp}~$\leq2$; see discussion in \S\ref{sec:prebursts}) 
observed with \xte/PCA or \sax/WFC
we measured the peak bolometric flux, incorporating the cross-instrument calibration factors given in equation \ref{eqn:pca_wfc_crosscal}. %:sax-xte-pflux}. %jc190918 \ref{eqn:pca_wfc_crosscal}.
Table \ref{table:peakfluxes} lists each source in the MINBAR catalog and the  burster type (as  in Table \ref{tab:bursters}). The third column lists the number of PRE bursts 
observed from the source. The fourth column lists the mean Eddington flux for each source, calculated by least-squares fitting the peak fluxes of the individual bursts weighted by their inverse-squared uncertainty.
For the WFC we incorporated corrections based on the comparison of bursts observed simultaneously with the PCA, as described in \S\ref{burstpcountxcal}. 
The uncertainty on this calculation is the standard deviation of all the burst peak fluxes for a source or, if only one burst was reported, the uncertainty of the peak flux of that burst. The fifth column in the Table lists distances measured by {\it Gaia}\/ 
\citep{Gaia2016, gaiadr2},
where available, taken from the catalog of \cite{bailer-jones18}, and marked with a G in the table column.
We found that 33 of our targets were either not detected by Gaia at all, and three more sources did not produce a parallax measurement. 25 sources were in fields so crowded we could not identify which of the stars was the LMXB of interest. In particular, the Galactic centre and globular cluster sources could generally not be unambiguously identified, and GX~17+2 was confused with a foreground star (e.g. \citealt{callanan02}). We thus obtained Gaia distances for 
13 % following jz comments; 2020 Jan
of the 73 sources in this analysis. %\hw{can you plot the burst-derived distances against the Gaia distances? -- dkg}

A further eleven LMXB systems are listed 
with measured distances: %jc190918
ten sources residing in globular clusters, whose distances can be determined from their Hertzsprung-Russell diagrams, 
and 4U~1608$-$52 \citep{guver10a}. These eleven objects are indicated by reference numbers in parentheses in the distance column.

We calculated approximate distances to the sources for hydrogen-rich and -poor material by assuming that peak flux is reached at the touchdown point and that the radius of the neutron star is 11.2\,km \citep{steiner18}, or upper limits to its distance by assuming that the brightest burst from it was 
a lower limit on
the Eddington luminosity. The results of these calculations are given in the sixth and seventh columns of the table. 

For four sources
(MXB 1730$-$335 or the Rapid Burster, GS~1826$-$24, IGR~17480$-$2446, and SLX~1735$-$269),
we include Eddington fluxes even though there are no PRE bursts in our sample (these sources are listed with a zero in the corresponding column of the Table, rather than an ellipsis). 
For the Rapid Burster we adopt $28\times 10^{-9}$\,erg\,s$^{-1}$\,cm$^{-2}$ \citep{bagnoli13}, and we here assume an uncertainty of of 25\%. %\hw{can you please check this? I can't find that value in that reference (I only recall that other paper that claimed a PRE burst from this source) -- dkg}. %This is correct, it's in the last paragraph of sect 3.2.1
The Terzan 5 source IGR~17480$-$2446 exhibited frequent non-radius expansion bursts, and we therefore estimate its Eddington flux the same way \cite{bagnoli13} did for the Rapid Burster: assuming it has a mass of $1.4\ M_\odot$ and a photosphere with hydrogen mass fraction $X=0.7$.
The prolific burster GS~1826$-$24 showed one PRE burst observed by {\it NuSTAR}, reaching a peak flux of $(40\pm 3)\times 10^{-9}$\,erg\,s$^{-1}$\,cm$^{-2}$ \citep{chenevez16}. 
Finally, a PRE burst from SLX~1735$-$269 was observed with JEM-X to reach $2.1\pm 0.4$~Crab \cite[2--30~keV; ][this burst is not present in MINBAR, see \S\ref{minbar}]{molkov05}. Since the Crab and a thermonuclear burst have very different spectra we match fluxes by first assuming a Crab spectrum of
$I(E)=9.59E^{-2.108}$\,photons\,s$^{-1}$\,cm$^{-2}$\,keV$^{-1}$ with interstellar absorption of $3.45\times 10^{21}$\,cm$^{-2}$\citep{willingale01}. The burst spectrum was taken to be a blackbody with temperature 2.5\,keV and an interstellar absorption column density of $1.5\times 10^{22}$\,cm$^{-2}$, and we adjusted the normalisation to give 2.1 Crab units over the 3--20\,keV range in \cite{molkov05}, which finally gives an Eddington flux of 
$(53\pm 21)\times 10^{-9}$\,erg\,s$^{-1}$\,cm$^{-2}$ in the 2--30\,keV range. We have increased the uncertainty to 40\% of the measurement to allow for uncertainties in this matching procedure.

 %Where distances are available, they generally agree with the ones estimated from the burst peak fluxes to within uncertainties, with the exception of 4U~0614+09 which seems to be a little more distant than the peak flux analysis suggests. 

\begin{figure}[ht]
	\includegraphics[width=\columnwidth]{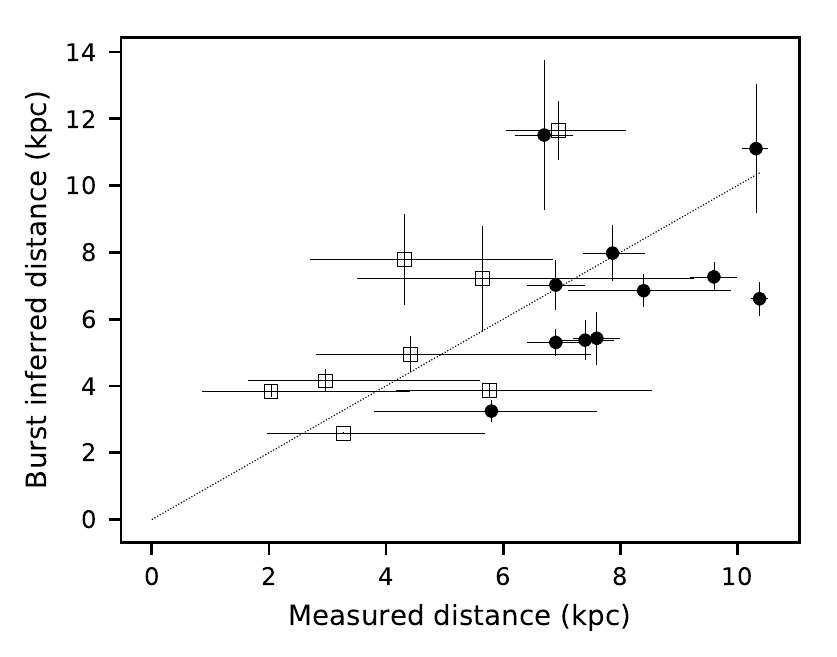}
	\caption{Distances to bursters inferred from bursts against their measured distances. Objects plotted with open squares have Gaia distance measurements, and those with filled circles are globular cluster or other measurements. The dotted line indicates equal inferred and measured distances.
	\label{fig:distancecomp} }
\end{figure}

In Figure \ref{fig:distancecomp} we plot the distances inferred from bursts against the independently-measured distances,
for bursters where both quantities are known. 
either from {\it Gaia\,} parallax or from the distance of the host globular cluster.
Parallactic distances are biased towards nearer objects,
as more distant objects are too faint optically
to provied a reliable parallax. 
The agreement is generally passable, but with some significant outliers from the 1:1 line. In the case of 4U~0614+09, the very high brightness of its single burst leads to a seemingly overprecise distance determination from the burst flux. We note that the distances from bursts neglect the possible effects of emission anisotropy, as well as possible variations in the Eddington luminosity based on the neutron star mass between different systems.

{\it Gaia}\/ distances, taken from \cite{bailer-jones18}, need to be carefully interpreted. Although the method of \cite{bailer-jones18} is more robust than simply assuming the distance is the inverse of the parallax, it is a probabilistic model based on assumptions regarding the distribution of matter in the Galaxy. It is unlikely that the LMXB population follows the stellar population. This may explain the discrepancy in the distance to 4U~1254-69 between the value inferred from {\it Gaia}\/ (whose DR2 lists a parallax of 0\farcs32$\pm0$\farcs21) and that inferred from burst peak fluxes or otherwise. There are five literature values for the latter that are all larger than 7.6 kpc \citep[e.g.,][]{motch87,zand03b,gambino17}, while the {\it Gaia}\/ distance is $3.18^{+3.16}_{-1.33}$~kpc.
We defer an in-depth study of the discrepant sources for a later paper.

%%%%%%%%%%%%%%%%%%%%%%%%%%%%%%%%%%%%%%%%%%%%%%%%%%%%%%%%%%%%%%%%
%%%%  Automatically generated. Do not edit.                 %%%%
%%%%%%%%%%%%%%%%%%%%%%%%%%%%%%%%%%%%%%%%%%%%%%%%%%%%%%%%%%%%%%%%
%table produced by get_avg_pflux.py at 2020-02-07 12:00:14
\startlongtable % AASTeX 6.3 environment
\begin{deluxetable*}{lcccccc}
\tabletypesize{\scriptsize}
\tablecaption{Mean Peak Fluxes and Estimated Distances from Bursts in the MINBAR Catalog
        \label{table:peakfluxes}
}
\tablehead{
\colhead{}   & Type            & \colhead{PRE}    & \colhead{$F_\textrm{Edd}$}                      & \colhead{Dist (kpc)}     & \multicolumn{2}{c}{Dist (kpc), inferred}\\
\colhead{Source} & & \colhead{Bursts} & \colhead{(10$^{-9}$\,erg\,cm$^{-2}$\,s$^{-1}$)} &  (Measured)           & \colhead{$X=0.7$} & \colhead{$X=0.0$} 
}
\colnumbers
\startdata
              4U 0513-40 &         GC &                        6 &            14.4 $\pm6.7$ & $10.32^{+0.20}_{-0.24}$ (1) &             8.5 $\pm1.5$ &            11.1 $\pm1.9$\\[3pt]
              4U 0614+09 &       ARSC &                        1 &           266.0 $\pm6.0$ & $3.27^{+2.42}_{-1.30}$ (G) &           1.99 $\pm0.02$ &           2.59 $\pm0.03$\\[3pt]
            EXO 0748-676 &       DEOT &                        5 &            46.5 $\pm4.3$ &            \textellipsis &             4.7 $\pm0.2$ &             6.2 $\pm0.3$\\[3pt]
             4U 0836-429 &          T &            \textellipsis &            \textellipsis & $3.18^{+2.25}_{-1.40}$ (G) &                   $<$6.9 &                   $<$9.0\\[3pt]
             2S 0918-549 &          C &                        5 &          119.1 $\pm14.4$ & $5.77^{+2.77}_{-1.60}$ (G) &             3.0 $\pm0.2$ &             3.9 $\pm0.2$\\[3pt]
             4U 1246-588 &          C &                        4 &          120.3 $\pm11.9$ & $2.03^{+2.37}_{-1.17}$ (G) &             3.0 $\pm0.1$ &             3.8 $\pm0.2$\\[3pt]
              4U 1254-69 &         DS &            \textellipsis &            \textellipsis & $3.18^{+3.16}_{-1.33}$ (G) &                   $<$6.0 &                   $<$7.9\\[3pt]
        SAX J1324.5-6313 & \textellipsis &            \textellipsis &            \textellipsis &            \textellipsis &                   $<$4.7 &                   $<$6.1\\[3pt]
              4U 1323-62 &          D &            \textellipsis &            \textellipsis &            \textellipsis &                   $<$5.2 &                   $<$6.8\\[3pt]
                 Cir X-1 &      ADMRT &            \textellipsis &            \textellipsis & $6.17^{+2.86}_{-1.96}$ (G) &                  $<$13.6 &                  $<$17.7\\[3pt]
             4U 1608-522 &       AOST &                       50 &          169.0 $\pm41.2$ & $5.80^{+1.80}_{-2.00}$ (2) &             2.5 $\pm0.3$ &             3.2 $\pm0.3$\\[3pt]
             4U 1636-536 &        AOS &                      140 &           72.5 $\pm18.8$ & $4.42^{+3.08}_{-1.63}$ (G) &             3.8 $\pm0.4$ &             5.0 $\pm0.5$\\[3pt]
           XTE J1701-462 &       TZR? &                        2 &            43.4 $\pm1.4$ &            \textellipsis &             4.9 $\pm0.1$ &             6.4 $\pm0.1$\\[3pt]
            MXB 1658-298 &       DEOT &                       13 &           17.0 $\pm15.9$ &            \textellipsis &             7.9 $\pm2.2$ &            10.2 $\pm2.9$\\[3pt]
             4U 1702-429 &         AO &                       79 &           88.7 $\pm45.0$ &            \textellipsis &             3.4 $\pm0.6$ &             4.5 $\pm0.8$\\[3pt]
              4U 1705-32 &          C &            \textellipsis &            \textellipsis &            \textellipsis &                   $<$4.4 &                   $<$5.7\\[3pt]
              4U 1705-44 &         AR &                       19 &           41.3 $\pm17.5$ &            \textellipsis &             5.0 $\pm0.8$ &             6.6 $\pm1.1$\\[3pt]
           XTE J1709-267 &         TC &            \textellipsis &            \textellipsis &            \textellipsis &                   $<$2.7 &                   $<$3.6\\[3pt]
           XTE J1710-281 &        DET &                        3 &             7.1 $\pm1.8$ &            \textellipsis &            12.2 $\pm1.3$ &            15.9 $\pm1.7$\\[3pt]
        SAX J1712.6-3739 &         TC &                        2 &           76.0 $\pm46.9$ &            \textellipsis &             3.7 $\pm0.8$ &             4.8 $\pm1.0$\\[3pt]
             2S 1711-339 &          T &            \textellipsis &            \textellipsis &            \textellipsis &                   $<$4.9 &                   $<$6.4\\[3pt]
         RX J1718.4-4029 &          C &                        1 &            47.2 $\pm6.2$ &            \textellipsis &             4.7 $\pm0.3$ &             6.1 $\pm0.4$\\[3pt]
         IGR J17191-2821 &         OT &            \textellipsis &            \textellipsis &            \textellipsis &                   $<$5.9 &                   $<$7.7\\[3pt]
           XTE J1723-376 &          T &            \textellipsis &            \textellipsis &            \textellipsis &                   $<$3.7 &                   $<$4.8\\[3pt]
              4U 1722-30 &        GAC &                       27 &           61.8 $\pm16.6$ & $7.40^{+0.50}_{-0.50}$ (3, 4, 5) &             4.1 $\pm0.5$ &             5.4 $\pm0.6$\\[3pt]
              4U 1728-34 &       AORC &                      496 &           94.0 $\pm35.9$ &            \textellipsis &             3.3 $\pm0.5$ &             4.4 $\pm0.7$\\[3pt]
            MXB 1730-335 &       TGDR &                        0 &            28.0 $\pm7.0$ & $7.87^{+0.56}_{-0.50}$ (5) &             6.1 $\pm0.6$ &             8.0 $\pm0.8$\\[3pt]
             KS 1731-260 &        OST &                       90 &           50.5 $\pm20.4$ &            \textellipsis &             4.6 $\pm0.7$ &             5.9 $\pm0.9$\\[3pt]
            SLX 1735-269 &         SC &                        0 &           52.9 $\pm21.0$ &            \textellipsis &             4.5 $\pm0.7$ &             5.8 $\pm0.9$\\[3pt]
             4U 1735-444 &       AR?S &                       27 &           34.2 $\pm22.0$ & $5.65^{+3.62}_{-2.14}$ (G) &             5.5 $\pm1.2$ &             7.2 $\pm1.6$\\[3pt]
           XTE J1739-285 &          T &            \textellipsis &            \textellipsis & $4.06^{+4.25}_{-2.44}$ (G) &                   $<$6.1 &                   $<$7.9\\[3pt]
            SLX 1737-282 &          C &                        1 &           68.1 $\pm12.4$ &            \textellipsis &             3.9 $\pm0.3$ &             5.1 $\pm0.4$\\[3pt]
             KS 1741-293 &          T &            \textellipsis &            \textellipsis &            \textellipsis &                   $<$4.2 &                   $<$5.4\\[3pt]
         GRS 1741.9-2853 &         OT &                        6 &            35.3 $\pm9.8$ &            \textellipsis &             5.5 $\pm0.6$ &             7.1 $\pm0.8$\\[3pt]
             1A 1742-294 & \textellipsis &                        3 &            37.7 $\pm1.3$ &            \textellipsis &             5.3 $\pm0.1$ &             6.9 $\pm0.1$\\[3pt]
        SAX J1747.0-2853 &         TS &                       18 &           52.7 $\pm31.4$ &            \textellipsis &             4.5 $\pm0.9$ &             5.8 $\pm1.2$\\[3pt]
         IGR J17473-2721 &          T &                        3 &          113.6 $\pm11.7$ &            \textellipsis &             3.0 $\pm0.1$ &             4.0 $\pm0.2$\\[3pt]
            SLX 1744-300 &         T? &                        4 &            13.7 $\pm3.2$ &            \textellipsis &             8.7 $\pm0.9$ &            11.4 $\pm1.1$\\[3pt]
                  GX 3+1 &         AS &                       54 &           53.3 $\pm15.2$ &            \textellipsis &             4.4 $\pm0.5$ &             5.8 $\pm0.7$\\[3pt]
         IGR J17480-2446 &       GOPT &                        0 &            36.1 $\pm9.0$ & $6.90^{+0.50}_{-0.50}$ (3, 4, 6) &             5.4 $\pm0.6$ &             7.0 $\pm0.7$\\[3pt]
             1A 1744-361 &      TA?DR &            \textellipsis &            \textellipsis &            \textellipsis &                   $<$7.0 &                   $<$9.1\\[3pt]
        SAX J1748.9-2021 &        TGA &                       12 &            38.0 $\pm6.1$ & $8.40^{+1.50}_{-1.30}$ (3, 4) &             5.3 $\pm0.4$ &             6.9 $\pm0.5$\\[3pt]
            EXO 1745-248 &       DGST &                        5 &           63.4 $\pm10.9$ & $6.90^{+0.50}_{-0.50}$ (3, 4, 6) &             4.1 $\pm0.3$ &             5.3 $\pm0.4$\\[3pt]
         IGR J17498-2921 &        OPT &                        1 &            51.6 $\pm1.6$ &            \textellipsis &             4.5 $\pm0.1$ &             5.9 $\pm0.1$\\[3pt]
              4U 1746-37 &        ADG &                        3 &             5.4 $\pm0.8$ &            \textellipsis &            13.9 $\pm1.0$ &            18.2 $\pm1.3$\\[3pt]
        SAX J1750.8-2900 &       A?OT &                        4 &            54.3 $\pm6.1$ &            \textellipsis &             4.4 $\pm0.2$ &             5.7 $\pm0.3$\\[3pt]
            GRS 1747-312 &       DEGT &                        3 &            13.4 $\pm7.3$ & $6.70^{+0.50}_{-0.50}$ (3, 4, 6) &             8.8 $\pm1.7$ &            11.5 $\pm2.2$\\[3pt]
         IGR J17511-3057 &        OPT &            \textellipsis &            \textellipsis &            \textellipsis &                   $<$4.1 &                   $<$5.4\\[3pt]
        SAX J1752.3-3138 &          T &            \textellipsis &            \textellipsis &            \textellipsis &                   $<$6.8 &                   $<$8.9\\[3pt]
        SAX J1753.5-2349 &          T &            \textellipsis &            \textellipsis &            \textellipsis &                   $<$4.4 &                   $<$5.7\\[3pt]
         IGR J17597-2201 &          D &                        3 &            15.7 $\pm0.8$ &            \textellipsis &             8.2 $\pm0.2$ &            10.7 $\pm0.3$\\[3pt]
        SAX J1806.5-2215 &          T &            \textellipsis &            \textellipsis &            \textellipsis &                   $<$4.8 &                   $<$6.2\\[3pt]
             2S 1803-245 &        TAR &            \textellipsis &            \textellipsis &            \textellipsis &                   $<$4.2 &                   $<$5.4\\[3pt]
        SAX J1808.4-3658 &       OPRT &                       11 &          230.2 $\pm26.3$ &            \textellipsis &             2.1 $\pm0.1$ &             2.8 $\pm0.1$\\[3pt]
           XTE J1810-189 &          T &                        1 &            54.2 $\pm1.8$ &            \textellipsis &             4.4 $\pm0.1$ &             5.7 $\pm0.1$\\[3pt]
        SAX J1810.8-2609 &         TO &                        2 &           111.3 $\pm7.2$ &            \textellipsis &             3.1 $\pm0.1$ &             4.0 $\pm0.1$\\[3pt]
   XMMU J181227.8-181234 &          T &                        1 &             2.4 $\pm0.3$ & $14.00^{+2.00}_{-2.00}$ (7) &            20.9 $\pm1.2$ &            27.2 $\pm1.5$\\[3pt]
           XTE J1814-338 &        OPT &            \textellipsis &            \textellipsis &            \textellipsis &                   $<$8.6 &                  $<$11.3\\[3pt]
              4U 1812-12 &         AC &                       18 &          203.1 $\pm40.1$ &            \textellipsis &             2.3 $\pm0.2$ &             3.0 $\pm0.3$\\[3pt]
                 GX 17+2 &        RSZ &                       14 &            14.6 $\pm5.0$ &            \textellipsis &             8.5 $\pm1.2$ &            11.1 $\pm1.5$\\[3pt]
        SAX J1818.7+1424 & \textellipsis &            \textellipsis &            \textellipsis &            \textellipsis &                   $<$5.9 &                   $<$7.7\\[3pt]
             4U 1820-303 &      AGRSC &                       65 &           60.5 $\pm22.6$ & $7.60^{+0.40}_{-0.40}$ (3, 4, 8) &             4.2 $\pm0.6$ &             5.4 $\pm0.8$\\[3pt]
        SAX J1828.5-1037 &          S &            \textellipsis &            \textellipsis &            \textellipsis &                   $<$5.8 &                   $<$7.6\\[3pt]
              GS 1826-24 &          T &                        0 &            40.0 $\pm3.0$ &            \textellipsis &             5.1 $\pm0.2$ &             6.7 $\pm0.2$\\[3pt]
             XB 1832-330 &         GC &                        1 &            33.8 $\pm4.5$ & $9.60^{+0.40}_{-0.40}$ (3, 4, 9) &             5.6 $\pm0.3$ &             7.3 $\pm0.4$\\[3pt]
                 Ser X-1 &        ARS &                        7 &           29.4 $\pm13.8$ & $4.31^{+2.54}_{-1.61}$ (G) &             6.0 $\pm1.0$ &             7.8 $\pm1.4$\\[3pt]
       HETE J1900.1-2455 &        OIT &                        7 &           123.9 $\pm8.6$ &            \textellipsis &             2.9 $\pm0.1$ &             3.8 $\pm0.1$\\[3pt]
                 Aql X-1 &     ADIORT &                       17 &          103.3 $\pm19.6$ & $2.97^{+2.64}_{-1.32}$ (G) &             3.2 $\pm0.3$ &             4.2 $\pm0.3$\\[3pt]
             XB 1916-053 &        ADC &                       12 &            30.6 $\pm3.5$ &            \textellipsis &             5.8 $\pm0.3$ &             7.6 $\pm0.4$\\[3pt]
           XTE J2123-058 &        TAE &            \textellipsis &            \textellipsis &            \textellipsis &                  $<$12.3 &                  $<$16.0\\[3pt]
                 M15 X-2 &       GCR? &                        2 &            40.8 $\pm7.2$ & $10.38^{+0.15}_{-0.15}$ (1) &             5.1 $\pm0.4$ &             6.6 $\pm0.5$\\[3pt]
                 Cyg X-2 &         ZR &                        8 &            13.1 $\pm2.3$ & $6.95^{+1.16}_{-0.91}$ (G) &             8.9 $\pm0.7$ &            11.6 $\pm0.9$\\[3pt]
        SAX J2224.9+5421 & \textellipsis &            \textellipsis &            \textellipsis &            \textellipsis &                   $<$6.0 &                   $<$7.9\\[3pt]
\enddata
\tablerefs{1. \cite{WatkinsEtAl2015}, 2. \cite{guver10a}, 3. \cite{gccat96}, 4. \cite{Harris2010}, 5. \cite{ValentiEtAl2010}, 6. \cite{ValentiEtAl2007}, 7. \cite{goodwin19b}, 8. \cite{heasley00}, 9. \cite{ChaboyerEtAl2000}, G. \cite{Gaia2016, gaiadr2, bailer-jones18}}
\end{deluxetable*}

\subsection{Burst recurrence times, $\tau$ and $\alpha$-values}
\label{sec:trec}

We calculated the separation $t_{\rm sep}$ for each pair of bursts from each source. 
Bursts that were observed by more than one instrument were counted as a single event, and we quote the same separation (and recurrence time, where available). 

We also attempted to constrain the recurrence time $\Delta t$, by two separate methods.
First, we identified pairs of bursts for which uninterrupted observations were available for any instrument (or combination) over the burst interval. We allowed a maximum gap between observations of 10~s, which is sufficient to rule out that bursts with typical durations have been missed.
Because of the regular interruptions in the low-Earth orbits of \sax\ and \xte, 
the 2-min. gaps between \igr\ pointings, 
and the low duty cycle even for all three instruments combined (typically 2\%; see \S\ref{subsec:obs_summ}) for observations for all instruments, we found few examples of bursts with uninterrupted data over intervals longer than 1~hr. 

Consequently, we also identified bursts occurring periodically, so that the recurrence time $\Delta t$ could be inferred even when intervening bursts were missed in data gaps.
Here we selected subsamples of bursts occurring within a day (or up to a few days), assigned a trial ``burst order'' based on the shortest separation between any two bursts, and perform a linear fit to estimate the steady recurrence time. We adjusted the order for individual events to try to minimise the residuals compared to a model with a constant recurrence time, and checked that any predicted burst times which were not observed, fell within data gaps. We then reported the best-fit recurrence time (and error) for the entire group. 
The same approach was used to identify the ``reference'' bursts reported by \cite[]{gal17a}.
This exercise was straightforward for sources exhibiting regular bursts, like GS~1826$-$24 \cite[which was in its hard state over the entire period covered by the MINBAR observations; cf. with][]{chenevez16}, but such behaviour is scarce for most other sources.
In total, we report the recurrence time of 693 bursts.

For those bursts where the recurrence time was measured or inferred, and the fluence $E_b$ and persistent flux $F_{\rm per}$ were also measured, we calculated the ratio $\alpha$ of the burst to persistent fluxes:
\begin{equation}
    \alpha = \frac{\Delta t\,F_{\rm per}c_{\rm bol}}{E_b} \label{eq:alpha}
\end{equation}
\added{where
$c_{\rm bol}$ is the
bolometric correction estimating the ratio of bolometric flux to that in the common 3--25~keV band (see \S\ref{bolcor}). }
The $\alpha$-value is understood to be a measure of the relative efficiency of accretion and thermonuclear burning, i.e. 
\begin{equation}
    \alpha \propto \frac{Q_{\rm grav}}{Q_{\rm nuc}}
\end{equation}
where $Q_{\rm grav}=c^2z/(1+z)\approx GM_{\rm NS}/R_{\rm NS}$ is the specific accretion yield, and $M_{\rm NS}$, $R_{\rm NS}$ and $z$ are the mass, radius, and surface redshift for the neutron star, respectively. The nuclear burning yield $Q_{\rm nuc}$ depends primarily upon the fuel composition; for mean hydrogen fraction $\bar{X}$, $Q_{\rm nuc}\approx(1.35+6.05\bar{X})$~MeV/nucleon \cite[]{goodwin19a}.
We note that the measured $\alpha$ values as defined in equation \ref{eq:alpha} can only be related to the neutron-star parameters if the anisotropy of the persistent and burst emission is also taken into account (see \S~\ref{bolcor}).

\begin{figure}[ht]
\includegraphics[width=\columnwidth]{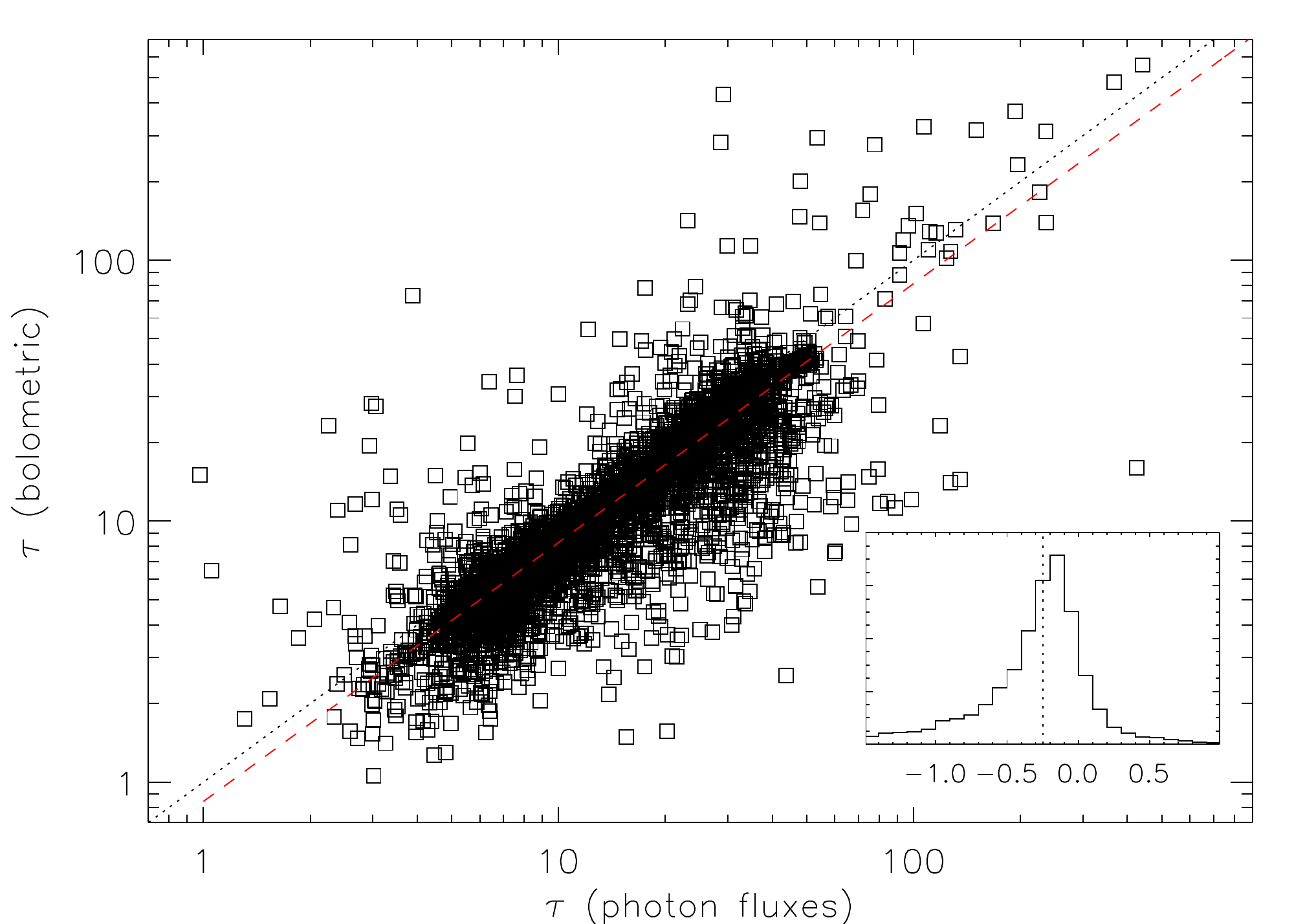}
\caption{
Comparison of 4282 $\tau$ values (ratio of burst fluence to peak flux) measured from the bolometric and photon fluxes. Note the relatively good average correspondence of the two values; the line of best-fit is overplotted ({\it dashed red line}). The inset shows the distribution of the fractional variation between the two quantities, demonstrating that the bolometric value is systematically lower than the value calculated from the photon fluxes, by a factor of 0.250.
\label{fig:tau-fit}}
\end{figure}

We measured the $\tau$ value for each burst, which is defined as the ratio of the fluence to the peak flux
(following \citealt{vppl88}; see also G08). Bolometric peak fluxes (and fluences) were only available for the bursts from \sax/WFC and \xte/PCA, so for the other bursts we calculated the ratio of the photon fluence and peak flux, as determined in \S\ref{sec:lcs:modeling}.
For those bursts where both measures are available, the quantities are reasonably consistent; see Fig. \ref{fig:tau-fit}. The $\tau$ value determined from the bolometric fluence and flux is systematically lower than the value from the photon fluence and flux, by a factor of 
0.250.
We carried out a linear fit to the logarithms of the two quantities, deriving a best-fit relationship of 
\begin{equation}
    \log\tau_b = (-0.165\pm0.005) + (0.9902\pm0.0018) \times \log\tau_p
    \label{eq:tau-fit}
\end{equation}
where $\tau_b$ and $\tau_p$ are the $\tau$-values calculated from the bolometric and photon flux quantities, respectively.
We then adopted the bolometric $\tau_b$ value for those 4283 bursts where it could be measured, and for the remainder, the value of $\tau_p$, corrected via the expression in equation \ref{eq:tau-fit}.

We note that since the quantities $\tau$ and $\alpha$ are both based on ratios of fluxes measured by the same instrument, they should be independent of any variations in the absolute calibration of the instruments (cf. with \S\ref{sec:crosscal}).

\subsection{Bolometric corrections and the estimated accretion rate}
\label{bolcor}
In order to estimate the accretion rate 
at the time of each burst,
we calculated for each observation the $\gamma$-parameter \cite[]{vppl88}. This quantity is normally defined as the ratio of the persistent flux $\fper$ to the mean peak flux of the Eddington-limited bursts from a given source. We describe in \S\ref{fluxEdd} how we identify Eddington-limited bursts, and how the average peak flux of these events $\langle F_{\rm Edd} \rangle$ is measured for each source. We adopted several corrections to this calculation in an attempt to improve the accuracy of the $\gamma$ values for MINBAR.
We calculated $\gamma_i$ for each observation $i$ as
\begin{equation}
 \gamma_i = 1.7\frac{c_{\rm bol}F_{{\rm per},i}}{\langle F_{\rm Edd} \rangle}
    \frac{\xi_p}{\xi_b}
\end{equation}
where 
$F_{{\rm per},i}$ is the persistent flux measured in the 3--25~keV band,
$\langle F_{\rm Edd} \rangle$ is the average Eddington flux adopted for the source, and 
$\xi_p$, $\xi_b$ are the model-predicted persistent and burst flux anisotropy factors.
We describe each of these corrections below.

First, we corrected for the systematic differences between the measured fluxes for the different instruments, as described in \S\ref{sec:crosscal}, normalising to the \xte/PCA.
The persistent flux  measured by the latter instrument may include contributions from multiple active sources in the field. In such cases, we adopt the observation-averaged flux value, but flag the corresponding entry to indicate the possibility (see \S\ref{minbar-obs}).

Second, we note that the persistent fluxes are calculated over the common instrumental band of 3--25~keV, while the burst fluxes are bolometric, estimated from the spectral fit parameters. Thus, we corrected each of the persistent flux values by 
a bolometric correction factor, chosen as follows. 
We calculated bolometric corrections for selected persistent spectra measured by \xte\/ in the MINBAR sample, partially following the approach of G08.
We chose representative, long-exposure observations for selected sources and carried out combined fits of each PCU spectrum (as described above) along with HEXTE spectra above 15~keV. We set the upper energy limit for each HEXTE spectrum 
at the maximum  to which the source could be detected (typically 40--80~keV). 
For observations best-fit already with Comptonization components, we used PCA data alone (since those data were sufficient to constrain the spectral turnover at a few times the plasma temperature $kT_e$) to fit the spectra and estimate the bolometric correction.

We fit the broadband spectra with a Comptonization continuum component attenuated by neutral absorption,  also for some observations with a Gaussian component representing fluorescent Fe K$\alpha$ emission around 6.4~keV. In {\sc xspec} we generated an idealized response using the {\tt dummyrsp} command, covering the energy range 0.1--200~keV with 200 logarithmically spaced energy bins,  and  integrated the model flux over this range. We  calculated the bolometric correction $c_{\rm bol}$ for each observation as the ratio between the 0.1--200 and 
3--25~keV fluxes measured from the broadband spectral fits. We did not correct for the neutral absorption. 
The error on the bolometric correction was estimated as the standard deviation of the derived correction over the active PCUs, where more than one was active. %\ek{what if 1 or 2 active PCUs?} 

\begin{deluxetable}{lrcc}
\tablecaption{Bolometric correction values adopted for different sources. 
Entries with no data for the number of corrections have only a single estimate
\label{tab:bolcorr} }
\tablecolumns{4}
\tablewidth{0pt}
\tablehead{ & \colhead{No.} & \colhead{No. of} & \\
\colhead{Source} & \colhead{bursts} & \colhead{corrections} & \colhead{$c_{\rm bol}$} }
\startdata
4U 0513$-$40              &   35 & \nodata & $1.47\pm0.02$ \\
EXO 0748$-$676            &  357 &  4 & $1.6\pm0.3$ \\
4U 0836$-$429             &   78 & \nodata & $1.82\pm0.02$ \\
4U 1254$-$69              &   34 &  3 & $1.30\pm0.15$ \\
4U 1323$-$62              &   99 &  3 & $1.65\pm0.05$ \\
Cir X-1                 &   14 &  4 & $1.12\pm0.06$ \\
4U 1608$-$522             &  145 &  4 & $1.59\pm0.13$ \\
4U 1636$$-$$536             &  664 & 44 & $1.51\pm0.12$ \\
XTE J1701$-$462           &    6 &  2 & $1.44\pm0.07$ \\
MXB 1658$-$298            &   27 & 17 & $1.32\pm0.05$ \\
4U 1702$-$429             &  278 & 11 & $1.4\pm0.3$ \\
4U 1705$-$44              &  267 &  7 & $1.51\pm0.15$ \\
XTE J1709$-$267           &   11 &  3 & $1.45\pm0.05$ \\
XTE J1710$-$281           &   47 & \nodata & $1.42\pm0.13$ \\
IGR J17191$-$2821         &    5 & \nodata & $1.36\pm0.04$ \\
XTE J1723$-$376           &   12 & \nodata & $1.05\pm0.02$ \\
4U 1728$-$34              & 1169 & 43 & $1.40\pm0.15$ \\
MXB 1730$-$335            &  126 & 33 & $1.30\pm0.05$ \\
KS 1731$-$260             &  366 &  6 & $1.62\pm0.13$ \\
4U 1735$-$444             &   71 & 10 & $1.37\pm0.12$ \\
XTE J1739$-$285           &   43 & \nodata & $1.30\pm0.06$ \\
SAX J1747.0$-$2853        &  113 & \nodata & $1.93\pm0.06$ \\
IGR J17473$-$2721         &   61 &  3 & $1.6\pm0.5$ \\
SLX 1744$-$300            &  303 &  4 & $1.45\pm0.14$ \\
GX 3+1                  &  201 &  2 & $1.458\pm0.008$ \\
IGR J17480$-$2446         &  303 & 34 & $1.21\pm0.02$ \\
EXO 1745$-$248            &   25 &  7 & $1.8\pm0.3$ \\
SAX J1748.9$-$2021        &   46 & 15 & $1.43\pm0.08$ \\
4U 1746$-$37              &   37 &  5 & $1.33\pm0.07$ \\
SAX J1750.8$-$2900        &   24 &  2 & $1.338\pm0.008$ \\
GRS 1747$-$312            &   21 & \nodata & $1.34\pm0.04$ \\
SAX J1806.5$-$2215        &    9 & \nodata & $1.30\pm0.05$ \\
GX 17+2                 &   43 &  9 & $1.35\pm0.10$ \\
4U 1820$-$303             &   67 &  2 & $1.45\pm0.17$ \\
GS 1826$-$24              &  454 & 13 & $1.66\pm0.11$ \\
Ser X-1                 &   55 & 15 & $1.45\pm0.08$ \\
Aql X-1                 &   96 &  7 & $1.65\pm0.10$ \\
XB 1916$-$053             &   36 & \nodata & $1.37\pm0.09$ \\
XTE J2123$-$058           &    6 &  2 & $1.35\pm0.06$ \\
Cyg X-2                 &   70 & 54 & $1.41\pm0.05$ \\
\hline
non-pulsar mean & & & $1.42\pm0.17$ \\
\hline
SAX J1808.4-3658        &   12 & \nodata & $2.14\pm0.03$ \\
XTE J1814-338           &   28 & \nodata & $1.86\pm0.03$ \\
\hline
pulsar mean & & & $2.00\pm0.2$ \\
\enddata
\end{deluxetable}

\begin{figure}[h]
	\includegraphics[width=\columnwidth]{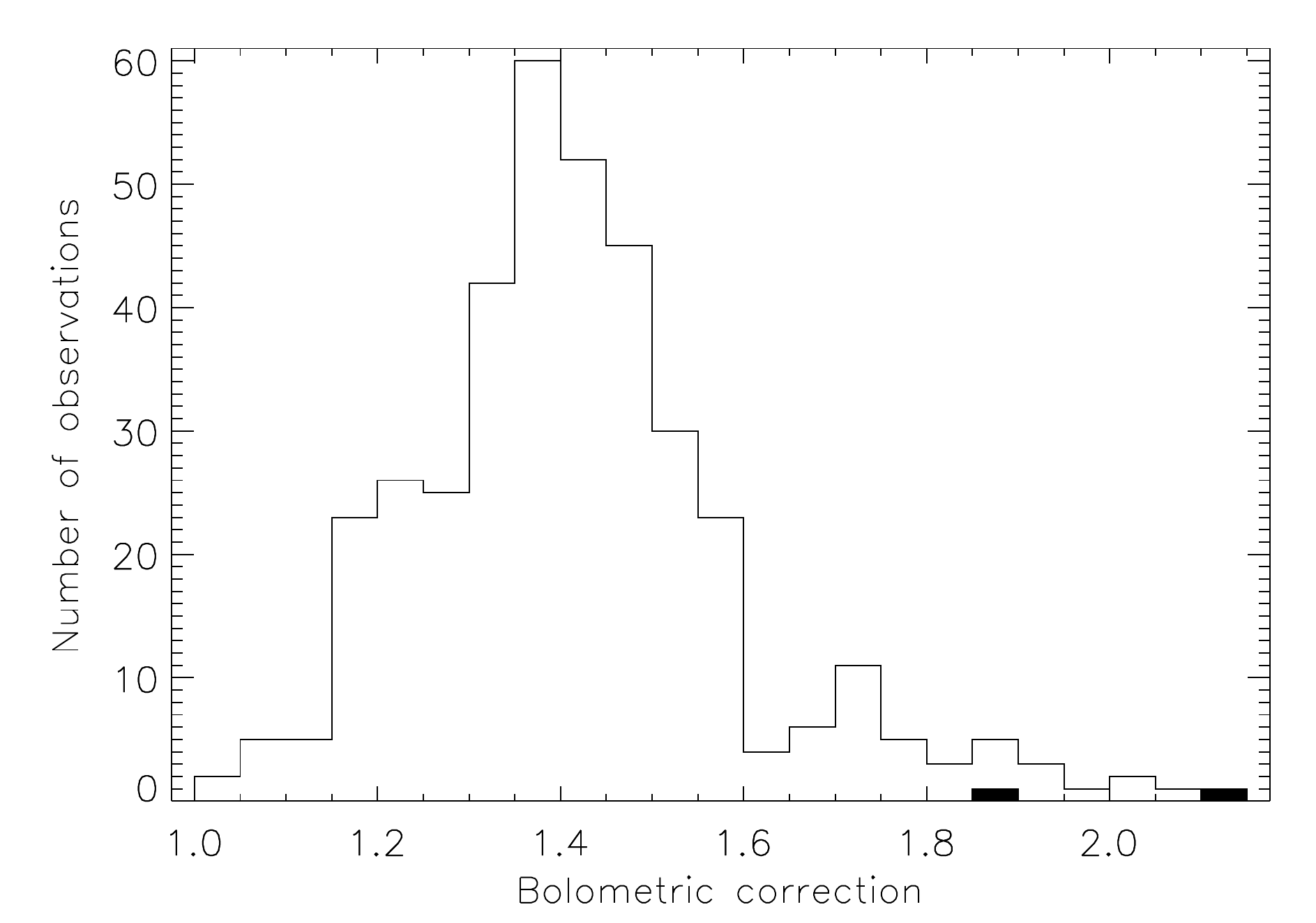}
	\caption{Distribution of bolometric corrections derived for selected 
	% \ek{(why these are selected? why not give it for all? why pulsars/non-pulsars as a group?)} 
	% they are selected because they were the ones that were done! (couldn't do them all, as yet). Pulsars and non-pulsars tend to have different values, hence the grouping - see the figure. - dkg
	\xte/PCA observations including bursts, over the MINBAR sample. The shaded region corresponds to the bolometric corrections for the accretion-powered millisecond pulsars (SAX~J1808.4$-$3658 and XTE~J1814$-$338).
	\label{fig:bcdist} }
\end{figure}

Our sample of bolometric corrections includes observations of 24 bursting sources as analysed by G08, 
augmented by observations of 
2 % as at 2018 Sep 9
additional sources (Table \ref{tab:bolcorr}). The values of $c_{\rm bol}$  varied between 1.05--2.14, depending upon the spectral shape (Fig. \ref{fig:bcdist}).

Where a bolometric correction was not available for the observation containing an individual burst, we adopted one of two sets of averages. % \ek{which sets are you talking about? pulsars/non-pulsars?}
If any bolometric correction estimates for that source was available (i.e. if the source was one of those listed in Table \ref{tab:bolcorr}), then we adopted the mean of those values.
In the absence of an observation or source specific measurement, we adopted the overall mean of 
$c_{\rm bol} = 1.4\pm0.3$
for the non-pulsing sources, and
$2.00\pm0.14$
for the pulsars (not including intermittent pulsars HETE~J1900.1-2455, Aql X-1 and SAX~J17498.9$-$2021). In that case we set the uncertainty on the bolometric correction attribute in the table to zero (column {\tt bce}; see \S\ref{minbar}).
The likely error introduced is thus no more than about $40$\%. 

The third additional factor introduced compared to the treatment of G08 is the adopted composition for the Eddington-limiting atmosphere. The available evidence suggests that even sources that accrete mixed H/He typically exhibit radius-expansion bursts that reach the Eddington limit for pure He material (\citealt{gal06a}; see also \citealt{bult19b}). Adopting the peak Eddington fluxes without correction would underestimate the accretion rate by a factor of 1.7, corresponding to the %composition jc190918
ratio of Eddington luminosities between mixed H/He ($X=0.7$) and pure He ($X=0$). Thus, we multiply the $\gamma$-values by an additional factor of 1.7.

Fourth, it is known that the accretion disk and mass donor intercept and reflect some fraction of the burst and persistent flux, such that the flux measured by a distant observer may be more or less than the isotropic value. The degree of enhancement (or reduction) of the flux is generally parameterised as $\xi_{\rm b,p}$ (where the subscript b indicates burst emission, and p persistent), with
\begin{equation}
L_{b,p} = 4\pi d^2\xi_{b,p}F_{b,p}
\end{equation}
\cite[e.g.][]{fuji88,ls85}.
Although the inclination values for the bursters in our sample are generally poorly constrained (except for the dipping sources), we may adopt the most likely value assuming an isotropic distribution (i.e. $P(i)\propto \cos(i)$).
For the non-dippers, which we assume $i<75^\circ$, the median expected value is 
$50^\circ$; 
for the dippers 
(EXO~0748$-$676, 4U~1254$-$69, 4U~1323$-$62, Cir~X$-$1,
UW~CrB, % missing from list; added 25.3.19
MXB~1658$-$298, XTE~J1710$-$281, MXB~1730$-$335, EXO~1745$-$248, 1A~1744$-$361, 4U~1746$-$37, GRS~1747$-$312, IGR~J17597$-$2201, GX~13+1, and
XB~1916$-$053)
and the eclipsing source
XTE~J2123$-$058
we adopt the median value for an isotropic distribution with $i>75^\circ$,
$i=82^\circ$ \cite[cf. with $75^\circ$ for EXO 0748$-$676;][]{parmar86}.
All these sources consistently exhibit dips when active; we note that one additional source in our sample, Aql~X-1, has shown intermittent dips in \xte\/ observations \cite[]{gal16a}. While the intermittent dipping activity suggests the inclination in that system might be intermediate between the non-dippers and dippers, 
dynamical constraints from measurements of absorption features attributed to the companion suggest instead an inclination $<47^\circ$ \cite[]{mata17}.
Thus,
we group it with the non-dippers and adopt $i=50^\circ$.

We calculate the anisotropy corrections from  models\footnote{
The models assume that the accretion disk extends to the NS surface,
which may not be the case in the ``low hard'' state, when the disk is thought to be interrupted above the surface \cite[e.g.][]{done07}. The disk may also be temporarily interrupted during a burst even in the ``high soft'' state, due to Poynting-Robertson effects \cite[]{fragile20}. Thus, the true value of the anisotropy corrections may be systematically different from those we assume.} of the burst and persistent emission by \cite{he16}, as 
$(\xi_b, \xi_p)$ = (0.898, 0.809) 
for the non-dippers, and 
$(\xi_b, \xi_p)$ = (1.639, 7.27) % for i=82 deg
for the dippers. Since $\gamma$ is the ratio of the persistent flux to the burst peak flux, we multiply by the ratio $\xi_p/\xi_b$, which is 0.90 for the non-dippers, and 
4.43 
for the dippers. 

We expect that 
$\gamma_i$ will be approximately proportional to $\dot{m}/\dot{m}_{\rm Edd}$, but we acknowledge that the approximate nature of these corrections likely introduces some error, and also that there may be other sources of systematic error, for example changing radiation efficiency as a function of source and/or accretion rate \cite[cf. with][]{gal18a}.

\subsection{Spectral colours and the $S_\text{z}$ diagram}
\label{sec:colour}

\begin{figure}[h!t]
	\includegraphics[width=\columnwidth]{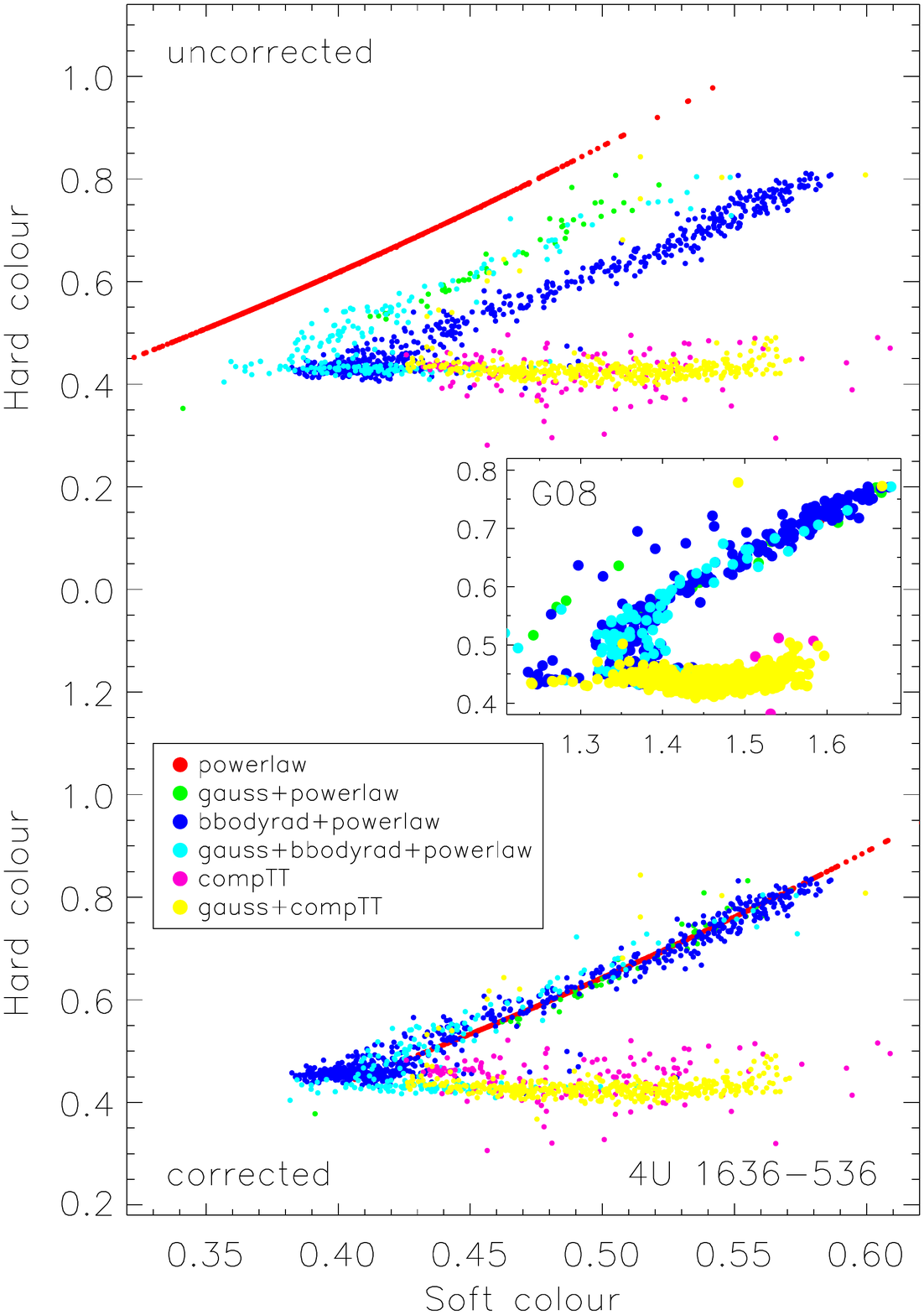}
	\caption{Comparison of the ``raw'' (uncorrected) spectral colours calculated from the best-fit spectral model ({\it top panel}), with the corrected values ({\it bottom panel}). 
	Each point represents the averaged colours over an observation, with the spectral model indicated by the colour. 
	Note the displacement  in the colour-colour tracks for observations with different spectral models.
	The corrections to the colours for each model are as given by Table \ref{tab:col_corr}.
	The inset shows the corresponding (instrumental) colours for the source, derived by G08.
	\label{fig:col_corr} }
\end{figure}

As with the flux measurements, we seek to correct the measured spectral colours (calculated as described in \S\ref{colours}), for any systematic variation arising from the instruments. Our choice of defining the colours based on the integrated model fluxes (rather than X-ray counts) is intended to correct for this kind of variation, but unfortunately we introduce a different issue, related to the spectral model adopted. This issue is illustrated in Fig. \ref{fig:col_corr}, which shows the distribution of ``raw'' spectral model colour values for the best-observed source, 4U~1636$-$536. 

\begin{figure*}[h!t]
	\includegraphics[width=\textwidth]{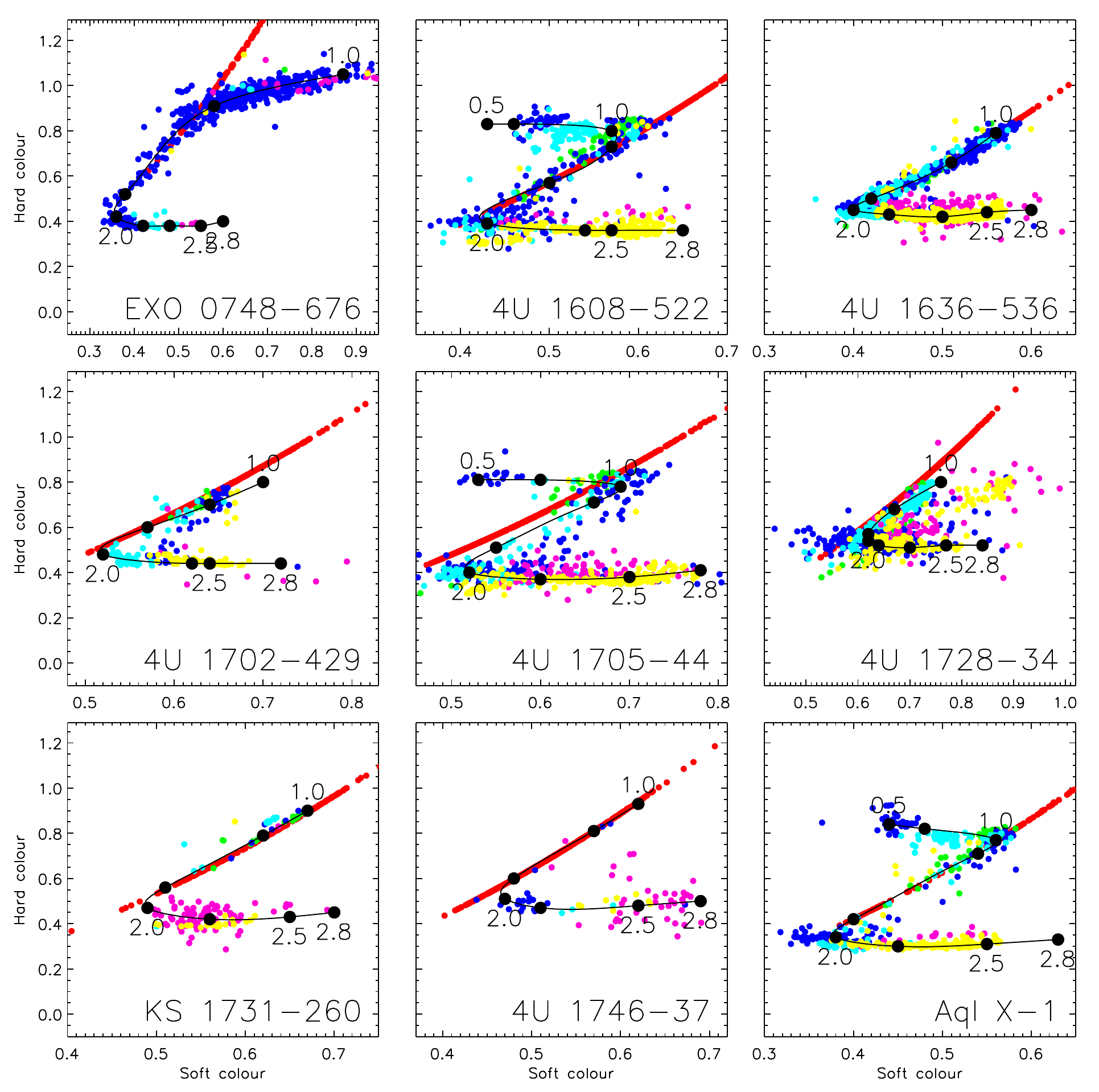}
	\caption{Colour-colour diagrams for selected sources from the MINBAR observation sample, chosen based on a large number of observations spanning a wide range of persistent spectral states. Within each panel, each  symbol represents the average over a single observation, with the colour indicating the adopted spectral model (colours as for Fig. \ref{fig:col_corr}). The colour measurements are corrected as described in \S\ref{sec:colour}, excluding EXO~0748$-$676.
	The loci by which the position in the colour-colour diagram (the $S_\text{z}$ parameter) is determined, is overplotted in each pane ({\it black symbols and lines}). Key $S_\text{z}$ values are marked.
	\label{fig:col_all} }
\end{figure*}

We adopted small corrections to the soft and hard colours, designed to align the different tracks in the colour-colour diagram for the uncorrected (``raw'') values (Fig. \ref{fig:col_corr}, top panel). 
The corrections are intended to align with the model adopted for the highest signal-to-noise observations, the {\tt gauss+compTT} model.
The corrections for the other model choices are listed in Table \ref{tab:col_corr}.
While the offsets were determined visually from inspection of the colour-colour diagram for 4U~1636$-$536, we checked that they also provided adequate corrections for the other sources. 
The sole exception was EXO~0748$-$676, for which the bulk of the observations were fit with a powerlaw or blackbody+powerlaw, and for which the overlap of the observations in the colour-colour space was best {\it without} the correction derived for 4U~1636-536. We speculate that the discrepancy for EXO~0748$-$676 is related to the high system inclination; it alone, amongst the sources for which a colour-colour diagram could be plotted, exhibits ``dips'' and eclipses once each orbital period, indicative of high system inclination \cite[]{parmar86}.
We then applied these corrections to all the colours for each of the 8 sources with well-defined colour-colour diagrams (excluding EXO~0748$-$676), as shown in Fig. \ref{fig:col_all}.

\begin{deluxetable}{lcc}
\tablecaption{Corrections to spectral colours (for sources excluding EXO~0748$-$676; see \S\ref{sec:colour}) as a function of adopted spectral model
  \label{tab:col_corr}
}
\tablewidth{0pt}
\tablehead{
 \colhead{Spectral model}
 & \colhead{Soft colour}
 & \colhead{Hard colour}
}
\startdata
{\tt bbodyrad+powerlaw }       & +0.000 & +0.025 \\
{\tt compTT  }                 & +0.000 & +0.025 \\
{\tt gauss+bbodyrad+powerlaw } & +0.025 & +0.000 \\
{\tt gauss+powerlaw }          & +0.050 & +0.025 \\
{\tt powerlaw }                & +0.100 & +0.025 \\
\enddata
\end{deluxetable}

We note that the corrected colours adopted for the observations in MINBAR
appear to provide a sharper 
delineation between the ``soft'' and ``hard'' tracks (traditionally referred to as ``island'' and ``atoll'', respectively) compared to colours derived from the counts (e.g. G08). This contrast is  illustrated in Fig. 
\ref{fig:col_corr}, 
which compares the MINBAR colour track for 4U~1636$-$536 against the data available from the G08 catalog (inset).

\begin{figure*}[h!t]
	\includegraphics[width=\textwidth]{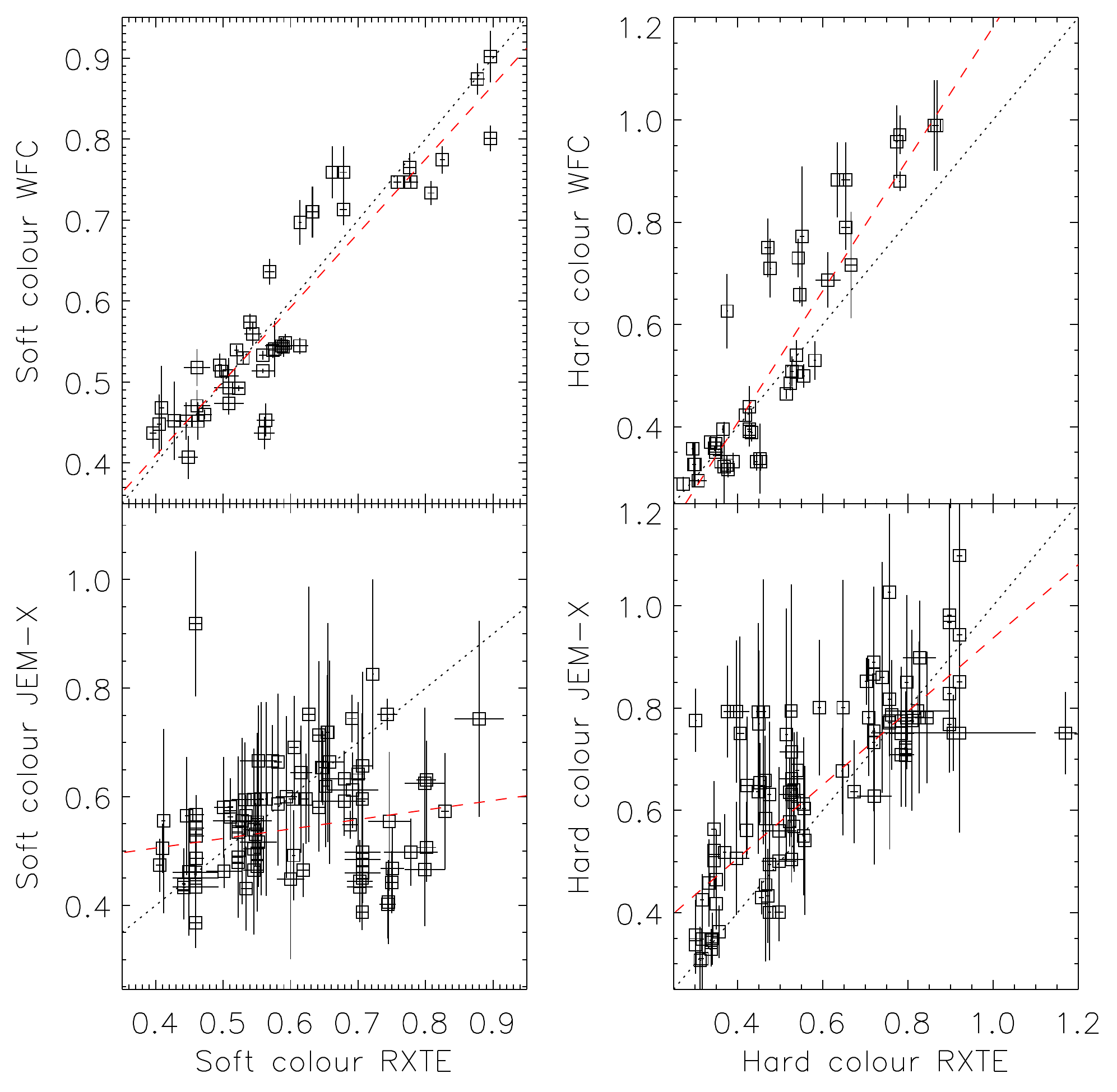}
	\caption{Comparison of soft ({\it left panels}) and hard ({\it right}) colours measured by different instruments, in overlapping observations. The top row of plots shows the comparison between \sax/WFC and \xte/PCA, while the bottom row shows the comparison between \igr/JEM-X and \xte/PCA.
	The best-fit linear relation for each pair of measurements is overplotted ({\it dashed red line}); the fit parameters are listed in Table \ref{tab:crosscal}.
	\label{fig:col_cross} }
\end{figure*}

As with the other quantities for which we establish cross-correlation relations between pairs of instruments, we also compared the soft and hard colours. In Fig. \ref{fig:col_cross} we plot colours measured independently with \sax/WFC and \igr/JEM-X, against those measured by \xte/PCA, for overlapping observations (following the approach adopted in \S\ref{sec:crosscal}).

For most combinations we find a reasonable correlation, with the lines of best fit not deviating overmuch from the 1:1 ideal. The exception is for the soft colour measured by JEM-X, which has a line of best fit with slope $0.18\pm0.10$. Inspection of Fig. \ref{fig:col_cross} suggests the correspondence is not universally as poor as suggested by this value; the colours agree reasonably well on average in the range 0.4--0.65, but a group of observations with soft colour in the range 0.7--0.8 measured by PCA, instead have low colour values in the range 0.4--0.5 with JEM-X. In any case, the primary discriminator of spectral state is the hard colour, which (on average) is much better correlated between the two instruments.
We list the correlation coefficients for each pair of parameters also in Table \ref{tab:crosscal}.

Following G08, we also attempted to specify a quantity parameterising the position of each observation within the colour-colour diagram, referred to as $S_\text{z}$. We took the approach of defining a track which followed the shape of the colour-colour diagram for each source, with 1--2 vertices defined at points where the track changes direction. The example for 4U~1636$-$536 shown in Fig. 
\ref{fig:col_all}, top-right panel,
is anchored by the vertices at $S_\text{z}=1$ and 2, defined, respectively, as the maximum extent of the ``island'' track to large values of the hard colour; and the transition between the hard ``island'' and soft ``banana'' branch.
Generally only the PCA observations were sufficient to precisely determine the spectral colours, so we prioritised sources with many detections with that instrument. This parameterisation was, furthermore, only possible for those sources where there were sufficient high-signal observations covering a range of spectral states. Notable exceptions include 4U~0513$-$40, the accretion-powered pulsars SAX~J1808.4$-$3658 and HETE~J1900.1$-$2455 (with their consistently hard spectra), and the Z-sources Cyg~X-2 and Cir~X-1 (with their consistently soft spectra).
Ultimately we defined the $S_\text{z}$ value for 9 sources: 
EXO~0748$-$676, 4U~1608$-$522, 4U~1635$-$536, 4U~1702$-$429, 4U~1705$-$44, 4U~1728$-$34, KS~1731$-$260, 4U~1746$-$37 and Aql~X-1 (Fig. \ref{fig:col_all}).
These values are stored in the {\tt s\_z} column of the observation table (see \S\ref{minbar-obs}). For each entry in the burst table we copied the $S_\text{z}$ value (along with the soft and hard colours) from the host observation through to the burst table (\S\ref{minbar}).

We also checked the $S_\text{z}$ values determined for MINBAR against the earlier values calculated by G08. 
We broadly find a good 1:1 correspondence of the values, although with some scatter about the line (typical RMS values of 0.1), and with larger deviations notably for observations fit with power law models alone.

\section{The MINBAR sample}
\label{minbar}

Data release 1 of MINBAR consists of \ubursts\ unique bursts, detected within \observations\ observations of \minbarsources\ burst sources 
made
between February 8, 1996, and May 3, 2012.

The complete MINBAR catalog consists of 
four tables:
\begin{itemize}
\item {\tt sources} lists all the known burst sources, and relevant properties, as described in \S\ref{sec:sources};
\item {\tt minbar} listing properties of each analysis of each burst observed independently by each instrument;
\item {\tt minbar-osc} giving the timing properties of \xte/PCA bursts from those sources which have exhibited burst oscillations, as described in \S\ref{osc_table}.
\item {\tt minbar-obs} listing properties and analysis results from each separate observation of each burst source, as described in \S\ref{minbar-obs}
\end{itemize}
The source table is provided in FITS format, and the other tables in ASCII.

The burst table contains analysis results for every \xte/PCA,  \sax/WFC and \igr/JEM-X burst  from the  sources described in \S\ref{sec:sources}, detected in the observations making up the observation table (\S\ref{minbar-obs}).
As we list each detected event separately, and 
28
bursts were detected by more than one instrument (see \S\ref{crosscal:bursts}), the burst table includes \bursts\ entries.

The source, burst and observation tables are also available via a web interface\footnote{\url{http://burst.sci.monash.edu}}. 
The complete list of observations from which the MINBAR sample was drawn (including non-detections) is only available via the web interface.
The sample of bursts that is queried via the web interface includes 
approximately 1600 
additional events not provided in the ASCII tables; these include 
type-II events (from the Rapid Burster, observed with the \xte/PCA), 
and events initially identified as burst candidates, but subsequently rejected due to the lack of evidence for cooling, or identification with other mechanisms. 
The selection criterion to retrieve only the events also found in the ASCII table is {\tt type=1}. 
\added{We also provide the burst lightcurves and the time-resolved spectroscopic analysis results via the web interface. }

Similarly, the observation table queried via the web interface also includes observations in which the target (or any other source in the FOV, in the case of \xte/PCA) is not detected above our significance threshold; 
\deleted{duplicate entries, added in error;}
and observations from sources in which bursts have not been detected by any of the instruments analysed here, earlier than the cutoff date. The selection criteria to retrieve only the events also found in the ASCII observations table is {\tt \deleted{flag>=0,}sig>=3}.

\subsection{Table format}
\label{subsec:minbar}

The burst table columns are listed in Table \ref{minbar-table}. Below we describe in more detail how the column entries relate to the analysis in \S\ref{sec:analysis}.

% moved in response to the proof Q5 to ensure correct numbering
\setcounter{magicrownumbers}{-45}
\begin{deluxetable*}{ccccp{7cm}}
\tablewidth{480pt}
\tabletypesize{\scriptsize}
\tablecaption{Burst table columns, formats and description
  \label{minbar-table}}
\tablehead{
& \colhead{Web table} & \colhead{ASCII table} \\
\colhead{Column} & \colhead{attribute} & \colhead{Format} & \colhead{Units} &
\colhead{Description}  }
\startdata
\rnum & {\tt name} & A23 & & Likely burst origin \\ % name
\rnum & {\tt instr} &  A3 & &  Instrument label \\ % instr 
\rnum & {\tt obsid} & A20 & & Observation ID \\ % obsid
\rnum & {\tt time} & F11.5 & d & Burst start time (MJD UT)  \\ % time   
\rnum & {\tt entry} & I4 & & MINBAR burst ID \\
\rnum & {\tt entry\_obs} & I6 & & MINBAR observation ID in which this burst falls \\ 
\rnum & {\tt bnum} & I3 & & Order of the event within the observation \\
\rnum & {\tt xref} & I3 & &  Burst ID in external catalog (G08 or C17) \\
\rnum   & {\tt mult} & I1 & &       Number of MINBAR instruments which detected this event \\ % mult  
\rnum & {\tt angle} &   F6.2 & arcmin & Angle between the source position and pointing axis \\ % angle
\rnum & {\tt vigcorr} & F5.3 & & Vignetting correction factor \\ % vigcorr
\rnum & {\tt sflag} & A11 & & Data quality/analysis flags \\ % sflag 
\rnum & {\tt rexp} &    A1  & &  Photospheric radius expansion flag\\ % rexp; convert from integer
\rnum & {\tt rise} &    F5.2 & s & Rise time \\ % rise
\rnum & {\tt tau} &    F5.1 & s & Ratio of fluence to peak flux, $\tau=E_b/F_{\rm peak}$ \\ % tau 
\rnum & {\tt taue} &   F5.1 & s & Uncertainty on $\tau$ \\ % % taue 
\rnum & {\tt dur} & F6.1 & s & Burst duration \\ % dur
\rnum & {\tt dure} & F6.1 & s & Uncertainty on burst duration \\ % dure
\rnum & {\tt edt} & F6.1 & s & Exponential decay timescale \\ % edt 
\rnum & {\tt edte} & F7.3 & s & Uncertainty on exponential decay timescale \\ % edte
\rnum & {\tt tdel} &    F8.1 & hr & Time since previous burst from this source \\ % tdel
\rnum & {\tt trec} &    F7.1 & hr & Inferred recurrence time $T_{\rm rec}$ \\ % trec; no error?
\rnum & {\tt perflx} &  F6.3 & $10^{-9}\ \epcs$ & Persistent 3--25~keV flux prior to the burst, $F_{\rm per}$ \\ % perflx
\rnum & {\tt perflxe} & F5.3 & $10^{-9}\ \epcs$ &  Uncertainty on persistent flux \\ % perflxe
\rnum & {\tt alpha} &   F6.1 & &  Ratio of integrated persistent flux to burst fluence, $\alpha$ \\ % alpha
\rnum & {\tt alphae} &  F6.1  & & Uncertainty on $\alpha$ \\ % alphae
\rnum & {\tt bc} &      F5.3 & & Bolometric correction adopted for persistent flux \\ % bc
\rnum & {\tt bce} &     F5.3 & & Uncertainty on bolometric correction \\ % bce
\rnum & {\tt gamma} &   F6.4 & & Ratio of persistent flux to peak PRE burst flux, $\gamma$ \\ % gamma
\rnum & {\tt sc} &      F6.3 & &  Soft colour \\ % sc
\rnum & {\tt hc} &      F6.3 & & Hard colour \\ % hc
\rnum & {\tt s\_z} &     F6.3 & & Position on colour-colour diagram, $S_\text{z}$ \\ % s_z
\rnum & {\tt pflux} &   F6.2 & count~s$^{-1}$~cm$^{-2}$ &  Peak photon flux \\ % pflux
\rnum & {\tt pfluxe} &   F5.2 & count~s$^{-1}$~cm$^{-2}$ & Uncertainty on peak photon flux \\ % pfluxe
\rnum & {\tt fluen} &   F8.3 & count~cm$^{-2}$& Integrated photon flux \\ % fluen
\rnum & {\tt fluene} &   F7.3 & count~cm$^{-2}$& Uncertainty on integrated photon flux \\ % fluene
\rnum & {\tt bpflux} &  F6.2 & $10^{-9}\ \epcs$ & Bolometric peak flux $F_{\rm peak}$ \\ % bpflux
\rnum & {\tt bpfluxe} &  F5.2 & $10^{-9}\ \epcs$ &  Uncertainty on bolometric peak flux \\ % bpfluxe
\rnum & {\tt kT} & F4.2 & keV & Blackbody temperature $kT$ at burst peak \\ % kT      
\rnum & {\tt kTe} & F4.2 & keV & Uncertainty on $kT$ at burst peak \\ % kTe    
\rnum & {\tt rad} &     F6.1 & ${\rm km}/10\,{\rm kpc}$ & Blackbody normalisation at burst peak \\ % rad
\rnum & {\tt rade} &    F5.1 & ${\rm km}/10\,{\rm kpc}$ &  Uncertainty on blackbody normalisation at burst peak \\ % rade
\rnum & {\tt bfluen} & F6.4 & $10^{-6}\ \epc$ & Bolometric fluence (integrated bolometric flux) $E_b$\\ % bfluen
\rnum & {\tt bfluene} & F6.4 & $10^{-6}\ \epc$ & Uncertainty on bolometric fluence \\ % bfluene
\rnum & {\tt refs} & A20 & &  References for the burst \\
\enddata
\end{deluxetable*}

\paragraph{1. Likely burst origin} ({\tt name} in the web table) The adopted origin for the burst.
For the imaging instruments, the origin can be determined unambiguously, except for  pairs of close sources 
(see \S\ref{saxobs} and \S\ref{jemxobs}). 
For \xte/PCA, where the FOV covers more than one source, the origin is assigned by matching the observed properties of the burst with the known source behaviour, following G08.

\paragraph{2. Instrument label} ({\tt instr}) The instrument label is encoded as a three-character string. The first two characters correspond to the satellite and instrument,  
i.e. 
\begin{itemize}
\item {\tt XP}: \xte/PCA
\item {\tt IJ}: \igr/JEM-X
\item {\tt SW}: \sax/WFC
\end{itemize}
The third character corresponds to the camera number (for the WFC and JEM-X; see \S\ref{obs-sax} and \S\ref{obs-jemx},  respectively). 
For JEM-X observations later in the mission, both instruments were active; these are indicated by instrument code {\tt IJX}, and the provided attributes are an average over the results for the two cameras individually 
(see \S\ref{obsanl}).
For the PCA, the third character encodes the number of PCUs active, with the possible values listed in Table \ref{pca-code-table}. 

\begin{deluxetable}{cl|cl}
\tablewidth{0pt}
\tablecaption{\xte/PCA instrument codes
  \label{pca-code-table}}
\tablehead{
\colhead{Label} & \colhead{PCUs active} & \colhead{Label} & \colhead{PCUs active} }
\startdata
0 & 0 &    h & 0, 1, 3 \\
1 & 1 &    j & 0, 2, 3 \\   
2 & 2 &    k & 1, 2, 3 \\
3 & 3 &    o & 0, 1, 4 \\
4 & 4 &    q & 0, 2, 4 \\
b & 0, 1 & r & 1, 2, 4 \\
c & 0, 2 & u & 0, 3, 4 \\
d & 1, 2 & v & 1, 3, 4 \\
f & 0, 3 & x & 2, 3, 4 \\
g & 1, 3 & l & 0, 1, 2, 3 \\
i & 2, 3 & s & 0, 1, 2, 4 \\
m & 0, 4 & w & 0, 1, 3, 4 \\
n & 1, 4 & y & 0, 2, 3, 4 \\
p & 2, 4 & z & 1, 2, 3, 4 \\
t & 3, 4 & a & 0, 1, 2, 3, 4 \\
e & 0, 1, 2 \\
\enddata
\end{deluxetable}

\paragraph{3. Observation ID}
({\tt obsid}) 
The identifier for each observation is specified by the instrument's science team. For \sax\/, this attribute corresponds to the observation period ('OP') which identifies a contiguous observation with a constant pointing.

For \igr, each entry in the observation table corresponds to a science window each with a unique observation ID. 
This attribute is a 12-digit number  of the  form {\tt  RRRRPPPPSSSF}, 
where
{\tt RRRR} is the revolution number of the S/C as defined from perigee passage;
{\tt PPPP} is the pointing number within the revolution (reset
      to {\tt 0000} when the revolution number increments;
{\tt SSS} is the subdivision number, 
beginning at {\tt 001} and 
resetting on each new pointing; and
{\tt F} is the type identifier of the science window, with allowed values of
      0 (``pointing''), 1 (``Slew''), and 2 (``Engineering'').
For the observations included in the sample here, we selected only the ``Pointing'' type ({\tt F=0}).

For \xte, the observation ID is of the form {\tt  NNNNN-TT-VV-SS[X]}
where:
{\tt NNNNN} is the five-digit proposal number assigned by the guest observer facility (GOF);
{\tt TT} is a two-digit target number,
which may be zero if there was only one target for the proposal;
{\tt VV} is the two-digit viewing number, assigned by GOF, which tracks the number of scheduled visits (epochs) for each target;
{\tt SS} is the two-digit sequence number used for identifying different pointings that make up the same viewing (if the 
viewing was further split into more than one interval);
and 
{\tt X} the optional 15th character, which when present, indicates:
{\tt S}       ``raster'' scan observation or
{\tt R}       ``raster'' grid observation.

We caution that, for some bursts detected by \xte/PCA, the burst may actually occur in the slew before or after the observation, in which case the corresponding  dataset on the archive will be labeled with an additional {\tt -A} or {\tt -Z}. Furthermore, some longer ($>8$~hr) observations are split into multiple obsids, labeled with an extra digit ({\tt -0}, {\tt -1} etc), which include the FITS data actually covering the burst.

\paragraph{4. Burst start time } ({\tt time}) The burst start time, in MJD UT, as defined in \S\ref{sec:lcs:modeling}.

\paragraph{5. MINBAR burst ID} ({\tt entry}) The unique identifier for each burst in the MINBAR sample.
The ordering of this identifier is arbitrary, based primarily on the history of burst assembly.

\paragraph{6. MINBAR observation ID} ({\tt entry\_obs}) The unique identifier of the observation in the observation table (see \S\ref{minbar-obs}) in which this burst was detected.

\paragraph{7. Order of the event within the observation } ({\tt bnum}) The ranking in time order of this event in the entire observation, irrespective of the origin. For \sax\/ and \igr, the ranking includes each burster in the FOV; 
the ranking may be incidentally out of time order. Additionally, for observations covering the Rapid Burster, the order is determined including type-II events, which are otherwise not part of the MINBAR sample. 

\paragraph{8. Burst ID in external catalog} ({\tt xref}) This attribute is the corresponding entry value in the catalog of bursts detected with \xte/PCA \cite[]{bcatalog}, or with \igr/JEM-X \cite[]{chelov17} 

\paragraph{9. Number of MINBAR instruments which detected the event} ({\tt mult}) There 
are \burstdupes\
bursts detected simultaneously by more than one instrument (see \S\ref{sec:crosscal}). For these events, we set this attribute to 2. For other bursts, it is 1.

\paragraph{10. Angle between the source position and the pointing axis} ({\tt angle}) 
Generally the angle (and the corresponding vignetting correction) will be identical to that for the host observation (see \S\ref{minbar-obs}), but may vary (for example, in the case of \xte\/ observations which include multiple sources in the FOV, or for which the pointing is not constant during the observation).

\paragraph{11. Vignetting correction factor} ({\tt vigcorr}) The factor assumed in the analysis by which the count rate and other quantities are scaled to take into account the instrumental vignetting (see \S\ref{sec:burst-lightcurves}), as for the observation table (\S\ref{minbar-obs}) 

\paragraph{12. Data quality/analysis flags} ({\tt sflag}) Indicates a number of sub-optimal situations for the data analysis, as described in Table \ref{burst-analysis-flag-table}. 

\begin{deluxetable*}{ccp{12cm}}
\tablecaption{Analysis flags relevant to MINBAR bursts
  \label{burst-analysis-flag-table}}
\tablehead{
\colhead{Label} & \colhead{Instrument} & \colhead{Description} }
\startdata
- & all & No significant analysis issues \\
a & PCA & The burst was observed during a slew, and thus offset from the source position.; fluxes and fluence have been scaled by $1/(1-\Delta\theta)$ \\ % flag +1
b & PCA & The observation was offset from the source position; flux and fluence have been adjusted via setting the source position for response matrix generation \\ % flag +2
c & PCA & The origin of the burst is uncertain; the burst may have been from another source in the field of view. If the origin is not the centre of the FOV, the flux and fluence have been adjusted by calculating the response for the assumed source position \\ % flag +4
d & PCA, WFC & Buffer overruns (or some other instrumental effect) caused gaps in the high time resolution data, affecting the time-resolved spectroscopic analysis \\ % flag +8
e & PCA, WFC & The burst was so faint that only the peak flux could be measured, and not the fluence or other parameters; or, alternatively, that the burst was cut off by the end of the observation, so that the fluence is an underestimate \\ % flag +16
f & PCA & An extremely faint burst or possibly problems with the background subtraction, resulting in no time-resolved spectral fit results \\ % flag +32
g & all & The full burst profile was not observed, so that the event can be considered an unconfirmed burst candidate. Typically in these cases the initial burst rise is missed, so that the measured peak flux and fluence are lower limits only. Also includes long bursts observed with \igr/JEM-X spanning multiple science windows (observations) \\ % flag +64
h & PCA & High-time resolution modes don't cover burst, preventing any time-resolved spectroscopic results and oscillation search % flag +128 %jc190918
\enddata
\end{deluxetable*}

\paragraph{13. Photospheric radius-expansion flag} ({\tt rexp}) This attribute indicates the presence of photospheric radius expansion (PRE), 
determined as described in \S\ref{sec:prebursts}.
The possible values are 2.0 (1.0), indicating confirmed presence (absence); a value in the range $(1.0,2.0)$, specifying the probability $p_i={\tt rexp}-1$ (according to a machine-learning classification scheme) that the burst exhibits radius-expansion; 3.0, indicating marginal evidence; -1.0, indicating insufficient data to assess.

\paragraph{14. Rise time} ({\tt rise}) %, risee}) 
The burst rise time (in seconds) estimated from the lightcurve analysis, as described in \S\ref{sec:lcs:modeling}.

\paragraph{15 \& 16. Ratio of fluence to peak flux, $\tau$} ({\tt tau, taue}) 
This quantity is a measure of the burst timescale, $\tau = E_b/F_{\rm peak}$ \cite[following][]{vppl88}, and the estimated uncertainty, as calculated in \S\ref{sec:trec}.

\paragraph{17 \& 18. Burst duration} ({\tt dur, dure}) The approximate duration of the burst, and its uncertainty (see \S\ref{sec:lcs:modeling}).

\paragraph{19 \& 20. Exponential decay timescale} ({\tt edt, edte}) The decay timescale and uncertainty (in seconds) for an exponential fit to the intensity lightcurve (see \S\ref{sec:lcs:modeling}).

\paragraph{21. Time since previous burst} ({\tt tdel}) The elapsed time $t_{\rm sep}$ in hours since the previous burst from this source (see \S\ref{sec:trec}).
This attribute is zero for the earliest burst from each source present in the MINBAR sample.

\paragraph{22. Inferred recurrence time} ({\tt trec}) The recurrence time in hours inferred for the burst. This quantity may be 
shorter than
the elapsed time since the previous bursts, in cases where we infer a steady recurrence time (with undetected bursts falling in data gaps; see \S\ref{sec:trec}). % \dg{should really also have an error}

\paragraph{23 \& 24. Pre-burst persistent flux} ({\tt perflx, perflxe}) The estimated persistent flux and uncertainty immediately prior to the burst.
For \xte/PCA and \igr/JEM-X this value is identical to that measured for the entire observation, 
but for \sax/WFC we estimate fluxes from spectra extracted over shorter intervals, as described in \S\ref{saxtrse}.
For some bursts, the persistent emission is undetectable; we flag these cases by setting the uncertainty to $-1$, in which case the provided value is the estimated $3\sigma$ upper limit.

\paragraph{25 \& 26. Ratio of integrated persistent flux to burst fluence, $\alpha$} ({\tt alpha, alphae}) This quantity is calculated as $\alpha=\Delta t F_{\rm per}c_{\rm bol}/E_b$
depends on the inferred recurrence time $\Delta t$ (column 22) as well as the persistent flux $F_{\rm per}$ (column 23) and the bolometric burst fluence, $E_b$ (column 43), and also incorporates the bolometric correction factor $c_{\rm bol}$ (column 27). % changed from 30 -- jc20181123

\paragraph{27 \& 28. Bolometric correction adopted for persistent flux} ({\tt bc, bce}) The estimated correction factor $c_{\rm bol}$ (and uncertainty) by which the 3--25~keV persistent flux needs to be multiplied for the best estimate of the bolometric flux.
Where the error ({\tt bce}) is zero, the value adopted is the mean over all other measurements for that source (if any are available), or the mean over all sources of the same class, as described in \S\ref{bolcor}. % \dg{does this mean that in these 'zero error cases' no error propagation is done for $c_{bol}$ in the $\alpha$  calculation? --cs200120}

\paragraph{29. Ratio of persistent flux to peak PRE burst flux, $\gamma$} ({\tt gamma}) The ratio of the estimated bolometric persistent flux to the average Eddington flux from the source % \notitie{if observed -- jc} 
(from Table \ref{table:peakfluxes}, where available); after \cite{vppl88}.
We adopt the average persistent flux for the host observation, taken from the observation table (see \S\ref{minbar-obs}), rather than the {\tt perflx} value (column 23, see above). The $\gamma$ value for the burst is thus identical to that for the host observation.
The $\gamma$ value also takes into account the bolometric correction (specific to the observation or source, where available) and the best-guess correction for the system anisotropy, as described in \S\ref{bolcor}.

\paragraph{30 \& 31. Soft \& hard spectral colour} ({\tt sc, hc}) The soft and hard spectral colours calculated over the entire observation, as described in \S\ref{colours}; these attributes are duplicated from the host observation in the observation table (\S\ref{minbar-obs}). 

\paragraph{32. Position on colour-colour diagram $S_\text{z}$} ({\tt s\_z}) This attribute is also calculated from the observation table, and is copied here.

\paragraph{33 \& 34. Peak photon flux} ({\tt pflux, pfluxe}) The peak photon flux and uncertainty, calculated from the count rate rescaled by the adopted instrumental effective area (see \S\ref{sec:lcs:modeling}).

\paragraph{35 \& 36. Integrated photon flux} ({\tt fluen, fluene}) The integrated photon flux over the burst duration. This quantity is expected to be approximately proportional to the bolometric fluence.

\paragraph{37 \& 38. Bolometric peak flux} ({\tt bpflux, bpfluxe}) The estimated peak bolometric flux of the burst, based on the parameters determined from time-resolved spectroscopy (columns 37--44 are only present for sufficiently bright bursts observed with \xte/PCA and \sax/WFC).

\paragraph{39 \& 40. Blackbody temperature at burst peak} ({\tt kT, kTe}) The best-fit value of the blackbody temperature $kT$ and its uncertainty, in keV, for the spectrum with maximum bolometric flux.

\paragraph{41 \& 42. Blackbody normalisation at burst peak} ({\tt rad, rade}) The square root of the best-fit value of the blackbody normalisation and its uncertainty, for the spectrum with maximum bolometric flux, in units of $({\rm km}/10\,{\rm kpc})$.
For some bursts, the 
radius could not be constrained; we flag these cases by setting the uncertainty to $-1$, in which case the provided value is the estimated $3\sigma$ upper limit.

\paragraph{43 \& 44. Bolometric fluence} ({\tt bfluen, bfluene}) The integrated bolometric flux over the entire burst duration, in units of $10^{-6}\ \epc$, calculated as described in \S\ref{trsanl}. 

\paragraph{45. References for the burst} ({\tt refs}) Here we indicate prior analyses in the literature which included or focussed on this event. 
The list of references may not be complete. 
References are numbered, and may be matched with the list below: 

1. \cite{kuul03a};   2. \cite{zand14a};   3. \cite{zand17b};   4. \cite{kuul10};   5. \cite{zand10a};   6. \cite{zand14b};   7. \cite{chelov05};   8. \cite{aranzana16};   9. \cite{zand05a};  10. \cite{corn02b};  11. \cite{jon01};  12. \cite{zand11a};  13. \cite{zand08a};  14. \cite{piro97};  15. \cite{bhatt07a};  16. \cite{zand03b};  17. \cite{barnard01};  18. \cite{linares10};  19. \cite{corn03a};  20. \cite{stroh98b};  21. \cite{mill99};  22. \cite{mill00};  23. \cite{gil02};  24. \cite{muno02b};  25. \cite{gal06a};  26. \cite{lyu16};  27. \cite{bhatt06a};  28. \cite{jonk04};  29. \cite{homan07};  30. \cite{wij01};  31. \cite{wij01c};  32. \cite{mss99};  33. \cite{zand05b};  34. \cite{ford98};  35. \cite{assa01};  36. \cite{cocchi98};  37. \cite{cocchi99b};  38. \cite{kuul09c};  39. \cite{kaptein00};  40. \cite{marshall99b};  41. \cite{chenevez07};  42. \cite{brandt06a};  43. \cite{mgl00};  44. \cite{suleimanov11b};  45. \cite{franco01};  46. \cite{vs01};  47. \cite{gal03b};  48. \cite{stroh97b};  49. \cite{stroh98};  50. \cite{stroh96};  51. \cite{falanga06};  52. \cite{gbm3cat};  53. \cite{fox01};  54. \cite{guerr99};  55. \cite{muno00};  56. \cite{bazz97};  57. \cite{brandt05};  58. \cite{zand02};  59. \cite{stroh97};  60. \cite{cocchi99a};  61. \cite{chakraborty12};  62. \cite{werner04};  63. \cite{zand98b};  64. \cite{brandt06b};  65. \cite{chenevez11b};  66. \cite{hartog03};  67. \cite{kuul00};  68. \cite{chenevez06};  69. \cite{zand99b};  70. \cite{ferrigno11};  71. \cite{jonk00b};  72. \cite{gal04a};  73. \cite{nat99};  74. \cite{bazz97b};  75. \cite{kaaret02};  76. \cite{zand03};  77. \cite{zand03a};  78. \cite{li18b};  79. \cite{cocchi99c};  80. \cite{zand99c};  81. \cite{chelov07a};  82. \cite{chelov07b};  83. \cite{chenevez12a};  84. \cite{corn07b};  85. \cite{muller98b};  86. \cite{zand98c};  87. \cite{zand01};  88. \cite{gal06c};  89. \cite{chak03a};  90. \cite{bhatt06f};  91. \cite{bhatt06d};  92. \cite{delmonte08b};  93. \cite{nat00b};  94. \cite{ubert98};  95. \cite{fiocchi09};  96. \cite{stroh03a};  97. \cite{watts05};  98. \cite{cocchi00};  99. \cite{kuul02a}; 100. \cite{zand04a}; 101. \cite{cocchi01c}; 102. \cite{clock99}; 103. \cite{ubert97}; 104. \cite{gal03d}; 105. \cite{kong00}; 106. \cite{zand98}; 107. \cite{kajava17a}; 108. \cite{zhang98}; 109. \cite{1916burst}; 110. \cite{tomsick99}; 111. \cite{ts98}; 112. \cite{smale01}; 113. \cite{smale98b}; 114. \cite{ts02}

\subsection{Burst sample completeness}
\label{completeness}

\begin{figure}
	\begin{center}
	\includegraphics[width=\columnwidth]{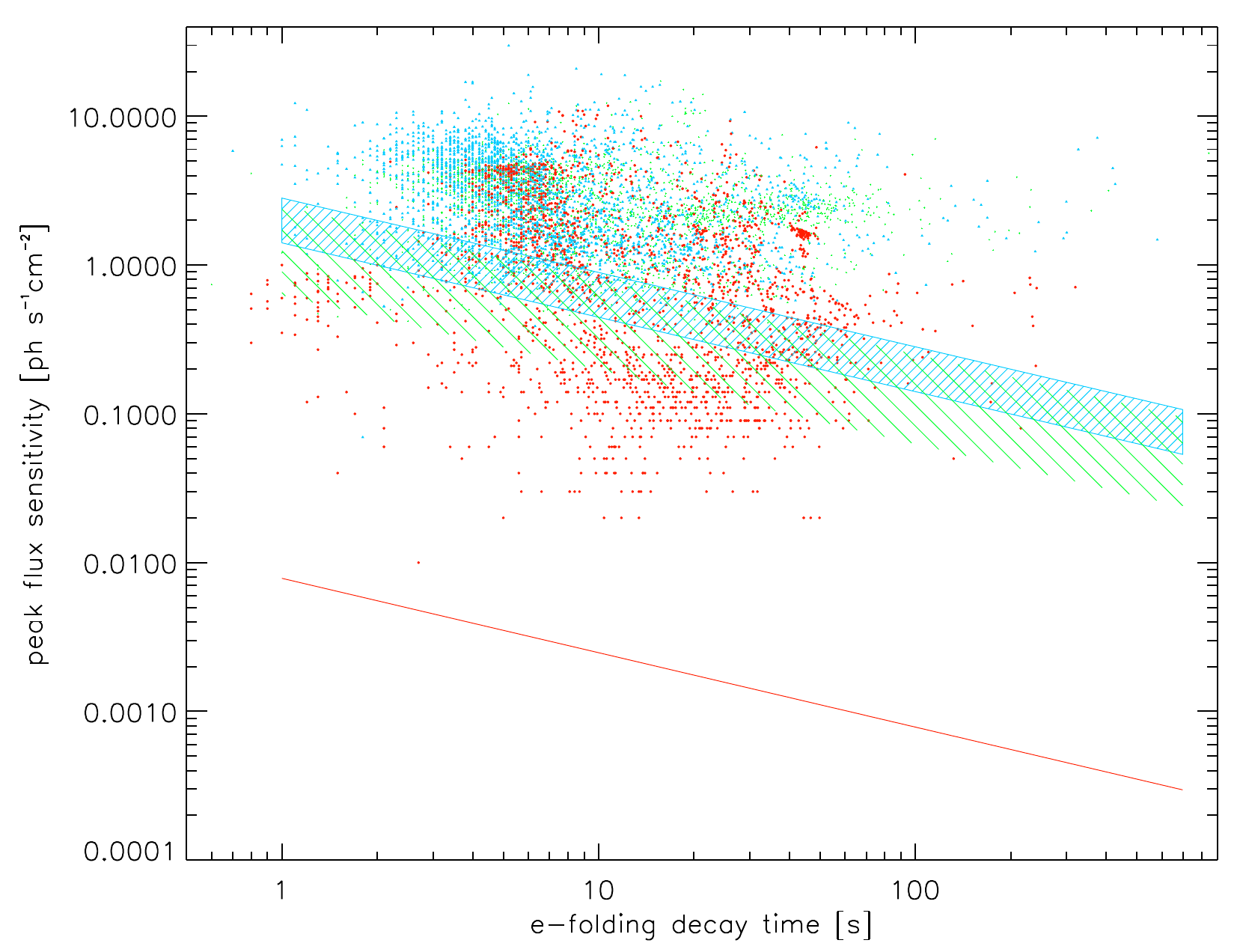}
	\end{center}
	\caption{
	% \dg{This is a new plot, but I can't get a good pdf version. Can you please try to get one from the ps file? Thx - jz111019}
	% it looks OK to me; I cropped some whitespace. what I usually do is generate an .pdf file and include that, then the pdf version will be generated automatically. See other figure examples -- dkg (2019 Nov)
	Estimated sensitivities of the three instruments employed in MINBAR in terms of the peak flux detection threshold, plotted as a function of burst $e$-folding decay time, compared to the properties of the detected bursts.
	% I tried to shorten this caption a bit and reduce duplication with the text --- dkg (2019 Sep)
	% The best signal-to-noise ratio is obtained for fast-rise-exponential-decay burst light curves by integrating the signal up to 1.2 times the e-folding decay time and this has been employed in this calculation. 
	The estimated sensitivities are shown for \xte/PCA ({\it red solid line}), \igr/JEM-X ({\it blue hatched region}) and \sax/WFC ({\it green hatched region}). The hatched areas indicate the variation in sensitivity across the FOV of the latter two instruments, which is about a factor of 2 for JEM-X and a factor of 4 for WFC.  %\jz{range looks bigger for the JEM-X region --- dkg} \dg{Correct, and this is explained in the new text..? - jz}% \jz{can you explain quantitatively how  these regions are defined? -- dkg} \dg{Done. OK? - jz_sep19}. 
	% Also shown are the coordinates of the 
	Each burst in MINBAR is plotted, with colour indicating the instrument (PCA {\it red}; JEM-X {\it blue}: and WFC {\it green}), with the horizontal position from the best-fit $e$-folding decay time 
	% (even if an exponential function is a bad fit; see \S\ref{sec:lcs:modeling}), 
	and the vertical position given by the measured 
	% \notitie{(3--25 keV) -- jc190918} 
	% actually no - this is the photon flux from the lightcurves, which is not from the common energy band --- dkg (2019 Nov)
	peak flux. Note that the sensitivities are only first order estimates, because they vary considerably from observation to observation; see the text for more details. 
	%\cs{Celia, you wrote "green and blue are confused in the printed document"; do you mean that they are swapped, or that they are difficult to separate from one another? --- dkg}
		\label{fig:burst_sensitivity} }
\end{figure}

The degree of completeness of our sample depends on both the selection of observations that comprise our search scope (see \S\ref{minbar-obs}), but also the probability of unambiguously detecting each burst within each observation.
We illustrate the relative sensitivity of each instrument to bursts in Fig. \ref{fig:burst_sensitivity}, 
as a function of the duration of the burst (expressed in $e$-folding decay time). 
The sensitivity depends on the detailed time profile of the burst, the non-burst noise level, and other observing conditions. % However, 
The best sensitivity (as plotted in the Figure) is achieved for
a fast-rise exponential decay function in the photon count rate domain; the source position on the optical axis of the instrument; constant non-burst noise level; and the optimal time interval over which the signal is accumulated, i.e. 
from the burst start to 1.25 times the $e$-folding decay time. In the cases of the wide field-of-view instruments, WFC and JEM-X, we plotted the sensitivity for two extremes in the noise level. 
The 
vertical extent of the regions for WFC and JEM-X is determined by the range of sensitivity across the FOV; this range is narrower for JEM-X 
due to the filtering of JEM-X data for most bursts within only the central $5^\circ$-radius of the $6.6^\circ$-radius field of view (see \S~\ref{jemxburstid}).
One should note that the sensitivity limit drawn for the PCA in Fig.~\ref{fig:burst_sensitivity} does not take into account the high persistent fluxes of some sources and, therefore, may be underestimating the true detection limit for some incidental bursters.

Also shown in Fig.~\ref{fig:burst_sensitivity} are all bursts in MINBAR. While bursts detected with WFC and JEM-X hover just above the theoretical sensitivity curves, those detected with PCA are well above that, indicating that PCA covers for each burster the full range of burst peak fluxes.
This figure suggests that 
the PCA observations are sufficiently sensitive to detect the faintest thermonuclear bursts that occur, although for very faint events it becomes a challenge to confirm a thermonuclear origin based on time-resolved spectroscopy.

There are a number of instances which might result in bursts occurring during the observation intervals of the three instruments, being overlooked by our search strategy.
First, the burst may simply be too faint, or observed at too large an angle from the instrument aimpoint. In such cases it is challenging to confirm the presence of weak bursts, except where there is other corroborating evidence for the events. 
Such evidence may include the detection of the event by an instrument other than the three used for this sample, or a predicted event based on a series of events with a regular recurrence time.

Second, the good time intervals over which our light curves are extracted may not encompass the entire period in which a particular source is observed (and in which bursts may be detected).
It is possible that different choices for the criteria defining the good-time intervals, and/or longer-term variations in the data extraction algorithm arising from software version changes,  may result in 
slightly different 
observation intervals which either exclude previously detected events or reveal new, previously overlooked bursts.

Third, there were a number of instrumental issues that prevented some data being analysed for the MINBAR sample.
For JEM-X, some of the early data from the mission was taken in a (now deprecated) ``restricting imaging mode'', which is no longer supported by the available versions of the OSA software, and we cannot produce light curves (or spectra; see \S\ref{minbar-obs}) for 114 ScWs between \igr\/ revolutions 30 (2003-01-12) and 163 (2004-02-14). 
Notable events that are affected by this issue include the long burst from SLX~1735$-$269 on MJD~52897.733 \cite[see][]{molkov05,zand10a}. 

We performed a number of tests to ensure the completeness of the data.
First, we cross-matched the events seen in each instrument, with any overlapping observations by the other instrument.
This cross-check confirmed the detection of \burstdupes\ events seen in more than one instrument, which we adopt for the purposes of cross-calibrating the instruments as described in \S\ref{crosscal:bursts}.

Second, we compared our detected sample with other samples from the same data, as reported in the literature. 
For example, 
\cite{chelov17} list 2201 events detected by JEM-X and IBIS/ISGRI through 2015 January.  1925 of these events fall within the interval adopted for the MINBAR sample, and we find matches in the MINBAR sample for 
1467
Most of the matched bursts agree in the start time to within $<100$~s, but a few events have offsets of up to 5~min. Additionally, 13 events in the other sample from MXB~1730$-$335 are flagged as type-II in MINBAR and hence excluded from our list (although these events are available via the web interface).
The \burstsigr\ events in MINBAR detected by JEM-X implies that there are more than a thousand additional events in our sample compared to that of \cite{chelov17}. 
Even so, we tried to assess below why some events in the other sample were not identified by our analysis.

Many of the missing events are labeled ``ISGRI'' in the \cite{chelov17} sample, and so it is possible they were detected only in the wider field of view of that instrument.
Of the remaining events, 
one (their 
\#803) is 
attributed to a different source within the FoV 
a type-II event from MXB~1730$-$335 not included in MINBAR).
Another event (their \#786) is the continuation of \#785, a long burst from SLX~1737-282 (MINBAR \#5608) which spans two science windows.
19
fall within observations that were not included as part of MINBAR (see \S\ref{minbar-obs}).
Just one of these science windows was taken in the ``restricted'' imaging mode that was unavailable for analysis for the MINBAR sample (see \S\ref{jemxobs}).

Third, we analysed selected groups of bursts observed close together in time, to determine whether they were consistent with a regular recurrence time. Where the predicted time of a burst fell within an observation, but where the burst search found no candidates, we double-checked the lightcurve to confirm the burst absence.
In four cases this search resulted in additional bursts being identified in WFC observations.

We conclude that the MINBAR sample is essentially complete for those observations that are included in the search, and down to the level where the faintness of the bursts (and/or the data quality; see below) makes it difficult to confirm the presence of bursts in low signal-to-noise data. 
We further discuss the completeness of the observation sample in \S\ref{minbar-obs}.

\subsection{Burst demographics}
\label{sec:burst_summary}

\begin{figure*}[h!t]
	\includegraphics[width=\textwidth]{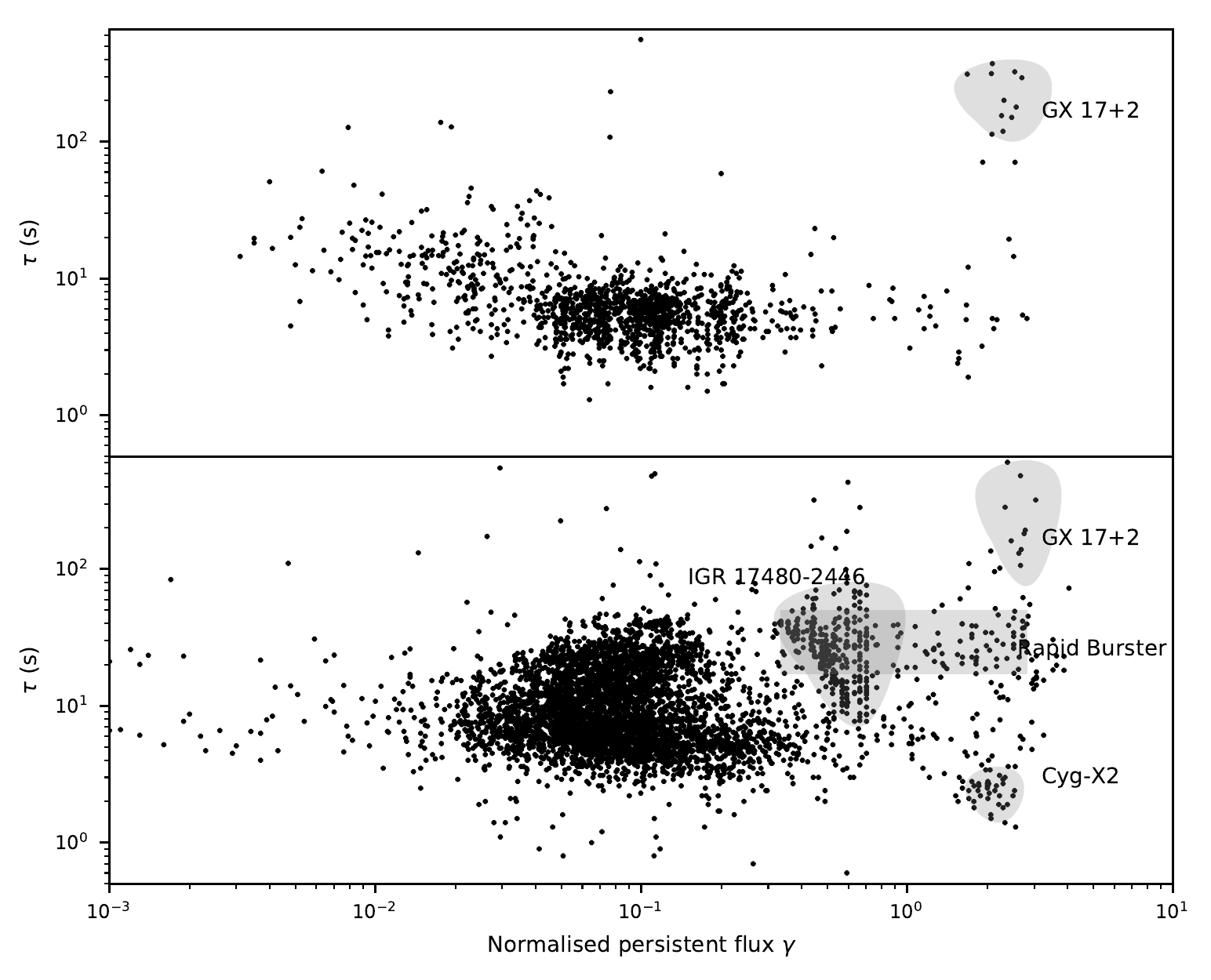}
	\caption{Burst timescale $\tau$ against $\gamma$ for PRE ({\it top panel}) and non-PRE ({\it bottom panel}) bursts. The burst timescale $\tau$ is calculated as the ratio of the fluence to peak flux, i.e. $\tau=E_b/F_{\rm peak}$; $\gamma$ is a proxy for the accretion rate, as a fraction of the Eddington value. 
	The majority of bursts are observed around $\gamma\approx0.1$, which is partially a sampling effect determined by the burst rate, which typically peaks close to this value (see \S\ref{sec:burstrates}).
	% \notitie{Should we say something about cases $\gamma>1$? - jz111019} \jz{can you suggest something?}
	At the highest accretion rates, $\gamma\gtrsim1$, we find the well-known anomalous cases GX~17+2 and Cyg~X-2; numerical models predict that burning should be stable, so that no bursts would be observed. Distinctly different behaviour at roughly the same accretion rate is observed for the slowly-rotating transient IGR~J17480$-$2446. 
	% \dg{same as with the four selected sources, I would maybe highlight the regions occupied in the lower plot by the GS~1826 bursts. Also, it might be interesting to plot differently bursts for which PRE character is measured differently from those for which PRE is only inferred, and see which regions of the plot is occupied by one or another sample of bursts} - Celia
	% I don't think 1826 falls in a distinct region of parameter space, like the highlighted sources. Re. the PRE bursts, what are you trying to show here? --- dkg (2020 Feb)
	\label{fig:tau_v_gamma} } 
\end{figure*}

We summarise the MINBAR burst sample in a plot showing the burst timescale $\tau$ against the inferred accretion rate $\gamma$ (as a fraction of the Eddington rate; Fig. \ref{fig:tau_v_gamma}).  We divide the sample into radius expansion bursts (\texttt{rexp}$>$\rexpthresh; {\it top panel}) and all other bursts ({\it bottom panel}). 
The density of bursts in any given region of the $\gamma$-$\tau$ parameter space is a consequence of both the typical burst rate (see \S\ref{sec:burstrates}) and 
the typical time that sources spend in that range of accretion rates (see \S\ref{subsec:obs_summ}).
Several atypical sources can be identified, and are marked with grey patches. These are the strongly accreting GX~17+2 and Cyg~X$-$2, the Rapid Burster, and IGR~J17480$-$2446. 

Several trends are immediately apparent.
Most 
radius expansion bursts 
occur at an accretion rate corresponding to $\gamma\approx0.1$, with  burst timescale $\tau\approx5$~s. A slight downward trend is also apparent,
with $\tau$ becoming shorter as the accretion rate increases (cf. \citealt{vppl88, murakami80b}). This trend can be understood as a faster onset to ignition as the accretion rate increases, leading to a smaller fluence and hence smaller $\tau$ value since all these bursts are limited to roughly the same peak luminosity (the Eddington value). 
At the lowest accretion rates, where the cool fuel layers allow a substantial reservoir to accumulate prior to ignition, we find the longest bursts with the most intense radius-expansion, including ``intermediate-duration'' events (see \S\ref{sec:superexpansion}).

Non-PRE bursts also occur 
predominantly 
around 
$\gamma\approx0.1$ and with $\tau\approx5$~s,
but extend 
to a second locus with  higher $\tau$ values, up to the $\tau\approx 20$~s region. 
The almost bimodal distribution of timescales for non-radius expansion bursts seen at $\gamma\approx0.1$ may be identified with long, relatively infrequent bursts characteristic of $rp$-process burning (exemplified by those bursts observed in the hard state of GS~1826$-$24; e.g. \citealt{zand17b}), occurring at roughly the same accretion rate (albeit in different sources) as short-duration, weak events, likely made up of a significant fraction of short waiting time bursts.
This feature was already apparent in 
G08, but is more pronounced with the additional bursts in the MINBAR sample.

The similar location of the Rapid Burster and IGR~J17480$-$2446 on this plot is 
suggestive; IGR~J17480$-$2446 has a slow rotation rate of 11\,Hz \citep{papitto11}, slower than any other known burster, and there are compelling, though indirect, reasons for supposing the Rapid Burster to likewise be a slow rotator \citep{bagnoli13}.

A few sources, for instance GX~17+2 and Cyg~X-2, appear to exhibit super-Eddington luminosity. Although the uncertain bolometric correction may play a role, so that the accretion onto the neutron star is actually below the Eddington limit (see \S \ref{bolcor}) it is also thought possible that 
accretion at a few times the Eddington rate may occur (e.g., \citealt{balu10}).

In the sections below we discuss additional aspects of the sample, including the range of burst rates found over the included sources;
and 
the range of peak fluxes and burst temperatures in the bursts with time-resolved spectroscopic analyses.

\subsection{Burst rates}
\label{sec:burstrates} % not to be confused with the earlier section sec:burstrate

\begin{figure}
	\begin{center}
	\includegraphics[width=\columnwidth]{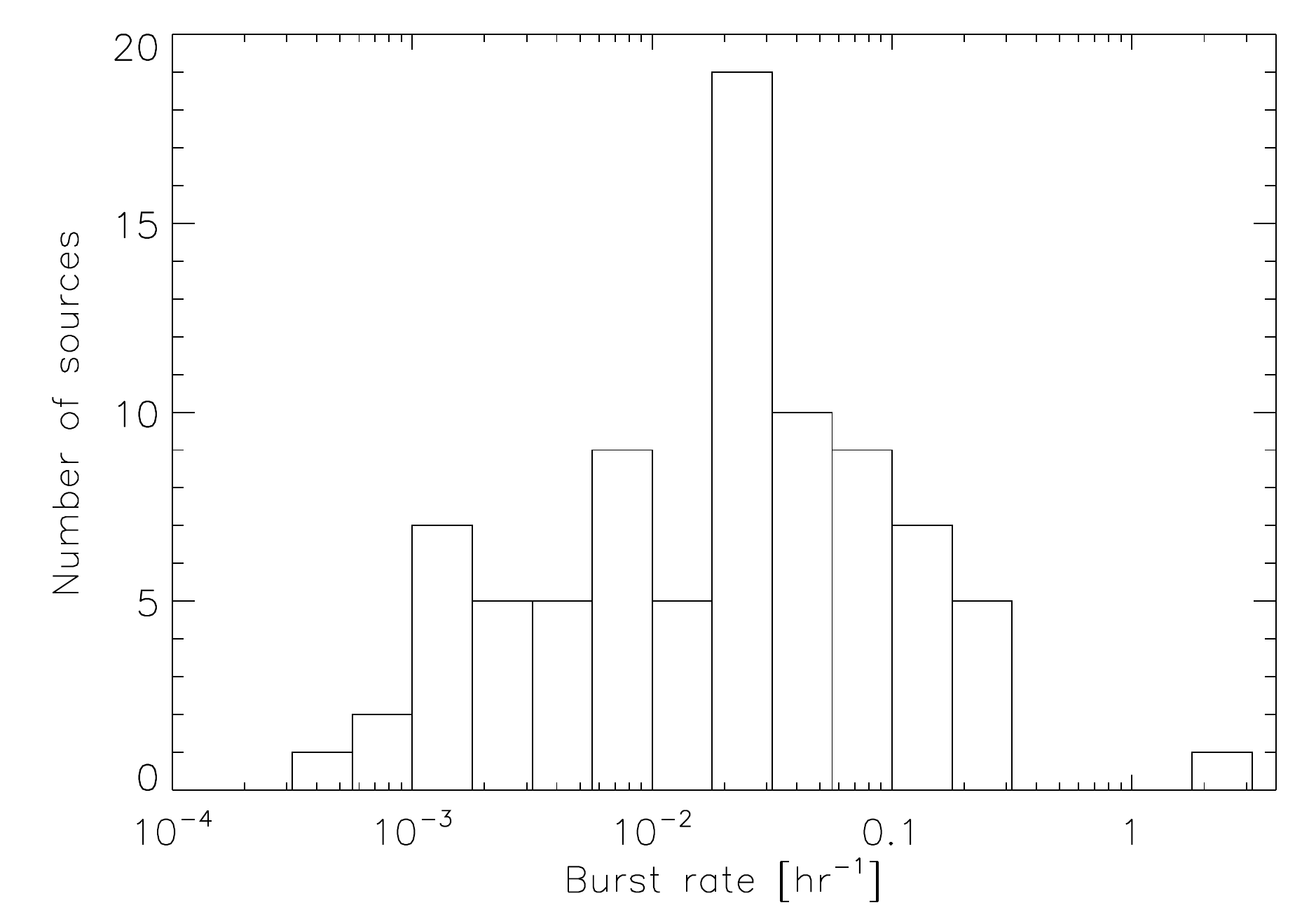}
	\end{center}
	\caption{Distribution of average burst rates for 85 sources in the MINBAR sample. The rates are calculated using the estimated total exposure time for each source when it was active (i.e. the flux was above our detection threshold in any of the three instruments). The most frequent burster in the sample by almost an order of magnitude is IGR~J17480$-$2446 (Terzan 5 X-2), at 1.86~hr$^{-1}$ on average during its 2010 outburst.
	\label{fig:burst_rates} }
\end{figure}

The large size of the MINBAR sample provides a unique opportunity to compare burst rates over a large number of sources. By combining the data from the observation sample (see \S\ref{minbar-obs}) we calculated the average burst rate for each source over all the observations present in MINBAR (Table \ref{tab:bursters}).
However, this quantity is a lower limit on the actual rate, because (for transients) it includes intervals where the source was quiescent. Thus, we calculated the rate while active, including only those observations where the source was detected at $3\sigma$ significance or better. The resulting distribution of burst rates is shown in Fig. \ref{fig:burst_rates}.

\begin{figure}
	\begin{center}
	\includegraphics[width=\columnwidth]{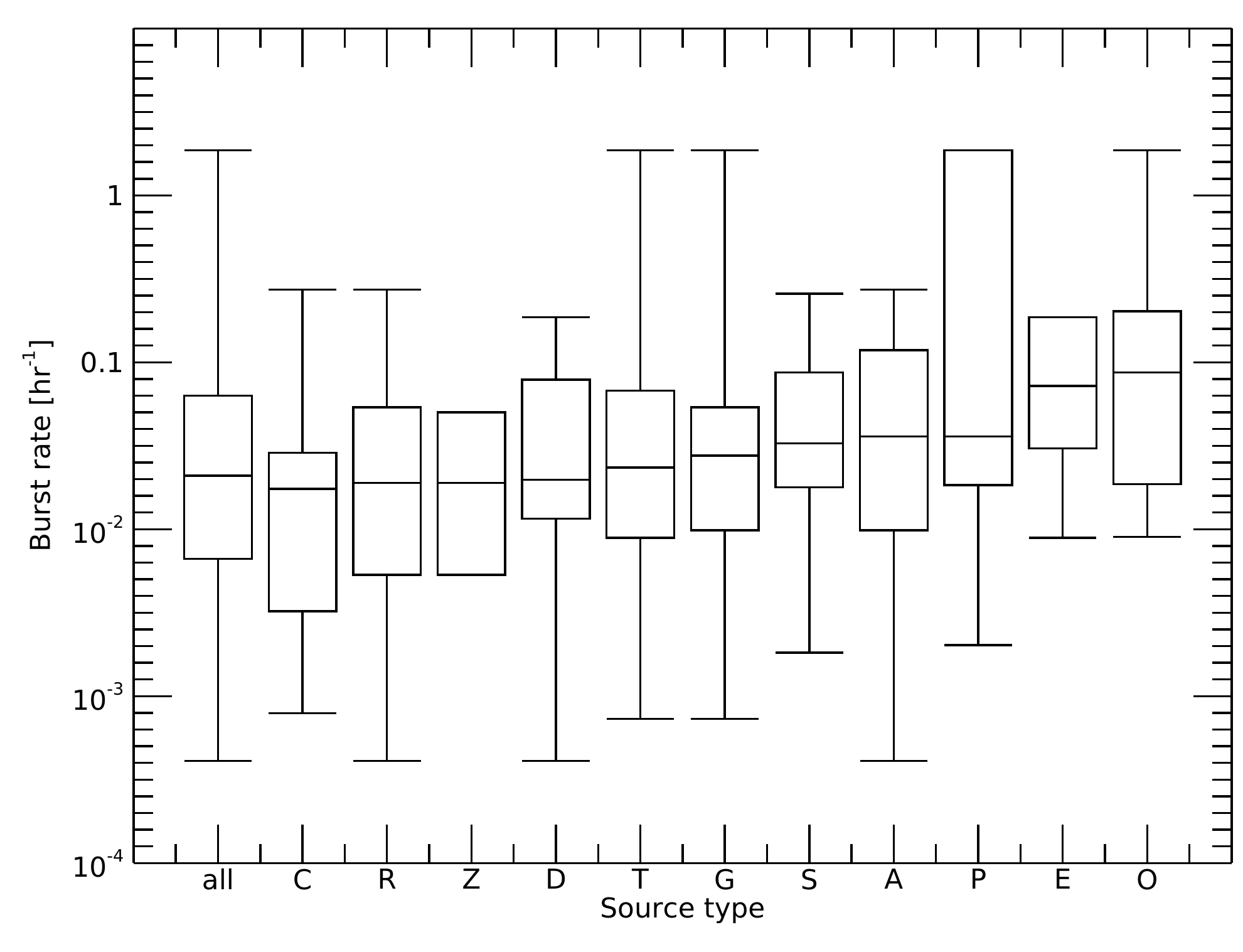}
	\end{center}
	\caption{Distribution of average burst rates for all the  sources in the MINBAR sample, grouped by source type. We display each distributions as a standard ``box-whisker'' style, with the error bars showing the minimum and maximum, the box giving the 25th and 75th percentile, and the median value indicated by the horizontal line. The leftmost symbol includes all sources (i.e. the same distribution as in Fig. \ref{fig:burst_rates}); the source types are sorted by increasing median burst rate. For their meaning, see note a in Table~\ref{tab:bursters}.
	\label{fig:rates_by_type} }
\end{figure}

Although there is a remarkably wide (4 orders of magnitude) range of rates over the source sample, we note that this range likely arises primarily from the range of accretion rates at which the sources were observed. We  find  more modest variation in the mean burst rates per source type (Fig. \ref{fig:rates_by_type}).
The set of burst sources with the lowest median rate are the ultracompact binaries (type 'C'), which is broadly consistent with both the observations of typically low accretion rates from ultracompacts \cite[e.g.][]{zand07}, as well as expectations from theoretical ignition models, since the weak contribution from hydrogen burning will tend to delay burst ignition. Conversely, the highest burst rates on average are found for the sources with burst oscillations, which may be a selection effect. As not all bursts exhibit burst oscillations, a high burst rate favours the observation of many bursts and hence burst oscillation detection.

One notable contribution that can likely bias the measured burst rates to higher values are the presence of 
much shorter recurrence times of order a few minutes, which have been measured in 
several sources 
\cite[e.g.][]{boirin07a}. These events can occur as soon as a few minutes after the previous event, and occur in groups of up to 4; both aspects of which are inconsistent with theory. 
In the MINBAR sample, 493 bursts have recurrence times of less than 1~hr, and come in multiples of up to four events, from 25 sources; the shortest measured separation is 3.88~min 
for a pair of bursts from 4U~1705$-$44 detected with \sax/WFC (MINBAR IDs \#1550, 1551).
\cite{keek10} carried out a systematic analysis of a subset of these events, with 136 recurrence times from 15 sources, drawn from the 3387 bursts from PCA and WFC data that made up MINBAR at the time.  
In the full MINBAR sample, 
half of the events arise from a single source, IGR~J17480$-$2446, which shows behaviour distinct from all the others. During its 2010 outburst, the only one detected to date, the burst rate from this system increased steadily with accretion rate, up to the point the bursts were replaced by mHz X-ray oscillations \cite[e.g.][]{linares12a}. Although this behaviour is similar to what is predicted by numerical models as the accretion rate approaches the Eddington value, only IGR~J17480$-$2446 behaves in this manner, which may be a consequence of its unusually slow rotation period.

Of the sources contributing to the remaining 239 ``episodic'' short waiting time bursts, 12 have measured orbital periods, of $\approx2$~hr or longer, and none are confirmed ultracompacts, with just one candidate, XMMU~J181227.8$-$181234 \cite[]{goodwin19b}.
\cite{keek10} estimated the fraction of these bursts at 30\%,
for the persistent flux range in which most such events are observed. 
Within the full MINBAR sample, and excluding the ultracompact candidates and IGR~J17480$-$2446, we find instead a fraction of approximately 4\%. This fraction is likely an underestimate, because some weak secondary (and tertiary) bursts would likely be missed in the lower-sensitivity JEM-X and WFC observations.

This predominance of H-rich accretors suggests that hydrogen-burning processes play a crucial role in creating short recurrence times. As far as the neutron star spin frequency is known, these sources all spin fast at over 500~Hz. Rotationally induced mixing may explain burst recurrence times of the order of 10~min. Short recurrence time bursts generally occur at all mass accretion rates where normal bursts are observed, but for individual sources the short recurrence times may be restricted to a smaller interval of accretion rate. 
Recent numerical simulations explain this phenomenon as due to reignition of left-over hydrogen mixed into the ashes layer \citep{keek17a}.

For 14 sources, where we have more than $\approx100$ bursts per source, we can also measure the burst rate as a function of accretion rate, as described in \S\ref{sec:burstrate}.

\subsection{Burst peak flux and peak temperature}

\begin{figure}
	\begin{center}
	\includegraphics[width=\columnwidth]{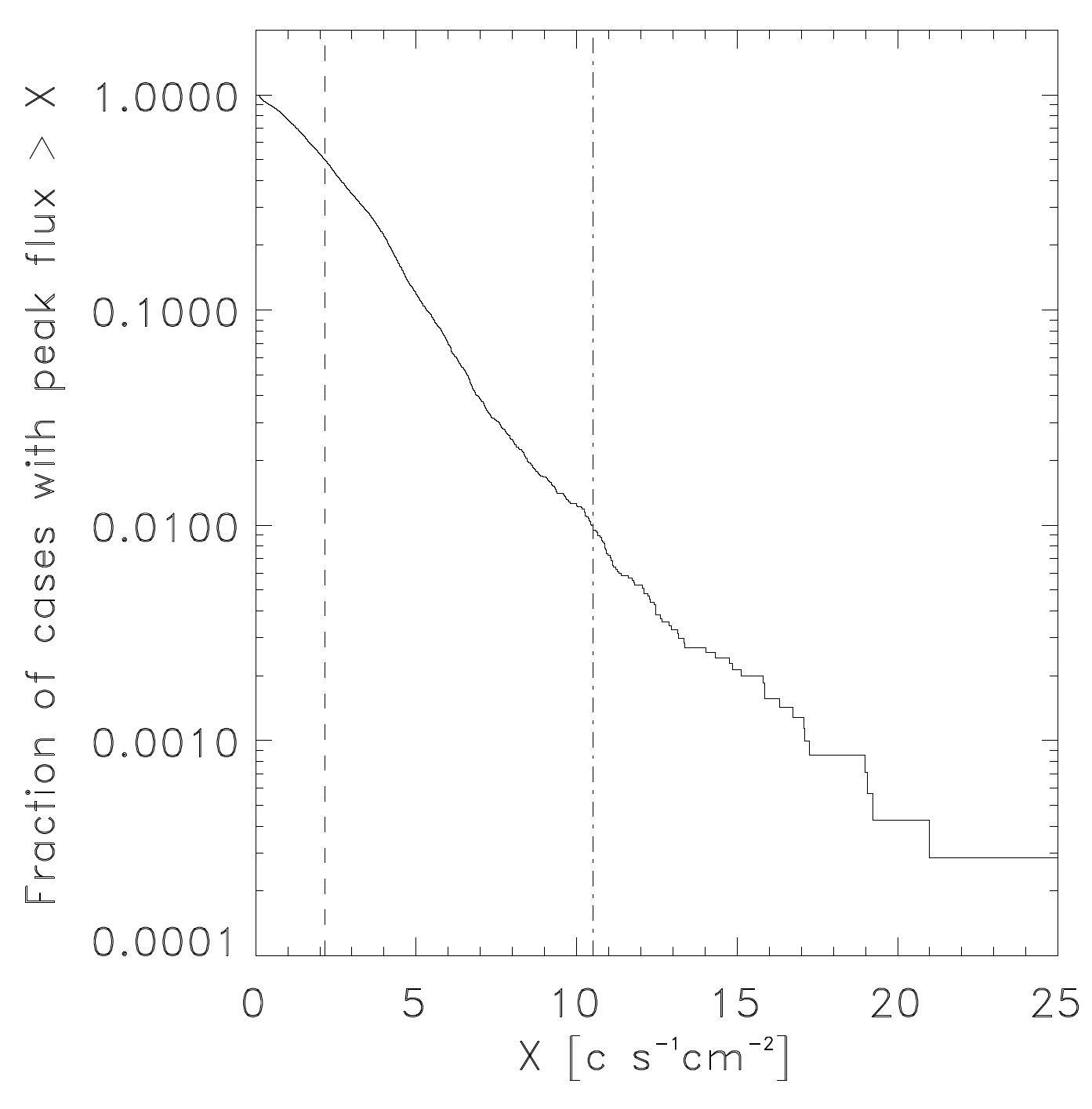}
	\end{center}
	\caption{Cumulative distribution of observed peak photon flux. The dashed line indicates the 50\% mark and the dot-dashed line the 1\% mark. %\jz{suggest putting some vertical lines indicating boundaries referred to in the text, \& perhaps also  median and 68\% boundaries --- dkg} \dg{Done the two lines, 50\% = median line. I think 68\% is not necessary - jz13112019}
	\label{fig:pfcumulative} }
\end{figure}

\begin{figure}
	\begin{center}
	\includegraphics[width=\columnwidth]{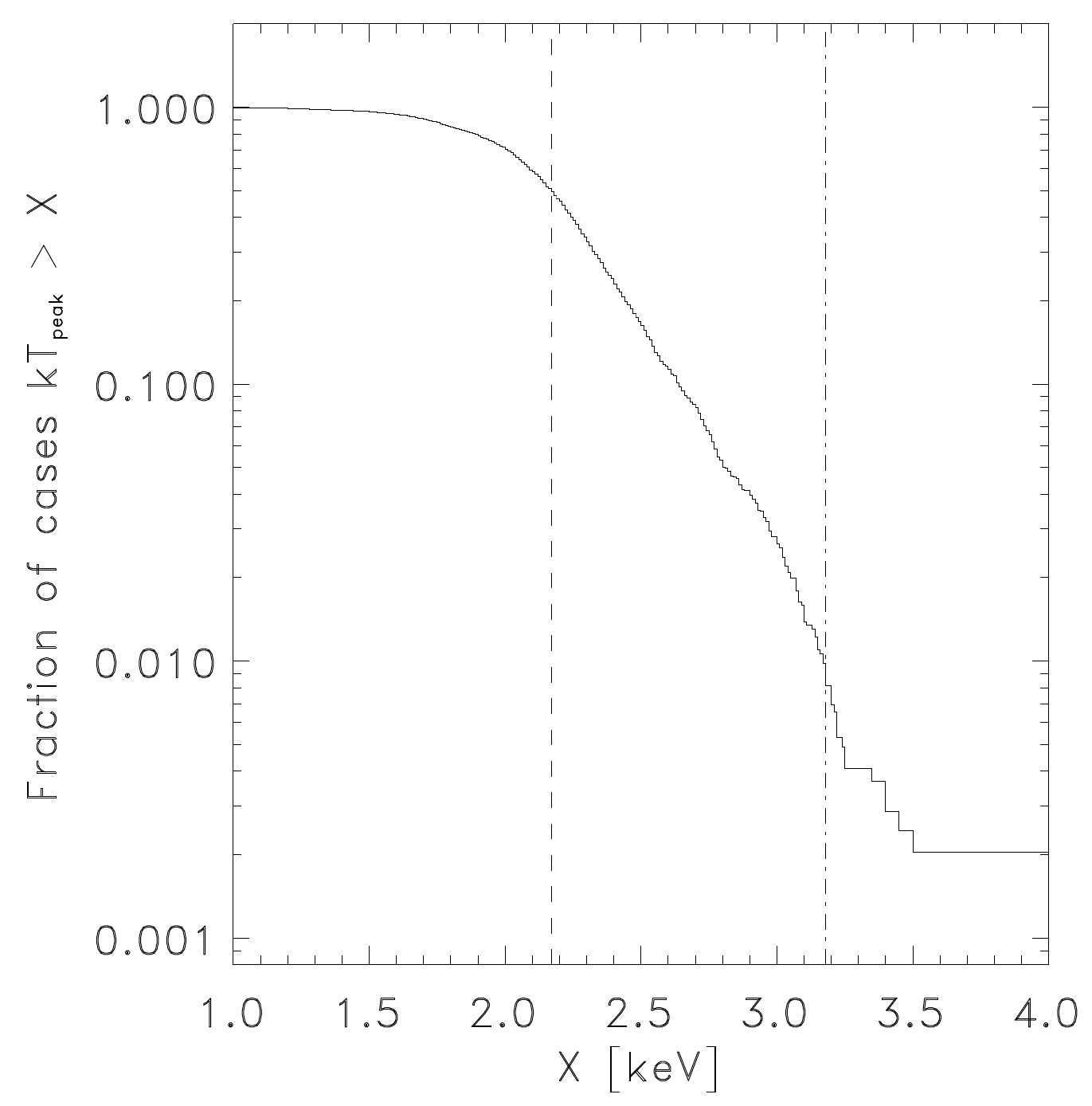}
	\end{center}
	\caption{Cumulative distribution of peak burst k$T$ in keV resulting from time-resolved spectroscopy with a blackbody model, after selecting only those cases for which the 1$\sigma$ uncertainty is smaller than 0.2 keV.  The dashed line indicates the 50\% mark and the dot-dashed line the 1\% mark.
	\label{fig:ktecumulative} }
\end{figure}

The observed burst peak photon fluxes range up to 21~count~s$^{-1}$cm$^{-2}$ (see Fig.~\ref{fig:pfcumulative}) which is equivalent to 10~Crab units\footnote{As explained in \S~\ref{obs-jemx}, 1 Crab unit represents the photon flux of the Crab nebula plus pulsar in the same bandpass, and translates to a 3--25 keV flux of $3\times10^{-8}\epcs$ for a power law of photon index -2.1 and absorption due to cold interstellar matter with $N_{\rm H}=3\times10^{21}$~cm$^{-2}$.}. Half of all bursts are brighter than 1~Crab, and 1\% of all bursts are brighter than 5~Crab. Burst peak energy fluxes have a dynamic range of a factor of 2$\times 10 ^3$, between $2\times10^{-10}$ and $4\times10^{-7}$~erg~s$^{-1}$cm$^{-2}$ (see Fig.~\ref{fig:burst_sensitivity}), although there are relatively few bursts
below $1\times10^{-9}$ erg~s$^{-1}$cm$^{-2}$. The smallest peak fluxes, measured with the most sensitive of the three instruments (\xte/PCA), are an order of magnitude above the sensitivity limit and, therefore, appear to probe the true minimum peak flux, at least for known X-ray bursters in our Galaxy (see also \S\ref{sec:future}). 

The cumulative distribution of the peak temperatures as measured in the time-resolved burst spectrum with a blackbody model 
(for the PCA and WFC bursts only) is shown in Fig.~\ref{fig:ktecumulative}. There are hardly any bursts with peak temperatures cooler than 1~keV. This limit 
may be % dkg (2020 Feb)
due to the low-energy cutoff of the bandpass of all employed instruments of $\approx2$~keV. 69\% of the peak temperatures in our sample are between 2 and 3~keV, while 28\% are between 1 and 2~keV. Just 2\% are higher than 3~keV. The highest temperature is about 3.5~keV which is marginally consistent (to within the spectral fit uncertainties) with 3~keV. This 
limit is robust, because the bandpasses of all instruments allow the measurement of temperatures that are a factor of roughly 3 higher.
This limit is naturally explained as the maximum temperature on the surface of a NS before the radiation pressure becomes so large that the photosphere will leave the NS surface and the temperature drops again, and it is called the Eddington temperature 
\cite[see e.g.][]{lew93}.

We discuss additional results derived from the MINBAR sample, and also suggestions for future research directions, in \S\ref{sec:discussion}.

\subsection{Comparison with G08}
\label{sec:g08comp}

G08 published a catalog of 1187 PCA X-ray bursts that are part of the 
\burstsxte\ 
PCA bursts in MINBAR, excluding 
5 % by my calculation --- dkg (2019 Dec)
events that were discarded 
as unlikely to be type-I (thermonuclear) X-ray bursts\footnote{These are events numbered \#53 and 94 in G08, from EXO~0748$-$676; \#2 from 4U~0919$-$54; \#47 from 2E~1742.9$-$2929; and \#30 from 4U~1746-37. The latter event, on 2004 Nov 8 at 15:46:15.168 UTC, is roughly coincident with GRB20041108C detected by KONUS, see \url{http://gcn.gsfc.nasa.gov/konus_2004grbs.html} }. 
Here we compare the results with the previous analysis for several parameters, to test the robustness of the MINBAR values. 

\begin{figure}
	\includegraphics[width=\columnwidth]{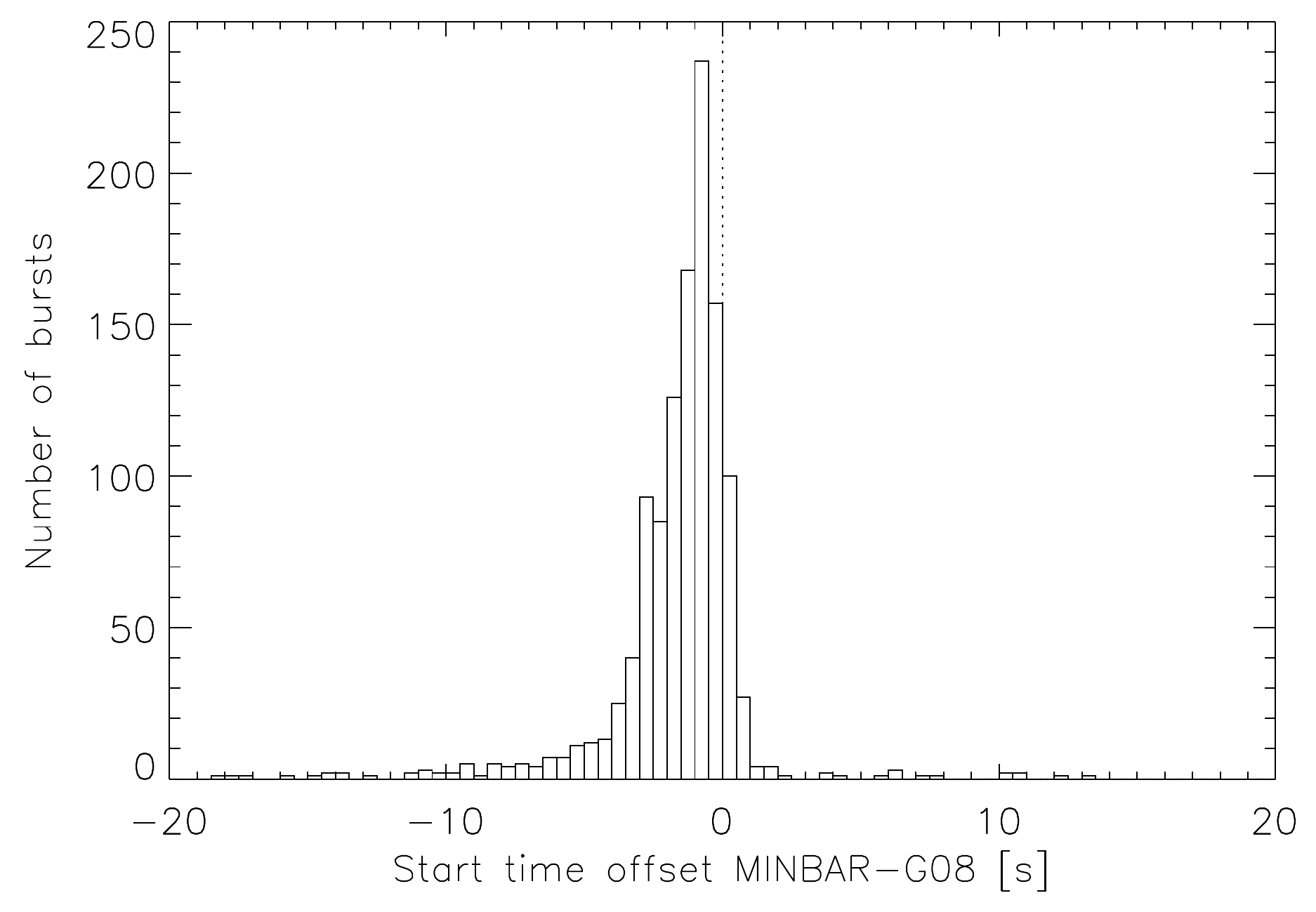}
	\caption{Offset between start time for 1170 \xte/PCA bursts common to both MINBAR and the sample of G08, and for which the start time was covered by the data (i.e. excluding bursts flagged ``g''; Table \ref{burst-analysis-flag-table}). The median offset is -1.1~s. 
	% \jz{Seven bursts have an absolute offset larger than 20~s --- dkg (2019 Dec)} \dg{you mean 10 s, right? See my email to you of Dec 18, 2019 - jz060120}
	% still trying to sort this out... dkg (2020 Feb)
	\label{fig:comp_time}}
\end{figure}

We first compared the start time of the bursts, by calculating the offset between the time determined for MINBAR and in G08. The distribution of residuals is  skewed towards negative values, with the MINBAR start times typically  earlier (by $\approx1.1$~s) than for  G08  (Fig. \ref{fig:comp_time}). This offset may be understood as arising from the different definition of start times, with the G08 values also relying on the bolometric flux measurements rather than the instrumental lightcurves (see \S\ref{sec:burst-lightcurves}).

We next compared the peak flux and fluence values. We can directly compare the peak count rate values provided by G08 with the peak (photon) flux calculated for MINBAR, once the effective area correction is taken into account (see \S\ref{sec:area}).
The agreement is good (within 10\%) for 
about half of the bursts, 
but the MINBAR values are $\approx6$\% smaller on average. 
We attribute this offset to the fact that in G08 the peak flux was determined from light curves with a time binning of 0.25~s while for the new analysis we used 1.0~s.
Variability on time scales shorter than 1~s, including statistic fluctuations, 
will tend to result in systematically higher intensities in G08, by approximately the measured fraction.
For the remainder of the bursts, where the values were discrepant at $>10$\%, inspection of a few tens of these events indicates
that these bursts had an incorrect value for the number of active PCUs in G08. 
We note that this has no effect on the spectroscopic analysis in G08, because that analysis relies on a different algorithm. 

\begin{figure}
	\includegraphics[width=\columnwidth]{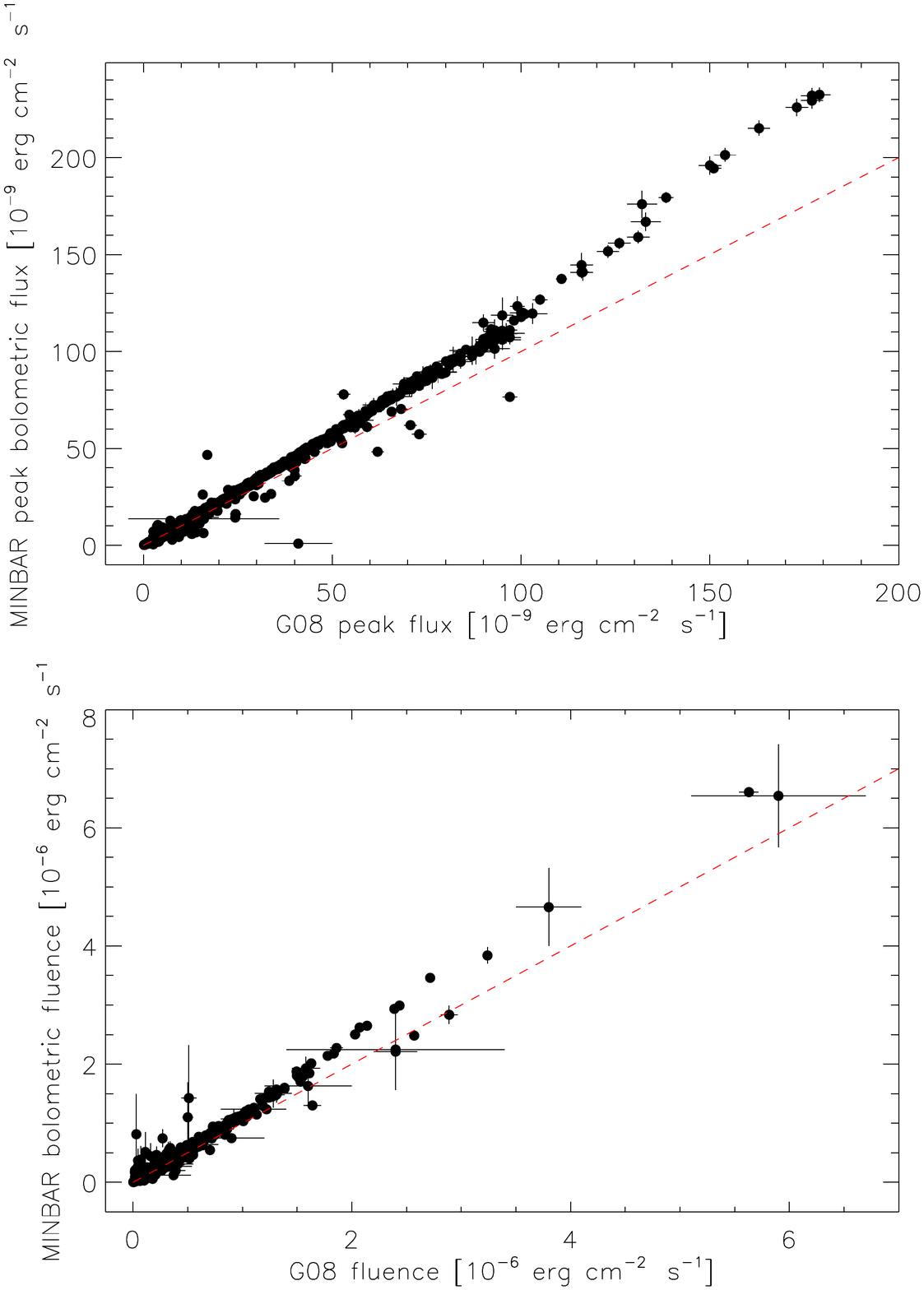}
	\caption{Comparison of burst bolometric peak flux ({\it top panel}) and fluence ({\it bottom panel}) measured for 1140 bursts common to both MINBAR and G08. The top panel compares the MINBAR bolometric peak fluxes to the equivalent measurements from G08. The red dashed line is 1:1, and the bias towards higher values for the MINBAR measurements demonstrates the effect of the deadtime correction in the MINBAR sample.
	The bottom panel compares the bolometric fluence from MINBAR against the equivalent measurements from G08. Other details are as for the upper panel.
	\label{fig:comp_flux_fluen}}
\end{figure}

G08 also measured the peak flux from the bolometric flux measurements from time-resolved spectra, while in MINBAR we quote values both from the instrumental lightcurves and the bolometric flux measurements including the effects of deadtime correction.
We find that the peak flux and fluence values calculate from the lightcurves for MINBAR correlate well with the values from G08.
We also compared the bolometric peak flux and fluence values, for those bursts where the lightcurve was observed with PCA in full (i.e. excluding bursts with flags e, f, g or h; see  Table \ref{burst-analysis-flag-table}). The values for the remaining 1140 bursts are highly correlated as expected since the two analyses are based on the same spectral extraction, but the effect of including the deadtime correction (see \S\ref{xtetrse}) clearly biases the MINBAR bolometric peak fluxes and fluences to higher values (Fig. \ref{fig:comp_flux_fluen}).

\begin{figure}
	\begin{center}
    \includegraphics[width=\columnwidth]{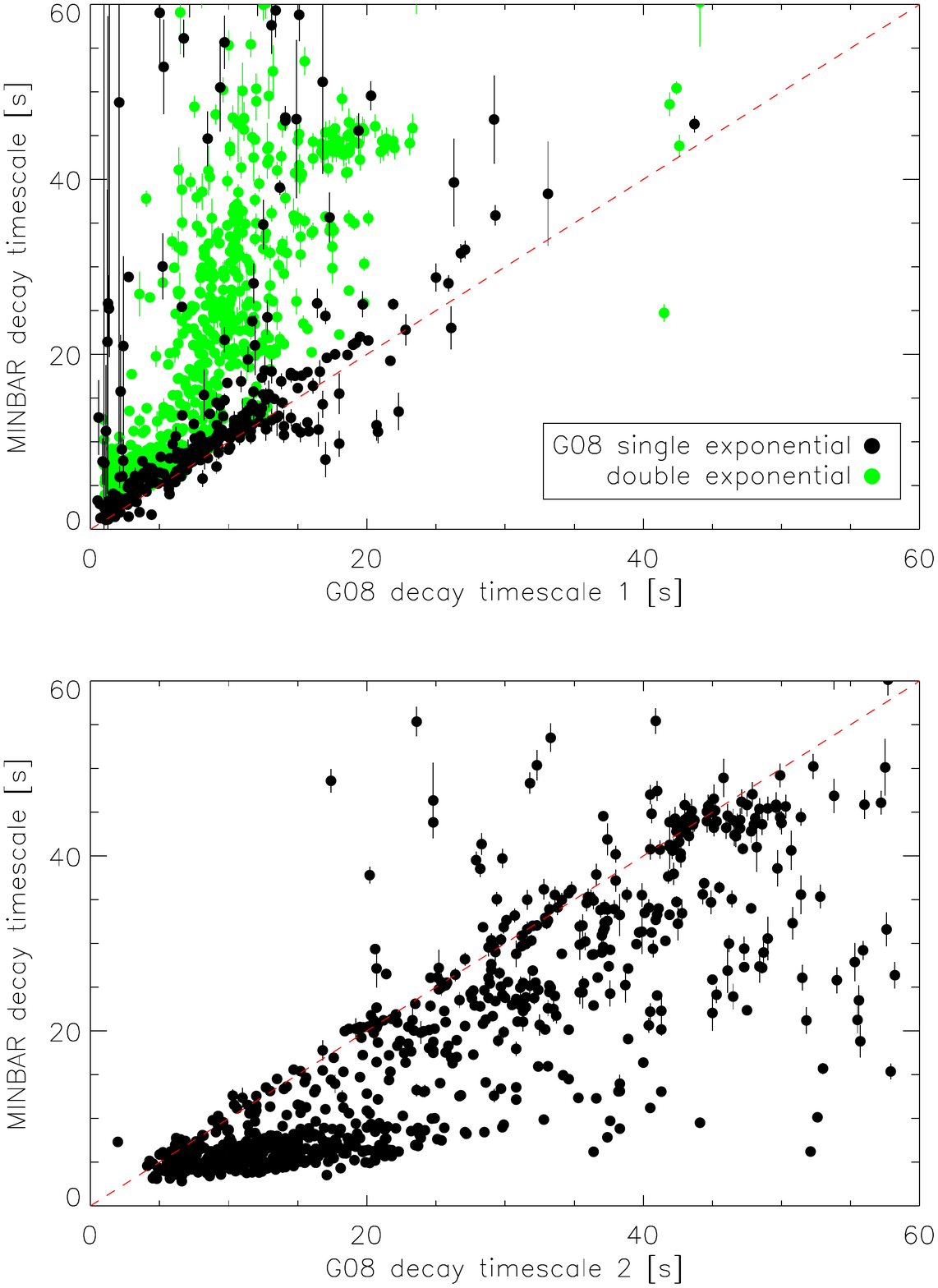}
	\end{center}
	\caption{Comparison of decay time scale from lightcurve fits to MINBAR data and those in G08. In the top panel we compare the MINBAR decay timescale to the first decay timescale quoted by G08, separately for bursts with a single or double exponential. We also compare the MINBAR timescale with the second exponential timescale in G08, where present ({\it bottom panel}).
	\label{fig:comp_timescale}}
\end{figure}

We also compared the burst timescales, via the $\tau$ value and exponential decay timescales. The $\tau$ values were very similar since
% corrected in proof
% (for the PCA) bursts
(for the PCA bursts) 
they are based on the same measurements in both samples. However, the single exponential decay timescale in MINBAR provides a  simpler description of the burst decay than the double exponential provide by G08.
For those bursts in G08 described purely by a single exponential, the agreement between the decay timescales is good (Fig. \ref{fig:comp_timescale}, top panel). However, the MINBAR values substantially overestimate the first decay timescale, for those bursts in G08 that were fitted with a double (broken) exponential curve. In contrast, the MINBAR timescale significantly {\it under}-estimates the second exponential timescale (Fig. \ref{fig:comp_timescale}, bottom panel). This pattern can be understood with the MINBAR value as being an average over the typically more complex decay revealed by the high signal-to-noise PCA measurements.

In summary, we find our analysis results to be highly consistent with G08 once differences in the data analysis procedures and minor errors in the earlier sample are taken into account.

\section{Burst oscillations}
\label{osc_table}

Here we describe how the burst oscillation analysis described in \S\ref{burstosc} is presented in the MINBAR sample.

\begin{deluxetable*}{cccp{10cm}}
\tabletypesize{\scriptsize}
\tablecaption{Burst oscillation table columns, formats and description
  \label{tab:oscillations}}
\tablehead{
\colhead{Column} & \colhead{Format} & \colhead{Units} &
\colhead{Description}  }
\setcounter{magicrownumbers}{-20}
\startdata
\rnum & A23 & & Burster name \\
\rnum &  A3 & &  Instrument label \\ % instr 
\rnum & A15 & & Observation ID \\ % obsid
\rnum & I4 & & MINBAR burst ID \\ 
\rnum & I6 & & MINBAR observation ID \\
\rnum & F13.3 & MET & Start time of the bin with the most significant signal \\ % (relative to the burst start) \\
\rnum & F5.2 & s & Duration of the bin \\
\rnum & I2 & & Number of time bins exceeding the count threshold into which the burst was divided \\
\rnum & F7.1 & counts$\,{\rm s^{-1}\,PCU}^{-1}$ & Background rate estimated from the pre-burst emission \\
\rnum & I1 & & Detection flag on burst oscillation; 1 for detection, 0 otherwise \\ % "<" or blank
\rnum & A1 & & Phase during which oscillation was detected (peak phase=90\% maximum); n = none, r = rise, p = peak, t = tail  \\
\rnum & I1 & & Detection criterion by which the highest-power signal was selected\\ % not true -- dkg
\rnum & I1 & & Flag for signal found in the first bin after the burst start time; 1=yes, 0=no\\% 1=yes, 0=no
\rnum & F5.2 & \% rms & Amplitude of detected burst oscillation (or limit for non-detection) \\
\rnum & F5.2 & \% rms & Lower error on amplitude \\
\rnum & F5.2 & \% rms & Upper error on amplitude \\
\rnum & F5.2 & \% rms & 3$\sigma$ upper limit on amplitude for bursts without detected oscillation signals\\
\rnum & F5.1 & Hz & Frequency of the selected signal \\
\rnum & F5.1 & & Signal power of the detected burst oscillation \\
\rnum & F5.1 & & Measured power of the most-significant detected signal \\
\enddata
\end{deluxetable*}

\subsection{Table format}
\label{osc_table_format}
Below we list the table columns, units, and the format in the ASCII file. See also Table~\ref{tab:oscillations}. % Column \ref{xreftest} is copied from the burst table (testing the row number cross-references here)

\paragraph{1--3 Burster name, instrument label and observation ID}  attributes are identical to the corresponding columns in the burst table (see \S\ref{minbar}).

\paragraph{4 \& 5 MINBAR burst and observation ID} identify the burst ID in the MINBAR table (see \S\ref{minbar}), as well as the host observation in the observation table (\S\ref{minbar-obs}), respectively.

\paragraph{6 \& 7. Time range for bin} specified via the time in MET seconds corresponding to the start of the bin, and the bin duration in s. 

\paragraph{8. Number of time bins exceeding the count threshold} $N_{\rm t}$

\paragraph{9. Background rate} $C_{\rm B}$ per PCU measured over the time range 20--5~s prior to the burst

\paragraph{10. Detection flag} $=1$ for a detection, or $=0$ for no detection (in which case columns 14--16 are limits)

\paragraph{11. Burst phase for detection} (r)ise, (p)eak, (t)ail or (n)one

\paragraph{12. Detection criterion} by which the time bin identified as having the most significant signal was selected; 1: single bin, not in the first second;
                        2: single bin, signal in the first second;
                        3--5: double time-frequency bin;

\paragraph{13. First bin flag} $=1$ if the signal was found in the first time bin following the start

\paragraph{14--16. Amplitude of signal and uncertainty} given as \%rms, with the $1\sigma$ lower and upper bounds, respectively

\paragraph{17. Upper limit (3$\sigma$) on amplitude for bursts without detected oscillation signals} given as \%rms.

\paragraph{18. Frequency of the signal} to within a Hz.

\paragraph{19. Signal power of the detected oscillation} $Z_s^2$ (see \S~\ref{burstosc})

\paragraph{20. Power of the most-significant signal} $Z_m^2$ (see \S~\ref{burstosc})

\subsection{Burst oscillation summary}

\label{osc_summary}
\begin{deluxetable*}{lccccccccl}
\tablecaption{Analysed burst oscillation sources.
\label{tab:osc_summary}}
\tablecolumns{10}
\tablewidth{0pt}
\tablehead{\colhead{Source} & \colhead{Type\tablenotemark{\footnotesize{a}}} & \colhead{$\nu_\text{spin}$} & \colhead{Number} & \multicolumn{5}{l}{Bursts with oscillations\tablenotemark{\footnotesize{b}}}  & \colhead{Ref.}\\
\colhead{} & \colhead{} & \colhead{(Hz)} & \colhead{of bursts} & \colhead{Total (fraction)} & \colhead{PRE} & \colhead{Rise} & \colhead{Peak} & \colhead{Tail}  & \colhead{}} 
\startdata
IGR J17511$-$3057  & OPT   & 245 & 10  & 10 (1.00) & 0 & 1  &  3  &  6 & [1]\\
IGR J17191$-$2821  & OT    & 294 & 5   & 3 (0.60) & 0 & 0  &  0  &  3 & [2]\\
XTE J1814$-$338    & OPT   & 314.4\tablenotemark{\footnotesize{c}} & 27  & 27 (1.00)& 1 & 1  &  20 &  6 & [3] \\
4U 1702$-$429      & AO    & 329 & 49  & 35 (0.71)& 0 & 5  &  5  &  25 & [4] \\
4U 1728$-$34       & ACOR  & 363 & 169 & 49 (0.29) & 19& 14 &  14 &  21 & [5] \\
HETE J1900.1$-$2455& IOT   & 377 & 7   & 1  (0.14) & 1 & 0  &  0  &  1 & [6] \\
SAX J1808.4$-$3658 & OPRT  & 401 & 8   & 8  (1.00) & 7 & 7  &  0  &  1 & [7]\\
KS 1731$-$260      &  OST  & 524 & 27  & 6  (0.22)& 4 & 3  &  1  &  2 & [8,9] \\
SAX J1810.8$-$2609 & OT     & 532 & 6   & 1  (0.17) & 1 & 0  &  0  &  1 & [10] \\
Aql X$-$1          & ADIORT& 550 & 71  & 8  (0.11) & 6 & 2  &  2  &  4 & [11] \\
EXO 0748$-$676     & DEOT  & 552 & 145 & 2  (0.01)& 1 & 1  &  0  &  1 & [12] \\
MXB 1658$-$298     & DEOT  & 567 & 24  & 3  (0.13) & 3 & 1  &  0  &  2 & [13] \\
4U 1636$-$536      & AOS   & 581 & 347 & 82 (0.24) & 62& 31 &  7  &  43 & [14,15]\\
GRS 1741.9$-$2853  & OT    & 589 & 2   & 0  (0.00)\tablenotemark{\footnotesize{d}}& 0 & 0  &  0  &  0 & [16]\\
SAX J1750.8$-$2900 & A?OT  & 601 & 6   & 1  (0.17)& 1 & 1  &  0  &  0 & [17,18]\\
4U 1608$-$522      & AOST  & 620 & 47  & 8  (0.17)& 6 & 3  &  1  &  4 & [18,19]\\
\enddata
\tablenotetext{a}{Source type as listed in Table \ref{tab:bursters}}
\tablenotetext{b}{We list in how many bursts in our sample oscillations were detected, and specify for the bursts with oscillations how many of those were PRE bursts (flagged with 2) and in which phase of the burst the strongest signal was found.}
\tablenotetext{c}{The burst oscillation frequency of XTE J1814$-$338 has been found to be very stable at a frequency of 314.4 Hz \citep{stroh03a}. 
We have set $\nu_0$ for this source to the known oscillation frequency of 314.4 Hz, to ensure that signals that would otherwise fall between the bins are not missed.}
\tablenotetext{d}{We did not detect any burst oscillations in the bursts of GRS 1741.9$-$2853 included in this sample. Burst oscillations are however detected in other bursts from this source \citep{stroh97} that are eliminated from this search because they met our elimination criteria.}
\tablerefs{1. \cite{altamirano10c}, 2. \cite{altamirano10b}, 3. \cite{stroh03a}, 4. \cite{mss99}, 5. \cite{stroh96}, 6. \cite{watts09b}, 7. \cite{chak03a}, 8. \cite{smith97}, 9. \cite{muno00}, 10. \cite{bilous18}, 11. \cite{zhang98}, 12. \cite{gal10a}, 13. \cite{wij01}, 14. \cite{stroh98b}, 15. \cite{stroh02b}, 16. \cite{stroh97}, 17. \cite{kaaret02}, 
18. 
G08, 
19. \cite{hartman03}}
\end{deluxetable*}

\begin{figure*}
	\includegraphics[width=\textwidth]{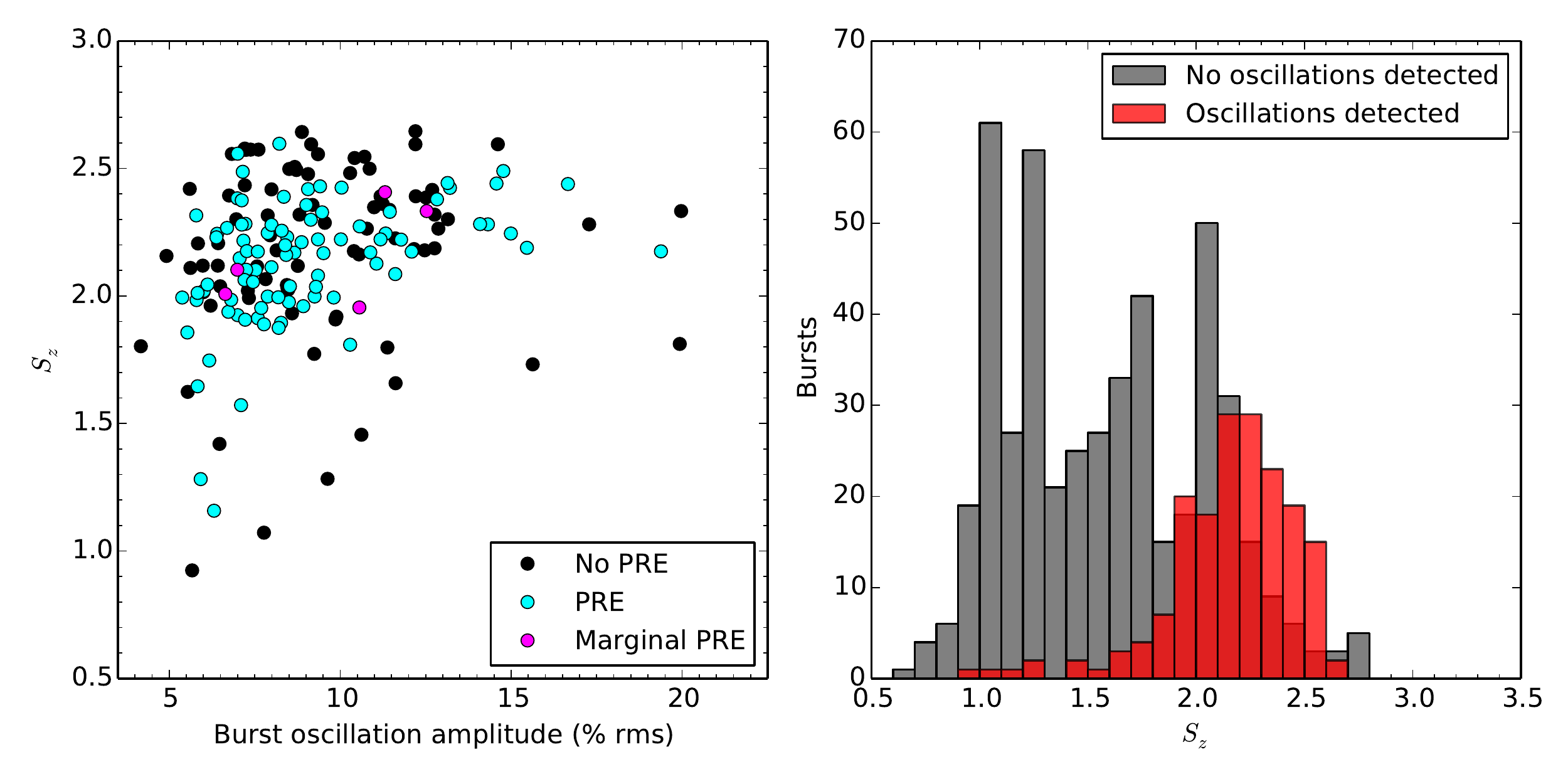}
	\caption{Relation between burst oscillations and $S_\text{Z}$ for all sources in Table \ref{tab:osc_summary} combined. Left: $S_\text{z}$ value at the time of the burst as function of burst oscillation amplitude of the strongest oscillation signal. This figure combines the results from all bursts (from all sources) with detected oscillation signals. Bursts without detected oscillations are omitted, as are bursts for which the $S_\text{z}$ could not be determined. Colours indicate whether the burst showed photospheric radius expansion (PRE), \S~\ref{sec:superexpansion}. Right: histograms of $S_\text{z}$ values for bursts with detected oscillations (red) and bursts without detected oscillations (grey).} 
	\label{fig:bo_sz} 
\end{figure*}

\begin{figure}
	\includegraphics[width=\columnwidth]{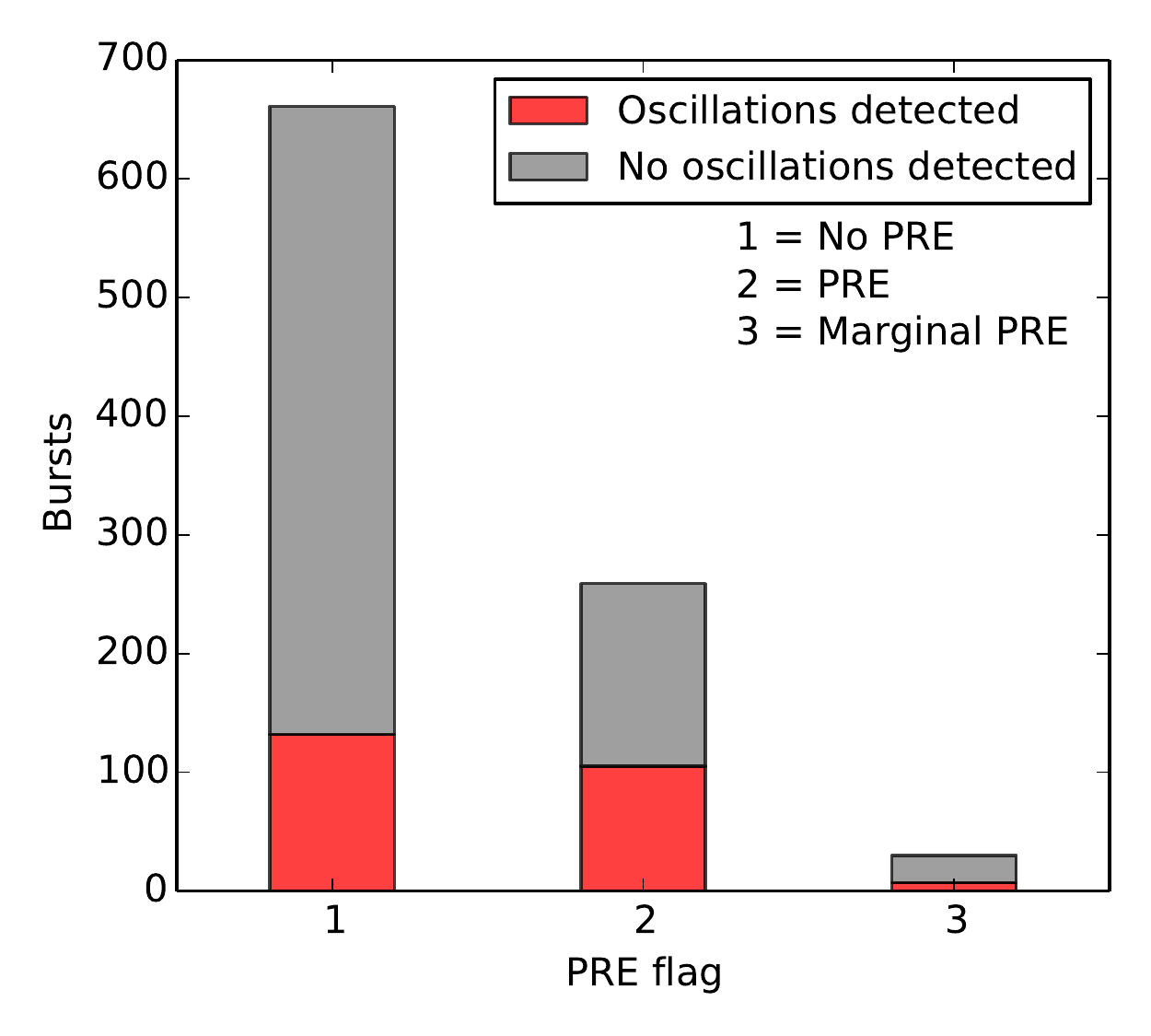}
	\caption{Stacked histograms of PRE flag for bursts from the burst oscillation sample with detected oscillations (red) and bursts from the sample without detected oscillations (grey). Note the much higher fraction of detections in the bursts with PRE flag {\tt rexp > 1}}
	% \lo{There are bursts with fractional PRE flags between 1 and 2. Maybe leave no space between balken?}} 
	\label{fig:bo_PRE}  
\end{figure}
\begin{figure*}
	\includegraphics[width=\textwidth]{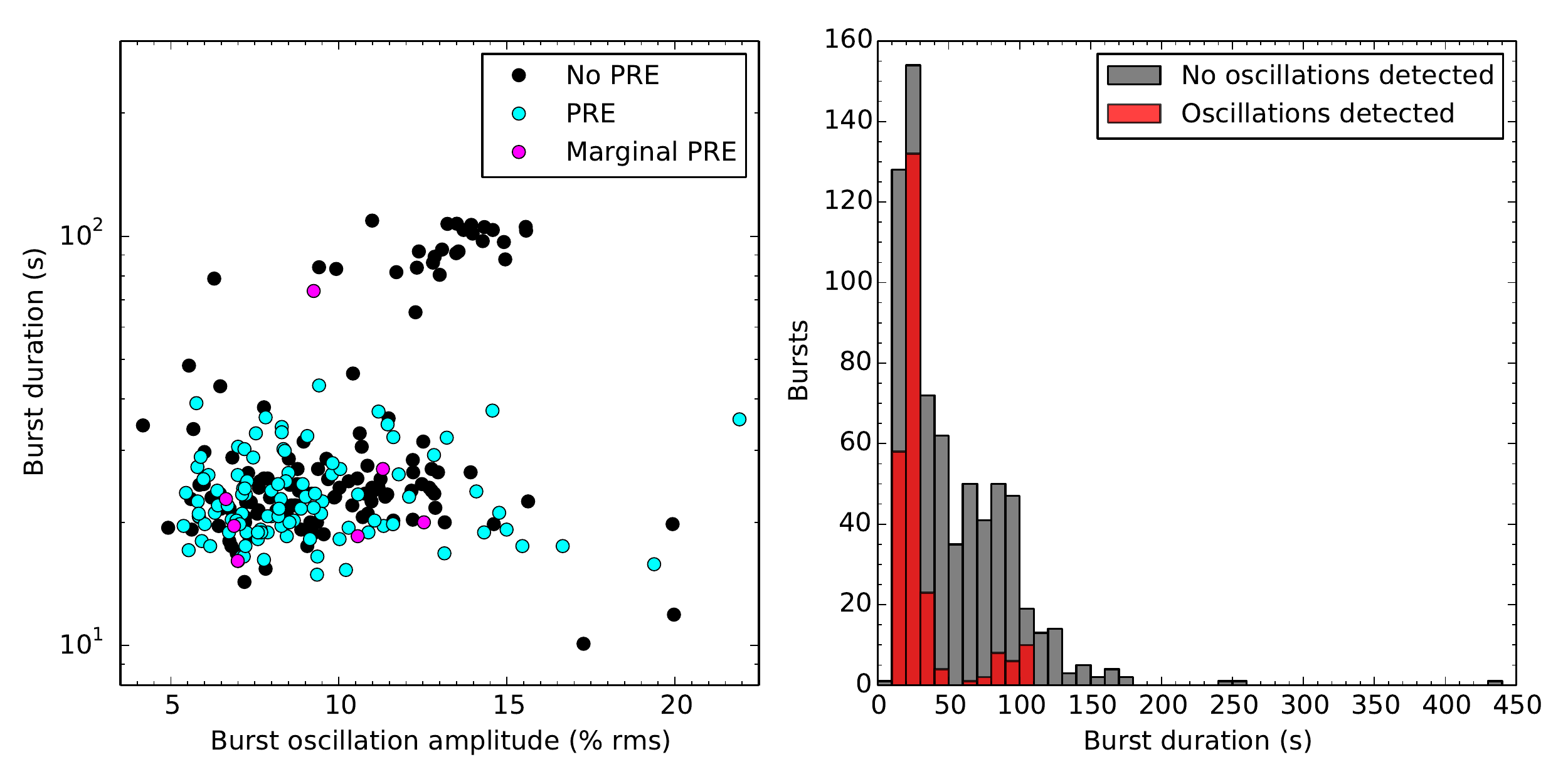}
	\caption{Relation between burst oscillations and burst duration for all sources in Table \ref{tab:osc_summary} combined. Left: Burst duration as function of burst oscillation amplitude of the strongest oscillation signal. Bursts without detected oscillations are omitted. Right: histograms of burst duration for bursts with detected oscillations (red) and without detected oscillation signals (grey). }
	\label{fig:bo_duration} 
\end{figure*}

In total, we have detected burst oscillations in 244 out of 950 bursts observed with \textit{RXTE} from 16 different sources. Table \ref{tab:osc_summary} summarizes per source the fraction of bursts in which oscillations were detected and in which phase of the burst the strongest detection signal was found. 
\cite{ootes17a}
found from analysis of a subset of the bursts presented here (694 vs. 950 bursts from burst oscillation sources) that the detectability of burst oscillations increases with $S_\text{z}$, as was also found in 
\cite{muno04a} and G08.
This correlation from the present analysis is shown Fig. \ref{fig:bo_sz} in which we plot histograms of the $S_\text{z}$ value for the bursts with and without oscillations (right panel). In this figure, we also plot in the left panel the $S_\text{z}$ value as function of burst oscillation amplitude (of the strongest signal). This shows that at low $S_\text{z}$ we detect oscillations with low amplitudes, while at higher $S_\text{z}$ we detect oscillations with both low and high amplitudes. The left panel of Figure \ref{fig:bo_sz} also indicates which of the bursts with oscillations show photospheric radius expansion. There is no apparent relation from this figure between $S_\text{z}$, burst oscillation amplitude, and PRE.

Next, we compare the detectability of burst oscillations and the detected oscillation amplitude of the strongest oscillation signal per burst to other burst properties presented in this paper. First of all, we find that burst oscillations are detected more often in bursts with PRE than without PRE (see Fig. \ref{fig:bo_PRE}). This result has previously been found to be correlated with spin frequency and burst type 
(\citealt{muno01,muno04a}, G08).
In relation to this correlation, bursts are more often found to show oscillations in bursts with higher peak fluxes; 
whilst at the same time the bursts with the highest peak fluxes 
also tend to 
experience PRE. Additionally, burst oscillations are found more often in bursts with short rise times ($\lesssim 3$ s) and short duration ($\lesssim 40$ s, Fig. \ref{fig:bo_duration}) which again coincides with those bursts that show most often a PRE-phase. Fig. \ref{fig:bo_duration} also shows a group of bursts with detected oscillations which have a burst duration $\gtrsim 70$ s. All but one of the bursts in this group are from XTE~1814$-$338 (all bursts from the non-intermittent accretion-powered pulsars seem to have burst oscillations in every burst, irrespective of the properties of the bursts; and XTE~1814$-$338 happens to have rather long bursts). 
We find no correlations between oscillation detectability and burst fluence or burst separation time. We find no relationship between the burst oscillation amplitude (of the strongest signal per burst) and any properties of the bursts in which they occur (except for the $S_\text{z}$ value). 

\section{Observation sample}
\label{minbar-obs}

The observation table contains information about  public \xte/PCA,  \sax/WFC and \igr/JEM-X observations of the burst sources described in \S\ref{sec:sources}, based on the selection criteria defined in  \S\ref{xteobs}, \S\ref{saxobs} and \S\ref{jemxobs}.
As we describe in \S\ref{minbar}, the completeness of the burst sample depends critically on the completeness of the observation sample. While our data selection criteria were designed to include every observation of known burst sources by each of the contributing instruments, we did identify some missing observations, notably for \igr/JEM-X, which resulted in a few missed bursts. 
With roughly 20 of the $\approx2000$ events in the burst sample of \cite{chelov17} missing for this reason, we estimate that the JEM-X observation sample is likely around 99\% complete. We plan to address the issue of these missing observations in future data releases.

In addition to the criteria for the individual instruments (as described in \S\ref{sec:source-data}), we filtered our analysis results for the accompanying ASCII table to list only observations in which the source was detected (based on the average count rate) at $3\sigma$ significance, or higher;
or where at least one burst was detected, even when the persistent emission was below our detection threshold.
The full set of observations, including those where no source is detected, is included in the sample available through the web interface.

After the selection for the observations where a source was detected at the $3\sigma$ level or higher, we retained observations from \xte\/ totalling \expxte.
The total exposure accumulated with the WFC for sources detected at $3\sigma$ significance or higher
is \expsax.  
The accumulated exposure for observations with significant detections by JEM-X is \expigr. 

The observations table includes a combined total of \observations\ 
PCA, WFC and JEM-X observations.

\subsection{Table format}

The observation table columns are listed in Table \ref{minbar-obs-table}. Below we describe in more detail how the column entries relate to the analysis in \S\ref{sec:analysis}.

% moved in response to the proof Q5 to ensure correct numbering
\startlongtable % AASTeX 6.3 command
\begin{deluxetable*}{ccccp{6cm}}
\tablewidth{480pt}
\tabletypesize{\scriptsize}
\tablecaption{Observation table columns, formats 
and description
  \label{minbar-obs-table}}
\tablehead{
& \colhead{Web table} & \colhead{ASCII table} \\
\colhead{Column} & \colhead{attribute} & \colhead{Format} & \colhead{Units} &
\colhead{Description}  }
\setcounter{magicrownumbers}{-45}
\startdata
\rnum & {\tt name} & A23 & & Source name \\
\rnum & {\tt instr} & A6 & & Instrument label \\
\rnum & {\tt obsid} & A20 & & Observation ID \\
\rnum & {\tt entry} & I6 & & MINBAR observation ID \\ 
\rnum & {\tt sflag} & A3 & & Data quality/analysis flags \\
\rnum & {\tt tstart} & F11.5 & MJD & Observation start time \\
\rnum & {\tt tstop} & F11.5 & MJD & Observation stop time \\
\rnum & {\tt exp} & I6 & s & Total exposure \\
\rnum & {\tt angle} & F7.2 & arcmin & Off-axis angle \\
\rnum & {\tt vigcorr} & F5.3 & & Vignetting correction factor \\
\rnum & {\tt nburst} & I3 & & Number of (type-I) bursts in the observation \\
\rnum & {\tt count} & F8.3 & count~cm$^{-2}$~s$^{-1}$ & Background-subtracted mean rate for target source \\
\rnum & {\tt counte} & F8.3 & count~cm$^{-2}$~s$^{-1}$ & Uncertainty\tablenotemark{a} on mean rate  \\
\rnum & {\tt sig} & F6.1 & & Detection significance for this observation \\
\rnum & {\tt flux} & F6.3 & $10^{-9}\ \epcs$ & Mean flux over the observation \\
\rnum & {\tt fluxe} & F6.3 & $10^{-9}\ \epcs$ & Estimated uncertainty\tablenotemark{a} on mean flux \\
\rnum & {\tt gamma} & F6.4 & & $\gamma$ ratio of persistent flux to mean peak flux of radius-expansion bursts \\
\rnum & {\tt sc} &  F6.3 & &     Soft colour \\
\rnum & {\tt hc} &  F6.3 & &     Hard colour \\
\rnum & {\tt s\_z} & F6.3 & & $S_\text{z}$ value,  giving position in the colour-colour diagram \\
\rnum & {\tt model} & A30  & & Spectral model ({\sc XSpec} syntax)  \\
\rnum & & E9.3 & & Spectral index $\Gamma$ of power law (where present) \\
\rnum & & E9.3 & & Uncertainty\tablenotemark{a} on spectral index $\Gamma$ \\
\rnum & & E9.3 & photons~keV$^{–1}$~cm$^{–2}$~s$^{–1}$ at 1~keV & Normalisation of power law (where present) \\
\rnum & & E9.3 & photons~keV$^{–1}$~cm$^{–2}$~s$^{–1}$ at 1~keV & Uncertainty\tablenotemark{a} on power-law normalisation \\
\rnum & & E9.3 & keV & Temperature $kT$ of blackbody component (where present) \\
\rnum & & E9.3 & keV & Uncertainty\tablenotemark{a} on blackbody temperature \\
\rnum & & E9.3 & $(R_{\rm km}/d_{\rm 10\ kpc})^2$ & Normalisation of blackbody component (where present) \\
\rnum & & E9.3 & $(R_{\rm km}/d_{\rm 10\ kpc})^2$ & Uncertainty\tablenotemark{a} on blackbody normalisation \\
\rnum & & E9.3 & keV & Input soft photon (Wien) temperature $T_0$ of Comptonisation component (where present) \\
\rnum & & E9.3 & keV & Uncertainty\tablenotemark{a} on Comptonisation input  temperature $T_0$\\
\rnum & & E9.3 & keV & Plasma temperature $kT$ of Comptonisation component (where present) \\
\rnum & & E9.3 & keV & Uncertainty\tablenotemark{a} on Comptonisation plasma temperature \\
\rnum & & E9.3 &  & Plasma optical depth $\tau_C$ of Comptonisation component (where present) \\
\rnum & & E9.3 &  & Uncertainty\tablenotemark{a} on Comptonisation optical depth  \\
\rnum & & E9.3 &  & Normalisation of Comptonisation component (where present) \\
\rnum & & E9.3 &  & Uncertainty\tablenotemark{a} on Comptonisation normalisation  \\
\rnum & & E9.3 & keV & Line energy for Gaussian component (where present) \\
\rnum & & E9.3 & keV & Uncertainty\tablenotemark{a} on Gaussian line energy \\
\rnum & & E9.3 & keV & Line width $\sigma$ for Gaussian component (where present) \\
\rnum & & E9.3 & keV & Uncertainty\tablenotemark{a} on Gaussian line width \\
\rnum & & E9.3 & photons~cm$^{-2}$~s$^{-1}$ & Normalisation for Gaussian component (where present) \\
\rnum & & E9.3 & photons~cm$^{-2}$~s$^{-1}$ & Uncertainty\tablenotemark{a} on Gaussian line normalisation \\
\rnum & {\tt chisqr} & F5.2 & & Mean reduced $\chi^2$ of spectral fits \\
\rnum & {\tt chisqre} & F5.2 & & Standard deviation of reduced $\chi^2$ from spectral fits, where more than one spectrum is fit 
\enddata
\tablenotetext{a}{Uncertainties are at the 1$\sigma$ (68\%) confidence level}
\end{deluxetable*}

\paragraph{1. Burster name} ({\tt name} in the web table) The target for the observation.
For the imaging instruments, we present analysis results for lightcurves and spectra extracted for each burst source within the FOV. 
For \xte/PCA, we list the source closest to the aimpoint in the case of multiple sources within the FOV, and/or the only active source within the FOV, as determined from contemporaneous ASM measurements (see \S\ref{obs-xte}).

\paragraph{2--3. Instrument, observation ID} these attributes are identical to the corresponding columns in the burst table (see \S\ref{minbar}).

\paragraph{4. MINBAR observation ID} ({\tt entry}) The unique numeric identifier for each observation in the MINBAR sample.

\paragraph{5. Analysis flags} ({\tt sflag}) Indicates a number of sub-optimal situations for the data analysis, as described in Table \ref{analysis-flag-table}. % \dg{note alternate format in web table}

\begin{deluxetable*}{ccp{10cm}}
\tablecaption{Analysis flags relevant to MINBAR observations
  \label{analysis-flag-table}}
\tablehead{
\colhead{Label} & \colhead{Instrument} & \colhead{Description} }
\startdata
- & all & No significant analysis issues \\
a & PCA & Multiple sources active in the field, but sources other than the named source contribute negligible flux \\ % 1
b & PCA & Multiple sources active in the field and sources other than the named origin contribute non-negligible flux \\ % 2 
c & PCA & Multiple sources active in the field and no information is available about the relative intensities \\ % 4
d & all & Could not constrain persistent flux in the spectral fit; flux value is $3\sigma$ upper limit \\
e & PCA & Standard filtering left no good times \\ % 16
f & PCA, JEM-X & No Standard-2 mode data, or no spectrum available \\ % 32
g & PCA & No FITS data available in archive % 128
\enddata
\end{deluxetable*}

\paragraph{6 \& 7. Observation start and end times} ({\tt tstart, tstop}) The nominal extent of each observation, in MJD (UT). Data may not be continuous throughout the interval, due to occultations, passages through regions of high particle flux, or other instrumental factors.

\paragraph{8. Total exposure} ({\tt exp}) The total on-source time for the observation in seconds, taking into account the data gaps.
We note that the treatment for different instruments is slightly different here, with the table entries corresponding to PCA and WFC observations typically spanning multiple satellite orbits, during which the target sources are not consistently visible. 
The exposure for these instruments thus is less than the observation time span (i.e. the difference between the start and stop times; columns 6 \& 7). 
For JEM-X however, each observation corresponds to a science window (SCW) which is (typically) an uninterrupted observation interval, so that the exposure is (approximately) the same as the observation time span.

\paragraph{9. Off-axis angle} ({\tt angle}) The angle (in arcmin) between the instrument aimpoint and the source position.

\paragraph{10. Vignetting correction factor} ({\tt vigcorr}) The factor describing 
the detector efficiency compared to a source located at the aimpoint.

We generated a separate response matrix for each observation factoring the position of the source within the FOV, so this attribute 
approximately
takes into account the 
decrease in instrumental sensitivity moving away from the aimpoint.

\paragraph{11. Number of (type-I) bursts detected in the observation} ({\tt nburst}) This is the  number of bursts 
from the source associated with this entry, detected in the observation. For PCA, which lacks the capability to discriminate between different sources in the field of view, there may be additional bursts from other sources. 
For fields containing MXB~1730$-$335 (the Rapid Burster) there additionally may be (many) type-II events, which are not included in MINBAR.
There may also be
additional weakly-significant candidates which could not be confirmed as bursts.

\paragraph{12 \& 13. Photon flux and error} ({\tt count, counte}) The background-subtracted  count rate (and $1\sigma$ uncertainty) in units of counts~cm$^{-2}$~s$^{-1}$ averaged over the entire observation. 
For JEM-X observations where both cameras are operational, we average over JEM-X 1 and 2, and adopt the empirical effective area of 64~cm$^{2}$ appropriate for the persistent emission, determined in \S\ref{sec:area}.
For the PCA, we give the count rate per active PCU, and adopt the effective area determined as for JEM-X, of 1400~cm$^{2}$.

\paragraph{14. Detection significance } ({\tt sig}) The estimated detection significance for this source in the observation. This is calculated as the source photon flux divided by the uncertainty. 
We only include observations in the table where the detection is at least at the (estimated) $3\sigma$ level, although this quantity is not always available for instrumental reasons. We also include any observations in which a burst has been detected.

\paragraph{15 \& 16. Mean persistent flux  for the observation} ({\tt flux, fluxe}) This attribute is the integrated 
flux $F_p$ and uncertainty in units of $10^{-9}\ \epcs$, based on the spectral model given in column 23, and the best-fit spectral parameters in columns 24--45. 
Note that for some observations, the signal-to-noise is insufficient to constrain the flux. These observations are flagged as ``d'' (Table \ref{analysis-flag-table}), and the flux provided is instead the estimated 3-$\sigma$ upper limit.

\paragraph{17. The $\gamma$-value} ({\tt gamma}) 
The ratio of the estimated bolometric persistent flux to the average Eddington flux from the source (from Table \ref{table:peakfluxes}, where available), as described in \S\ref{bolcor}. 

\paragraph{18 \& 19. The soft \& hard spectral colours}  ({\tt sc, hc}) These attributes parametrise the shape of the persistent spectrum, and are derived from the best-fit spectral model, as described in \S\ref{colours}.

\paragraph{20. The $S_\text{z}$ parameter}  ({\tt s\_z}) This attribute quantifies the position on the colour-colour diagram, for those sources with observations spanning sufficient range of spectral shapes to describe it (see \S\ref{colours}).

\paragraph{21. The spectral model}  ({\tt model}) This column specifies the spectral model adopted for the persistent spectrum, in {\sc xspec} format \cite[]{xspec, xspec12}. See \S\ref{obsanl} for a description of how the spectral models were chosen. 
Columns 22--43 list the spectral parameters corresponding to the adopted model, with columns 22--25 describing the power-law component, where present; 26--29 the blackbody component; 30--37 the Comptonisation component; and 38--43 the Gaussian component.
In the online web interface, each set of parameters is listed as attributes {\tt par1}, {\tt par1e}, {\tt par2}, {\tt par2e} and so on, with {\it par1} corresponding to the $N_H$ value, and the remaining parameters present in order depending upon the choice of spectral model.
Where no spectral information was available, or no good fit could be obtained 
this attribute (and the subsequent spectral parameter attributes below) is blank.

\paragraph{Power law component}

\paragraph{22 \& 23. Power law spectral photon index $\Gamma$ and uncertainty} For those observations with a power-law component, we list here the best-fit spectral photon index and uncertainty.

\paragraph{24 \& 25. Power law normalisation} The best-fit normalisation at 1~keV and uncertainty of the power-law component, where present.

\paragraph{Blackbody component}

\paragraph{26 \& 27. The blackbody temperature} For those observations with a blackbody component, we list in these columns the best fit temperature $kT$ and uncertainty in keV.

\paragraph{28 \& 29. The blackbody normalisation } The best fit normalisation and uncertainty for the blackbody, where present.

\paragraph{Comptonisation component}

\paragraph{30 \& 31. The Comptonisation component seed photon temperature} For those observations with a Comptonisation continuum component, we list in these columns the best-fit seed photon (Wien) temperature, k$T_0$ and uncertainty, in keV.

\paragraph{32 \& 33. The Comptonisation plasma temperature} The best-fit plasma temperature $kT$ and uncertainty. This attribute (and the optical depth $\tau_C$, below) are measured with a fixed ``geometry'' flag for the {\tt compTT} component of 1.0, corresponding to the default ``disk'' geometry %\dg{fill in here}.

\paragraph{34 \& 35. The Comptonisation optical depth} The best-fit optical depth $\tau_C$ for scattering for those observations including a Comptonisation component.

\paragraph{36 \& 37. The Comptonisation component normalisation } The best-fit normalisation and uncertainty of the {\tt compTT} component, where present.

\paragraph{Gaussian component}

\paragraph{38 \& 39. The centroid energy of the Gaussian} For those observations with a Gaussian component (simulating Fe K$\alpha$ emission around 6.4--6.7~keV), we list here the best-fit line centroid energy (and uncertainty). 

\paragraph{40 \& 41. The Gaussian width} The best-fit standard deviation $\sigma$ and uncertainty of the Gaussian component, where present.

\paragraph{42 \& 43. The Gaussian normalisation} The best-fit normalisation of the Gaussian component and estimated uncertainty, where present.

\paragraph{44 \& 45. The fit statistic} ({\tt chisqr, chisqre}) The reduced $\chi^2_\nu$ ($\equiv \chi^2/\nu$, where $\nu$ is the number of degrees of freedom in the fit). Where more than one spectrum was used for a simultaneous fit (e.g. for the case of the \xte/PCA where multiple PCUs were operational) we list the mean $\chi^2_\nu$ and the standard deviation.

\subsection{Observation summary}
\label{subsec:obs_summ}

\begin{figure}[h]
	\includegraphics[width=\columnwidth]{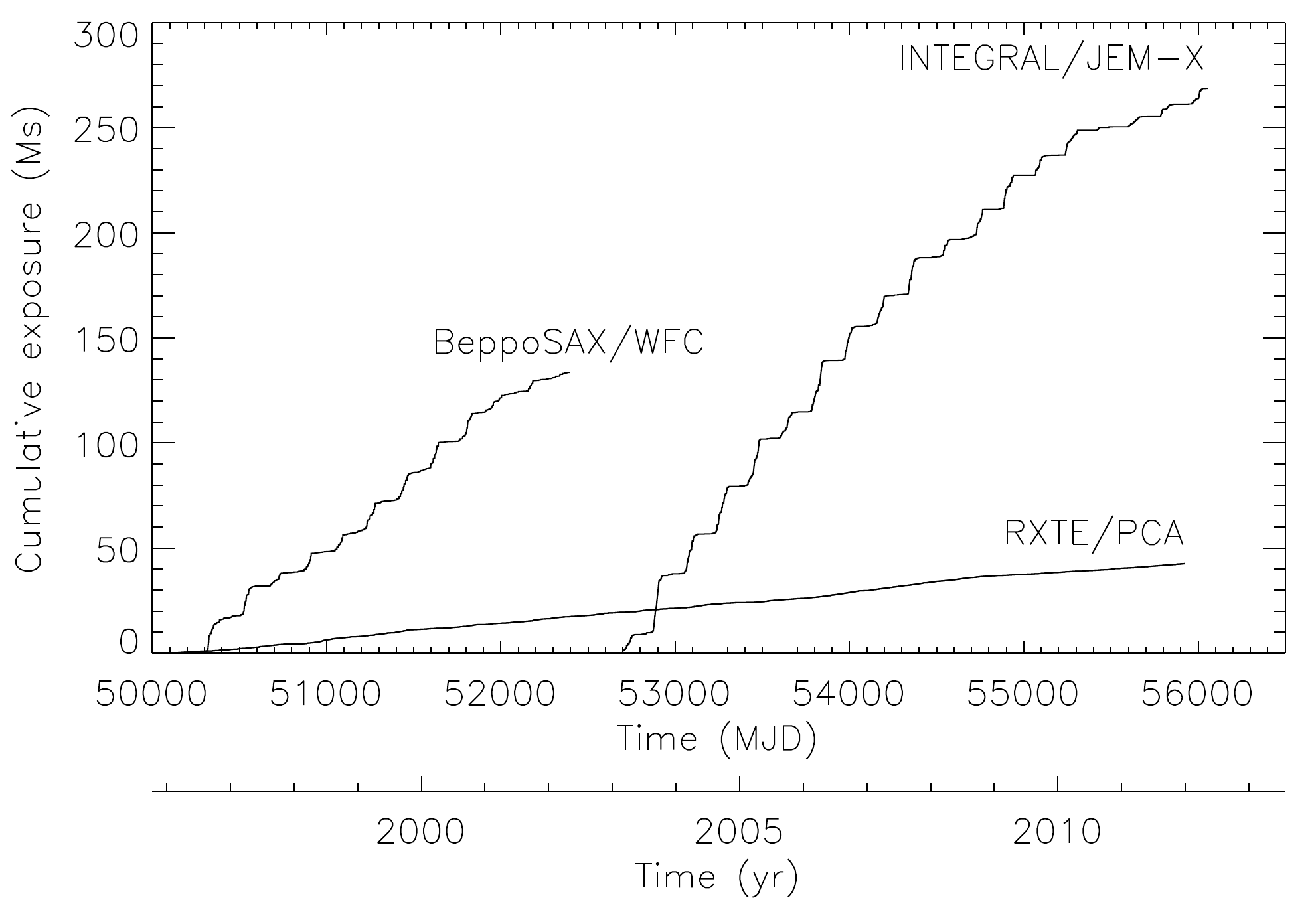}
	\caption{Cumulative exposure for each of the three missions comprising the MINBAR observation sample. 
	% \dg{need to update when you complete the JEM-X sample through rev. \pubdatacutoffrev\ (MJD \pubdatacutoffmjd).} 
	\label{fig:cum_exp} }
\end{figure}

The total exposure over all the sources was \expsax, \expigr, and \expxte\ for \sax/WFC, \igr/JEM-X and \xte/PCA, respectively. The cumulative exposure over the history of each mission evolved as shown in Figure \ref{fig:cum_exp}.
The 6-monthly ``steps'' visible in the curves for \sax\/ and \igr\/ are likely related to the semi-annual periods of visibility of the Galactic centre. The exposure for \xte\/ increases at a lower, although more steady rate over the mission lifetime.

The concentration of sources around the Galactic centre, and the corresponding observational focus on that area, results in a strong dependence of total exposure on angular distance from the centre. Most sources within $5^\circ$ of the Galactic centre have accumulated 15~Ms of total exposure. 
For sources more than $5^\circ$ away, 1--10~Ms is more typical.

\begin{figure}[h]
	\includegraphics[width=\columnwidth]{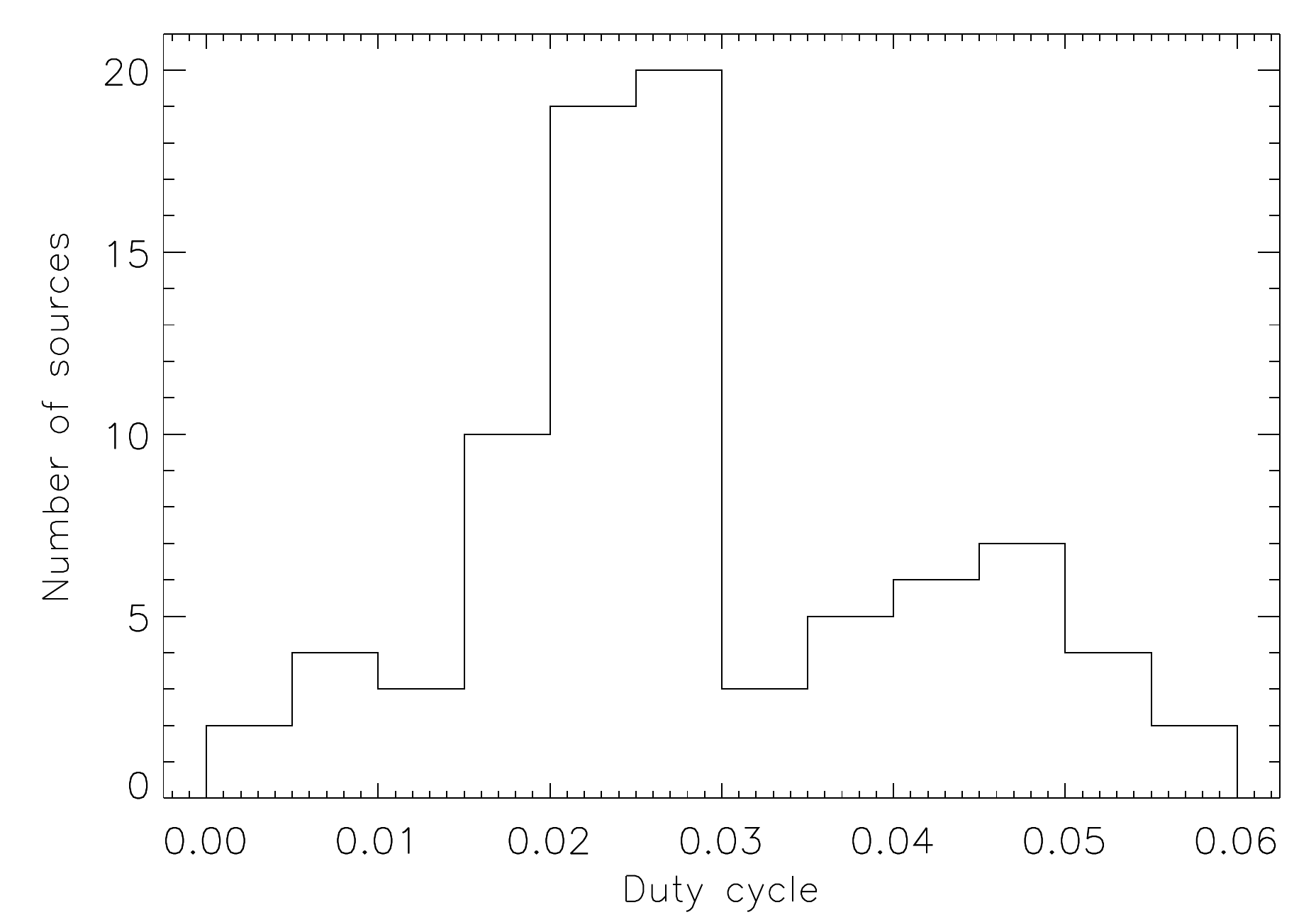}
	\caption{Duty cycle for each source contributing to MINBAR, calculated as the combined exposure divided by the timespan over which observations were made.
	\label{fig:duty_cycle} }
\end{figure}

We calculated the duty cycle for each source contributing to MINBAR, by merging the good time intervals from each instrument and calculating the overall combined exposure. We then divided this value by the total time span of the observations. The distribution of duty cycles was double-peaked, with sources clustering around $\approx2.5$\% or $\approx4.5$\% (Fig. \ref{fig:duty_cycle}). The higher peak corresponds to the Galactic centre sources, which had generally higher exposure.
We note that the mean duty cycle for the combined set of MINBAR observations, of 2.9\%, was substantially higher than the average for the individual instruments, at 1.2\% (\xte/PCA), 0.6\% (\sax/WFC) and 2.2\% (\igr/JEM-X).

\begin{figure}[h]
	\includegraphics[width=\columnwidth]{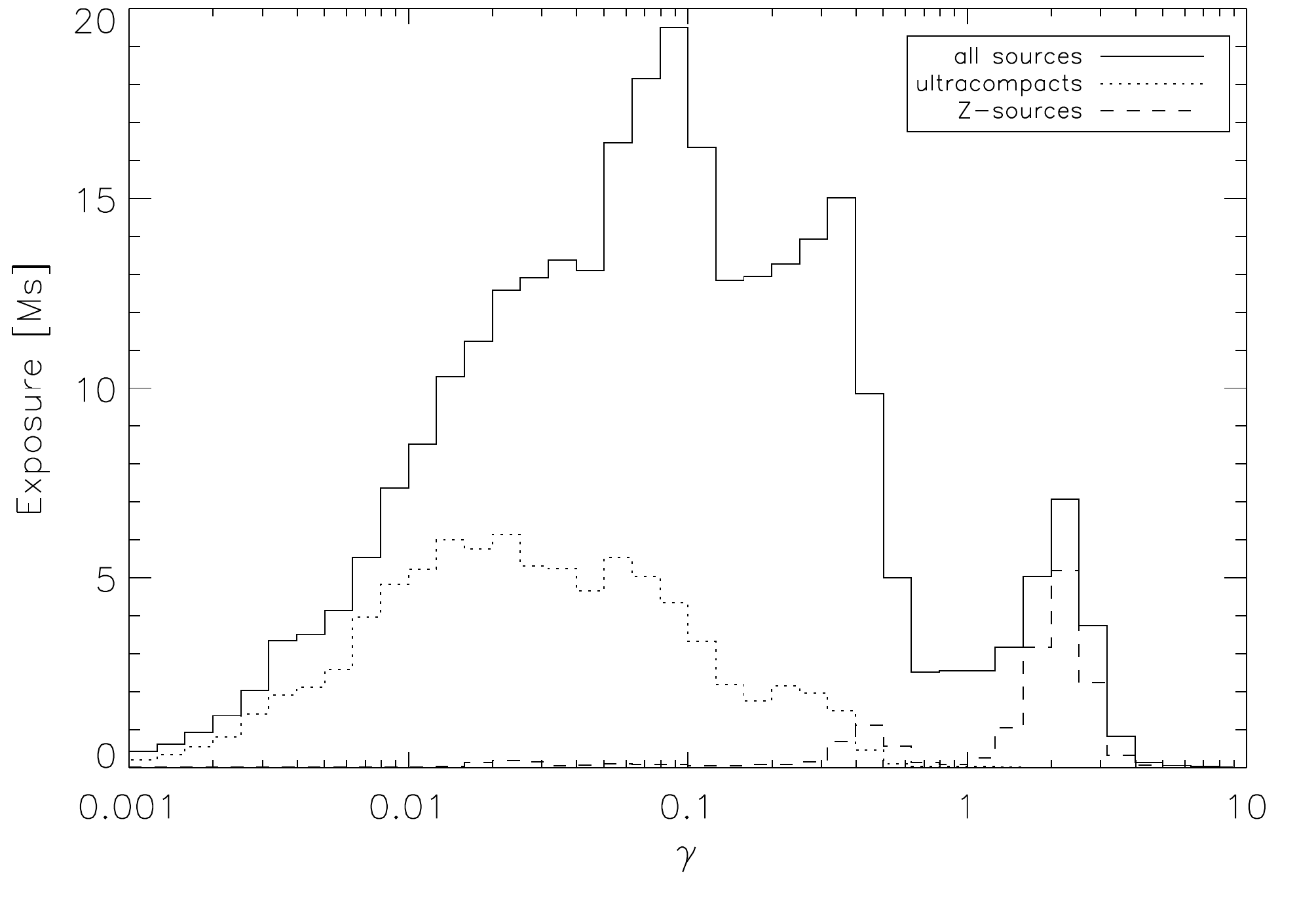}
	\caption{Exposure as a function of $\gamma$-value (proportional to the accretion rate in units of $\dot{m}_{\rm Edd}$), both for the entire MINBAR sample, and the subsamples comprised of the ultracompact sources (and candidates), and the Z-sources (see \S\ref{sec:sources}) 
	% \dg{need to update with new $\gamma$-values as of 2018 Sep}.
	% done 2019 Dec 17 -- dkg
	\label{fig:gammadist} }
\end{figure}

% added in proof, Q1
We show the exposure as a function of $\gamma$-value in Figure \ref{fig:gammadist}. This quantity is the ratio of the persistent flux $F_p$ to the average peak flux of radius-expansion bursts (for those sources where they are observed; see \S\ref{fluxEdd}). We adopted $\gamma$ as a measure of the accretion rate, in units of the Eddington value (see \S\ref{bolcor}).
We find that the highest exposure is accumulated at $\gamma\approx0.1$, corresponding to an inferred accretion rate of around $0.1\dot{m}_{\rm Edd}$. This accretion rate is (perhaps not coincidentally) also where the bursts have their highest density (see Fig. \ref{fig:tau_v_gamma}). The inferred accretion rate ranges over almost two orders of magnitude higher and lower. The lower range, 0.01--$0.1\dot{m}_{\rm Edd}$, is typically where the ultracompact sources fall, while the highest values, $\gamma>1$, are dominated by the Z-sources.

\begin{figure}[h]
	\includegraphics[width=\columnwidth]{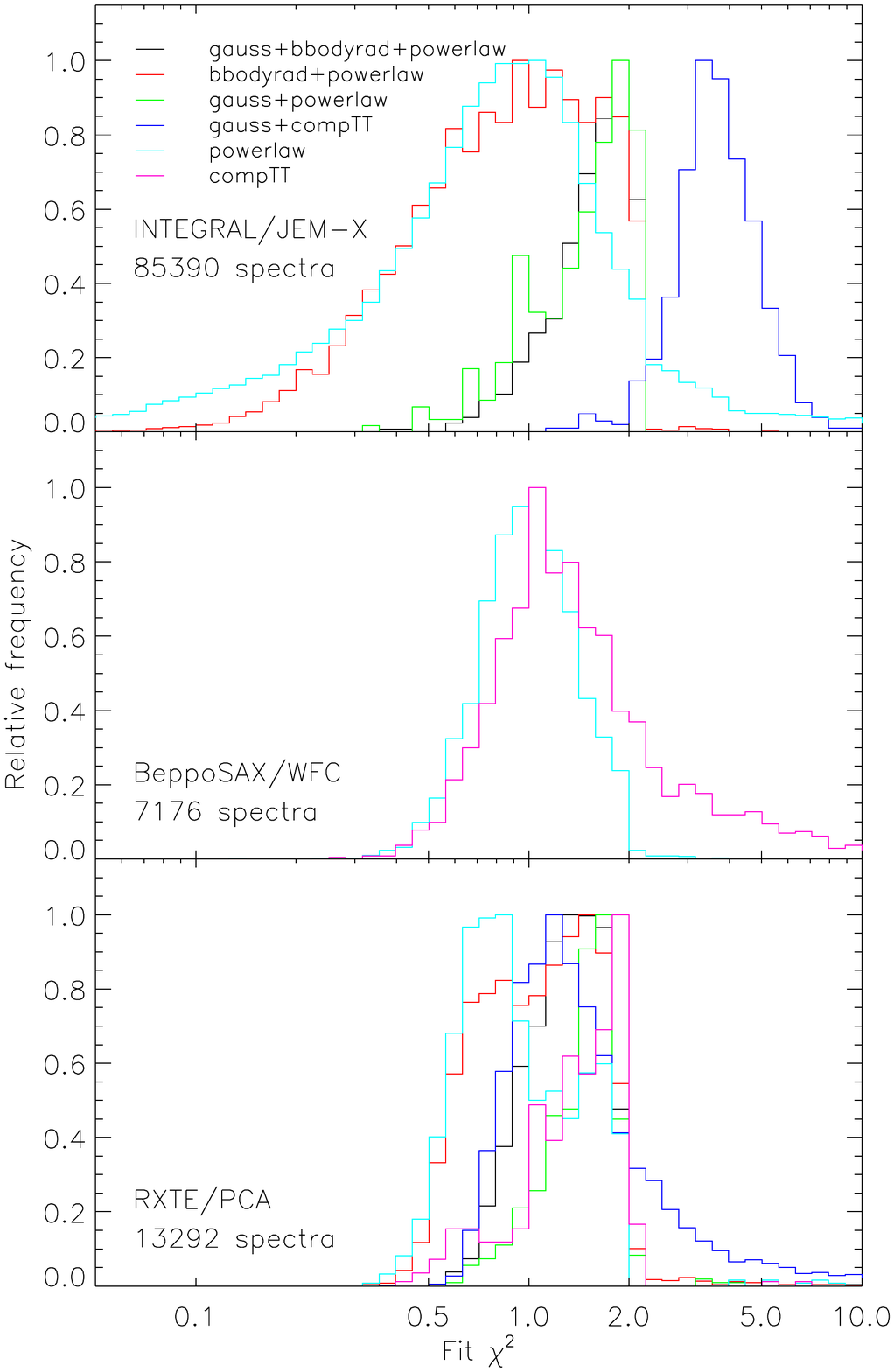}
	\caption{Distributions of reduced $\chi^2$ for persistent spectral fits to the observations in the MINBAR sample. Each panel shows the results from one of the instruments, and we break down the distributions into each model combination.
	\label{fig:specsum} }
\end{figure}

We carried out spectral fits for 
105858 % total(num) from review_fits.pro, 2019 Jun 17
individual observations, excluding those observations for which the persistent spectrum was not available, or where the source was so faint that the best-fit flux value was consistent with zero. We summarise the fit statistics in Fig. \ref{fig:specsum}. 
The fitting approach was slightly different for each instrument, resulting in a variety of different breakdowns against the range of spectral models adopted (Table \ref{tab:spec_summ}; see also \S\ref{obsanl}).

For \igr/JEM-X and \xte/PCA, we fit initially with a {\tt powerlaw} component alone, and successively added components for cases where the reduced $\chi^2_\nu$ value indicated a poor fit ($\chi^2_\nu>2$). 
For \xte/PCA, this approach yielded fit statistics that were in the majority less than this threshold, but with a not-insignificant tail at higher values, particularly for the ``apex'' model {\tt gauss+comptt} chosen for the highest signal-to-noise observations.

For \igr/JEM-X, the {\tt powerlaw} and {\tt bbodyrad+powerlaw} models resulted in broad distributions of $\chi^2_\nu$ centred around 1, indicating a good fit on average. Some {\tt powerlaw} fits yielded  $\chi^2_\nu$ values in excess of 2; the majority of these fits were for the lowest signal-to-noise spectra ({\tt sig}$<8$), for which the {\tt powerlaw} spectral index $\Gamma$ was frozen at 2. A smaller number of observations best fitted with {\tt gauss+powerlaw} exhibited a distribution of $\chi^2_\nu$ rising towards the threshold of 2. 
For the apex model, which was also the model chosen for the majority of the spectra, the $\chi^2_\nu$ values were distributed around a mode in the range 3--4, suggesting substantial systematic contributions to the 
spectral bin variations. 

For \sax/WFC, a choice of either {\tt powerlaw} or {\tt compTT} continuum were chosen, depending upon which model provided the best fit. The distribution of the resulting $\chi^2$ values for both models were centred around 1, but a significant fraction of the fits (particularly for the {\tt compTT} models) had much higher values, suggesting that additional components may be required.

\begin{deluxetable*}{rcccc}
\tablecaption{Summary of spectral fits by model and instrument
  \label{tab:spec_summ}
}
\tablewidth{0pt}
\tablehead{
   \colhead{Model} & \colhead{\igr/JEM-X} & \colhead{\sax/WFC} & \colhead{\xte/PCA} & \colhead{Total}
}
\startdata
{\tt powerlaw} with frozen $\Gamma=2$ & 53276 & 2847  & 143 & 56266  \\
{\tt powerlaw} & 77866 &  5047 &  1033 & 83946\\
{\tt gauss+powerlaw}  & 314 & \nodata &   450 & 764\\
{\tt bbodyrad+powerlaw} &  6051 & \nodata &  4959 & 11010\\
{\tt gauss+bbodyrad+powerlaw} &   601 & \nodata &  1859 & 2460\\
{\tt compTT} &     2 &  2129 &   403 & 2534 \\
{\tt gauss+compTT} &   556 & \nodata &  4588 & 5144 \\
\hline
Total &   85390 & 7176 &   13292  & 105858\\
\enddata
\tablecomments{The total number of observations with spectral fits is less than the total number of observations (\observations), because 
of a range of analysis issues preventing spectral fits; primarily, missing spectral files for JEM-X observations (flag ``f''; Table \ref{analysis-flag-table})
}
\end{deluxetable*}

\section {Discussion}
\label{sec:discussion}

The MINBAR sample of thermonuclear (type-I) X-ray bursts is the largest yet assembled, and provides an unprecedented overview of the diverse phenomenon of thermonuclear bursts. 
By combining the extensive observations of the wide-field instruments on \sax\/ and \igr, with the high-sensitivity and high timing resolution offered by the \xte/PCA, we present complementary views that incorporate detailed information down to millisecond time resolutions coupled with good statistics for rare events in many tens of burst sources.
The provision, for the first time, of a companion
observation catalog (not previously available for other large burst studies) offers the prospect of improved understanding of how the accretion flow affects the surface burning, as well as providing critical data on long-term accretion behaviour, and timescale and patterns for variation in spectral shapes.

In this section we briefly summarise the principal conclusions arising from the assembly and study of the MINBAR sample, and provide some suggestions for future directions both with this sample and future observations.

\subsection{Burst rate}
\label{sec:burstrate}  % not to be confused with the later section sec:burstrates

The pattern of variation in burst rate as a function of accretion rate for 
selected sources with large ($>100$) numbers of bursts in MINBAR
supports the classification of sources into two main groups, thanks to substantially improved statistics provided for individual sources. For the first group, with typical members ultracompact candidates or with relatively slow ($\lesssim300$~Hz) rotation speed, the burst rate appears to increase steadily with accretion rate, to the point (in at least one source) where the burning transitions instead to quasi-stable mHz oscillations, similar to theoretical predictions. 
In the second group, typified by those fast rotators ($\gtrsim300$~Hz), the burst rate reaches a maximum at some intermediate accretion rate, typically one tenth (or lower) of the Eddington rate. Above that accretion rate, the burst rate {\it decreases} with increasing accretion rate. 

Although this behaviour has been observed before \cite[e.g.][G08]{corn03a} the detail provided via the MINBAR sample offers the most detailed 
view of this dichotomy, and also provides evidence that 
the accretion rate at which the burst rate reaches a maximum is anti-correlated with the spin frequency \cite[]{gal18a}. 
While this result remains perplexing, an explanation may be emerging based on the variation of ignition latitude with accretion rate \cite[e.g.][]{cavecchi17}. Even if this explanation is not correct, it seems more clear than ever that the effects of rotation on the ignition of thermonuclear bursts cannot be ignored.

\subsection{Accretion emission changes during bursts}
    
    The 
    % excellent 
    high signal-to-noise pre-burst emission  and time resolved burst spectra provided by the MINBAR catalog, particularly for the bursts observed by \xte/PCA, has enabled in-depth studies of the influence of bursts on the persistent emission 
    % \ek{(refer also Degenaar et al. SSRv review?}. 
    \cite[e.g.][]{deg18a}.
    Typically, the persistent emission is found to increase during the early stages of bursts, by a factor of several. Initially identified in analysis of the PCA bursts contributing to MINBAR \citep{worpel13a}, the effect has been confirmed by other instruments, including a joint \xte/{\it Chandra}\/ observation of SAX~J1808.4$-$3658 \cite[]{zand13a}. Accounting for this effect typically leads to a significant improvement in the fit statistic $\chi^2_\nu$ for the majority of time-resolved burst spectra, although 
    % not yet to the point of formal consistency.
    it may not be formally required for individual bursts.
    
    This enhanced persistent emission occurs for both PRE and non-PRE bursts \cite[]{worpel15}, though its intensity varies more erratically, and the improvement in $\chi^2$ is not as good %\hw{can you elaborate on what you mean by ``chaotic''? -- dkg}
    in the former, perhaps due to the changing structure of the photosphere during radius expansion. The cause of persistent emission enhancement is still not understood. Reflection of surface nuclear burning off the accretion disc \cite[e.g.][]{zand13a,keek17b}, and temporarily enhanced accretion rate induced by radiation drag \cite[e.g.][]{walker89,walker92,ml93}, have been suggested as possible causes. \cite{peille14} find that 
    % No! corrected by dkg 2020 Apr
    % burst 
    kHz quasi-periodic
    oscillations in 4U~1636$-$536 and 4U~1608$-$522 are suppressed for several tens of seconds after bursts, suggesting that the inner accretion disc is indeed significantly affected.% \hw{ref? -- dkg}.
    \added{The most recent modelling suggests that the response of the accretion disk to a burst may be  complex, involving a number of effects \cite[]{fragile20}. }
    
     In the 10--20~keV range there is evidence of spectral hardening later in the burst (e.g. \citealt{vpd90,kuul02a}) but at higher ($\geq 30$\,keV) energies, this pattern is apparently reversed; the influx of burst luminosity often causes a {\it reduction} of hard X-ray photons (e.g. \citealt{mc03,ji13, ji14, ji15, kajava17b}). This effect has been attributed to the rapid cooling of an extended corona. These changes may be related to the accretion state of the source, and the temperature relevant to coronal cooling may also vary from source to source \cite[e.g.][]{fragile18}.
Future MINBAR data releases, perhaps incorporating bursts observed by {\it NuSTAR}\/ (with its improved sensitivity at high energies), may enable a more thorough and systematic investigation of such spectral changes.

\subsection{Cooling after bursts}
\label{sec:cooling}
    The cooling tails of type-I X-ray bursts are a potent diagnostic of the layers of the neutron stars above and just below the ignition layer. This domain offers some interesting physics, where electrons are partly degenerate and where photons contribute significantly to the pressure and heat capacity. Traditionally, the cooling tails, expressed in either units of photon or energy flux, were modeled with an exponential decay function. That model is often satisfactory for the first 90\% of the decay, but 
    % not further down. 
    no later.
    
    % Cumming \& Macbeth (2004, ApJ 603, L37) 
    \cite{cm04} 
    and \cite{cumming06} introduced a more physically-based decay function for superbursts, consisting of a broken power-law function. The first shallow power law represents the phase when the heat wave is traveling from the ignition depth to the photosphere, the second steep power law all times beyond.  
    
    \cite{zand14a} extended this work to ordinary (helium-fueled) X-ray bursts from ultracompact X-ray binaries, in which 
    % hydrogen is deficient and 
    no nuclear burning due to hydrogen burning (rp-process) is expected that may extend into the cooling phase. The 37 X-ray bursts for this study were extracted from the MINBAR database (excluding the superburst from 4U~1636$-$536 on MJD~51962.70296). 
The study found for all bursts that the cooling tails for 99\% down from the peak flux could better be modeled with a single power law function than an exponential decay function. In fact, the single power law function is simpler than what is actually expected from theory,
which predicts
a changing power law index as the result of a changing dominance of different contributors (ions, electrons, photons) to the heat capacity with temperature. The data shows singular power laws with decay indices of 1.3--2.5, but peaking at 1.8 which is the value expected when electrons determine the heat capacity.
    
    % need to add the burst refs for kuuttila17 also?
    \cite{zand17b} extended this work to 1254
    X-ray bursts from the MINBAR database\footnote{Note that their table C1 lists the MINBAR burst ID ({\tt entry} attribute of the {\tt minbar} table; see \S\ref{minbar}) for all but 26 of those bursts; the published version of the MINBAR table now includes those additional events, which can be identified by time. % \jz{do you want to update the table with the missing values? --- dkg} \dg{do you mean update a published table? I think that is not possible..? - jz13112019} ,% 2019 Sep
    },
    including hydrogen-rich sources. The analysis method was accommodated to filter out the contribution from $rp$-burning to the cooling tail. This order-of-magnitude larger sample provided a confirmation of the earlier study and provided for the first time statistical data for the $rp$-process in X-ray bursts. All bursts selected for these two studies were PCA bursts, because only those provided enough statistical quality to probe the cooling below 10\% of the peak flux. Regarding bursts from WFC and JEM-X, we remark that usually an exponential decay fits the data just as well as a power law decay.
    
    \cite{kuuttila17} also performed a study of cooling tails in 540 bursts from the PCA sample, but followed a different approach whereby they attempted to measure the changes in the power law index as expected from theory. They did not allow for an $rp$-process component. 
    % \jz{conclusions? -- dkg} \dg{I think this text speaks for itself..? - jz13112019} % 2019 Sep

\subsection{Bursts during transient outbursts}

Several notable transient outbursts occurred during the period covered by the MINBAR observation.
\cite{chenevez11b}
describe the bursting behaviour of the transient source IGR~J17473$-$2721 during a six-month long outburst in 2008, which seemed to be triggered by the occurrence of a burst. 
The entire outburst was well covered by several instruments,
and spanned a
range of accretion luminosities between 1\% and 20\% of Eddington. A total of 
61 
bursts were observed throughout the outburst, among which, one occurred simultaneously in both JEM-X and PCA. % Two particularly interesting results can be stressed: 1) 

This outburst was notable for a wide range of bursting behaviour, with
seven distinct phases identifiable, seemingly covering 
several of the regimes understood theoretically \cite[e.g.][]{gal17b}.
Additionally, the transition between some pairs of states seemed to occur
at accretion rates 10 times 
higher 
than 
predicted by theory.

The burst 
rate 
dropped when the accretion rate reached 15\% of Eddington, shortly before the peak of the outburst which was accompanied by 
a sudden persistent spectral change from the ``hard'' to ``soft'' state. The burst activity resumed after one month, when the accretion rate returned below 5\% of Eddington, thus demonstrating a hysteresis of burst rate vs. accretion rate. 

We note that similar burst intermissions have been observed from other bursting transients %\jc{refs? --- dkg}
(e.g. EXO~1745-248; 
G08).
One interpretation is the stabilization of the thermonuclear burning at high temperatures due to the heating of the neutron star crust by accretion, and the subsequent thermal relaxation of the crust delaying the resumption of unstable burning after the accretion rate reached back the level at which the burning stabilized.
A similar effect is also observed following superbursts, when heating of the envelope instead by carbon burning is inferred to cause stable burning of the H and He fuel, so that bursting ceases. A subset of the data comprising MINBAR was employed to derive the strongest limit ($<15$~d) so far on burst quenching by superbursts 
\cite[]{keek12a}.
The value here is a limit because the incomplete coverage means that earlier bursts may have occurred but not been observed. Even so, this limit is shorter than that observed for the transients, 29~d (for IGR~J17473$-$2721) and 39~d (for 4U~1745$-$248).

Bursts from another transient, IGR~J17254$-$3257, were observed occurring at slightly different accretion rates but with markedly different durations. \cite{chenevez07} compare two bursts seen by JEM-X while the source was at a low accretion rate. The first burst observed from this source (MINBAR \#4806, on MJD~53052.82221)
was short, at an accretion rate $<0.5\%$ of Eddington, thus consistent with helium burning triggered by hydrogen instability 
\cite[][; case 3]{fhm81}.
Another burst,
\#6229  on MJD~54009.301122 
was observed 
at a comparable accretion rate 
with a duration of 15~min, 
typical of the cooling of a thick fuel layer, here interpreted as helium produced by hydrogen burning at low accretion rate \citep{pbt07}. However, IGR~J17254$-$3257 (= 1RXS~J172525.5$-$325717) is an ultracompact X-ray binary candidate 
% fixed duplicate in proofs, Q11
% \citep{zjm07} 
\citep{zand07} % \jc{doesn't seem to be the correct reference; can you check?  --- dkg}
from which only H-poor accretion is expected. In such case, a more likely interpretation of the long burst would be the burning of a thick layer of pure helium slowly accreted from the degenerate companion onto the NS surface 
\cite[e.g.][]{cumming06}.

\subsection{Rare and unusual bursts}
\label{sec:peculiar_trends}
    With such a large sample, rare events are detected,
    % picked up. Rare can 
    here referring both to 
    % a particular source which is seen to burst only a few times 
    sources with very low burst rates (e.g., SAX~J1324.5$-$6313, 4U~1705$-$22, SLX~1732$-$304, Swift~J1749.4$-$2807, SAX~J2224.9+5421) but also to bursts with extraordinary characteristics (peak fluxes and temperatures, durations, unusual time profiles). 
    % The development of MINBAR has resulted in a few publications about such rare events.
    
    \cite{chenevez06} discuss an unusual event from the regular burster GX~3+1 detected by JEM-X (MINBAR \#5309, on MJD~53248.78684), 
    % detected an X-ray burst 
    that appeared initially as a common short (10-s timescale) burst,
    % from this source, 
    with a brief Eddington-limited phase; but that was followed by a 30-min long tail. It is not clear what caused this long tail: cooling of a very thick layer (while ignition must have been at a shallow depth) or prolonged hydrogen burning due to a layer that remained hot for a long time.
    
    2S~0918$-$549 is a persistently accreting ultracompact X-ray binary with only 7 bursts in MINBAR,
    % . Two times a burst was detected that is
    two with durations in the ``intermediate'' range. One of these bursts (\#1798, on MJD~50357.88531) was detected with the WFC \citep{zand05a} and one, \#3663 on MJD~54504.12698, with the PCA \citep{zand11a}. 
The latter event has a 
burst 
time scale of $\tau=139$~s, 
an Eddington-limited phase of 70~s and, most importantly, it shows $\sim50$\%-amplitude variations 2--3~min after the onset.% just like the WFC bursts reported in 2005. 
The WFC burst also shows strong variations in the tail. Apart from these two bursts, a handful more such bursts from other sources have been reported from the 
instruments contributing to the
MINBAR sample 
    %
    % \citep{molkov05,zand2008} 
    \citep{molkov05,zand08a}
    % \jz{please check the 2008 reference -- dkg} \dg{It is correct. This paper, among other stuff, deals with WFC bursts from 1246-588, one of which shows modulations in the tail - jz_sep19}
    %
    or from other instruments (e.g., 
% fixed duplicate in proofs, Q11
% \citealt{degenaar13a,degenaar18,zand19}),
\citealt{degenaar13a,deg18a,zand19}), 
and they all seem to be associated to intermediate-duration bursts with long PRE phases, the most powerful and energetic He powered bursts. It 
    % is believed 
    has been suggested
    that the variations are due to an accretion disk that is strongly disturbed, both dynamically and radiatively, by the powerful and explosive burst.

Another example of an intermediate-long burster is represented by the ultracompact X-ray binary candidate SLX~1737$-$282. Indeed, only long bursts, lasting more than 15 minutes, have so far been detected from this source at low accretion rate ($\sim0.5$\% Eddington) by the WFC \citep{zand02} and JEM-X \citep{falanga08}. They are all interpreted as resulting from the unstable burning of a thick pure helium layer slowly accreted from an H-poor stellar companion.

We note that the MINBAR sample omits one of the long bursts observed by JEM-X, from SLX 1735-269 on MJD~52897.73280 \cite[]{molkov05}, due to the issue with unavailability of certain data modes as described in \S\ref{jemxobs}.

\subsection{Superexpansion}
\label{sec:superexpansion}
    Some 20\% 
    % \dg{update this with the actual fraction in MINBAR} % 2019 Sep
    % Correct, see sec. sec:prebursts
    of all bursts exhibit photospheric radius expansion, due to nuclear fluxes reaching the Eddington limit (G08). The  expansion is usually 
    % measured to be rather mild 
    modest, with expansion factors of just a few. However, there is a subset of these events, perhaps 1\% of all bursts, where the expansion is much larger, with factors reaching 10$^2$. This phenomenon is referred to as ``superexpansion'' \citep{zand10a}. 
    
    During superexpansion bursts the photosphere expands and cools so much that the thermal radiation moves out of the X-ray band during the most extreme expansion. 
    % and the appearance is created of a precursor 
    The lightcurves of such bursts often feature a ``precursor'' followed by a dropout caused by the low blackbody temperatures. 
    %\ek{(possibly refer to Lewin et al. paper [1984ApJ...277L..57L] where precursors are 
    %described for the first time, in the context of PRE bursts)}. PRE is already discussed in section 5.3
    Eventually, the photosphere returns to the neutron star and the rest of the burst (in fact, most) can be observed in X-rays.
    
    Superexpansion 
    % must be due to a nuclear power 
    bursts have energies that are substantially larger than 
    % in bursts with mild photospheric expansion. 
    typical PRE bursts. 
    % This can be combination of a larger amount of fuel being ignited and faster nuclear burning processes. 
    \cite{zand14b} assembled a catalog of 39 of these events \cite[based on an earlier sample of 32;][]{zand10a}, 
    % 35 from the MINBAR sample and 4 from the literature. 
    33 present in the MINBAR sample (two missing include the superburst observed with the PCA from 4U~1820$-$303, and the long burst from SLX 1735-239 on MJD~52897), and four others from the literature.
    % burst_ref db has 26 bursts from zand10a (which has a total of 32 bursts); excludes 
    % - the superburst from 1820-30; 
    % - the long burst with JEM-X from SLX 1735-239 (MJD 52897.733)
    % and 2 from zand14b --- dkg (2019 Sep)
    % here's the 33 matches between table A1 of zand14a and MINBAR, generated with the following IDL code:
    It turns out that all these bursts are from hydrogen-deficient ultracompact X-ray binaries 
    % (inducing dominance of the fast 3$\alpha$ helium burning) 
    with low average mass accretion rates leading to cooler fuel layers and, therefore, larger ignition depths and larger amounts of fuel being ignited per burst.
    
    The superexpansion observed in relatively short X-ray bursts from 4U 1820-303 \citep{zand12a} poses somewhat of a puzzle in this respect, because the ignition depth is relatively shallow. We suspect this may be explained by a difference in He abundance in the fuel layer. 
    %\jz{why? -- dkg} % 2019 Sep
    Superexpansion observed in superbursts may partly be due to the effects from a shock wave propagating from the carbon ignition layer \citep{weinberg07,keek12b}.

    \cite{zand14b} discuss unusual events from 4U~0614+091 (see also \citealt{kuul10}) and 2S 0918-549 (the same bursts as discussed in 
    % the previous paragraph
    \S\ref{sec:peculiar_trends}).  Both bursts show precursors of extremely short duration, namely a few tens of milliseconds. Furthermore, these bursts during the precursors show fluxes that surpass the well-measured Eddington flux by a factor of about 2. This is interpreted as nova-like shells expanding at mildly relativistic speeds of a few tenths of the speed of light. Due to the brevity of the precursors, such a phenomenon can only be detected with a high detector area instrument such as the PCA and these are the only two events for which this has ever been detected. The brevity points to very fast flame speeds on the neutron star surface, which must be induced by a detonation instead of the more common deflagration, or by some kind of auto-ignition regime when the temperature distribution across the neutron star surface just prior to ignition is extraordinarily uniform and ignition conditions are supercritical everywhere on the surface.

\subsection{Narrow spectral features}

Constraining the equation of state (EOS) of neutron star matter and diagnosing
the composition of the neutron star photosphere provide strong motivation
to search for narrow features in X-ray burst spectra 
\cite[e.g.][]{waki84,nak88,magnier89,cott02}.
Results have been tentative so far. While early results
concerned absorption lines, the measurements of fast neutron star
spins starting in the nineteen nineties 
(\citealt{stroh97b}; see also \citealt{watts12a})
drowned most of the hope for that because Doppler
smearing washes away the signal 
\cite[e.g.][]{baubock13}.

Investigations in this area
received new impetus with the theoretical prediction
that absorption edges instead of lines might yield strong imprints in
the spectrum 
\cite[]{wbs06}.
Observational follow-up of such features
\cite[]{zand10a,kajava17a,li18b}
has resulted in strong detections of absorption features in \xte/PCA
spectra of 5 bright PRE bursts (MINBAR burst identification numbers \#2254, 2705, 2994, 3301 and the superburst
of 4U 1820-303) 
with optical depths between 0.1 and 3 and edge energies between 5 and 12 keV. These two parameters vary strongly within bursts. 

Although the significance of these features is strong,
the poor spectral
resolution of the PCA precludes a convincing verification of the
typical absorption edge profile. Their application to the NS EOS is
also limited because only singular edges have been detected in each
burst, resulting in an uncertainty in the identification of the
responsible atom and ionization state and, thus, in an uncertainty in
the rest wavelength.

More recent measurements with {\it NICER}\/ also in a PRE burst observed from 4U~1820$-$303, provide evidence for multiple narrow spectral lines in the range 1--3~keV \cite[]{stroh19a}. The inferred redshift of $1+z=1.046$ is likely too low to indicate emission at the neutron star surface, and might also include contributions from blueshift arising from a wind. The principle challenge for future observations of such features arise primarily in detecting the bursts, since they tend to occur episodically and unpredictably.

\subsection{Model-observation comparisons}
\label{sec:model-obs}

A key motivation for combining the burst observations from different satellites was to increase the number of unambiguous measurements of burst recurrence times.
One of the principal difficulties of studying bursts with satellite-based instruments is the ambiguity that arises from the regular interruptions due to the (typically $\approx90$~min) low-Earth orbits. The maximal duty cycle of $\approx60$\% means that even for the most intense observations, there is a high probability of missing intervening bursts, introducing substantial uncertainty for the recurrence times. This issue can be circumvented for sources with highly regular bursts (so that the occurrence of intervening events can be inferred even if not observed), but such behaviour is unexpectedly rare. 

Unfortunately, the rather low duty cycle of observations 
($\approx 2.9$\% on average; see \S\ref{subsec:obs_summ})
limits the efficacy of the MINBAR sample in this regard. Despite the many thousands of observations from different sources with each instrument, only 
68
observations with \xte/PCA occurred with appreciable overlap (as defined in \S\ref{sec:area}) with either of the other two instruments. Furthermore, only 
\burstdupes\ 
bursts were observed with two instruments simultaneously (see \S\ref{crosscal:bursts}). Dedicated, long-duration observations of prolific bursts sources with instruments in much wider orbits offering uninterrupted coverage \cite[e.g. EXO~0748$-$676 with {\it XMM-Newton};][]{boirin07a} may offer a more effective way of gathering measurements of burst recurrence times. 

Nevertheless, \cite{gal17a} selected several key sources representing different ignition cases, and identified closely-spaced groups of bursts to infer the recurrence times. \cite{gal17a} chose the ``textbook'' or ``clocked'' burster, GS~1826$-$24, which has been a focus of many studies to date. \cite[e.g.][]{clock99,gal03d,heger07b,zand09a}. %\dg{Please include in the bib file the reference to Ubertini 1999 1999ApJ...514L..27U}. %was already cited somewhere under a different label
\cite{gal17a} selected observations that were not subject to episodes when the spectrum apparently softened, such that significant contributions to the X-ray flux fell below the low-energy threshold for our instruments \cite[3~keV;][]{thompson08}. \cite{gal17a} combined observations from the different instruments so as to measure the recurrence time without the ambiguity of missing bursts, and identified three ``reference'' epochs over a range of accretion rates to serve as comparison data for numerical models.

These data have already been the subject of attempts to constrain the system properties (including fuel composition, distance, and neutron star mass and radius) with {\sc MESA} 
\cite[]{mesa15} and {\sc Kepler} \cite[]{woos04,johnston18}. The {\sc MESA} studies indicate that the accretion rate for the source was substantially higher than would be expected given the measured persistent flux \cite[]{meisel18a}, and also provides support to the goal of constraining individual nuclear reactions in the $rp$-process chains \cite[]{meisel19}. A different approach was taken by \cite{johnston19}, who precomputed large grids of {\sc Kepler} models and used an interpolation scheme to efficiently probe the parameter space, including neutron star mass and radius. Although these efforts are still at a relatively preliminary stage, and key information (such as the relative agreement of the two models) are as yet unexplored in detail, the prospects for both astrophysical and nuclear physics constraints seem promising.

\cite{gal17a} also selected trains of He-rich bursts, including from the accretion-powered millisecond pulsar SAX~J1808.4$-$3658 during its 2002 October outburst, and the 11-min binary 4U~1820$-$303. While these two sources both show H-poor bursts, SAX~J1808.4$-$3658 likely accretes H-rich material which is exhausted by steady burning prior to ignition \cite[]{goodwin19c}, while 4U~1820$-$303 cannot accommodate a H-rich donor in its very close orbit, and so must accrete material with likely hydrogen fraction no more than 0.1 \cite[e.g.][]{cumming03}. The sample also includes a high-quality lightcurve of a superburst observed by \xte/PCA from 4U~1636$-$536, which is not included in MINBAR.

\subsection{Future work}
\label{sec:future}

There are a number of directions that future studies utilising the MINBAR sample may develop. 
First, there is an extensive sample of bursts observed by JEM-X after our cutoff date of \pubdatacutoffdate. It would be a relatively straightforward exercise to extend our burst search and analysis procedures to those data, and further increase the sample. We plan to make our basic data analysis procedures available so that this analysis can  be extended to these and other data 
One limitation of the JEM-X data at the present time is the unavailability of time-resolved spectroscopy (see \S\ref{igrtrse}). 
OSA version 11.0, released in 2018 Oct, 
may make this analysis feasible in the near future,  
but likely only for bright, long bursts observed close to on-axis. 

Second, there is the prospect of adding bursts observed with other detectors similar to those used for the present sample, including 
{\it EXOSAT}/ME \citep[e.g.,][]{damen90}, {\it Ginga}/LAC \citep[e.g.,][]{vpd90}, and {\it ASTROSAT}/LAXPC \citep[e.g.,][]{beri19}. 
A slightly more challenging goal would be to add data for other types of instruments, including {\it Swift}/XRT and BAT \citep[e.g.,][]{zand19}, {\it XMM-Newton} \citep[e.g.,][]{boirin07a}, {\it Chandra}\/ \citep[e.g.,][]{zand13a} and {\it NICER}
% fixed duplicate in proofs, Q11
% \citep[e.g.,][]{bult19}
\citep[e.g.,][]{bult19b}
The difficulty with these latter instruments is that the bandpass typically only goes up to $\approx10$~keV (although may extend well below 3~keV; or above 15~keV for {it BAT}), and the shape of the instrumental response will be very different from the currently-included MINBAR instruments, offering additional challenges to the instrumental cross-calibration (see \S\ref{sec:area}).

The benefit of adding data from other instruments may be limited, given the much smaller number of bursts typically accumulated (at most a few hundred, compared to the thousands detected by the three instruments contributing to MINBAR). The substantial effort of adapting the analysis to a new instrument (and mitigating any instrument-specific analysis effects that arise) thus may not provide sufficient return, in terms of large increases in the overall sample size. On the other hand, the possibility of characterising the properties of bursts and persistent emission at energies below 2~keV may make this exercise worthwhile.

One question where the low-energy data may play a critical role is 
the 
intrinsic lower limit to the burst peak flux (or fluence) distribution (see \S \ref{sec:burst_summary}). 
That is, 
what is the minimum amount of fuel that can be ignited in a burst? 
One possibility (about which much speculation has been made) is that only part of the NS surface is ignited (which could happen, for example, if the flame front collapses). If the NS is spinning and the spin axis is not aligned with the line of sight, and the burning region is not centered on a rotation pole, an oscillating flux should emerge. This yields a reduction of the peak flux when measured over time scales larger than the spin period which is quite possible because rotation periods are often in the order of milliseconds. The maximum reduction, ignoring GR effects, is a factor of 2 for an amplitude of 100\% (e.g., \citealt{ms14}).

A second method to 
achieve lower 
peak flux is to reduce the peak temperature. To diminish the peak energy flux by a factor of 10$^2$ while leaving the whole NS emitting, it would suffice to reduce the peak temperature by a factor of roughly 100$^{1/4}$=3.1, so somewhat less than 1~keV. If this is the predominant reason for the minimum observed in our sample, this implies that peak burst temperatures smaller than about 1~keV are not present which is actually the case (Fig.~\ref{fig:ktecumulative}). However, the capability to measure such low temperatures in our sample is limited because it goes hand in hand with low statistical significances due to the low-energy cutoff of the bandpass of 
$\approx2$~keV
(see above).

A better prospect to measure peak temperatures lower than 1~keV is provided by other instruments that are sufficiently sensitive 
at low energies
to measure faint bursts. This goal is only possible with {\it NICER}, {\it Chandra}\/ and {\it XMM-Newton}\/ for the nearest bursters (e.g., \citealt{zand13a}). This limitation is shown by measurements of burst tails, when the temperature may drop below 1~keV (see e.g., \citealt{zand08a}).
This behaviour may be related to the presence of a boundary layer between the NS and the accretion disk with similar temperatures and emission areas as the NS surface  \cite[e.g.][]{vpl86}, which precludes the practical measurement of lower temperatures.

A third method to reduce the peak flux of a burst is by the obscuration of (part of) the NS by a circumstellar medium with a medium optical thickness and/or with such a temperature that most
atoms are ionized \cite[e.g.][]{gal08a}. This medium may be structure on the surface of the accretion disk. The obscuration may not have a clear spectral dependence if the medium is highly ionized and the obscuration occurs
through mostly Thompson scattering. 
We conclude that pursuing faint bursts is an interesting subject for a future study with the MINBAR database.

As we suggest in \S\ref{sec:model-obs}, a more targeted approach 
for future observational programs
may offer a better return on the analysis investment. For studies of burst recurrence times, long-duration observations of sources with high burst rates by instruments in long orbits may be a better source of detailed information than the (generally) low-duty cycle observations analysed here. Such observations also exist in the archives \cite[e.g.][]{haberl87,boirin07a,kong07}, and could be combined with newer observations for a comprehensive study. 

For intermediate-duration bursts and superbursts \citep[see][for a recent observational review of superbursts]{zand17c}, dedicated samples of those events with uniform analysis procedures could provide stronger constraints on the cooling behaviour and crust properties than for individual sources. Assembly of a catalog (partially overlapping with MINBAR) has been under way for some years, with involvement by some of the authors of this paper. 
As with any such directions, the MINBAR sample offers a critical broad overview of the bursting process which can inform further in-depth studies, for example by identifying priority targets or spectral states for new observations.

Determining burst luminosities has historically been difficult for sources not located in globular clusters because the distances to the sources cannot easily be measured. This situation has begun to change, with the arrival of parallax measurements from the Gaia satellite. We expect that future {\it Gaia}\/  data releases will provide more precise and accurate distance measurements, and increase the number of sources for which distances can be determined. We also aim in future MINBAR releases to refine our knowledge of burster positions such that more objects can be unambiguously matched to the {\it Gaia}\/  catalog. Nonetheless it seems that burst peak flux analyses will remain useful distance indicators for the foreseeable future.

\section{Conclusion}
\label{sec:conclusion}

We have assembled the largest sample of thermonuclear (type-I) X-ray bursts yet available, from three long-duration missions featuring proportional counter detectors covering a common energy range. 
We have developed and applied common analysis procedures and investigated in detail the properties of the sample.
These data and analysis results offer a uniquely comprehensive overview of the burst phenomenology, but even so, the sample does not encompass the full range of observed behaviour, since additional archival observations from other instruments are available, and new observations are continually being taken.
In the previous section we describe several possible extensions to the sample; we plan to provide the key data analysis routines to the community to allow others to contribute to this sample.
We hope that these data, and the related tools, will provide an invaluable resource to the community long into the future.

\acknowledgments

This research has been supported under the Australian Academy of Science's Scientific Visits to Europe program, and the Australian Research Council's Discovery Projects (project DP0880369) and Future Fellowship (project FT0991598) schemes. 
The research leading to these results has received funding from the European Union’s Horizon 2020 Programme under AHEAD project (grant agreement n. 654215).
Partly based on observations with \igr, an ESA project with instruments and science data centre funded by ESA member states (especially the PI countries: Denmark, France, Germany, Italy, Switzerland, Spain) and with the participation of Russia and the USA.
This work has made use of data from the European Space Agency (ESA)
mission {\it Gaia} (\url{https://www.cosmos.esa.int/gaia}), processed by
the {\it Gaia} Data Processing and Analysis Consortium (DPAC,
\url{https://www.cosmos.esa.int/web/gaia/dpac/consortium}). Funding
for the DPAC has been provided by national institutions, in particular
the institutions participating in the {\it Gaia} Multilateral Agreement.
JZ acknowledges the strong support from NWO-I/SRON.
HW acknowledges support by the German DLR under Fkz 50 OR 1405, Fkz 50 OR 1711, and Fkz 50 OR 1814. 
LO acknowledges support from an NWO Top Grant, Module 1 (PI Rudy Wijnands).  
ALW acknowledges support from ERC Starting Grant No.639217 CSINEUTRONSTAR. 
The authors thank S. Brandt and C.A. Oxborrow for their help processing JEM-X data.

% added software citations; see
% https://journals.aas.org/policy-statement-on-software
\software{
HEASoft \cite[]{ftools95},
\igr\/ Off-line Science Analysis (OSA),
XSpec \cite[]{xspec12}, % no DoI
IDL, %
% Astropy \citep{http://dx.doi.org/10.1051/0004-6361/201322068},
Matplotlib \cite[]{matplotlib07} % \citep{http://dx.doi.org/10.1109/MCSE.2007.55}}
}

% {\it Facilities:}
\facility{RXTE},  \facility{BeppoSAX},  \facility{INTEGRAL}.

\clearpage %all that overlapping text makes me irrationally annoyed-hw

\bibliography{all,newbib} %Duncan, can you merge these? I cannot edit all.bib
\bibliographystyle{apj}

\clearpage

\end{document}